\magnification=1200 \hsize=5 true in \vsize=7.5 true in
\baselineskip=3 ex

\def\secskp{\vskip 1.5 true cm}
\def\VE{\vfill\eject}
\hoffset 1.5 true cm
\voffset 2 true cm

\def\sp{\hskip .5 cm}
\def\skp{\vskip .6 cm}
\def\secskp{\vskip 1.5cm}
\def\sc{\scriptstyle}

\def\qed{\vrule height6pt width3pt depth 0pt}
\def\br{\overline}
\def\sc{\scriptscriptstyle}
\def\zbar{\bar z} 
\def\ftil{\tilde f}  
\def\hhat{\hat h}  \def\fhat{\hat f}
\def\zbar{\bar z}  
\def\zs{{\rm Z\hskip-.6ex Z}}
\def\zz{{\rm Z\hskip-.7ex Z}}
\def\rl{{\rm I\!R}} 
\def\cx{{\,\rm C\hskip-1.1ex l}\,}
\def\cxs{{\,\rm l\hskip-1ex C}}
\def\inr{\int_{-\infty}^\infty}
\def\inrr{\int_{\, {\rm I\!R}^2}}
 
\def\inrs{\int_{\, {\rm I\!R}^s}}      
 
\def\incs{\int_{\ {\rm l\hskip-1ex C}^s}}
\def\phs{{2\pi i\nu t}}  \def\tpi{{2\pi i}}

\def\has{h_{a,s}}
\def\ap{|a|^{1/2}}  \def\am{|a|^{-1/2}}

\def\rightheadline{\tenrm\hfil {\sl 1. Coherent--State
Representations}\hfil\folio}  
\def\leftheadline{\tenrm\folio\hfil {\sl G. Kaiser}\hfil}

\def\sp{\hskip .5 cm}
\def\skp{\vskip .5 cm}
\def\secskp{\vskip 1.5 cm}

\def\sst{\scriptstyle}
\def\sc{\scriptscriptstyle}
\def\qed{\vrule height6pt width3pt depth 0pt}
\def\sdp{{\bigcirc\hskip-1.7ex s\ }}
\def\del{{\sqcup\hskip-1.6ex\sqcap\ }}
\def\dd{\,\partial{\kern-1.3ex^{^\leftrightarrow}}\!\!}
\def\dir{\partial\kern -1.2ex /\,}
\def\dirt{\partial\kern -1.1ex /\,}
\def\psl{p\kern -1.0ex /\,}
\def\pslt{p\kern -.9ex /\,}
\def\br{\overline}
\def\ve{\varepsilon}
\def\w{\wedge}
\def\dx{\widehat{dx}}
\def\dy{\widehat{dy}\,}
\def\zbar{{\bar z}} 
\def\ftil{\tilde f}  
\def\ftd{\tilde\phi} 
 \def\gbar{\bar g}
\def\hhat{\hat h}  \def\fhat{\hat f}  \def\ghat{\hat g}
\def\zs{{\rm Z\hskip-.6ex Z}}
\def\zz{{\rm Z\hskip-.7ex Z}}
\def\rl{{\rm I\!R}} 
 
\def\cx{{\rm C\hskip-1.1ex l}\,}
\def\cxs{{\,\,\rm l\hskip-1ex C}}
\def\inr{\int_{-\infty}^\infty}
\def\inrr{\int_{\, {\rm I\!R}^2}}
 
\def\inrs{\int_{\, {\rm I\!R}^s}}    
 
\def\incs{\int_{\ {\rm l\hskip-1ex C}^s}}
\def\ino{\int_{\Omega_m} d\tilde p\ }
\def\inop{\int_{\Omega_m^+}d\tilde p\ }

\def\ins{\int_\sigma d\sigma \,}
\def\insp{\int_{\sigma_+} d\sigma \,}
\def\insp{\int_{\sigma_+} d\sigma \,}
\def\insm{\int_{\sigma_-} d\sigma \,}
\def\inds{{ 1\over 2\pi i}\int_{-\infty}^\infty\,{ d\tau \over \tau -i}\,}
\def\om{{\Omega _m}}
\def\omp{{\Omega _m^+}}
\def\omm{{\Omega _m^-}}
\def\ol{{\Omega _\lambda }}
\def\olp{{\Omega _\lambda ^+}}
\def\olm{{\Omega _\lambda ^-}}
\def\tb{{\cal T}}
\def\tp{{{\cal T}_+}}
\def\tm{{{\cal T}_-}}
\def\bl{{B_\lambda }}
\def\blp{{B_\lambda^+ }}
\def\blm{{B_\lambda^- }}
\def\vp{\overline{V }_+}
\def\inp{\int_{\rl^n} d^np\,\,}
\def\inpv{(2\pi )^{-s-1} \,\int_{\overline{ V}_+} dp\,\,}
\def\fin{\phi _{\rm in}}
\def\fout{\phi _{\rm out}}

\def\einp{e^+_{{\rm in}, z}}
\def\einm{e^-_{{\rm in}, z}}
\def\eoutp{e^+_{{\rm out}, z}}
\def\eoutm{e^-_{{\rm out}, z}}
\def\gret{G_{\rm ret}}
\def\gadv{G_{\rm adv}}
\def\intf{\int_{\cal K} d\mu (f)\, }
\def\kc{{\cal K}}
\def\hc{{\cal H}}

\def\h{{{1\over 2}}}

\def\mt{\mapsto}
\def\ap{{\alpha +1}}

\def\has{H_\alpha^*}

\def\l{\langle\, }
\def\r{\,\rangle}

\def\cap{\chi ^{\alpha+1} }

\def\sp{\hskip .5 cm}
\def\skp{\vskip .6 cm}
\def\secskp{\vskip 1.5cm}
\def\qed{\vrule height6pt width3pt depth 0pt}
\def\br{\overline}
\def\sc{\scriptscriptstyle}
\def\zbar{\bar z} 
\def\ftil{\tilde f}  
\def\hhat{\hat h}  \def\fhat{\hat f}
\def\zbar{\bar z}  
\def\zs{{\rm Z\hskip-.6ex Z}}
\def\zz{{\rm Z\hskip-.7ex Z}}
\def\rl{{\rm I\!R}} 
\def\cx{{\,\rm C\hskip-1.1ex l}\,}
\def\cxs{{\,\rm l\hskip-1ex C}}
\def\inr{\int_{-\infty}^\infty}
\def\inrr{\int_{\, {\rm I\!R}^2}}
 
\def\inrs{\int_{\, {\rm I\!R}^s}}      
 
\def\incs{\int_{\ {\rm l\hskip-1ex C}^s}}
\def\phs{{2\pi i\nu t}}  \def\tpi{{2\pi i}}

\def\has{h_{a,s}}
\def\ap{|a|^{1/2}}  \def\am{|a|^{-1/2}}

\def\sp{\hskip .5 cm}
\def\skp{\vskip .5 true cm}
\def\secskp{\vskip 1.5 true cm}
\def\sc{\scriptscriptstyle}
\def\qed{\vrule height6pt width3pt depth 0pt}
\def\sdp{{\bigcirc\hskip-1.7ex s\ }}
\def\del{{\sqcup\hskip-1.6ex\sqcap\ }}
\def\dd{\,\partial{\kern-1.3ex^{^\leftrightarrow}}\!\!}
\def\br{\overline}
\def\w{\wedge}
\def\dx{\widehat{dx}}
\def\dy{\widehat{dy}\,}
\def\zbar{{\bar z}} 
\def\ftil{\tilde f}  
 \def\gbar{\bar g}
\def\hhat{\hat h}  \def\fhat{\hat f}
\def\zs{{\rm Z\hskip-.6ex Z}}
\def\zz{{\rm Z\hskip-.7ex Z}}
\def\rl{{\rm I\!R}} 
\def\cx{{\rm C\hskip-1.1ex l}\,}
\def\cxs{{\rm l\hskip-1ex C}}
\def\inr{\int_{-\infty}^\infty}
\def\inrr{\int_{\, {\rm I\!R}^2}}
 
\def\inrs{\int_{\, {\rm I\!R}^s}}      
 
\def\incs{\int_{\ {\rm l\hskip-1ex C}^s}}
\def\ino{\int_{\Omega_m} d\tilde p\ }
\def\inop{\int_{\Omega_m^+}d\tilde p\ }

\def\ins{\int_\sigma d\sigma \,}
\def\omp{{\Omega _m^+}}
\def\omm{{\Omega _m^-}}
\def\olp{{\Omega _\lambda ^+}}
\def\olm{{\Omega _\lambda ^-}}
\def\tb{{\cal T}}
\def\tp{{\cal T}_+}
\def\tm{{\cal T}_-}
\def\bl{{B_\lambda }}
\def\blp{{B_\lambda^+ }}
\def\blm{{B_\lambda^- }}

\def\z{(z)}
\def\x{(x)}

\def\rightheadline{\tenrm\hfil {\sl 3. Complex Spacetime}\hfil\folio} 
\def\leftheadline{\tenrm\folio\hfil {\sl G. Kaiser}\hfil}

\def\mb{{\bar \mu }}

\def\mt{\mapsto}
\def\h{{1 \over 2}}
\def\d{\partial}
\def\db{\bar\partial}
\def\t{(t)}
\def\x{(x)}
\def\p{(p)}
\def\z{(z)}
\def\sp{\hskip .5 cm}
\def\skp{\vskip .5 true cm}
\def\secskp{\vskip 1.5 truecm}

\def\sst{\scriptstyle}
\def\sc{\scriptscriptstyle}
\def\qed{\vrule height6pt width3pt depth 0pt}
\def\sdp{{\bigcirc\hskip-1.7ex s\ }}
\def\del{{\sqcup\hskip-1.6ex\sqcap\ }}
\def\dd{\,\partial{\kern-1.3ex^{^\leftrightarrow}}\!\!}
\def\dir{\partial\kern -1.2ex /\,}
\def\dirt{\partial\kern -1.1ex /\,}
\def\psl{p\kern -1.0ex /\,}
\def\pslt{p\kern -.9ex /\,}
\def\br{\overline}
\def\ve{\varepsilon}
\def\w{\wedge}
\def\dx{\widehat{dx}}
\def\dy{\widehat{dy}\,}
\def\zbar{{\bar z}} 
\def\ftil{\tilde f}  
\def\ftd{\tilde\phi} 
 \def\gbar{\bar g}
\def\hhat{\hat h}  \def\fhat{\hat f}  \def\ghat{\hat g}
\def\zs{{\rm Z\hskip-.6ex Z}}
\def\zz{{\rm Z\hskip-.7ex Z}}
\def\rl{{\rm I\!R}} 
\def\st{\rl^{s+1}}
\def\cx{{\rm C\hskip-1.1ex l}\,}
\def\cxs{{\rm l\hskip-1ex C}}
\def\inr{\int_{-\infty}^\infty}
\def\inrr{\int_{\, {\rm I\!R}^2}}
 
\def\inrs{\int_{\, {\rm I\!R}^s}}      
 
\def\incs{\int_{\ {\rm l\hskip-1ex C}^s}}
\def\ino{\int_{\Omega_m} d\tilde p\ }
\def\inop{\int_{\Omega_m^+}d\tilde p\ }

\def\ins{\int_\sigma d\sigma \,}
\def\insp{\int_{\sigma_+} d\sigma \,}
\def\insp{\int_{\sigma_+} d\sigma \,}
\def\insm{\int_{\sigma_-} d\sigma \,}
\def\inds{{ 1\over 2\pi i}\int_{-\infty}^\infty\,{ d\tau \over \tau -i}\,}
\def\om{{\Omega _m}}
\def\omp{{\Omega _m^+}}
\def\omm{{\Omega _m^-}}
\def\ol{{\Omega _\lambda }}
\def\olp{{\Omega _\lambda ^+}}
\def\olm{{\Omega _\lambda ^-}}
\def\tb{{\cal T}}
\def\tp{{{\cal T}_+}}
\def\tm{{{\cal T}_-}}
\def\bl{{B_\lambda }}
\def\blp{{B_\lambda^+ }}
\def\blm{{B_\lambda^- }}
\def\vp{\overline{V }_+}
\def\inp{(2\pi )^{-s-1} \,\int_{\rl^{s+1}} dp\,\,}
\def\inpv{(2\pi )^{-s-1} \,\int_{\overline{ V}_+} dp\,\,}
\def\fin{\phi _{\rm in}}
\def\fout{\phi _{\rm out}}

\def\einp{e^+_{{\rm in}, z}}
\def\einm{e^-_{{\rm in}, z}}
\def\eoutp{e^+_{{\rm out}, z}}
\def\eoutm{e^-_{{\rm out}, z}}
\def\gret{G_{\rm ret}}
\def\gadv{G_{\rm adv}}
\def\intf{\int d\mu (f)}
\def\intfp{\int_{{{\cal K}}_0'} d\mu (f)\, }
\def\kc{{\cal K}}
\def\kco{{\cal K}_0}
\def\kcp{{\cal K}_0'}
\def\hc{{\cal H}}

\def\r{\rangle\, }
\def\l{\,\langle}

\def\rightheadline{\tenrm\hfil {\sl 4.  Quantized Fields}\hfil\folio} 
\def\leftheadline{\tenrm\folio\hfil {\sl G. Kaiser}\hfil}

\def\mb{{\bar \mu}}

\def\mt{\mapsto}
\def\h{{1 \over 2}}
\def\d{\partial}
\def\db{\bar\partial}
\def\t{(t)}
\def\x{(x)}
\def\z{(z)}
\def\sp{\hskip .5 cm}
\def\skp{\vskip .5 true cm}
\def\secskp{\vskip 1.5 truecm}

\def\sst{\scriptstyle}
\def\sc{\scriptscriptstyle}
\def\qed{\vrule height6pt width3pt depth 0pt}
\def\sdp{{\bigcirc\hskip-1.7ex s\ }}
\def\del{{\sqcup\hskip-1.6ex\sqcap\ }}
\def\dd{\,\partial{\kern-1.3ex^{^\leftrightarrow}}\!\!}
\def\dir{\partial\kern -1.2ex /\,}
\def\dirt{\partial\kern -1.1ex /\,}
\def\psl{p\kern -1.0ex /\,}
\def\pslt{p\kern -.9ex /\,}
\def\br{\overline}
\def\ve{\varepsilon}
\def\w{\wedge}
\def\dx{\widehat{dx}}
\def\dy{\widehat{dy}\,}
\def\zbar{{\bar z}} 
\def\ftil{\tilde f}  
\def\ftd{\tilde\phi} 
 \def\gbar{\bar g}
\def\hhat{\hat h}  \def\fhat{\hat f}  \def\ghat{\hat g}
\def\zs{{\rm Z\hskip-.6ex Z}}
\def\zz{{\rm Z\hskip-.7ex Z}}
\def\rl{{\rm I\!R}} 
\def\st{\rl^{s+1}}
\def\cx{{\rm C\hskip-1.1ex l}\,}
\def\cxs{{\rm l\hskip-1ex C}}
\def\inr{\int_{-\infty}^\infty}
\def\inrr{\int_{\, {\rm I\!R}^2}}
 
\def\inrs{\int_{\, {\rm I\!R}^s}}      
 
\def\incs{\int_{\ {\rm l\hskip-1ex C}^s}}
\def\ino{\int_{\Omega_m} d\tilde p\ }
\def\inop{\int_{\Omega_m^+}d\tilde p\ }

\def\ins{\int_\sigma d\sigma \,}
\def\insp{\int_{\sigma_+} d\sigma \,}
\def\insp{\int_{\sigma_+} d\sigma \,}
\def\insm{\int_{\sigma_-} d\sigma \,}
\def\inds{{ 1\over 2\pi i}\int_{-\infty}^\infty\,{ d\tau \over \tau -i}\,}
\def\om{{\Omega _m}}
\def\omp{{\Omega _m^+}}
\def\omm{{\Omega _m^-}}
\def\ol{{\Omega _\lambda }}
\def\olp{{\Omega _\lambda ^+}}
\def\olm{{\Omega _\lambda ^-}}
\def\tb{{\cal T}}
\def\tp{{{\cal T}_+}}
\def\tm{{{\cal T}_-}}
\def\bl{{B_\lambda }}
\def\blp{{B_\lambda^+ }}
\def\blm{{B_\lambda^- }}
\def\vp{\overline{V }_+}
\def\inp{(2\pi )^{-s-1} \,\int_{\rl^{s+1}} dp\,\,}
\def\inpv{(2\pi )^{-s-1} \,\int_{\overline{ V}_+} dp\,\,}
\def\fin{\phi _{\rm in}}
\def\fout{\phi _{\rm out}}

\def\einp{e^+_{{\rm in}, z}}
\def\einm{e^-_{{\rm in}, z}}
\def\eoutp{e^+_{{\rm out}, z}}
\def\eoutm{e^-_{{\rm out}, z}}
\def\gret{G_{\rm ret}}
\def\gadv{G_{\rm adv}}
\def\intf{\int_{\cal K} d\mu (f)\, }
\def\kc{{\cal K}}
\def\hc{{\cal H}}

\def\has{H_\alpha^*}

\def\l{\langle\, }
\def\r{\,\rangle}

\def\cap{\chi ^{\alpha+1} }

\def\sp{\hskip 1.0 ex}
\def\skp{\vskip 0.2ex}

\nopagenumbers

\vskip 2cm

\centerline{\bf Quantum Physics, Relativity, and Complex Spacetime:}
\centerline{\bf Towards a New Synthesis}
\vskip .2cm

\centerline{\bf Gerald Kaiser}

\centerline{Department of Mathematics}

\centerline{University of Massachusetts--Lowell}

\centerline{kaiser@wavelets.com\  $\bullet$\  http://wavelets.com}

\vskip 1cm

\centerline{\bf ABSTRACT}

\noindent
The positivity of the energy in relativistic quantum mechanics implies that wave functions can be continued analytically to the forward tube $ {\cal T}_+$ in complex spacetime. For Klein-Gordon particles, we interpret ${\cal T}_+$ as an extended (8D) classical phase space containing all 6D classical phase spaces as symplectic submanifolds. The evaluation maps $e_z: f\to f(z)$ of wave functions on ${\cal T}_+$ are relativistic coherent states reducing to the Gaussian coherent states in the nonrelativistic limit. It is known that no covariant probability interpretation exists for Klein-Gordon particles in real spacetime because the time component of the conserved "probability current" can attain negative values even for positive-energy solutions. We show that this problem is solved very naturally in complex spacetime, where $|f(x-iy)|^2$ is interpreted as a probability density on all 6D phase spaces in ${\cal T}_+$ which, when integrated over the "momentum" variables y, gives a conserved spacetime probability current whose time component is a positive regularization of the usual one. Similar results are obtained for Dirac particles, where the evaluation maps $e_z$ are spinor-valued relativistic coherent states. For free quantized Klein-Gordon and Dirac fields, the above formalism extends to n-particle/antiparticle coherent states whose scalar products are Wightman functions. The 2-point function plays the role of a reproducing kernel for the one-particle and antiparticle subspaces.

\vskip .5 cm
\noindent
Originally published as a book in 1990 by North-Holland, Amsterdam
\noindent
\copyright \hskip 2pt Gerald Kaiser 2003. All rights reserved.

\VE

\baselineskip=2.5 ex 
\noindent {\bf Unified field Theory}
\bigskip

\sl 

\noindent In the beginning there was Aristotle \hfill\break 
And objects at rest tended to remain at rest \hfill\break 
And objects in motion tended to come to rest \hfill\break 
And soon everything was at rest \hfill\break 
And God saw that it was boring. \hfill\break 
\skp
\noindent Then God created Newton \hfill\break 
And objects at rest tended to remain at rest \hfill\break 
But objects in motion tended to remain in motion \hfill\break 
And energy was conserved and momentum was conserved  \hfill\break 
\hbox{\hskip.5cm} and matter  was conserved \hfill\break 
And God saw that it was conservative. \hfill\break 
\skp
\noindent Then God created Einstein \hfill\break 
And everything was relative \hfill\break 
And fast things became short \hfill\break 
And straight things became curved \hfill\break 
And the universe was filled with inertial frames \hfill\break 
And God saw that it was relatively general  \hfill\break 
\hbox{\hskip.5cm} but some of it was especially relative. \hfill\break 
\skp
\noindent Then God created Bohr \hfill\break 
And there was the $\cal P$rinciple \hfill\break 
And the principle was  $\cal Q$uantum \hfill\break 
And all things were quantized \hfill\break 
But some things were still  $\cal R$elative \hfill\break 
And God saw that it was confusing. \hfill\break 
\skp
\noindent Then God was going to create  $\cal M$ \hfill\break 
And  $\cal M$ would have unified \hfill\break 
And $\cal M$ would have fielded a theory \hfill\break
And all would have been one \hfill\break 
But it was the seventh day \hfill\break 
And God rested \hfill\break
And objects at rest tend to remain at rest.

\bigskip\skp

\hbox{\hskip2cm Adapted from a poem  by Tim Joseph}\rm

\VE

\countdef\pageno=0\pageno=-2

\def\leftheadline{\tenrm\folio\hfil} 
\def\rightheadline{\hfill\folio}

\headline={\ifodd\pageno\rightheadline\else\leftheadline\fi}

\VE

\baselineskip=3.5 ex 
\centerline {\bf  CONTENTS } 
\vskip 1cm
\parindent=0pt

{\bf Preface  } \dotfill iv
\skp
{\bf Suggestions to the Reader } \dotfill  vii

\skp
{\bf Chapter 1.  Coherent--State Representations } 

\sp 1.1. Preliminaries \dotfill 1

\sp 1.2. Canonical coherent states  \dotfill 7

\sp 1.3. Generalized frames and resolutions of unity   \dotfill 14

\sp 1.4. Reproducing--kernel Hilbert spaces    \dotfill 22

\sp 1.5. Windowed Fourier transforms   \dotfill 26

\sp 1.6. Wavelet transforms  \dotfill 33

\skp

{\bf  Chapter 2. Wavelet Algebras and Complex Structures}

This chapter is not essential, hence omitted to save space; the same material is available at  http://wavelets.com/pdf/92Siam.pdf






\skp

{\bf Chapter 3. Frames and Lie Groups  }

\sp 3.1. Introduction  \dotfill 44

\sp 3.2. Klauder's group--frames  \dotfill 44

\sp 3.3. Perelomov's homogeneous $G$--frames  \dotfill 50

\sp 3.4. Onofri's's holomorphic  $G$--frames  \dotfill 57

\sp 3.5. The rotation group  \dotfill 74

\sp 3.6. The harmonic oscillator as a contraction limit  \dotfill 82

{\bf Chapter 4. Complex Spacetime  }

\sp 4.1. Introduction  \dotfill 89

\sp 4.2.  Relativity, phase space and quantization  \dotfill 90

\sp 4.3.  Galilean frames  \dotfill 99

\sp 4.4.  Relativistic frames  \dotfill 110

\sp 4.5.  Geometry and Probability  \dotfill 128

\sp 4.6.  The non--relativistic limit  \dotfill 142

\sp Notes  \dotfill 147

\skp

{\bf Chapter 5.  Quantized Fields }

\sp 5.1. Introduction  \dotfill 151

\sp 5.2. The multivariate Analytic--Signal transform  \dotfill 154

\sp 5.3. Axiomatic field theory and particle phase spaces  \dotfill 161

\sp 5.4. Free Klein--Gordon fields  \dotfill 180

\sp 5.5. Free Dirac fields  \dotfill 193

\sp 5.6. Interpolating particle coherent states   \dotfill 203

\sp 5.7. Field coherent states and functional integrals  \dotfill 208

\sp  Notes \dotfill 216

\skp

{\bf Chapter 6.  Further Developments}

\sp 6.1. Holomorphic gauge theory  \dotfill 218

\sp 6.2. Windowed X--Ray transforms: Wavelets revisited  \dotfill 227

\skp

{\bf References} \dotfill 237

\VE

\baselineskip=2.5 ex 

\parindent=4ex

\centerline{\bf PREFACE}
\skp

\noindent
The idea of complex spacetime as a
unification of spacetime and classical phase space, suitable as a possible
geometric basis for the synthesis of Relativity and quantum theory, 
first occured to me in 1966 while I was a physics graduate student 
 at the University of Wisconsin.    In 1971, during a seminar I gave
 at Carleton University in Canada,  it was  pointed out to me  that
the formalism I was developing   was related to the  coherent--state
representation,  which was then unknown to me.  This turned out to
be a fortunate circumstance, since many of the subsequent developments
have been inspired by ideas related to coherent states.  My main
interest at that time was to formulate relativistic coherent states.

In 1974,   I was struck by the appearance of  tube domains in  axiomatic
quantum field theory.   These domains result from the analytic continuation
of certain functions (vacuum expectaion values) associated with the theory
to complex spacetime, and powerful methods from the theory of several
complex variables are then used to prove important properties of these
functions in {\sl real\/}  spacetime.  However, the complexified spacetime
itself is usually not regarded as having any physical significance.
What intrigued me was the possibility that these tube domains  may, in fact,
have a direct physical interpretation as (extended) classical phase spaces.  If
so, this would give the idea of complex spacetime a firm physical foundation,
since in quantum field theory the  complexification is based on solid physical
principles.   It could also show the way to the
construction of relativistic coherent states.  These  ideas were
successfully worked out in 1975-76,  culminating in  a mathematics thesis
in  \/1977 at the University of Toronto entitled   ``Phase--Space Approach
to Relativistic Quantum Mechanics.''

Up to that point, the theory could only describe free particles.
The next  goal was to see how interactions could be added.  Some progress in
this direction   was made in 1979-80, when a natural way was found to
extend gauge theory to complex spacetime.  Further progress came during
my sabbatical in 1985-86, when a method was developed for extending
quantized fields themselves (rather than their vacuum expectation values) to
complex spacetime.  These ideas have so far produced no ``hard''
results,  but I  believe that they are on the right path.

Although much work remains to be done,  it seems to me that enough
structure is now in place to justify writing a book.  I hope that this volume
will be of interest to researchers in theoretical and mathematical physics,
mathematicians interested in the structure of fundamental physical theories
and assorted graduate students searching for new directions.  Although the
topics are fairly advanced, much effort has gone into making the book
self--contained and  the subject matter accessible to someone with an
understanding of the rudiments of quantum mechanics and functional
analysis.  

A novel feature of this book, from the point of view of mathematical
physics,  is the special attention given to ``signal analysis'' concepts,
especially  time--frequency localization and the new idea of {\sl
wavelets.\/}   It turns out that relativistic coherent states are similar to
wavelets, since they undergo a Lorentz contraction in the direction of
motion.   I have learned  that engineers struggle with many of the same
problems as physicists, and  that  the interplay between ideas from
quantum mechanics  and signal analysis can be very helpful to both camps. 
For that reason,  this book may also be of interest to engineers and
engineering students.

 \skp
The contents of the book are as follows.  In chapter 1  the
simplest examples of coherent states and time--frequency localization are
introduced, including the original ``canonical'' coherent states, windowed
Fourier transforms and  wavelet transforms.  A generalized notion of
frames is defined which includes the usual (discrete) one as well as
continuous resolutions of unity, and the related concept of  reproducing
kernels is discussed. 

In chapter 2 a new, algebraic approach to orthonormal bases of wavelets
is formulated.  An operational calculus is developed which simplifies the
formalism considerably and provides insights into its symmetries.  This is
used to find a {\sl complex structure\/}  which explains the 
symmetry between the low--  and the high--frequency filters in wavelet
theory.   In the usual formulation, this symmetry is clearly evident but
appears to be accidental.  Using this structure, complex wavelet
decompositions are considered which are analogous to analytic
coherent--state representations.

In chapter 3  the concept of generalized coherent states based on
Lie groups and their homogeneous spaces is reviewed.  Considerable
attention is given to holomorphic (analytic) coherent--state representations,
which result from the possibility of Lie  group complexification. The
rotation group provides a simple yet non--trivial proving ground for these
ideas, and the resulting construction  is known as the ``spin
coherent states.''   It is then
shown that the group associated with the Harmonic oscillator  is a
 weak contraction  limit (as the spin $s\to\infty$) of the rotation group and,
correspondingly, the canonical coherent states are  limits of the  spin
coherent states.  This explains why the canonical coherent states transform
naturally under the dynamics generated by the harmonic oscillator.

In chapter 4, the interactions between  phase space, quantum
mechanics   and Relativity are studied.   The main ideas of the phase--space
approach to relativistic quantum mechanics are
developed for free particles, based on the relativistic coherent--state
representations developed in my  thesis.   It is shown that  such
representations admit a covariant probabilistic interpretation,  a feature
absent in the usual spacetime theories.  In the non--relativistic  limit, the
representations are seen to ``contract'' smoothly to representations of the
Galilean group which are  closely related to the canonical coherent--state
representation.  The Gaussian weight functions in the latter 
are seen to emerge  from the geometry of the mass hyperboloid.

In chapter 5, the formalism is extended to quantized fields. The basic tool
for this  is the {\sl Analytic--Signal transform,\/} which can be applied to an
arbitrary function on $\rl^n$ to give a function on $\cx^n$ which, although
not in general analytic, is ``analyticity--friendly'' in a certain sense.  It is
shown that even the most general fields satisfying the Wightman axioms
generate a complexification of spacetime which may be interpreted as an
extended classical phase space for certain special states associated with the
theory.   Coherent--state representations  are developed for free
Klein--Gordon and Dirac fields, extending the results of
chapter 4.   The analytic Wightman two--point functions play the role of
reproducing kernels.   Complex--spacetime densities of observables such as
the energy, momentum, angular momentum and charge current are seen to
be 
 {\sl regularizations\/}  of their counterparts in real spacetime.    In
particular, Dirac particles do not undergo their usual Zitterbewegung.  The
extension to complex spacetime separates, or {\sl polarizes,\/}  the
positive-- and negative--frequency parts of free  fields, so that Wick
ordering becomes unnecessary.   A  functional--integral representation  is
developed for  quantized fields which combines the coherent--state
representations for particles (based on a finite number of degrees of
freedom) with that for fields (based on an infinite number of degrees of
freedom).

In chapter 6 we give a brief account of some  ongoing work, beginning with
a  review of the idea of holomorphic gauge theory.   Whereas in
real spacetime it is not possible to derive gauge potentials and gauge fields
from a (fiber)  metric, we show how this can be done in complex spacetime. 
Consequently, the analogy between General Relativity and gauge theory
becomes much closer in complex spacetime than it is in real spacetime. In
the ``holomorphic'' gauge class, the relation between the (non--abelian)
Yang--Mills field and its potential becomes linear due to the cancellation of
the non--linear part which follows from an integrability condition.   Finally,
we come full circle by generalizing the  Analytic--Signal transform   and
pointing out that  this generalization is a higher--dimensional version of the
wavelet transform which is, moreover, closely related to various classical
transforms \/ such as the Hilbert, Fourier--Laplace and Radon transforms.
\skp

I am deeply grateful to G.~Emch for his continued help
and encouragement over the past ten years, and to  John Klauder and
Ray Streater for having  read  the manuscript carefully and made many
invaluable comments, suggestions and corrections.  (Any remaining errors
are, of course, entirely my responsibility.)  I also  thank D.~Buchholtz, 
F.~Doria,  D.~Finch,   S.~Helgason,  I.~Kupka,  Y.~Makovoz,  J.~E. Marsden, 
M.~O'Carroll,  L.~Rosen, M.~B.~Ruskai and R.~Schor   for miscellaneous
important assistance and moral support at various times.   Finally,  I am
indebted to L.~Nachbin,  who first invited me to write this volume in 1981
(when I was not prepared to do so) and again in 1985 (when I was), and
who arranged for a tremendously interesting visit to Brazil in 1982. {\sl
Quero tamb\'em agradecer a todos os meus colegas Brasileiros!\/}

\vskip 2cm

\centerline{\bf Suggestions to the Reader} 
\skp

 The reader primarily interested in the phase--space approach to
relativistic quantum theory  may on first reading skip chapters 1--3 and
read only chapters 4--6, or even just chapter 4 and either chapters 5 or 6,
depending on interest. These chapters form a reasonably self--contained
part of the book. Terms defined in the previous chapters, such as ``frame,''
can be either ignored or looked up using the extensive index.  The index also
serves partially as a glossary of frequently used symbols.  The reader
primarily interested in signal analysis,  time--frequency localization and
wavelets, on the other hand, may read chapters 1 and 2 and skip directly to
sections 5.2 and 6.2.  The mathematical reader unfamiliar with the ideas of
quantum mechanics is urged to begin by reading section 1.1, where some
basic notions are developed, including the  Dirac notation used throughout
the book.

\VE

\countdef\pageno=0\pageno=1

\def\leftheadline{\tenrm\folio\hfil {\sl 
1. Coherent--State Representations}\hfil}  
\def\rightheadline{\hfill\folio}

\headline={\ifodd\pageno\rightheadline\else\leftheadline\fi}
\def\be{$$}\def\ee{$$}\def\ba{$$\eqalign}\def\a{&}  
\def\eno{\eqno(}\def\ea{$$}
\nopagenumbers

\centerline{\bf Chapter 1}\skp
\centerline{\bf COHERENT--STATE REPRESENTATIONS}
\vskip 3 cm
\noindent {\bf  1.1. Preliminaries}
\skp

 \noindent  In this section we establish some notation and conventions which
will be followed in the rest of the book.  We also give a little background on
the main concepts and formalism of non--relativistic  and relativistic
quantum mechanics, which should make this book accessible to
non--specialists.
\skp
\centerline{\sl 1. Spacetime and its Dual}

\noindent In this book we deal almost exclusively with flat spacetime, though
we usually let space be $\rl^s$ instead of $\rl^3$, so that spacetime becomes
$X=\rl^{s+1}$.  The reason for this extension is, first of all, that it involves
little cost since  most of the ideas to be explored here readily generalize to 
$\rl^{s+1}$, and furthermore, that it may be useful later.  Many models in
constructive quantum field theory  are based on two-- or three--
dimensional spacetime, and many currently popular attempts to unify
physics, such as string theories and Kaluza--Klein theories, involve
spacetimes of  higher dimensionality than four or (on the string
world--sheet) two--dimensional spacetimes.  An event $x\in X$ has
coordinates

\be
 x=(x^\mu )=(x^0,x^j),
\eno1)
 \ee
where $x^0\equiv t$ is the time  coordinate and $x^j$ are the space
coordinates.  Greek indices run from 0 to $s$, while latin indices run from 1
to $s$.  If we think of $x$ as a {\sl translation vector,\/} then $X$ is the
vector space of all translations in spacetime. Its {\sl dual\/} $X^*$ is the set
of all linear maps $k{:}\ X\to \rl$.  By linearity, the action of $k$ on $x$
(which we denote by $kx$ instead of $k(x)$)   can be written as

\be
 kx=\sum_{\mu =0}^sk_\mu x^\mu \equiv k_\mu x^\mu,
\eno2)
 \ee
where we adopt the Einstein summation convention of automatically
summing over repeated indices. Usually there is no relation between $x$
and $k$ other than the pairing $(x,k)\mapsto kx$. But suppose we are given
a scalar product on $X$,

\be
 x\cdot x'=g_{\mu \nu }x^\mu x'^\nu 
\eno3)
 \ee
where $(g_{\mu \nu })$ is a non--degenerate matrix.  Then each $x$ in $X$
defines a linear map $x^*{:}\ X\to\rl$  by $x^*(x')=x\cdot x'$, thus giving a
map $*{:}\ X\to X^*$, with 

\def\leftheadline{\tenrm\folio\hfil {\sl 1. Coherent--State
        Representations}\hfil} 
\def\rightheadline{\tenrm\hfil {\sl 1.1. Preliminaries}\hfil\folio}  

\headline={\ifodd\pageno\rightheadline\else\leftheadline\fi}

\be
 (x^*)_\nu\equiv x_\nu =g_{\mu \nu }x^\mu.
\eno4)
 \ee 
Since $g_{\mu \nu }$ is non--degenerate, it also defines a scalar product on
$X^*$, whose metric tensor is denoted by $g^{\mu \nu }$. The map $x\to x^*$
establishes an isomorphism between the two spaces, which we  use to
identify them.  If $x$ denotes a set of  inertial coordinates in free spacetime,
then the scalar product is  given by 

\be
 g_{\mu \nu}={\rm diag}(c^2, -1, -1, \cdots, -1)
 \ee  
where $c$ is the speed of light.  $X$, together with this scalar product, is
called {\sl Minkowskian\/}  or {\sl Lorentzian\/}  spacetime.

 It is often convenient to work in a single space rather than the dual pair $X$
and $X^*$.  Boldface letters will denote the spatial parts of vectors in $X^*$.
Thus $x=(t,-{\bf x} ), \ k=(k_0, {\bf k} )$ and 

\be
 x\cdot x'=c^2tt'-{\bf x} \cdot{\bf x'} \ \ \hbox{and}\  \ kx=k\cdot x^*=k_0t-
{\bf k} \cdot {\bf x},
 \eno5)
 \ee
where ${\bf x} \cdot{\bf x'}$ and ${\bf k} \cdot{\bf x}$ denote the usual
Euclidean inner products in $\rl^s$.
\skp
\centerline {\sl 2. Fourier Transforms}

\noindent The Fourier transform of a function $f{:}\ X\to \cx$
(which, to avoid analytical subtleties for the present, may be assumed to be a
Schwartz test function; see Yosida [1971]) is a function $\fhat{:}\ X^*\to \cx$
given by

\be
 \fhat(k)=\int_X dx\,e^{\tpi kx}f(x)
\eno6)
 \ee
where $dx\equiv dt\,d^s{\bf x} $ is Lebesgue measure on $X$.  $f$ can be
reconstructed from $\fhat$ by the inverse Fourier transform,  denoted by
$\check{\ }$ and given by

\be
 f(x)=\int_{X^*} dk\,e^{-\tpi kx}\fhat(k)\equiv (\fhat)\check{\,}(x),
\eno7)
 \ee
where $dk=dk_0\,d^s{\bf k} $ denotes Lebesgue measure on $X^*\approx
X$. Note that the presence of the $2\pi $ factor in the exponent avoids the
usual need for factors of $(2\pi )^{-(s+1)/2}$ or $(2\pi )^{-s-1}$ in front of
the integrals. Physically, $k$ represents a {\sl wave vector:\/} $k_0\equiv
\nu $ is a frequency in cycles per unit time, and $k_j$ is
a wave number in cycles per unit length.  Then the interpretation of
the linear map $k{:}\ X\to \rl$ is that $2\pi kx$ is the total
radian phase gained  by the plane wave 
$g(x')=\exp(-2\pi ikx')$ through the spacetime translation $x$, i.e. $2\pi k$
 ``measures''  the radian phase shift. 
Now in pre--quantum relativity, it was realized that the {\sl energy\/} $E$
combines with the {\sl momentum\/} {\bf p}   to form a vector $p\equiv
(p_\mu )=(E,{\bf p}  )$ in $X^*$. Perhaps the single most fundamental
difference between classical mechanics  and quantum mechanics  is   that in
the former, matter is  conceived to be made of ``dead sets'' moving in space
while in the latter, its microscopic structure is that of {\sl waves\/} descibed
by  complex--valued wave functions which, roughly speaking, represent its
distribution in space in probabilistic terms.   One important consequence of
this difference is that while in classical mechanics  one is free to specify
position and momentum independently, in quantum mechanics   a complete
knowledge of the distribution  in space, i.e. the wave function, determines
the distribution in momentum space via the Fourier transform. The classical
energy is re--interpreted as the frequency  of the associated wave
by Planck's {\sl Ansatz,}

\be
 p_0\equiv E=2\pi \hbar \nu
\eno8)
 \ee
where $\hbar$ is Planck's constant, and the classical momentum is
re--interpreted as the wave--number vector of the associated
wave by De Broglie's relation, 

\be
 {\bf p}  =2\pi \hbar {\bf k}.
\eno8')
 \ee
These two relations are unified in relativistic terms as $p_\mu =2\pi \hbar
k_\mu $.  Since a general wave function is a
superposition of plane waves, each with its own frequency and wave
number, the relation of  energy and momentum to the  the spacetime
structure is very different in quantum mechanics from what is was in
classical mechanics: They become operators on the space of wave functions:

\be
 (P_\mu f)(x)=\int_{X^*}dk\,p_\mu e^{-\tpi kx}\fhat(k)=
i\hbar{\partial \over{\partial x^\mu }}f(x),
\eno9)
 \ee
or, in terms of $x^*$,

\be
 P_0=i\hbar{\partial \over{ \partial t}}\  \ \hbox{and}\  \
P_k=-i\hbar{\partial \over{\partial  x_k}}.
 \eno9')
 \ee
This is, of course, the source of the uncertainty principle. In terms of
energy--momentum, we obtain the
``quantum--\-mechanical'' Fourier transform and its inverse,

\ba{
 \fhat(p)\a=\int_X dx\,e^{ipx/\hbar}f(x)\cr
f(x)\a=(2\pi \hbar)^{-s-1}\int_{X^*} dp\,e^{-ipx/\hbar}\fhat(p).
 \cr}
 \eno10)
 \ea

If  $f(x)$ satisfies a differential equation, such as the Schr\"odinger equation
or the Klein--Gordon equation, then $\fhat(p)$ is supported on an
$s$--dimensional submanifold $P$ of $X^*$ (a paraboloid or two--sheeted
hyperboloid, respectively) which can be parametrized by ${\bf p}  \in \rl^s$.
We will write the solution as

\be
 f(x)=(2\pi \hbar)^{-s}\int_P d\mu (p) \,e^{-ipx/\hbar} \fhat(p),
\eno11)
\ee
where $\fhat (p)$ is, by a mild abuse of notation, the ``restriction'' of $\fhat$
to $P$ (actually, $|\fhat(p)|^2$ is a density on $P$)
 and $d\mu (p)\equiv \rho (p)\,d^s{\bf p}  $ is an appropriate
invariant measure on $P$. For the Schr\"odinger equation $\rho (p)\equiv
1$, whereas for the Klein--Gordon equation, $\rho (p)=\,|\,p_0\,|\,^{-1} $.
 Setting $t=0$ then shows that $\fhat(p)$ is related to the initial wave
function by

\be
 f({\bf x} ,0)=\bigl( \rho (p)\fhat  \bigr)\check{\,}({\bf x}),
\eno12)
 \ee
where now ``$\ \check{}\ $'' denotes the the $s$--dimensional inverse
Fourier transform of the function $\fhat$ on $P\approx\rl^s$.

 We will usually work with ``natural units,'' i.e. physical units so
chosen that  $\hbar=c=1$.  However, when considering the non--relativistic
limit ($c\to\infty$) or the classical limit ($\hbar\to 0$), $c$ or $\hbar$ will
be re--inserted into the equations.
 \skp 

\centerline  {\sl 3. Hilbert Space}

\noindent Inner products in
Hilbert space will be linear in the {\sl second\/} factor and antilinear in the
first factor.  Furthermore, we will make some discrete use of Dirac's very
elegant and concise {\sl bra--ket\/} notation,  favored by physicists and often
detested or misunderstood by mathematicians.  As this book is aimed at a
mixed audience, I will now take a few paragraphs to review this notation
and, hopefully, convince mathematicians of its correctness and value.  When
applied to coherent--state representations, as opposed to representations in
which the position--or momentum operators are diagonal, it is perfectly
rigorous. (The bra--ket notation is problematic when dealing with
distributions, such as the generalized eigenvectors of position or momentum,
since it tries to take the ``inner products'' of such distributions.)

Let ${\cal  H}$ be an arbitrary complex Hilbert space with inner product
$\langle \cdot,\cdot\rangle$.  Each element $f\in {\cal  H}$ defines a
bounded linear functional \break
 $f^*{:}\ {\cal H}\to \cx$ by

\be
f^*(g)=\langle f, g\rangle. 
\eno13)
 \ee
The Riesz representation theorem guarantees that the converse is also true:
Each bounded linear functional $L: {\cal H}\rightarrow \cx$ has the form
$L=f^*$ for a unique $f\in {\cal  H}$.  Define the {\sl bra\/} $\langle f|$
corresponding to $f$ by

\be
 \langle f|\,=f^*: {\cal H}\rightarrow \cx.
\eno14)
 \ee
Similarly, there is a one-to-one correspondence between vectors
 $g\in {\cal  H}$ and linear maps

\be
 |g\rangle {:} \ \cx\rightarrow {\cal H}
\eno15)
 \ee
defined by 

\be
 |g\rangle(\lambda)=\lambda  g, \quad \lambda \in \cx,
\eno16)
 \ee
which will be called {\sl kets}.  Thus elements of ${\cal H}$ will be denoted
alternatively by $g$ or by $ |g\rangle$.  We may now consider the
composite map {\sl bra--ket}

\be
 \langle f|\,g\rangle\,{:}\ \cx\rightarrow \cx,
\eno17)
 \ee
given by 

\ba{
\langle f\,|\,g\rangle(\lambda )\a=f^*(\lambda g)=\lambda f^*(g)\cr
\a=\lambda \langle f,\,g\rangle.
\cr}
\eno18)
 \ea
Therefore the ``bra--ket'' map is simply the multiplication
by the inner product $\langle f,\,g\rangle$ (whence it derives its name). 
Henceforth we will identify these two and write $\langle f|\,g\rangle$ for
both the map and the inner product. The reverse composition

\be
 |g\rangle\langle f |  :{\cal H}\rightarrow {\cal H}
\eno19)
 \ee
may be viewed as acting on kets to produce kets:

\be
|g\rangle\langle f |\bigl(|h\rangle\bigr)=|g\rangle\bigl(\langle f |
h\rangle\bigr).
\eno20)
 \ee

To illustrate the utility of this notation,  as well as some of its
pitfalls, suppose that we have an orthonormal basis $\{g_n\}$ in ${\cal H}$. 
Then the usual expansion of an arbitrary vector $f$ in ${\cal H}$ takes the
form

\be
 f=\sum_n g_n\langle\, g_n\,|\,f\,\rangle\equiv \sum_n
\,|\,g_n\,\rangle\langle\, g_n\,|\,f\,\rangle,
 \eno21)
 \ee
from which we have the ``resolution of unity''

\be
 \sum_n \,|\,g_n\,\rangle\langle\, g_n\,|\,=I
\eno22)
 \ee
where $I$ is the identity on $\cal H$ and the sum converges in the strong 
operator topology.  If $\{h_n\}$ is a second orthonormal basis, the relation
between the expansion coefficients in the two bases is \be \langle\,
h_n\,|\,f\,\rangle=\langle\, h_n\,|\,If\,\rangle=\sum_m \langle\,
h_n\,|\,g_m\,\rangle \langle\, g_m\,|\,f\,\rangle. 
\eno23)
 \ee

In physics, vectors such as $g_n$ are often written as $\,|\,n\,\rangle$,
which can  be a source of great confusion for mathematicians. Furthermore,
functions in $L^2(\rl^s)$, say, are often written as $f(x)=\langle\,
x\,|\,f\,\rangle$,  with $\langle\, x\,|\,x'\,\rangle=\delta (x-x')$, as though
the $\,|\,x\,\rangle$'s formed an orthonormal  basis. This notation is very
tempting; for example, the Fourier transform is written as a ``change of
basis,''

\be
 \langle\, k\,|\,f\,\rangle=\int dx\,\langle\, k\,|\,x\,\rangle\langle\,
x\,|\,f\,\rangle
 \eno24)
 \ee
with the ``transformation matrix'' $\langle\, k\,|\,x\,\rangle=\exp (2\pi
ikx)$. One of the advantages of this notation is that it permits one to think of
the Hilbert space as ``abstract,'' with $\langle\,g_n\,|\,f\,\rangle, \/
\langle\, h_n\,|\,f\,\rangle, \langle\, x\,|\,f\,\rangle$ and $\langle\,
k\,|\,f\,\rangle$ merely different  ``representations'' (or ``realizations'') of
the same vector $f$.  However, even with the help of
distribution theory, this use of Dirac
notation is unsound, since it attempts to extend the
Riesz representation theorem to distributions by allowing inner products of
them. (The ``vector'' $\langle\, x\,|\,$ is a distribution which evaluates {\sl
test functions\/} at the point $x$; as such, $\,|\,x'\,\rangle$ does not exist
within modern--day distribution theory.)  We will generally abstain from
this use of the bra--ket notation.

Finally, it should be noted that the term ``representation'' is used in two
distinct ways: (a) In the above sense, where abstract  Hilbert--space vectors
are represented by functions in various function spaces, and (b) in
connection with groups, where the action of a group on a Hilbert space is
represented by  operators.

This notation will be especially useful when discussing frames, of which
coherent--state representations  are examples.

\secskp

\noindent {\bf 1.2. Canonical Coherent States}
\skp

\def\rightheadline{\tenrm\hfil {\sl 1.2. Canonical Coherent
States}\hfil\folio}

\noindent We begin by recalling the original coherent-state
representations \break (Bargmann [1961], Klauder [1960, 1963a, b], Segal
[1963a]). Consider a spinless non-relativistic particle in $\rl^s$ (or $s/3$
such particles in $\rl^3$), whose algebra of observables is generated by the
position operators $X_k$ and momentum operators $P_k, k=1, 2, \ldots s.$ 
These satisfy the ``canonical commutation relations''

\be
[X_k,X_l]=0,  \qquad  [P_k,P_l]=0, 
\qquad [X_k,P_l]=i\delta_{kl}I ,
 \eno1)
\ee
where $I$ is the identity operator.  
The operators $-iX_k$, $-iP_k$ and $-iI$ together  form a {\sl real\/}  Lie
algebra known as the {\sl Heisenberg algebra}, which is  irreducibly
represented on $L^2(\rl^s)$ by 

\be
 X_kf(x)=x_kf(x), \qquad P_kf(x)=-i{\partial \over{ \partial
x_k}}f(x),
  \eno2)
\ee
the {\sl Schr\"odinger representation.}  

 As a  consequence of the above commutation relations between
$X_k$ and $P_k$, the position and momentum of the particle obey the {\sl
Heisenberg uncertainty relations,\/} which can be derived simply as
follows.  The expected value, upon measurement, of an observable
represented by an operator $F$ in the state represented by a wave
function $f(x)$ with $\|f\|=1$ (where $\|\cdot\|$ denotes the norm in
$L^2(\rl^s)$) is given by

\be
 \langle\,F\,\rangle=\langle\, f\,|\,Ff\,\rangle.
\eno3)
 \ee
In particular,  the expected position-- and momentum coordinates of the
particle are $ \langle\, X_k\,\rangle$ and $\langle\, P_k\,\rangle$. The
uncertainties $\Delta _{X_k}$ and $\Delta _{P_k}$ in position and momentum
are given by the variances

\ba{
\Delta _{X_k}^2\a=\langle\, (X_k-\langle\, X_k\,\rangle)^2\,\rangle
   =\langle\, X_k^2\,\rangle-\langle\, X_k\,\rangle^2\cr
\Delta _{P_k}^2\a=\langle\, (P_k-\langle\, P_k\,\rangle)^2\,\rangle
   =\langle\, P_k^2\,\rangle-\langle\, P_k\,\rangle^2.
 \cr}
\eno4)
 \ea
Choose an arbitrary constant $b$ with units of area (square length) and
consider the operators 

\be
 A_k=X_k+ibP_k=x_k+b{\partial \over{\partial x_k}}.
  \eno5)
\ee
Notice that although $A_k$ is non--Hermitian, it is {\sl real } in the
Schr\"odin\-ger representation.  Let 

\be
\bar  z _k\equiv \langle\, f\,|\,A_kf\,\rangle=\langle\,
X_k\,\rangle+ib\langle\, P_k\,\rangle,
 \eno6)
 \ee
where $\bar z_k$ denotes the complex--conjugate of $z_k$.  Then for 
$\delta A_k\equiv A_k-\bar z _kI$ we have  $\langle\, \delta
A_k\,\rangle=0$ and

\be
 0\le \|\delta A_kf\|^2=\Delta _{X_k}^2+b^2\Delta _{P_k}^2-b.
\eno7)
 \ee
The right--hand side is a quadratic in $b$, hence the inequality for
all $b$ demands that the discriminant be non--positive,  giving the
uncertainty relations 

\be
 \Delta _{X_k}\Delta _{P_k}\ge {1\over 2}.
\eno8)
 \ee
Equality is attained if and only if $\delta A_kf=0$, which shows that the
{\sl only\/} minimum--uncertainty states are given by wave functions $f(x)$
satisfying the eigenvalue equations

\be
 A_kf=\bigl(x_k+b{\partial \over{x_k }}\bigr)f=\bar z _kf
\eno9)
 \ee
for some real number $b$ (which may actually depend on $k$)  and some $z
\in \cx^s$.  But square--integrable solutions exist only for $b>0$, and then
there is a unique solution (up to normalization) $\chi _z $ for each $z \in
\cx^s$. To simplify the notation, we now choose $b=1$. Then $A_k$ and
$A_k^*$ satisfy the  commutation relations 

\be
 [A_k, A_l]=0,\qquad [A_k, A_l^*]=2\delta _{kl}I,
\eno10)
 \ee
 and  $\chi _z$  is given by 
\be
\chi  _z (x')=N\exp[-\bar z ^2/4+\bar z \cdot x'-{x'}^2/2],
\eno11)
 \ee
where the normalization constant is chosen as $N=\pi ^{-s/4} $, so that
$\|\chi _z\|=1$ for $z=0$. Here $\bar z ^2$ is the (complex) inner product of
$\bar z $ with itself.  Clearly $\chi _z $ is in $L^2(\rl^s),$ and if $z
= {x-ip}$, then

\be
\langle X_k\rangle=x_k\ \hbox{and}\  \langle P_k\rangle=p_k  
  \eno12)
\ee
in the state given by $\chi _z $.  The vectors $\chi _z $ are
known as the {\sl canonical coherent states\/}. They occur naturally in
connection with the harmonic oscillator problem, whose Hamiltonian can be
cast in the form

\be
 H={1\over {2m}}\bigl(P^2+m^2\omega ^2X^2\bigr)={1\over2}m\omega
^2A^*\cdot A+{s\omega \over 2}
 \eno13)
 \ee
with 

\be
 A_k= X_k+iP_k/m\omega  
\eno14)
 \ee
(thus $b=1/m\omega $).  They have the remarkable property that if the
initial state is $\chi _z $, then the state at time $t$ is $\chi _{z(t)}
$ where $z (t)$ is the orbit in phase space of the
corresponding  {\sl classical\/} harmonic oscillator with initial data given
by $z$.  These states were discovered by Schr\"odinger himself [1926], at
the dawn of modern quantum mechanics.  They were further investigated by 
Fock [1928] in connection with quantum field theory  and by von Neumann
[1931] in connection with the quantum measurement problem. Although they
span the Hilbert space, they do not form a basis because they possess a high
degree of linear dependence, and it is not easy to find complete, linearly
independent subsets.  For this reason, perhaps, no one seemed to know quite
what to do with them until the early 1960's, when it was discovered that
what really mattered was not that they form a basis but what we shall call 
a {\sl generalized frame}.  This allows them to be used in generating a {\sl
representation\/} of the Hilbert space by a space of analytic functions, as
explained below.  The frame property of the coherent states (which will be
studied and generalized in the following sections and in chapter 3) was
discovered independently at about the same time by Klauder, Bargmann  and
Segal. Glauber  [1963a,b] used these vectors with great effectiveness to
extend the concept of optical coherence to the domain of quantum
electrodynamics, which was made necessary by the discovery of the laser. 
 He dubbed them ``coherent states,'' and the name stuck to the point of being
generic. (See also Klauder and Sudarshan [1968].)  Systems of vectors now
called ``coherent'' may have nothing to do with optical coherence, but there
is at least one unifying characteristic, namely their frame property (next
section).

The coherent-state representation  is now defined as follows:  Let $\cal
F$ be the space of all functions

\ba{
\tilde f (z )\a\equiv\langle \chi _z \,|\, f\rangle
=\int d^sx'\,\overline{\chi _z (x')}f(x')   \cr
\a=N\int d^sx'\,\exp[-z^2/2+z\cdot x'-{x'}^2/2]f(x') 	
  \cr}
\eno15)
\ea
where $f$ runs through $L^2(\rl^s)$.  Because the exponential decays
ra\-pidly in $x'$, $\tilde f$ is entire in the variable $z \in \cx^s.$ 
Define an inner product on $\cal F$ by

\be
\langle \tilde f\,|\, \tilde g\rangle_{\cal F}
\equiv\int_{\cxs^s}d\mu(z )\,\overline{\tilde f(z)} \tilde g(z),
  \eno16)
\ee
where $z \equiv x-ip $ and

\be
 d\mu(z )=(2\pi )^{-s}\exp(-\bar z\cdot  z/2)\,d^sx\,d^sp .
\eno17)
 \ee
 Then we have the
following  theorem relating the inner products in $L^2(\rl^s)$ 
and $\cal F$.

\skp  
\proclaim     Theorem 1.1.  Let $f,g\in L^2(\rl^s)$ and let $\tilde f, \tilde g$
be the  corresponding entire functions in ${\cal F}$. Then
 \par

\be
\langle \tilde f\,|\, \tilde g\rangle_{\cal F}=\langle f\,|\,  g\rangle_{L^2}.
  \eno18)
\ee
\skp
\noindent {\bf Proof.}   To begin with, assume that $f$ is in the Schwartz
space ${\cal S}(\rl^s)$ of ra\-pidly decreasing smooth test functions. For
$z =x-ip $, we have  
$$
\chi_z (x')=N\exp[-\bar z ^2/2+x^2/2-(x'-x)^2/2+ip\cdot x'], 
 $$
hence

\be
\tilde f(x-ip)
=N\exp[(x^2+p^2)/4+ipx/2]\bigl(\exp[-(x'-x)^2/2]f\bigr)\hat{\,}(p)
 \eno19)
\ee
where $\hat{\ }$ denotes the Fourier transform with respect to $x'$.  Thus by
Plancherel's theorem (Yosida [1971]), 

\ba{
 (2\pi )^{-s}\int d^sp\,\a\exp(-p^2/2)|\tilde f(x-ip)|^2 \cr
\a=N^2\exp(x^2/2)\int d^sx' \exp[-(x'-x)^2]\,|f(x')|^2.  
\cr} 
\eno20)
\ea
Therefore

\ba{
\int d\mu  \a  (z )\,|\tilde f(z )\,|^2\cr
\a=N^2\int d^sx\,
\int d^sx'\exp[-(x'-x)^2]\,|f(x')\,|^2 \cr
\a=\int d^sx' \,|\,f(x')\,|\,^2
\cr}
\eno21)
\ea
after exchanging the order of integration.  This proves that 

\be
\|\tilde f\|^2_{\cal F}=\|f\|^2_{L^2}
  \eno22)
\ee
for $f\in {\cal S}(\rl^s)$, hence by continuity also for arbitrary $f\in
L^2(\rl^s)$.  By polarization the result can now be extended from the
norms to the inner products. \qed
\skp

The relation $f\leftrightarrow \tilde f$ can
be summarized neatly and economically in terms of Dirac's  bra-ket
notation. Since 

\be
 \overline{\tilde f(z )}=\overline{\langle \chi_z |
f\rangle}=\langle f |\chi_z \rangle, 
\eno23)
 \ee
theorem 1 can be restated as

\be
 \int_{\ \cxs^s}d\mu (z )\,\langle f | \chi_z \rangle\langle
\chi_z  | g\rangle=\langle f | g\rangle.
 \eno24)
 \ee
Dropping the bra $\langle f | $ and ket $| g\rangle$, we have the
operator identity
 
\be
 \int_{\ \cxs^s}d\mu (z )\, | \chi_z \rangle\langle \chi_z  | =I,
\eno25)
 \ee
where $I$ is the identity operator on $L^2(\rl^s)$  and the integral
converges at least in the sense of the {\sl weak operator
topology,\/}\footnote*{As will be shown in a more general context in the
next section, under favorable conditions the integral actually converges in
the {\sl strong\/} operator topology.}  i.e. as a quadratic form.  In Klauder's
terminology, this is a {\sl continuous resolution of  unity.}  A general
operator $B$ on $L^2(\rl^s)$  can now be expressed as an integral operator
$\tilde B$ on $\cal F$ as follows:

\ba{
(\tilde B\ftil)(z ) \a\equiv (Bf)\tilde{\,}(z )\cr
\a=\langle\, \chi_z \,|\, Bf\,\rangle\cr
\a=\langle \,\chi_z \,|\, BIf\,\rangle\cr
\a=\int_{\ \cxs^s}d\mu (w )\, \langle\, \chi_z \,|\,B \chi_w \,
\rangle\langle \,\chi_w   \,|\, f\,\rangle\cr
\a\equiv  \int_{\ \cxs^s}d\mu (w )\, \tilde B(z ,\bar w)
\ftil(w ).
 \cr}
\eno26)
 \ea 
Particularly simple representations are obtained for the
basic po\-sition-- and momentum operators. We get

\ba{
X_k \chi_z (x')\a\equiv x'_k \chi_z (x')\cr
\a=\left({\partial \over{\partial \bar z _k}}+{\bar z_k\over
2}\right)\chi_z (x')\cr
\a \cr
 P_k \chi_z \a=-i(A_k-X_k)\chi_z \cr
\a=i\left({\partial \over{\partial \bar z_k }}  -{\bar z _k\over
2}\right)\chi_z, 
 \cr}
 \eno27)
 \ea
thus 

\ba{
\tilde X_k\ftil\a=\left({\partial \over{\partial z
_k}}+{z_k\over 2} \right)\ftil\cr
\a \cr
\tilde P_k\ftil\a=-i\left({\partial \over{\partial z_k }}  -{z
_k\over 2}\right)\ftil.
 \cr}
 \eno28)
 \ea
Hence $X_k$ and $P_k$ can be  represented as differential rather than
integral operators.

As promised, the continuous resolution of the identity makes it possible to
{\sl reconstruct} $f\in L^2(\rl^s)$ from its transform $\tilde f\in \cal F$:
\be
f=If=\int_{\cxs^s}d\mu (z )\,|\,\chi_z \,\rangle\langle\,
\chi_z \,|\,f\,\rangle,
 \eno29)
 \ee
that is,  
\be
 f(x')=\int_{\cxs^s}d\mu (z )\,\chi_z (x')\tilde f(z ).
\eno30)
 \ee

Thus in many respects the coherent states behave like a {\sl basis\/} for
$L^2(\rl^s)$.  But they differ from a basis in at least one important respect: 
They cannot all be linearly independent, since there are uncountably many of
them and $L^2(\rl^s)$ (and hence also $\cal F$) is
separable.  In particular, the above reconstruction formula can be used to
express $\chi_z $ in terms of all the $\chi_w  $'s:

\be
 \chi_z =\int_{\cxs^s}d\mu (w )\,|\,\chi_w  \,\rangle\langle\,
\chi_w \,|\,\chi_z \,\rangle. 
\eno31)
 \ee
In fact, since entire functions are determined by their values on some
discrete subsets $\Gamma$ of \/$\cxs^s$, we conclude that the corresponding
subsets of coherent states $\{ \chi_z \,|\,z \in \Gamma\}$
are already complete since for any function $f$ orthogonal to them all, 
$\ftil(z )=0$ for all $z\in \Gamma$ and hence $\ftil\equiv 0$,
which implies $f=0 $ a.e. For example, if $\Gamma$ is a regular lattice, a
necessary and sufficient condition for completesness 
is that $\Gamma$ contain at least one point in each Planck cell (Bargmann
et al., [1971]), in the sense  that the spacings $\Delta x_k$ and $\Delta p_k$
of the lattice  coordinates $z _k=x_k+ip_k$ satisfy $\Delta x_k\Delta p_k\le
2\pi \hbar\equiv 2\pi $.  It is no accident that this looks like the 
uncertainty principle but with the inequality going ``the wrong way.''  The
exact coefficient of $\hbar$ is somewhat arbitrary and depends
on one's definition of uncertainty; it is possible to define measures of
uncertainty other than the standard deviation.  (In fact, a preferable---but
less tractable---definition of uncertainty uses the notion of {\sl entropy,\/}
which involves all moments rather than just the second moment.  See
Bialynicki--Birula and Mycielski [1975] and Zakai [1960].) The intuitive
explanation is that if $\tilde f$ gets ``sampled'' at least once in every Planck
cell, then it is uniquely determined since the uncertainty principle limits the
amount of variation which can take place within such a cell.  Hence the set of
all coherent states is {\sl overcomplete.} We will see later that reconstruction
formulas exist for some discrete subsystems of coherent states, which makes
them as useful as the continuum of such states. This ability to synthesize
continuous and discrete methods in a single representation, as well as to
bridge quantum and classical concepts, is one more aspect of the appeal and
mystery of these systems.

\secskp

\noindent {\bf  1.3.  Generalized Frames and Resolutions of Unity }

 \skp
\noindent Let $M$ be a set and $\mu $ be a measure on $M$ (with an
appropriate $\sigma $--algebra of measurable subsets) such that $\{M,\mu
\}$ is a $\sigma $--finite  measure space.  Let $\cal H$ be a Hilbert space
and $h_m\in {\cal H}$ be a family of vectors indexed by $m\in M$.  \skp
\noindent {\bf Definition.}\sp    
 The set
\be
 {\cal H}_M\equiv \{h_m\,|\,m\in M\}
\eno1)
 \ee
  is a {\sl generalized frame\/} in ${\cal H}$ if
\item{1.}  the map $h{:}\ m\mapsto h_m$ is weakly measurable, i.e. for each
$f\in \cal H$ the function  $ \tilde f(m)\equiv \langle\, h_m\,|\,f\,\rangle$ is
measurable, and \item{2.} there exist constants $0<A\le B$ such that
\be
 A\|f\|^2\le\int_Md\mu(m)\, |\, \langle\, h_m\,|\,f\,\rangle\,|\,^2%
\le B\|f\|^2\qquad \forall f\in {\cal H}.
\eno2)
 \ee
\skp
\noindent  ${\cal H}_M$ is a {\sl frame} (see Young [1980] and Daubechies
[1988a]) in the special case when $M$ is countable and $\mu $ is the
counting measure on $M$ (i.e., it assigns to each subset of $M$ the number of
elements contained in it).  In that case, the above condition becomes

\be
  A\,\|f\|^2\le\sum_{m\in M}|\, \langle\, h_m\,|\,f\,\rangle\,|\,^2%
\le B\,\|f\|^2\qquad \forall f\in {\cal H}.
\eno3)
 \ee
We will henceforth drop the adjective ``generalized'' and simply speak of 
``frames.'' The above case where $M$ is countable will be refered to as a {\sl
discrete \/} frame.

If $A=B$, the frame ${\cal H}_M$ is called {\sl tight}.  The coherent states of
the last section form a tight frame, with $M=\cx^s, d\mu (m)=d\mu (z ),
h_m=\chi _z $ and $A=B=1$.  

Given a frame, let $T$ be the map taking vectors in ${\cal H}$ to functions on
$M$ defined  by

\be
 (Tf)(m)\equiv\langle\, h_m\,|\,f\,\rangle\equiv\ftil(m),\quad f\in {\cal H}. 
\eno4)
 \ee
Then the frame condition states that $Tf$ is square-integrable with respect
to $d\mu $, so that $T$ defines a map 

\be
T{:} \ {\cal H}\to L^2(d\mu), 
 \eno5)
 \ee
with

\be
A\,\|f\|^2\le\|Tf\|^2_{L^2(d\mu)}\le B\,\|f\|^2.  
\eno6)
 \ee
The frame property can now be stated in operator form as

\be
 AI\le T^*T\le BI,
\eno7)
 \ee
where $I$ is the identity on ${\cal H}$.  In bra-ket notation,

\be
 G\equiv T^*T=\int_M d\mu(m)\,|\,h_m\,\rangle\langle\, h_m\,|\,  ,
\eno8)
 \ee
where the integral is to be interpreted, initially,  as converging in the weak
operator topology, i.e. as a quadratic form.  For a measurable subset $N$ of
$M$, write 

\be
 G(N)\equiv \int_N d\mu(m)\,|\,h_m\,\rangle\langle\, h_m\,|.
\eno9)
 \ee

\def\rightheadline{\tenrm\hfil{\sl 1.3. Generalized Frames and
Resolutions of Unity}\hfil\folio} 

 \skp

\noindent {\bf Proposition 1.2.} \sp {\sl If the integral $G(N)$ converges in
the strong operator topology of ${\cal H}$ whenever $N$ has finite measure,
then so does the complete  integral representing $G=T^*T$.
}
\skp
\noindent {\bf Proof}\footnote*{I thank M. B.  Ruskai for suggesting this
proof.}. \sp Since $M$ is $\sigma $--finite, we can choose an increasing
sequence $\{M_n\}$ of sets of finite measure such that  $M=\cup_nM_n$.
Then the corresponding sequence of integrals $G_n$ forms a bounded (by
$G$) increasing sequence of Hermitian operators, hence converges to $G$ in
the strong operator topology (see Halmos [1967], problem 94). \qed\skp

If the frame is tight, then $G=AI$
and the above gives a resolution of unity after dividing by $A$.  For non-tight
frames, one generally has to do some work to obtain a resolution of unity. 
The frame condition means that $G$ has a bounded inverse, with 

\be
 B^{-1} I\le G^{-1}\le A ^{-1} I.
\eno10)
 \ee
Given a function $g(m)$ in $L^2(d\mu)$, we are interested in answering the
following two questions: (a) \  Is $g=Tf$ for some $f\in {\cal H}$?
(b) \ If so, then what is $f$?
In other words, we want to:
\skp 
\item{(a)} Find the range $\Re_T\subset L^2(d\mu)$  of the map $T$.
\item{(b)} Find a left inverse $S$ of $T$, which enables us to reconstruct $f$
from $Tf$ by $f=STf$.
\skp
Both questions will be answered if we can explicitly compute $G^{-1} $.  For
let 

\be
 P=TG^{-1} T^*{:}\ L^2(d\mu)\to L^2(d\mu). 
\eno11)
 \ee
Then it is easy to see that

\ba{
\a (a)\ P^*=P\cr
\a (b)\ P^2=P\cr
\a (c)\ PT=T.
 \cr}
 \eno12)
 \ea
It follows that $P$ is the orthogonal projection onto the range of $T$,

\be
P{:}\ L^2(d\mu)\to \Re_T,  
\eno13)
 \ee
for if $g=Tf$ for some $f$ in ${\cal H}$, then $P g =PTf=Tf=g$, and conversely
if for some $g$ we have $P g=g$, then $g=T(G ^{-1} T^*g)\equiv Tf$.  
Thus $\Re_T$ is a closed subspace of $L^2(d\mu)$  and a function $g\in
L^2(d\mu)$ is in $\Re_T$ if and only if 

\ba{
g(m)\a=(P g)(m)=\bigl (TG ^{-1} T^*g\bigr )(m)\cr
\a=\langle\, h_m\,|\,G ^{-1} T^*g\,\rangle_{\cal H} \cr
\a=\langle\, TG ^{-1} h_m\,|\,g\,\rangle_{L^2(d\mu)}\cr
\a=\int_Md\mu(m')\,\overline{\bigl(TG ^{-1} h_m\bigr)(m')}g(m')\cr
\a=\int_Md\mu(m')\,\overline{\langle\, h_{m'}\,|\,G%
^{-1}h_m\,\rangle} g(m')\cr 
\a=\int_Md\mu(m')\, \langle\, h_m\,|\,G ^{-1} h_{m'}\,\rangle g(m').
 \cr}
 \eno14)
 \ea
The function

\be
 K(m,m')\equiv\langle\, h_m\,|\,G ^{-1} h_{m'}\,\rangle
\eno15)
 \ee
therefore has a property similar to the Dirac $\delta $-function with respect
to the measure $d\mu$, in that it {\sl reproduces\/} functions in $\Re_T$.
But it differs from the $\delta $-function in some important respects.  For one
thing, it is  bounded by

\ba{
 \,|\,K(m,m')\,|\,\a=\,|\,\langle\, h_m\,|\,G ^{-1} h_m'\,\rangle\,|\,\cr
\a\le \|G ^{-1} \|\,\|h_m\|\,\|h_{m'}\|\cr
\a\le A ^{-1} \|h_m\|\,\|h_{m'}\|<\infty 
 \cr}
 \eno16)
 \ea
for all $m$ and $m'$. 
Furthermore, the ``test functions'' which $K(m,m')$ reproduces form a Hilbert
space and  $K(m,m')$ defines  an integral operator, not merely a distribution,
on $\Re_T$.  In the applications to relativistic quantum theory to be
developed later, $M$ will be a complexification of spacetime and $K(m,m')$
will be holomorphic in $m$ and antiholomorphic in $m'$.  

The Hilbert space $\Re_T$ and the associated function \break
$K(m,m')$ are an
example of an important structure called a {\sl reprodu\-cing--kernel Hilbert
space\/} (see Meschkowski [1962]), which is reviewed briefly in the next
section.
 $K(m,m')$ is called a {\sl reproducing kernel\/} for $\Re_T$.

We can thus summarize our answer to the first question by saying that a
function $g\in L^2(d\mu)$ belongs to the range of $T$ if and only if it
satisfies the {\sl consistency condition}

\be
 g(m)=\int_Md\mu(m')\, K(m,m')g(m').
\eno17)
 \ee
Of course, this condition is only useful to the extent that we have
information about the kernel $K(m,m')$ or, equivalently, about the operator
$G ^{-1} $.  The answer to our second question also depends on the
knowledge of $G ^{-1} $.  For once we know that $g=Tf$ for some $f\in {\cal
H}$, then

\be
 f=G ^{-1} Gf=G ^{-1} T^*\bigl(Tf\bigr)=G ^{-1} T^*g.
\eno18)
 \ee
Thus the operator

\be
 S=G ^{-1} T^*{:}\ L^2(d\mu)\to \cal H  
\eno19)
 \ee
is a left inverse of $T$ and we can reconstruct $f$ by

\ba{
f\a=Sg=G ^{-1} T^*Tf\cr
\a=G ^{-1} \int_Md\mu(m)\, |\,h_m\,\rangle\langle\, h_m\,|\,f\,\rangle\cr
\a=\int_Md\mu(m)\, G ^{-1} \,|\,h_m\,\rangle g(m).
 \cr}
 \eno20)
 \ea
This gives $f$ as a linear combination of the vectors

\be
 h^m\equiv G ^{-1} h_m.
\eno21)
 \ee
Note that

\be
 \int_Md\mu(m)\, |\,h^m\,\rangle\langle\, h^m\,|=G ^{-1} G G ^{-1} =G ^{-1}, 
\eno22)
 \ee
therefore the set 

\be
 {\cal H}^M\equiv\{h^m\,|\,m\in M\}
\eno23)
 \ee
is also a frame, with frame constants $0<B ^{-1} \le A ^{-1} $.  We will call 
${\cal H}^M$ the frame {\sl reciprocal \/} to ${\cal H}_M$.  (In Daubechies
[1988a], the corresponding discrete object is called the {\sl dual\/} frame,
but as we shall see below, it is actually a generalization of the concept of
reciprocal basis; since the term ``dual basis'' has an entirely different
meaning, we prefer ``reciprocal frame'' to avoid confusion.)

The above reconstruction formula is equivalent to the resolutions of unity in
terms of the pair ${\cal H}_M, \,{\cal H}^M$ of reciprocal frames:

\be
 \int_Md\mu(m)\, \,|\,h^m\,\rangle\langle\, h_m\,|=I=%
\int_Md\mu(m)\, \,|\,h_m\,\rangle\langle\, h^m\,|.
\eno24)
 \ee

\skp
\noindent {\bf Corollary 1.3. }  {\sl 
Under the assumptions of proposition 1.2,  the above resolutions of unity
converge in the strong operator topology of ${\cal H}$.
}
\skp
\noindent The proof is similar to that of proposition 1.2  and will not be
given. The strong convergence of the resolutions of unity is  important, since
it means that the reconstruction formula is valid {\sl within \/} ${\cal H}$
rather than just weakly.  Application  to $f=h_k$ for a
fixed $k\in M$ gives

\be
 h_k=\int_Md\mu(m)\, h_mK(m,k),
\eno25)
 \ee
which shows that the frame vectors $h_m$ are in general not linearly
independent.  The consistency condition can be understood as requiring the
proposed function $g(m)$ to respect the linear dependence of the frame
vectors.  In the special case when the frame vectors {\sl are\/} linearly
independent, the frames ${\cal H}_M$ and ${\cal H}^M$ both reduce to
{bases of ${\cal H}$.  If ${\cal H}$ is separable (which we assume it is), it
follows that $M$ must be countable, and without loss in generality we may
assume that $d\mu $ is the counting measure on $M$ (re--normalize the
$h_m$'s if necessary).  Then the above relation becomes

\be
h_k=\sum_{m\in M}h_mK(m,k), 
\eno26)
 \ee
and linear independence requires that $K$ be the Kro\-necker $\delta$: 
$K(m,k)\break =\delta _m^k$.  Thus when the $h_m$'s are linearly
independent, ${\cal H}_M$ and ${\cal H}^M$ reduce to a pair of {\sl
reciprocal bases\/} for ${\cal H}$.  The resolutions of unity become

\be
 \sum_{m\in M}|\,h_m\,\rangle\langle\, h^m\,|=I=%
\sum_{m\in M}|\,h^m\,\rangle\langle\, h_m\,|,
\eno27)
 \ee
and we have the relation

\be
 h_k=\sum_{m\in M}h^m\langle\, h_m\,|\,h_k\,\rangle\equiv%
\sum_{m\in M}h^mg_{mk}
\eno28)
 \ee
where

\be
 g_{mk}\equiv\langle\, h_m\,|\,h_k\,\rangle=\langle\, h^m\,|\,Gh_k\,\rangle
\eno29)
 \ee
is an infinite-dimensional version of the {\sl metric tensor\/}, which
mediates between covariant and contravariant vectors. (The operator $G$
plays the role of a {\sl metric operator.})\/  In this case, $\Re_T
=L^2(d\mu)\equiv \ell^2(M)$ and the consistency condition reduces to an
identity.  The reconstruction formula becomes the usual expression for $f$
as a linear combination of the (reciprocal) basis vectors. If we further
specialize to the case of a {\sl tight\/} frame, then $G=AI$ implies that

\be
 \langle\, h_m\,|\,h_k\,\rangle=A\,\delta _m^k\qquad\hbox{and}\qquad 
 \langle\, h^m\,|\,h^k\,\rangle=A^{-1} \delta _k^m,
\eno30)
 \ee 
so ${\cal H}_M$ and ${\cal H}^M$ become {\sl orthogonal\/} bases. 
Requiring $A=B=1$ means that  ${\cal H}_M={\cal H}^M$ reduce to  a
single orthonormal basis.

Returning to the general case, we may summarize our findings as follows:
${\cal H}_M,\ {\cal H}^M$, $K(m,m')$ and $g(m,m')\equiv \langle\,
h^m\,|\,Gh_{m'}\,\rangle$ are generalizations
of the concepts of basis, reciprocal basis, Kronecker delta and 
metric tensor to the infinite--dimen\-sion\-al case 
where, in addition, the requirement of linear
independence is dropped.  The point is that the all-important reconstruction
formula, which allows us to express any vector as a linear combination of
the frame vectors,
 survives under the additional (and obviously necessary) restriction
that the consistency condition be obeyed.   The  useful concepts of orthogonal and orthonormal
bases generalize to tight frames and frames with $A=B=1$, respectively.  We
will call frames with $A=B=1$ {\sl normal}. Thus normal frames are nothing
but resolutions of unity.

Returning to the general situation, we must still supply a way of computing
$G ^{-1} $, on which the entire construction above depends.  In some of the
examples to follow, $G$ is actually a multiplication operator, so $G^{-1} $ is
easy to compute.  If no such easy way exist, the following procedure may be
used.  From $AI\le G\le BI$ it follows that

\be
- {1\over 2}(B-A)I\le G-{1\over 2}(B+A)I\le {1\over 2}(B-A)I.
\eno31)
 \ee
Hence letting

\be
 \delta={B-A\over {B+A}}\qquad \hbox{and} \qquad  c={2\over{B+A} },
\eno32)
 \ee
we have

\be
 -\delta I\le I-cG\le \delta I.
\eno33)
 \ee
Since $0\le \delta <1$ and $c>0$, we can expand

\be
 G ^{-1} =c\bigl[I-(I-cG)\bigr] ^{-1} =c\sum_{k=0}^\infty(I-cG)^k
\eno34)
 \ee
and the series converges uniformly since

\be
 \|I-cG\|\le \delta <1.
\eno35)
 \ee
The smaller $\delta $, the faster the convergence.  For a tight frame, $\delta
=0$ and $cG=I$, so the series collapses to a single term $G ^{-1} =c$.  Then
the consistency condition becomes

\be
 g(m)=c \int_Md\mu(m')\, \langle\, h_m\,|\,h_{m'}\,\rangle g(m'),
\eno36)
 \ee
and the reconstruction formula simplifies to

\be
 f=c \int_Md\mu(m)\, h_m\,g(m).
\eno37)
 \ee

If $0<\delta \ll1$, the frame is called {\sl snug\/} (Daubechies [1988a]) and
the above formulae merely represent the first terms in the expansions. 
However, it is found in practice that under certain conditions, this first term
gives a  good approximation to the reconstruction for snug frames.  It
appears to be an advantage to have much linear dependence among the
frame vectors (precisely that which is impossible when dealing with bases!),
so the transformed function $\ftil (m)$ is ``oversampled.''  For such frames,
the oversampling seems to compensate for the truncation of the series.  A
good measure of the amount of linear dependence among the $h_m$'s is the
size of the  orthogonal complement $\Re_T^\perp$ of the range of $T$, which
is the null space ${\cal N}(T^*)$ of $T^*$ since

\ba{
g\in\Re_T^\perp\a\iff\langle\,g\,|\,Tf\,\rangle=0\qquad\forall f%
\in{\cal H}\cr
 \a\iff\langle\, T^*g\,|\,f\,\rangle=0\qquad \forall f\in {\cal H}\cr
\a\iff T^*g=0.
 \cr}
 \eno38)
 \ea

Suppose we are given some function $g(m)\in L^2(d\mu)$ and apply the
reconstruction formula to it blindly, without worrying whether the
consistency condition is satisfied.  That is, consider the vector 

\be
 f\equiv G ^{-1} T^*g
\eno39)
 \ee
in ${\cal H}$.  Since $g$ can be uniquely written as an orthogonal sum
$g=g_1+g_2$ where $g_1\in \Re_T$ and $g_2\in \Re_T^\perp ={\cal N}(T^*)$,
we find that

\be
 f=G ^{-1} T^*g_1
\eno40)
 \ee
where $g_1$ does satisfy the consistency condition.  This means that 
applying the reconstruction formula to an {\sl arbitrary\/} $g\in L^2(d\mu)$
results in a {\sl least--squares approximation\/} $f$ to a reconstruction, in
the sense that  the ``error'' $g_2\equiv g-g_1=g-Tf$ has a minimal norm in
$L^2(d\mu )$.

For example, suppose we only have information about $\ftil(m)$ for $m$ in
some subset  $\Gamma$ of $M$.   Let

\ba{
g(m)=\cases{\ftil(m)&if $m\in\Gamma$\cr
        0&if $m\notin\Gamma$.}
 \cr}
 \eno41)
 \ea
Then $g$ does not belong to $\Re_T$, in general, and $f_1\equiv G ^{-1}T^*g$
is the least--squares approximation to the (unknown) vector $f$ in view of
our ignorance.  A similar argument applies if $M$ is discrete and $f$ must
be reconstructed from $Tf$ by numerical methods.  Then we must confine
ourselves to a {\sl finite \/} subset $\Gamma$ of $M$.  The above
procedure then gives a least--squares approximation to $f$ by {\sl
truncating\/} the reconstruction formula to a finite sum over $\Gamma$.
\skp
A final note:  The usual argument in favor of using {\sl bases\/} rather than
overcomplete sets of vectors is that one desires a {\sl unique \/}
representation of functions as linear combinations of basis elements.  When
a frame is not a basis, i.e. when the frame vectors are linearly dependent,
this uniqueness is indeed lost since we may add an arbitrary function
$e(m)\in \Re_T^\perp$ to $\ftil(m)$ in the reconstruction formula
without changing $f$ ($\Re_T^\perp\ne \{0\}$ since the frame vectors are
dependent).  However, there is still a unique {\sl admissible\/ } coefficient
function, i.e. one satisfying the consistency condition.  Moreover, as we shall
see, it usually happens  in practice that the set $M$,  in addition to being  a
measure space, has some further structure, and the reproducing kernel
$K(m,m')$ preserves this structure.  For example, $M$ could be a
topological space and $K$ be continuous on $M\times M$, or $M$ could
be a differentiable manifold and $K$ be differentiable, or (as will happen in
our treatment of relativistic quantum mechanics) $M$ could be a complex
manifold and $K$ be holomorphic.  Furthermore, $K$ could exhibit a certain
 boundary-- or asymptotic behavior.  In such cases, these properties
are inherited by all the functions $\ftil(m)$ in $\Re_T$, and then of all
possible coefficient functions for a given $f\in \cal H$, there is only {\sl
one\/} which exhibits the appropriate behavior.  In this sense, uniqueness is
restored.  We will refer to frames with such additional properties as {\sl
continuous, differentiable, holomorphic, \/} etc.

In addition to properties such as differentiability or holomorphy, the
kernels $K$ we will encounter will usually have certain {\sl invariance\/}
properties with respect to some group of transformations acting on $M$. 
This, too, will induce a corresponding structure on the function space
$\Re_T$.

\secskp

 \noindent {\bf 1.4.  Reproducing--Kernel Hilbert Spaces}
\skp

\def\rightheadline{\tenrm\hfil{\sl 1.4. Reproducing--Kernel
Hilbert Spaces}\hfil\folio} 

\noindent The function $K(m,m')$ of section 1.3 is an example of a very
general structure called a {\sl reproducing kernel}, which we now review
briefly since it, too, will play an important role in the chapters to come. 
The reader interested in learning more about this fascinating subject may
consult Aronszajn [1950],  Bergman [1950], Meschkowski [1962] or Hille
[1972]. \skp
Suppose we begin with an  arbitrary set $M$ and a set of functions $g(m)$
on $M$ which  forms a Hilbert space $\cal F$ under some  inner
product $\langle\, \cdot\,|\,\cdot\,\rangle$.  In section 1.3, $M$ happened
to a measure space, $\cal F$ was $\Re_T$ and the inner product was that of
$L^2(d\mu)$.  But it is important to keep in mind that the exact form of the
inner product in $\cal F$ does not need to  be specified, as far as the
general theory of reproducing kernels is concerned.  Suppose we are given
a complex--valued function $K(m,m')$ on $M\times M$ with the following
two properties: \skp
\item{1.}  For every $m\in M$, the function $K_m(m')\equiv
K(m',m)$ belongs to $\cal F$.  
\item{2.} For every $m\in M$ and every $g\in
\cal F$, we have  
\be
g(m)=\langle\, K_m\,|\,g\,\rangle.
\eno1)
 \ee

\noindent Then ${\cal F}$  is called a {\sl reproducing--kernel Hilbert
space\/} and \break $K(m,m')$ is called its reproducing kernel.  Some
properties of $K$ follow immediately from the definition.  For example, 
since $K_m\in \cal F$, property (2) implies that

\be
 K(m',m)=K_m(m')=\langle\, K_{m'}\,|\,K_m\,\rangle.
\eno2)
 \ee 
Thus $K$ must satisfy

\ba{
\a (a)\quad  \br{K(m,m')}=K(m',m)\cr
\a (b)\quad  K(m,m)=\|K_m\|^2\ge 0\quad \forall m\in M\cr
\a (c)\quad |K(m,m')|\le \|K_m\|\,\|K_{m'}\| .
 \cr}
 \eno3)
 \ea

One of the most important  and useful aspects of reproducing--kernel Hilbert
spaces is
 the fact that the  kernel function itself virtually generates the whole
structure.  For example, the function $L(m)\equiv \|K_m\|$ dominates every
$g(m)$ in $\cal F$ in the sense that 

\be
 |g(m)|=\,|\,\langle\, K_m\,|\,g\,\rangle\,|\,\le \|K_m\|\,\|g\|=\|g\|L(m)
\eno4)
 \ee
by the Schwarz inequality. In particular,  all the functions in ${\cal F}$ inherit
the boundednes and growth  properties of $K$. If $K$ has singularities, then
 {\sl some\/}  functions in ${\cal F}$ have similar singularities.

From the above it follows that for fixed  $m_0\in M$,

\be
 \sup_{\|g\|= 1}|g(m_0)|=L(m_0),
\eno5)
 \ee
the supremum being attained by $g_{m_0}(m)=K_{m_0}(m)/L(m_0)$, and
this is, up to an overall phase factor, the only function which attains the
supremum.  This fact has an important interpretation when $M$ is the
classical phase space of some system, $\cal F$ represents the Hilbert space
of  the corresponding quantum  mechanical system and $|g(m)|^2$ is the
probability density in the quantum mechanical state $g$ of finding the
system in the classical state $m$.  The above inequality then shows that
the state which maximizes the probability of being at $m_0$ is uniquely
determined (up to a phase factor) as $g_{m_0}$.  In other words, $g_{m_0}$
is a wave packet which is in some sense optimally localized at $m_0$ in
phase space.

Another example of how the kernel function $K$ embodies the properties of
the entire Hilbert space $\cal F$ is the following:  If the basic set
$M$ has some additional structure, such as being a measure space
(as it was in the setting of frame theory) or a topological space or a
$C^k, C^\infty$, real--analytic or complex manifold, and if $K(m,m')$, as a
function of $m$, preserves this structure (that is, $K(m,m')$ is measurable,
continuous, $C^k, C^\infty$, real--analytic or holomorphic in $m$,
respectively), then every function $g(m)$ in $\cal F$ has the same property.
\skp
The question arises:  If we are given a Hilbert space ${\cal F}$ whose
elements are all functions on $M$, how do we know whether this space
posesses a reproducing kernel?  Clearly a {\sl necessary\/} condition for $K$
to exist is that

\be
\,|\, f(m)\,|\,=\,|\, \langle\, K_m\,|\,f\,\rangle\,|\,\le \,\|K_m\|\,\|f\|
\eno6)
 \ee
for all $f\in \cal F$ and all $m\in M$.  But this means that for every {\sl 
fixed\/} $m$, the map $E_m{:}\ {\cal F}\to \cx$ defined by

\be
 E_m(f)=f(m)
\eno7)
 \ee
is a bounded linear functional on ${\cal F}$.  $E_m$ is called the {\sl
evaluation map\/} at $m$.  By the Riesz representation theorem, every
bounded linear functional $E$ on ${\cal F}$ must have the form
$E(f)=\langle\,e\,|\,f\,\rangle$ for a unique vector $e\in \cal F$. Hence
there exists a unique $e_m\in \cal F$ such that    

\be
 f(m)=\langle\, e_m\,|\,f\,\rangle
\eno8)
 \ee
for all $f\in \cal F$.  Since also $f(m)=\langle\, K_m\,|\,f\,\rangle$, we
conclude that $e_m=K_m$. Thus it follows that given {\sl only \/} ${\cal F}$
such that all the evaluation maps $E_m$ are bounded (not necessarily
uniformly in $m$), we can {\sl construct\/}  a reproducing kernel by

\be
 K(m,m')=\langle\, e_m\,|\,e_{m'}\,\rangle.
\eno9)
 \ee
That is, {\sl ${\cal F}$ posesses a reproducing kernel if and only if all the
evaluation maps  are bounded.}
\skp
In all of the applications we will encounter, the set $M$ will have a structure
beyond those mentioned so far: it will be a {\sl Lie group\/}  $G$ or a
{\sl homogeneous space\/}  of $G$.  That is, each $g\in G$ {\sl acts\/}  on
$M$ as an invertible transformation preserving whatever other structure
$M$ may have such as continuity, differentiability, etc, and these
tranformations form the group $G$ under composition.  Let us denote the
action of $g$ on $M$ by $m\mapsto mg$ (i.e., $G$ acts from the {\sl
right\/}).  Then the operator

\ba{
T_g{:}\,{\cal F}\a\to {\cal F}\cr
(T_gf)(m)\a=f(mg)
 \cr}
 \eno10)
 \ea
gives a {\sl representation \/} of $G$ on ${\cal F}$, since
$T_gT_{g'}=T_{gg'}$.  From $f(m)=\langle\, K_m\,|\,f\,\rangle$ it
now follows that 

\be
 K_{mg}=T_g^*\,K_m,
\eno11)
 \ee
hence

\be
 K(m'g,mg)=\langle\, K_{m'g}\,|\,K_{mg}\,\rangle=\langle\, K_{m'}\,|\,
T_g\,T_g^*\,K_m\,\rangle.
\eno12)
 \ee
Therefore {\sl  the reproducing kernel $K(m',m)$ is invariant  under the
action of $G$ if and only if all the operators $T_g$ are unitary, i.e. the
representation  $g\mapsto T_g$ is unitary. }
\skp
The group $G$ usually appears in applications as a natural set of operations
such as translations in space and time (evolution), changes of reference
frame or coordinate system, dilations and frequency shifts (especially
useful in signal analysis), etc.  Invariance under $G$ then means that the
transformed objects (wave functions, signals) form a description 
of the system equivalent to the original one, i.e. that the transformation
changes nothing of physical significance.
\skp
Finally, let us note that the concept of generalized frame, as defined in the
previous section, is a special case of a reproducing--kernel Hilbert space. 
For given such a frame $\{h_m\}$, the function $K(m',m)=\langle\,
h_{m'}\,|\,G ^{-1}h_m\,\rangle$ is a reproducing kernel for the space
$\Re_T$ since  (a):

\be
 K_m(m')\equiv K(m',m)=\left( T(G ^{-1} h_m)\, \right)(m')
\eno13)
 \ee
shows that  $K_m=T(G ^{-1} h_m)$ belongs to $\Re_T$, and  (b):

\be
 \ftil(m)=\int_M d\mu (m')\,K(m,m')\ftil(m')=
\langle\, K_m\,|\,\ftil\,\rangle_{\Re_T}.
\eno14)
 \ee
\secskp

 \noindent {\bf 1.5.  Windowed Fourier Transforms  }

\def\rightheadline{\tenrm\hfil{\sl 1.5. Windowed Fourier Transforms}
\hfil\folio} 
 \skp
\noindent The coherent states of section 2 form a (holomorphic) frame.  I
now want to give some other examples of frames, in order to develop a
better feeling for this  concept.  The frames to be
constructed in this section turn out to be closely related to the coherent
states, but have a distinct ``signal processing'' flavor which will lend some
further depth to our understanding of phase-space localization.  

For simplicity, we begin with the study of  functions $f(t)$ of a
single real variable which, for motivational purposes, will be thought of as
``time.''  Although in the model to be built below $f$ is real-valued, we
allow it to be complex-valued since our eventual applications will be
quantum--mechanical.  The same goes for the window function $h$. 
The extension to functions of several variables, such as wave functions or
physical fields in spacetime, is straightforward and will be indicated later. 
We think of $f(t)$ as a ``time--signal,'' such as the voltage going into a
speaker or the pressure on an eardrum.  We are interested in the frequency
content of this signal.  The standard thing to do is to find its Fourier
coefficients  (if $f$ is periodic) or its Fourier transform (if it is not periodic). 
But if $f$ represents, say, a symphony, this approach is completely
inappropriate.  We are not interested in the {\sl total} amplitude 

\be
 \hat f(\nu )=\inr dt\,e^{2\pi i\nu t}f(t)
\eno1)
 \ee
in $f$ of each frequency $\nu $.  For one thing, both we and the musicians
would be long gone before we got to enjoy the music.  Moreover, the musical
content of the signal, though coded into $\hat f$, would be as inaccessible to
us as it is in $f(t)$ itself.  Rather, we want to analyze the frequency
content of $f$ in {\sl real\/} time.  At each instant $t=s$ we hear a
``spectrum'' $\ftil(\nu ,s)$.  To accomplish this, let us make a simple (though
not very realistic) linear model of our auditory system.  We will speak of the
``ear,'' though actually the frequency analysis appears to be performed
partly by the nervous system as well (Roederer [1975]).

The ear's output at time $s$ depends on the input $f(t)$ for $t$ in some
interval $s-\tau \le t\le s$, where $\tau $ is a lag--time characteristic of the
ear.  In analyzing $f(t)$ in this interval, the ear may give different weights
to different parts of $f(t)$.  Thus the signal to be analyzed for output at time
$s$ may be modeled as 

\be
 f_s(t)=\overline{h(t-s)}\,f(t)
\eno2)
 \ee
where $\overline{h(t-s)}$ is the weight assigned to $f(t)$.  (As noted above,
we are allowing $h$ to be complex--valued for future applications, though
here it should be real--valued;  the bar means complex conjugation.)
The function $h(t)$ is characteristic of the ear, with support in the interval
$-\tau \le t\le 0$.  Such functions are known in communication theory as
{\sl windows\/}.

Having ``localized'' the signal around time $s$, we now analyze its frequency
content by taking the Fourier transform:

\be
 \ftil(\nu ,s)=\hat f_s(\nu )=\inr dt\,e^{2\pi i\nu t}\,\overline{h(t-s)}\,f(t).
\eno3)
 \ee
This function, representing our ``dynamical spectrum,''  is called a {\sl
windowed Fourier transform\/} of $f$  (Daubechies [1988a]).  If the window
is flat ($h(t)\equiv 1$),  $\ftil(\nu ,s)$ reduces to the ordinary Fourier
transform.  On the other hand, if $f(t)$ is a ``unit impulse'' at $t=0$, i.e.
$f(t)=\delta (t)$, then  $\ftil(\nu ,s)=\overline{h(-s)}$.  Hence
$\overline{h(-s)}$ is the ``impulse response'' (Papoulis [1962]) of the ear. 
Note that

\be
 \ftil(0,s)=\inr  dt\,\overline{h(t-s)}\,f(t)
\eno4)
 \ee
is nothing but the convolution of $f$ with the impulse response.  Thus the
windowed Fourier transform is a marriage between the Fourier transform
and convolution. Letting

\be
 h_{\nu ,s}(t)=e^{-2\pi i\nu t}\,h(t-s),
\eno5)
 \ee
we have

\be
 \ftil(\nu ,s)=\langle\, h_{\nu ,s}\,|\,f\,\rangle,
\eno6)
 \ee
the inner product being in $L^2(\rl)$.  As our notation implies, we want to
make a frame indexed by the set

\be
 M=\{(\nu ,s)\,|\,\nu ,s\in\rl\}\approx\rl^2,
\eno7)
 \ee
that is, the {\sl time-frequency plane}. $M$ corresponds to the ``phase
space'' in quantum mechanics  in the sense that if $t$ were the position
coordinate, then $\nu$ would be the
wavenumber and $2\pi \hbar\nu$ would be the momentum. Since we
expect all times and all frequencies to be equally important, let us guess
that an appropriate measure on $M$ is the Lebesgue measure, $d\mu(\nu ,s)
=ds\,d\nu$. Now 

\be
 \ftil(\nu ,s)=\bigl(\bar h_sf\bigr)\hat{\,}(\nu)
\eno8)
 \ee
where $\bar h_s(t)=\overline{h(t-s)}$ is the translated window and
$\hat{\,}$ denotes the Fourier transform.  Thus for ``nice'' signals (say, in
the space  of \hfill\break
 Schwartz test functions), Plancherel's theorem gives

\ba{
\inrr ds\,d\nu\,\,|\,\ftil(\nu,s)\,|\,^2\a=\int ds\,\int d\nu\,\,|\,(\bar h_sf)%
                  \hat{\,}(\nu)\,|\,^2\cr
\a=\int ds\,\int dt\,\,|\,\overline{ h_s(t)}f(t)\,|\,^2\cr
\a=\int dt\,\,|\,f(t)\,|\,^2\int ds\,\,|\,h(t-s)\,|\,^2\cr
\a=\|h\|^2\,\|f\|^2,
 \cr}
 \eno9)
 \ea
where both norms are those of $L^2(\rl)$. This shows that the family of
vectors $h_{\nu,s} $ is indeed a tight frame, with  frame constants
\break $A=B= \|h\|^2$.   Now $h_{\nu,s} =\exp(-2\pi i\nu t)h_s(t)$ is
just a translation of $h_s$ in {\sl frequency}, that is, 

\be
 \hat h_{\nu,s} (\nu ')=\hat h_s(\nu '-\nu).
\eno10)
 \ee
Thus if the Fourier transform \ $\hat h(\nu)$ of the basic window
is concentrated near $\nu=0$ (which, among other things, means that $h(t) $
must be fairly smooth, since any discontinuities or sharp edges introduce
high-frequency components), then that of $h_{\nu,s}$ is concentrated near 
$\nu'=\nu$.  In other words, $h_{\nu,s} $ is actually a window in {\sl phase
space} or time--frequency.  The fact that these windows form a tight frame
means that signals may be equally well represented in time--frequency as
 in time.  The resolution of unity is 

\be
 \|h\|^{-2}\inrr ds\,d\nu\,\,|\,h_{\nu,s} \,\rangle\langle\, h_{\nu,s} \,|\,=I,
\eno11)
 \ee
and if the consistency condition is satisfied the reconstruction formula gives

\be
 f(t)=\|h\|^{-2}\inrr ds\,d\nu\,e^{-2\pi i\nu t}\,h(t-s)\,\ftil(\nu ,s).
\eno12)
 \ee
As in section 1.2, we denote by ${\cal H}_M$ the frame formed by the
$h_{\nu ,s} $'s.
\skp
Incidentally, the windowed Fourier transform is quite symmetrical with
respect to the interchange of time and frequency.  It can be re--written as

\ba{
\ftil(\nu ,s)\a=\left( \bar h_s\,f \right)\hat{\ }(\nu )\cr
\a=\left( \hat{\bar h}_s*\hat f \right)(\nu )\cr
\a=e^{-2\pi i\nu s}\int d\nu' \,e^{2\pi i\nu' s}\,
\overline{\hat h(\nu '-\nu )}\,\hat f(\nu '),
 \cr}
 \eno13)
 \ea
where $\hat h$ now plays the role of a window in frequency used to
localize $\hat f$.  As will be seen in the next chapter, the reason for this
symmetry is that  windowed Fourier transforms, like the canonical
coherent states, are closely related to the Weyl--Heisenberg
group, which treats time and frequency in a symmetrical fashion. (This is
rooted in symplectic geometry.)

\skp
The $h_{\nu,s} $'s are highly redundant. We want to find  discrete subsets of
them which still form a frame, that is we want  {\sl discrete subframes}. The
following construction is taken from Kaiser [1978c, 1984a]. Let
$T>0$ be a fixed time interval and suppose we ``sample'' the output signal
$\ftil(\nu ,s)$ only at times $s=nT$ where $n$ is an integer.  To discretize the
frequency as well, note that the localized signal $f_{nT}(t)=\bar h_{nT}(t)f(t)$
has compact support in the interval  $nT~-\tau\le~t\le~nT$, hence we can
expand it in a Fourier series

\be
 f_{nT}(t)=\sum_m e^{-2\pi imFt}\,c_{mn}
\eno14)
 \ee
where $F=1/\tau $ and

\ba{
c_{mn}\a=F\int_{nT-\tau }^{nT}dt\,e^{2\pi imFt} \,f_{nT}(t)\cr
\a=F\int_{nT-\tau }^{nT}dt\,e^{2\pi imFt}\,\br{h(t-nT)}\,f(t)\cr
\a=F\,\ftil(mF,nT).
 \cr}
 \eno15)
 \ea
The only problem with this representation of $f_{nT}$ is that it only holds in
the interval $nT-\tau \le t\le nT$, since $f_{nT}$ vanishes outside this
interval while the Fourier series is periodic. To force  equality at all times,
we multiply both sides by $\tau h_{nT}(t)$:

\ba{
\tau h_{nT}(t)\,f_{nT}(t)\a=\tau \,|h_{nT}(t)|^2\,f(t)\cr
\a=\sum_me^{-2\pi imFt}\,h_{nT}(t)\,\ftil(mF,nT)\cr
\a=\sum_mh_{mF,nT}(t)\,\langle\, h_{,mF,nT}\,|\,f\,\rangle.
 \cr}
 \eno16)
 \ea
Attempting to recover the entire signal rather than just pieces of it, we
now sum both sides with respect to $n$:

\be
 \tau \Bigl(\sum_n|h(t-nT)|^2\Bigr)f(t)=\sum_{n.m}h_{mF,nT}(t)%
\langle\, h_{nT,mF}\,|\,f\,\rangle.
\eno17)
 \ee
Recovery of $f(t)$  is possible provided that the sum on the left converges to
a function 

\be
 g(t)\equiv\tau \sum_n|h(t-nT)|^2
\eno18)
 \ee
which is bounded above and below by positive constants:

\be
 0<A\le g(t)\le B.
\eno19)
 \ee
In that case, we have 

\be
 \sum_{n,m}\,|\,\langle\, h_{mF,nT}\,|\,f\,\rangle\,|\,^2=\int dt\,g(t)|f(t)|^2
\eno20)
 \ee
and hence the subset 

\be
 {\cal H}_M^{T,F}\equiv\{h_{mF,nT}\,|\,m,n\in \zz\}
\eno21)
 \ee
forms a discrete subframe of ${\cal H}_M$ with frame constants $A$ and
$B$.  This subframe is in general not tight, and the operator $G$ is just
multiplication by $g(t)$, so finding $G ^{-1} $ presents no problem in this
case.  The reconstruction formula is 

\be
 f(t)=g(t) ^{-1} \sum_{n,m\in \zs}h_{mF,nT}(t)\ftil(mF,nT).
\eno22)
 \ee 
We are therefore able to recover the signal by ``sampling'' it in phase space
at time  intervals $\Delta t=T$ and frequency intervals $\Delta \nu =F$. 
What about the uncertainty principle? It is hiding in the condition that the
discrete subset $h_{mFnT}$ forms a frame!  For we cannot satisfy $g(t)\ge
A>0$ unless the supports of $h(t)$ and $h(t-T)$ overlap, which implies that 
$T\le \tau $, or $\Delta t\Delta \nu =T/\tau \le1$.  
In quantum mechanics, the radian frequency is related to the energy by
$E=2\pi \hbar \nu $,  so the above condition is

\be
\Delta t\,\Delta E\le 2\pi \hbar.
\eno23)
 \ee
This looks like the uncertainty principle going ``the wrong way.''  The
intuitive explanation for this has been discussed at the end of section 1.3. 

Notice that the closer we choose $T$ to $\tau $, the more
difficult it is for the window function to be smooth.  In the limiting case
$T=\tau $, $h(t)$ must be discontinuous if the above frame condition is to be
obeyed.  As noted earlier, this means that its Fourier transform $\hat h(\nu
)$  can no longer be concentrated near $\nu =0$, so the frequency resolution
of the samples  suffers.  In concrete terms, this means that whereas for
``nice'' windows $h(t)$ we may hope to get a good approximation to the
reconstruction formula by truncating the double sum after an appropriate 
finite number of terms, this can no longer be expected when $T\to\tau $. 
In other words, it pays to oversample!  ``Appropriate'' here means that
we cover most of the area in the time--frequency plane where $\ftil$ lives. 
Clearly, if the sampling is done only for $|n|\le N$, we cannot expect to
recover $f(t)$ outside the interval \break\hfill
 $-NT-\tau \le t\le NT$.  If
$\hat h(\nu )$ is nicely peaked around $\nu =0$, say with a spread of
$\Delta \nu $, and the
 signal $f(t)$ is (approximately) ``band--limited,''  so that $\hat f(\nu
)\approx 0$ for $|\nu |\ge W$, then we can expect to get a good
approximation to $f(t)$ by truncating the sum with $m\le M$,  where $M$ is
chosen to satisfy $MF>W+\Delta \nu $.  On the other hand, we may actually
not be interested in recovering the exact original signal $f(t)$.  If we are only
interested in
 a particular time  interval and a particular
frequency  interval (say, to eliminate some high--frequency noise), then
the appropriate truncation would include only an area in the
time--frequency plane which is slightly larger than our area of interest.
 If we sample a given signal $f$ only in a finite subset $\Gamma$ of our
lattice (as we are bound to do in practice) and then apply the reconstruction
formula, truncating the sum by restricting it to $\Gamma$,  then the result
is a least--squares approximation $f_1$ to the original signal.

\skp
On a philosophical note, suppose we are given an arbitrary signal without
any idea of the type of information it carries.  The choice of a window $h(t)$
then defines a {\sl scale\/} in time, given by $\tau $, and it is this scale that
distinguishes what will be perceived as frequency and what as time.
Thus, variations of $f(t)$ in time intervals much
smaller than $\tau $ are reflected in the $\nu $--behavior of $\ftil$, while
variations at scales much larger than $\tau $ survive in the
$s$--dependence of $\ftil$.  What an elephant perceives as a tone may
appear as  a rhythm to a mouse.  
\skp

Finally, I wish to compare the above reconstruction with the
Nyquist/Shannon sampling theorem (Papoulis [1962]), which gives a
reconstruction for band-limited signals ( $\hat f(\nu )\approx 0$ for $|\nu
|\ge W$) in terms of samples of $f(t)$ taken at times $t=n/2W$ for
integer $n$.  Note, first of all, that in the time--frequency reconstruction
formula above, it was not necessary to assume that $f(t)$ is band-limited.  In
fact, the formula applies to all square--integrable signals.  But suppose that 
$f(t)$ {\sl is\/} band--limited as above, and  choose a window $h(t)$ such
that $\hat h(\nu )$ is concentrated around an interval of width $\Delta \nu $
about the origin, with $\Delta \nu<W$.  Then we expect $\ftil(\nu ,s)\approx
0$ for $|\nu |\ge W+\Delta \nu $ .  In order to reduce the double sum in our 
reconstruction formula to a single sum over $n$ as in the Nyquist theorem,
choose  $F=W+\Delta \nu $.  Then $\ftil(mF,s)\approx 0$
whenever $m\ne 0$.  To apply the reconstruction formula, we must still
choose a time interval $T<\tau $. By the uncertainty principle, $\tau \Delta
\nu >1/2$. The above condition on $\Delta \nu $ therefore implies that
$\tau >1/2W$.  It would thus seem that we could get away with a
slightly larger sampling interval $T$ than the   Nyquist interval
$T_N=1/2W$.  Our reconstruction formula reduces to

 \be
 f(t)\approx g(t)^{-1} \sum_nh(t-nT)\,\ftil(0,nT).
\eno24)
 \ee
The smaller the ratio $\Delta \nu /W$, the better this approximation is
likely to be.  But a small  $\Delta \nu $ means a large $\tau $, hence the
samples $\ftil(0,nT)$ are smeared over a large time  interval.
\secskp

\noindent {\bf  1.6. Wavelet Transforms }

\def\rightheadline{\tenrm\hfil{\sl 1.6. Wavelet Transforms}\hfil\folio} 
 \skp

\noindent The frame vectors for windowed Fourier transforms were the
wave packets  \be
 h_{\nu,s }(t)=e^{-\phs}h(t-s).
\eno1)
 \ee
The basic window function $h(t)$ was assumed to vanish outside of the
interval $-\tau <t<0$ and to be reasonably smooth with no steep slopes, so
that its Fourier transform $\hhat(\nu )$ was also centered in a small interval
about the origin.  Of course, since $h(t)$ has compact support, $\hhat(\nu )$
is the restriction to $\rl$ of an entire function and hence cannot vanish on
any interval, much less be of compact support.  The above statement simply
means that $\hhat(\nu )$ decays rapidly outside of a small interval
containing the origin.  At any rate, the factor $\exp(-\phs)$ amounted to a
translation of the window in frequency, so that $ h_{\nu,s }$ was a
``window'' in the time--frequency plane centered about $(\nu,s)$.  Hence the
frequency components of $f(t)$ were picked out by means of rigidly
translating the basic window in both time and frequency.  (It is for this
reason that the windowed Fourier transform is associated to the
Weyl--Heisenberg group, which is exactly the group of all translations in
phase space amended with the multiplication by phase factors 
necessary to close the Lie algebra, as explained in chapter 3.)  Consequently,
$h_{\nu,s } $ has the same width $\tau $ for all frequencies, and the
number of wavelengths admitted for analysis is $\nu \tau $.  For low
frequencies with $\nu \tau \ll1$ this is inadequate since we cannot gain any
meaningful frequency information by looking at a small fraction of a
wavelength.  For high frequencies with $\nu \tau \gg1$, too many
wavelengths are admitted.  For such waves, a time-interval of duration $\tau
$ seems infinite, thus negating the sense of ``locality'' which the windowed
Fourier transform was designed to achieve in the first place.  This deficiency
is remedied by the {\sl wavelet transform.}   The window $h(t)$, 
called the {\sl basic wavelet,\/} is now {\sl scaled\/} to accomodate waves of
different frequencies. That is, for $a\ne 0$ let

\be
 h_{a,s}(t)=|a|^{-1/2}h\left({t-s}\over a\right ).
\eno2)
 \ee
The factor $|a|^{-1/2}$ is included so that

\be
 \|h_{a,s}\|^2\equiv \inr dt\,\,|\,h_{a,s}(t)\,|\,^2=\|h\|^2.
\eno3)
 \ee
The necessity of using negative as well as positive values of $a$ will become
clear as we go along. It will also turn out that $h$ will need to satisfy a
technical condition.  Again, we think of both $h$ and $f$ as real but allow
them to be complex.  The wavelet transform is now defined by

\be
 \ftil(a,s)=\langle\, \has\,|\,f\,\rangle=\inr dt\,%
|a|^{-1/2}\br{h\left({t-s}\over a\right )}f(t).
\eno4)
 \ee

Before proceeding any further, let us see how the wavelet transform
localizes signals in the time--frequency plane.  The localization in time is
clear: If we assume that $h(t)$ is concentrated near $t=0$ (though it will no
longer be convenient to assume that $h$ has compact support), then
$\ftil(a,s)$ is a weighted average of $f(t)$ around $t=s$ (though the weight
function need not be positive, and in general may even be complex).  To
analyze the frequency localization, we again want  to express $\ftil$ in
terms of the Fourier transforms of $h$ and $f$.  This is possible because,
like the windowed Fourier transform, the wavelet transform involves rigid
time--translations of the window, resulting in a convolution--like
expression.  The ``impulse response''  is now (setting $f(t)=\delta (t)$)

\be
 g_a(s)=|a|^{-1/2}\br{h(-s/a)},
\eno5)
 \ee
and we have

\be
 \ftil(a,s)=\bigl(g_a*f\bigr)(s)=\bigl(\hat g_a\fhat\bigr)\check{\,}(s),
\eno6)
 \ee
with

\be
 \hat g_a(\nu )=\inr dt\,e^\phs \am \br{h(-t/a)}=\ap \br{\hhat(a\nu )}.
\eno7)
 \ee
Later we will see that discrete tight frames can be obtained with certain
choices of $h(t)$ whose Fourier transforms have compact support in a
frequency interval interval $\alpha \le \nu \le \beta $. Such functions (or,
rather, the operations of convolutiong with them) are called {\sl bandpass
filters\/}  in communication theory, since the only frequency components
in $f(t)$ to survive are those in the ``band'' $[\alpha ,\beta ]$. Then the
above expression shows that $\ftil(a,s)$ depends only on the frequency
component of $f(t)$ in the band $\alpha /a\le \nu \le \beta /a$ (if $a>0$) or
$\beta /a\le \nu \le \alpha /a$ (if $a<0$). Thus frequency localization is
achieved by {\sl dilations\/} rather than translations in frequency space, in
contrast to the windowed Fourier transform.  At least from the point of view
of audio signals, this actually seems preferable since it appears to be
frequency ratios, rather than frequency differences, which carry meaning. 
For example, going up an octave is achieved by doubling the frequency.
(However, frequency differences do play a role in connection with beats and
also in certain non--linear phenomena such as difference tones; see
Roederer [1975].)  Let us now try to make a continuous frame out of the
vectors $h_{a,s} $.  This time the index set is 

\be
 M=\{(a,s)\,|\,a\ne 0, s\in\rl\}\approx\rl^*\times\rl
\eno8)
 \ee
where $\rl^*$ denotes the group of non-zero real numbers under
multiplication. $M$ is the {\sl affine group\/} of translations and dilations of
the real line, $t'=at+s$, and this fact will be recognized as being very
important in chapter 3.  But for the present we use a more pedestrian
approach to obtain the central results.  This will make the power and
elegance of the group--theoretic approach to be introduced later stand out
and be appreciated all the more.  At this point we only make the safe
assumption that the measure $d\mu $ on $M$ is invariant under time
translations, i.e. that

\be
 d\mu (a,s)=\rho (a)da\,ds
\eno9)
 \ee
where $\rho (a)$ is an as yet undetermined density on $\rl^*$.  Then, using
Plancherel's theorem,

\ba{
\int_M d\mu (a,s)\,|\,\ftil(a,s)\,|\,^2\a=\int_M\rho (a)\,da\,ds\, 
                     |\,(\hat g_a\fhat)\check{\ }(s)\,|\,^2\cr
\a=\int_{\rl^*}\rho (a) da\,\int_\rl d\nu
          \,|\,\hat g_a(\nu )\,|\,^2\,|\,\fhat(\nu)\,|\,^2\cr
\a=\int_{\rl^*}\rho (a) da\,\int_\rl d\nu\,|a|\,|\,\hhat(a\nu
          )\,|\,^2\,|\,\fhat(\nu) \,|\,^2\cr
 \a=\inr d\nu \,H(\nu )\,|\,\fhat(\nu) \,|\,^2\cr
\a=\langle\, \fhat\,|\,H\fhat\,\rangle_{L^2(\rl)} ,
 \cr}
 \eno10)
 \ea
where

\ba{
H(\nu )\a=\int_{\rl^*} \rho (a)\cdot\,|\,a\,|\,da\,|\,\hhat(a\nu )\,|\,^2\cr
\a=\int_0^\infty a da\,\left[ \rho (a)\,|\,\hhat(a\nu )\,|\,^2+
            \rho (-a)\,|\,\hhat(-a\nu )\,|\,^2\right].
 \cr}
 \eno11)
 \ea
The frame condition is therefore $A\le H(\nu )\le B$ for some positive
constants $A$ and $B$.  To see its implications, we analyze the cases $\nu
>0$ and $\nu <0$ separately.  If $\nu >0$, let $\xi =a\nu $. Then

\be
 H(\nu )=\nu ^{-2}\int_0^\infty\xi d\xi \,\left[
     \rho (\xi /\nu )\,|\,\hhat(\xi)\,|\,^2+
     \rho (-\xi /\nu )\,|\,\hhat(-\xi)\,|\,^2\right].
\eno12)
 \ee
If $\nu <0$, let $\xi =-a\nu $. Then

\be
  H(\nu )=\nu ^{-2}\int_0^\infty\xi d\xi \,\left[
     \rho (-\xi /\nu )\,|\,\hhat(-\xi)\,|\,^2+
     \rho (\xi /\nu )\,|\,\hhat(\xi)\,|\,^2\right],
\eno13)
 \ee
giving the same expression. Therefore the frame condition requires that

\be
 \rho (\xi /\nu )=O\bigl((\xi /\nu )^{-2}\bigr) \quad\hbox{as}\quad 
\xi /\nu \to\pm\infty,
\eno14)
 \ee
unless $\hhat(\nu )$ vanishes in a neighborhood of the origin.  Note that the
above expression for $H(\nu )$ shows that if both $\rho (a)$ and $\hhat(\nu
)$ vanish for negative arguments then $H(\nu )\equiv 0$ and no frame
exists.  Hence to support general complex-valued windows (such as
bandpass filters for a positive--frequency band), it is necessary to include
negative as well as positive scale factors $a$.

The general case, therefore, is that we get a (generalized) frame whenever
$\rho (a)$ and $h(t)$ are chosen such that $0<A\le H(\nu )\le B$ is satisfied. 
The ``metric operator'' $G$ and its inverse are given in terms of Fourier
transforms by

\be
 Gf=\bigl(H\fhat\bigr)\check{\ }\qquad\hbox{and}\qquad G^{-1} f=
\bigl(H^{-1} \fhat\bigr)\check{\ }.
\eno15)
 \ee
Since $G$ is no longer a multiplication operator in the time domain (as it was
in the case of the discrete frame we constructed from the windowed Fourier
transforms), the action of $G ^{-1} $ is more complicated. It is preferable,
therefore, to specialize to tight frames. This requires that $H(\nu )$ be
constant, so the asymptotic conditions on $\rho $ reduce to the requirement
that $\rho $ be piecewise continuous:

\be
 \rho (a)=\cases{c^+/a^2,&if $a>0$\cr
                 c^-/a^2,&if $a<0$\cr}
\eno16)
 \ee
where $c^+$ and $c^-$ are non-negative constants (not both zero).  Then for
$\nu >0$, 

\be
 H(\nu )=\int_0^\infty{d\xi \over \xi }\left[c^+\,|\,\hhat(\xi )\,|\,^2+
c^-\,|\,\hhat(-\xi )\,|\,^2\right]
\eno17)
 \ee
and for $\nu <0$,

\be
 H(\nu )=\int_0^\infty{d\xi \over \xi }\left[c^-\,|\,\hhat(\xi )\,|\,^2+
c^+\,|\,\hhat(-\xi )\,|\,^2\right].
\eno18)
 \ee
Thus $H(\nu )=A=B$ requires either that $\,|\,\hhat(\xi )\,|\,^2=
\,|\,\hhat(-\xi )\,|\,^2$ (which holds if $h(t)$ is real) or that $c^+=c^-$.
Since we want to accomodate complex wavelets, we assume the latter
condition.  Then we have

\be
 A=B=c^+\int_{\rl^*}{d\xi \over {|\xi|} }\ \bigl|\hhat(\xi )\bigr|^2.
\eno19)
 \ee
We have therefore arrived at the measure

\be
 d\mu (a,s)=c^+da\,ds/a^2
\eno20)
 \ee
for tight frames, which coincides with the measure suggested by group
theory (see chapter 3).  In addition, we have found that the basic
wavelet $h$ must have the property that

\be
 c_h\equiv\int_{\rl^*} {d\xi \over{|\xi |}}\,\bigr| \hhat(\xi) \bigl|^2<\infty.
\eno21)
 \ee
In that case, $h(t)$ is said to be {\sl admissible.} This condition is also a
special case of a group--theoretic result, namely that we are dealing with a
{\sl square--integrable representation\/} of the appropriate group (in this
case, the affine group $\rl^*\times\rl$).  To summarize, we have constructed
a continuous tight frame of wavelets $h_{a,s} $ provided the basic wavelet is
admissible. The corresponding resolution of unity is 

\be
 c_h ^{-1} \int_{\rl^*\times\rl}{da\,ds\,\over{a^2}}\,|\,h_{a,s}
\,\rangle\langle\, h_{a,s} \,|\,=I.
 \eno22)
 \ee
The associated reproducing--kernel Hilbert space $\Re_K$ is the space of
functions $(Kf)(a,s)=\langle\, h_{a,s} \,|\,f\,\rangle\equiv\ftil(a,s)$
depending on the scale parameter $a$ as well as the time coordinate $s$.  As
$a\to0$, $h_{a,s} $ becomes peaked around $t=s$ and 

\be
 \ftil\sim|a|^{1/2}cf(s)
\eno23)
 \ee
where $c=\int\br{h(u)}\,du$.  The transformed signal $\ftil$ is a
smoothed--out version of $f$ and $a$ serves as a {\sl resolution parameter.}

Ultimately, all computations involve a finite number of operations, hence as
a first step it would be helpful to construct a {\sl discrete\/} subframe of our
continuous frame.  Toward this end, choose a fundamental scale parameter
$a>1$ and a fundamental time shift $b>0$.  We will consider the discrete
subset of dilations and translations 

\be
 D=\lbrace (a^m,na^mb)\,|\,m,n\in\zz\,\rbrace\subset\rl^*\times\rl.
\eno24)
 \ee
Note that since $a^m>0$ for all $m$, only positive dilations are included in
$D$, contrary to the lesson we have learned above.  This will be remedied
later by considering $\br{h(t)}$ along with $h(t)$.  Also, $D$ is not a
subgroup of $\rl^*\times\rl$, as can be easily checked.  The wavelets
parametrized by $D$ are 

\be
 h_{mn} =a^{-m/2} \,h\left( t-na^mb\over{a^m} \right)
=a^{-m/2} h(a^{-m}t-nb).
\eno25)
 \ee

To see that this is exactly what is desired, suppose $\hhat(\nu) $ is
concentrated on an interval around $\nu =F$ (i.e., $\hhat$ is a band--pass
filter). Then $\hhat_{mn}$ is concentrated around $\nu =F/a^m$.  For given
integer $m$, the ``samples''

\be
 f_{mn} \equiv\langle\, h_{mn} \,|\,f\,\rangle,\quad n\in\zz
\eno26)
 \ee
therefore represent (in discrete ``time'' $n$) the behavior of that part of the
signal $f(t)$  with frequencies near $F/a^m$.  If $m\gg 1$, $f_{mn} $ will
vary slowly with $n$, and if $m\ll-1$, it will vary rapidly with $n$ (if $f(t)$
has frequency components with $\nu \sim F/a^m)$.  Now the time--samples
are separated by the interval $\Delta _mt=a^mb$, so the {\sl sampling rate}

\be
 R_m=1/a^mb
\eno27)
 \ee
is automatically adjusted to the frequency range $\nu \sim F/a^m$ of the
output signal $\{f_{mn} \,|\,n\in \zz\}$:  The high--frequency components
get sampled proportionately more often.  This is an example of a 
topic in mathematics which has recently attracted intense activity under the
banner of {\sl multiscale analysis\/} (Mallat [1987], Meyer [1986]) and which
is in fact closely related to the subject of wavelet transforms.

Returning to the construction of a discrete frame, Parseval's formula gives

\ba{
 f_{mn} \a=\langle\, h_{mn} \,|\,f\,\rangle=\langle\, \hat h_{mn}
                   \,|\,\fhat\,\rangle\cr
\a=\inr d\nu \ \br{k_{mn}(\nu )}\,\fhat(\nu )
\cr}
 \eno28)
 \ea
where $k_{mn}(\nu )\equiv \hat h_{mn} (\nu )$ is the Fourier transform of
$h_{mn}(t)$,

\be
 k_{mn} (v)=a^{m/2} \exp(\tpi na^mb\nu )\,\hhat(a^m\nu ).
\eno29)
 \ee
We now assume that $k(\nu )\equiv \hhat(\nu )$ vanishes outside the
interval 

\be
 I_0=\lbrace  F/a\le \nu \le Fa \rbrace,
\eno30)
 \ee
where $F>0$ is some fixed frequency to be determined below.  The width of
the ``band'' $I_0$ is $W_0=(a-a ^{-1} )F$.  Therefore the function
\break $\br{k(a^m\nu )} \fhat(\nu )$ is  supported on the compact interval

\be
 I_m=\left[  F/a^m,\, F/a^{m-1}\right]
\eno31)
 \ee
of width $W_m=W_0/a^m$, and we can expand it in a Fourier series in that
interval:

\be
 \br{k(a^m\nu )}\,\fhat(\nu )=\sum_n\exp(\tpi n\nu /W_m)\,c_{mn},
\eno32)
 \ee
where 

\be
 c_{mn}=W_m ^{-1} \inr d\nu \,\exp(-\tpi n\nu /W_m)\,\br{k(a^m\nu )}
           \,\fhat(\nu ).
\eno33)
 \ee
Comparing this with 

\be
 f_{mn} =\inr d\nu \,a ^{m/2} \,\exp(-\tpi \nu nba^m)\,\br{k(a^m\nu )}
\hat f(\nu )
\eno34)
 \ee
suggests that we choose $F$ so that $W_m=1/a^mb$, which gives 

\be
 F={a\over{(a^2-1)\,b}}
\eno35)
 \ee
and

\be
 c_{mn}=a^{m/2} b f_{mn} .
\eno36)
 \ee

The Fourier series representation above only holds in the interval $I_m$,
since the left--hand side vanishes outside this interval while the
right--hand side is periodic. To get equality for all frequencies and
reconstruct $f(t)$, multiply both sides by $k(a^m\nu )$ and sum over $m$:

\ba{
\left(  \sum_m\,|\,k(a^m\nu )\,|\,^2\right)\,\fhat(\nu )\a=
b\sum_{m,n}a^{m/2}\exp\left(\tpi na^mb\nu\right)\,k(a^m\nu )\,f_{mn}\cr
\a=b\sum_{m,n}k_{mn}(\nu )\,f_{mn}.
 \cr}
 \eno37)
 \ea
To have a frame we would need the series on the left--hand side to converge
to a function $\chi ^+(\nu )$ with

\be
 0<A\le \chi ^+(\nu )\equiv \sum_m\,|\,k(a^m\nu )\,|\,^2\le B
\eno38)
 \ee 
for some constants $A$ and $B$. But this is a priori impossible, since
$\hhat(\nu )$ is supported on an interval of positive frequencies and
$a^m>0$, so $\chi ^+(\nu )$ vanishes for $\nu \le 0$.   However, we can
choose $h(t),\, a$ and $b$ such that $\chi ^+(\nu )$ satisfies the frame
condition for $\nu >0$.  Negative frequencies will be taken care of by
starting with the complex--conjugate of the original wavelet.  We adopt the
notation 

\be
 h^+(t)\equiv h(t),\quad h^-(t)\equiv \br{h(t)}.
\eno39)
 \ee
Then the Fourier transforms $k^\pm$ of $h^\pm$ are related by

\be
k^-(\nu )=\br{k^+(-\nu )},
\eno40)
 \ee
hence $k^-(\nu )$ is supported on $-I_0=[-Fa,\,-F/a]$.  A similar argument to
the above gives

\be
 \left(  \sum_m\,|\,k^-(a^m\nu )\,|\,^2\right)\,\fhat(\nu )=
b\sum_{m,n}k^-_{mn}(\nu )\,f_{mn}
\eno41)
 \ee
and

\be
 \chi ^-(\nu )\equiv \sum_m\,|\,k^-(a^m\nu )\,|\,^2=\chi ^+(-\nu ).
\eno42)
 \ee
Hence if $\chi ^+$ satisfies the frame condition for $\nu >0$, then

\be
 0<A\le \chi ^+(\nu )+\chi ^-(\nu )\le B
\eno43)
 \ee
for all $\nu \ne 0$.  Since $\{0\}$ has zero measure in frequency space, the
frame condition is satisfied by the joint set of vectors 

\be
 {\cal H}^{a,b}_M=\lbrace k^+_{mn},\, k^-_{mn} \,|\,m,n\in\zz \rbrace.
\eno44)
 \ee
The metric operator 

\be
 G\equiv \sum_{\epsilon =\pm}\ \sum_{m,n\in\zs}\,|\,k^\epsilon
_{mn}\,\rangle \langle\, k^\epsilon _{mn}\,|\,
\eno45)
 \ee
is given by

\be
 (Gf)(t)=\left( (\chi ^+ +\chi ^-)\fhat \right)\check{\,}(t)
\eno46)
 \ee
and satisfies the frame condition $0<AI\le G\le B$.
Since $G$ is no longer a multiplication operator in the time domain (as was
the case with the discrete frame connected to the windowed Fourier
transform), the recovery of signals would be greatly simplified if the
frame was tight.  The following construction is borrowed from Daubechies
[1988a].  Let $F=a/(a^2-1)b$ as above and let $k$ be any non--negative
integer or $k=\infty$.  Choose a real--valued function $\eta \in C^k(\rl) $
(i.e., $\eta $ is $k$ times continuously differentiable) such that 

\be
 \eta (x)=\cases{0 &for $x\le 0$\cr
                \pi /2&for $x\ge 1$.\cr}
\eno47)
 \ee
(Such functions are easily constructed; they are used in differential
geometry, for example, to make partitions of unity; see Warner [1971].) 
Define $h(t)$ through its Fourier transform $k^+(\nu )$ by

\be
 k^+(\nu)=
\cases{\sin\left[\eta\left({\nu-F/a\over{F- F/a}}\right)\right] 
&for $\nu\le F$\cr\
{}&{}\cr
      \cos\left[\eta\left({\nu-F\over{aF-F}}\right)\right], &for $\nu \ge F.$}
\eno48)
 \ee

Note that $k^+(\nu )$ is $C^k$ since the derivatives of $\eta (x)$ up to order
$k$ all vanish at $x=0$ and $x=\pi /2$.  This means that the wavelets in the
frame we are about to construct are all $C^k$.  Also, $k^+$
vanishes outside the interval $I_0=[F/a,\,Fa]$.  The width of its support is
$W_0=(a-a ^{-1} )F$, and for each frequency $\nu >0$ there is a unique
integer $M$ such that  $ F/a<a^M\nu \le F$, hence also $F<a^{M+1}\nu \le aF$. 
Therefore, for $\nu >0$, 

\be
 \chi^+(\nu )=\sin^2\left[\eta\left({\nu-a^{-1}F\over{F-a^{-1}
F}}\right)\right]  + \cos^2\left[\eta\left({\nu-F\over{aF-F}}\right)\right]=1.
\eno49)
 \ee
Thus 

\be
 \chi ^+(\nu )=\cases{0&for $\nu \le 0$\cr
          1&for$\nu >0,$\cr}
\eno50)
 \ee
i.e., $\chi ^+(\nu )$ is the indicator function for the set  of
positive numbers.  It follows that $\chi ^-(\nu )$ is  the
indicator function for the negative reals, and

\be
\chi  ^+ + \chi  ^- =1 \quad\hbox{a.e.}
\eno51)
 \ee
This choice of $k^+$ and $k^-=\br{k^+}$ gives us a tight frame,

\be
 \sum_{\epsilon =\pm}\ \sum_{m,n\in \zs}\,|\,k^\epsilon _{mn}\,\rangle
\langle\, k^\epsilon _{mn}\,|\,=I.
\eno52)
 \ee

This frame is not a basis; if it were, it would have to be an orthonormal
basis since it is a normal frame, hence  the reproducing kernel would have
to be diagonal. But

\be
 K(\epsilon ,m,n;\ \epsilon ',m',n') \equiv \langle\, k^\epsilon _{mn}\,|\,
k^{\epsilon'} _{m'n'}\,\rangle
\eno53)
 \ee
does not vanish for $\epsilon '=\epsilon ,\,n'=n$ and $m'=m\pm 1$, due to
the overlap of wavelets with adjacent scales.  However, it  is
possible to construct orthonormal bases of wavelets which, in addition, have
some other surprising and remarkable properties.  For example, such bases
have been found  (Meyer [1985],  Lemarie and Meyer [1986]) whose Fourier
transforms, like those above, are $C^\infty$ with compact support and 
which  are, simultaneously, unconditional bases for all the spaces $L^p(\rl)$
with $1<p<\infty$ as well as all the Sobolev spaces  and some other popular
spaces to boot.  Similar bases were constructed in connection with quantum
field theory  (Battle [1987])  which are only $C^k$ for finite $k$ but, in
return, are better localized in the time domain (they have exponential
decay). The concept of multiscale analysis (Mallat [1987], Meyer [1986])
provided a general method for the construction and study of orthonormal
bases of wavelets.  This was then used by Daubechies [1988b] to construct
orthonormal bases of wavelets having {\sl compact\/}  support and
arbitrarily high regularity. 

The mere existence of such bases  has surprised analysts and made wavelets
a hot new topic in current mathematical research. They are also  finding
important applications in a variety of areas such as signal analysis,
computer science and quantum field theory.  They are the subject of the next
chapter, where a new, algebraic, method is developed for their study.

\VE

\def\leftheadline{\tenrm\folio\hfil {\sl 3. Frames and Lie Groups}\hfil} 
\def\rightheadline{\hfill\folio}

\headline={\ifodd\pageno\rightheadline\else\leftheadline\fi}
\def\be{$$}\def\ee{$$}\def\ba{$$\eqalign}\def\a{&}  
\def\eno{\eqno(}\def\ea{$$}

\centerline{\bf Chapter 3}\skp
\centerline{\bf FRAMES AND LIE GROUPS}
\vskip 3 cm
\noindent {\bf  3.1. Introduction }
\skp

\noindent Although we have not sought to exploit it until now, it is clear
that all our frames so far have been obtained with the aid of group
operations.  The frames associated with the canonical coherent states and
the windowed Fourier transform were built using
translations in phase space (Weyl--Heisenberg group), while the wavelet
frames used  translations and dilations (the affine group).  In this chapter
we look for a unifying pattern in these constructions based on group
theory.  We analyze  the foregoing constructions in turn, and draw
separate lessons from each.  It will be natural to work in reverse order.  The
affine group, which is, in some sense, the simplest, will lead us to the
general method.  Successive refinements will be suggested by the
windowed Fourier transform and the canonical coherent states.
\secskp

 \noindent {\bf 3.2. Klauder's Group--Frames}

 \skp
\noindent This was the first of the group--theoretic constructions, pioneered
by J. R. Klauder [1960, 1963a, 1963b], who was also the first to apply it to the
affine group $G=\rl^*\times \rl$ (Aslaksen and Klauder [1968, 1969]).  An
element $g=(a,s)$ of $G$ acts on the real line (``time'') by dilation followed by
translation:

\be
 gt\equiv (a,s)t=at+s.
\eno1)
 \ee
The group--composition law is given by 

\be
 g'gt=(a',s')(a,s)t=a'(at+s)+s'=(a'a,a's+s')t,
\eno2)
 \ee
hence $g ^{-1}=(a ^{-1},-s/a)$.  (The form of the composition law shows that
the subgroup of dilations {\sl acts\/} on the subgroup of translations, so that
G is the semidirect product of the two.)  The frame vectors for the wavelet
transform were obtained from a single ``basic wavelet'' vector $h$ by
applying the transformation 

\def\rightheadline{\tenrm\hfil {\sl 3.2. Klauder's Group--Frames}\hfil\folio}

\be
 h(t)\mapsto \,|\,a\,|\,^{-1/2}\,h\left( {t-s \over{a }} \right)\equiv 
\left( U(a,s) h\right)(t).
 \eno3)
 \ee
It it easy to see that with respect to the inner product in $L^2(\rl)$, 
\skp
\item{(a)} $\left(U(a,s)^*h 
\right)(t)=\,|\,a\,|\,^{1/2}\,h(at+s)=\left(U(a,s)^{-1} h\right)(t)$, and
\item{(b)} $U(a',s')U(a,s)h=U(a'a,a's+s')h$.
\skp
\noindent In terms of the group operations, this means that $U(g)^*=
U(g) ^{-1}$ and $U(g'g)=U(g') U(g) $, respectively.  That is, each $U(g) $ is a
{\sl unitary\/} operator (on $L^2(\rl)$ ), and $g\mapsto U(g)$ is a {\sl
representation\/} of $G$.  This means that $U$ is a {\sl unitary
representation\/} of $G$ on $L^2(\rl)$.  The theory of such representations
for general groups is a deep and highly developed subject, and is of
fundamental importance in quantum mechanics, as was realized by
Hermann Weyl and others long ago (Weyl [1931]).  In the broadest
sense,  group representations amount to a  vast generalization of the  
exponential function (think of the map $z\mapsto e^{az}$ from $\cx$ to
$\cx^*$), and unitary group representations generalize the map  $x\mapsto
e^{iax}$ from $\rl$ to the unit circle.

A representation $U$ of a group $G$ on a Hilbert space $\cal H$
is said to be {\sl reducible\/} if $\cal H$ has a non--trivial closed subspace 
$S$ (i.e., $S\ne\{0\}$ and $S\ne \cal H$) which is {\sl invariant\/} under
$U$ (i.e., $ U(g) S\subseteq S$ for every $g\in G$).  If $U$ is unitary and $S$
is invariant, then clearly so is its orthogonal complement $S^\perp$. If no
such $S$ exists, then $U$ is said to be {\sl irreducible}.  It can be shown that the above
representation of $\rl^*\times \rl$ on $L^2(\rl)$ is, in fact, irreducible. 

 The method to be described below assumes that we begin
with a given irreducible unitary representation $U$ of a given group $G$ on
a given Hilbert space $\cal H$. In addition, we assume that the  group is a
{\sl Lie group}, meaning that it has a differetiable structure such that it is
essentially determined (in local terms) by its Lie algebra of left-- (or right--)
invariant vector fields.  (See Helgason [1978], Varadarajan [1974] or Warner
[1971] for background on Lie groups.)  The affine group and the
Weyl--Heisenberg group are examples of Lie groups, as is $\rl^n$ (under
vector addition as the group operation).

Given a general setup ($U$, $G$, $\cal H$ ) as above, choose an arbitrary
non--zero vector $h$ in $\cal H$.  (Klauder dubbed $h$ a ``fiducial
vector''; for the affine group, this was the ``basic wavelet''.)  For every $g\in
G$ define the vector 

\be
 h_g=U(g) h.
\eno4)
 \ee
Since $U$ is unitary, $\|h_g\|=\|h\|$.  The $h_g$'s are {\sl covariant\/} under
the action of $G$ on $\cal H$, i.e.,

\be
 U(g') h_g=U(g') U(g) h=U(g'g) h=h_{g'g}.
\eno5)
 \ee
Hence the set of all finite linear combinations (span) of $h_g$'s is invariant
under $U$, and therefore so is its closure $S$.  Since $U$ is irreducible and
$S\ne \{0\}$, it follows that $S=\cal H$.  This means that every vector
in $\cal H$ can be approximated to arbitrary precision by finite linear
combinations of $h_g$'s---a good beginning, if one is ultimately  interested
in reconstruction!  To build a frame from the $h_g$'s, we need a measure on
$G$.  Now every Lie group has an essentially unique (up to a constant
factor) {\sl left--invariant\/} measure, which we will denote by $d\mu$. 
This means that if $E$ is an arbitrary (Borel) subset of $G$, and if
$g_1E\equiv \{g_1g\,|\,g\in E\}$ is its {\sl left translate\/} by $g_1\in G$,
then

\be
 \mu (g_1E)\equiv \int_{g_1E}d\mu (g)=\mu (E)\equiv \int_E d\mu (g).
\eno6)
 \ee
In local terms, $d\mu (g_1g)=d\mu (g)$ for fixed $g_1$.  [Similarly, there
exists a {\sl right--invariant\/} measure $d\mu _R$ on $G$, which is in
general different from $d\mu$;  if $d\mu _R$ is proportional by a constant
to $d\mu$, the group $G$ is called {\sl unimodular}. Everything we do below
can be repeated, with obvious modifications, using $d\mu _R$, provided  that
$h_g$ is redefined as $U(g ^{-1} )h$.] 

 To find $d\mu $ for the affine group,
for example, we write it in the form of a density,

\be
 d\mu (a,s)=\rho (a,s)\,da\wedge ds ,
\eno7)
 \ee
where $da\wedge ds$ is a differential  2--form  denoting the area element
in $\rl^2 \supset G$. ($da\wedge ds$ becomes a  positive measure upon
choosing an {\sl orientation\/} in $\rl^2$ and orienting all subsets
accordingly;  see Warner [1971].)  Then, for fixed $(a_1, s_1)\in \rl^*\times
\rl$,

\ba{
d\mu ((a_1, s_1)(a,s))\a=\rho (a_1a, a_1s+s_1)\,d(a_1a)\wedge
                          d(a_1s+s_1)\cr
\a=a_1^2\rho (a_1a, a_1s+s_1)\,da\wedge ds,
 \cr}
 \eno8)
 \ea
hence left--invariance implies

\be
 a_1^2\rho (a_1a, a_1s+s_1)=\rho (a,s).
\eno9)
 \ee
Setting $a_1=1/a$ and $s_1=-s/a$ then shows that

\be
 d\mu (a,s)=da\wedge ds/a^2,
\eno10)
 \ee
where we have chosen the normalization $\rho (1,0)=1$.  This is precisely the
measure we obtained earlier using a more pedestrian approach. 

Returning to the general case, consider the {\sl formal\/}  integral 

\be
 J=\int_G d\mu (g)\,|\,h_g\,\rangle\langle\, h_g\,|\,.
\eno11)
 \ee
We want to show that (a) $J$ converges, in some sense, and (b) $J$ is a
multiple of the identity, thus giving a tight frame in the generalized sense
defined in chapter 1.  Before worrying about convergence, let us formally
apply $U(g_1)$ from the left:

\ba{
U(g_1)J\a=\int_G d\mu (g)\,U(g_1) \,|\,h_g\,\rangle\langle\, h_g\,|\, \cr
\a=\int_G d\mu(g) \,|\,h_{g_1g}\,\rangle\langle\, h_g\,|\, \cr
\a=\int_G d\mu(g') \,|\,h_{g'}\,\rangle\langle\, h_{g_1 ^{-1} g'}\,|\, \cr
\a= \int_G d\mu(g') \,|\,h_{g'}\,\rangle\langle\,h_{g'}\,|\,U(g_1 )\cr
\a=JU(g_1),
 \cr}
 \eno12)
 \ea
where we have used the left--invariance of the measure and 

\be
 \langle\, h_{g_1^{-1} g'}\,|\,=\left(\,|\, h_ {g_1^{-1} g'}\,\rangle\right)^*
=\left(U(g_1 ^{-1} ) \,|\,h_{g'}\,\rangle \right)^*
=\langle\, h_{g'}\,|\,U(g_1),
\eno13)
 \ee
which follows from unitarity.

That is, $J$ commutes with every representative $U(g) $ of the group.  By
Schur's Lemma (Varadarajan [1974]), it follows from the irreducibility of $U$
that $J$ is a multiple of the identity operator on $\cal H$.  Since $J$, if it
converges, is a positive operator, we arrive at the desired frame condition

\be
 \int_G d\mu(g) \,|\,h_g\,\rangle\langle\, h_g\,|\,=cI,
\eno14)
 \ee
where $c$ is a positive constant which can taken as unity by the
appropriate normalization of $d\mu $.  Returning to the formal integral
defining $J$, the above argument shows that {\sl if\/} the integral converges
in some sense, it must converge to $cI$, hence define a bounded operator. (Of
course $c$ could be infinite!)  Thus a {\sl necessary\/} condition for
convergence in the {\sl weak\/} sense (i.e., as a quadratic form) is that

\ba{
\langle\, h\,|\,J\,h\,\rangle\a=\int_G d\mu(g)\, \langle\, h\,|\,h_g\,\rangle
\langle\, h_g\,|\,h\,\rangle\cr
\a=\int_G d\mu(g) \,\,|\,\langle\, h\,|\,U(g)\,|\,h\,\rangle\,|\,^2 <\infty.
 \cr}
 \eno15)
 \ea
We have already encountered this condition in the special case of the affine
group, where it was called the {\sl admissibility condition\/} for $h$.  The
same terminology is used in the present, general setting.  The above
condition depends both on the representation $U$ and the choice of $h$.  If
it is satisfied for at least {\sl one\/} non--zero vector $h$, the
representation $U$ is called {\sl square--integrable} and $h$ is called {\sl
admissible}.  It turns out that the existence of an admissible
(non--zero) vector is also {\sl sufficient\/} for the weak convergence of the
integral $J$.  The following theorem is due to Carey [1976], and Dufflo and
Moore [1976]; $G$ can be any locally compact topological group, in
particular any Lie group. 
\skp

\bf Theorem 3.1. \sl   Let $U$ be a square--integrable unitary irreducible
representation of $G$ on a Hilbert space $\cal H$.  Then there exists a
unique self--adjoint (in general unbounded) operator $C$ on $\cal H$ such
that:

\item{(a)}  The domain of $C$ coincides with the set of all admissible
vectors.

\item{(b)}  If $h_1$ and $h_2$ are admissible, then for all $f_1$ and $f_2$
in $\cal H$ we have 

\be
 \int_G d\mu(g) \,\langle\, f_1\,|\,U(g)\,h_1\,\rangle\langle\,
U(g)\,h_2\,|\,f_2\,\rangle=\langle\, Ch_2\,|\,Ch_1\,\rangle
\langle\, f_1\,|\,f_2\,\rangle.
 \eno16)
 \ee

\item{(c)}  If $G$ is unimodular, then $C$ is a multiple of the identity.  
\skp\rm

Choosing $h_1=h_2\equiv h$ now leads to the earlier resolution of unity,
with $c=\|Ch\|^2$.  As usual, an  arbitrary vector $f$ in $\cal H$ can be
``presented'' as a function $\ftil(g)\equiv \langle\, h_g\,|\,f\,\rangle$ on
$G$, from which $f$ may be reconstructed as a linear combination of $h_g$'s
weighted by the reproducing kernel $K(g',g)=\langle\,
h_{g'}\,|\,h_g\,\rangle$.  

Note that the basic wavelet $h$ corresponds to 

\be
 \tilde h(g)=\langle\,h_g\,|\,h\,\rangle=\br{\langle\, h\,|\,U(g)h\,\rangle},
\eno17)
 \ee
so the admissibility condition is nothing but the requirement that $\tilde h$
belong to $L^2(d\mu )$. 

A potentially interesting generalization of this scheme is actually
possible.  By the foregoing theorem we can use two {\sl distinct\/}
admissible vectors: the vector $h_2$ is used to {\sl analyze\/} $f$, i.e.
present it as $\ftil(g)=\langle\, (h_2)_g\,|\,f\,\rangle$, while the vector $h_1$
is then used to {\sl synthesize\/} $f$ from $\ftil(g)$.  There is  no
fundamental reason to use the same ``wavelets'' for analysis and synthesis,
provided only that they overlap in the sense that

\be
 \langle\, Ch_2\,|\,Ch_1\,\rangle\ne0.
\eno18)
 \ee
Absorbing the reciprocal of this constant into the group measure, the map
$T{:}\ f \mapsto \ftil$ is an isometry from $\cal H$ onto its range $\Re_T$.

Due to the ``covariance'' of the $h_g$'s with respect to the
action of $G$, the representation of $G$ on $\Re_T$ acquires the simple {\sl
geometric\/} form

\ba{
 \left( \tilde U(g_1)\ftil \right)(g) \a
\equiv \langle\, h_g\,|\,U(g_1)f\,\rangle\cr
\a=\langle\, h_{g_1^{-1} g}\,|\,f\,\rangle\cr
\a=\ftil(g_1^{-1} g),
 \cr}
 \eno19)
 \ea
showing that $G$ acts on $\Re_T$ by merely translating the variable in the
base space $G$.  The realization of $f$ by $\ftil$ and $U$ by $\tilde U$ as
above is called the {\sl coherent--state representation \/} determined by
the pair $(U, h)$.  We will also refer to the frame $\{\,h_g\,|\,g\in G\}$ as the
{\sl group--frame\/} ($G$--frame) associated with $(U,h)$.  From a purely
mathematical point of view, one of the attractions of this scheme is that
although we  started with an arbitrary representation of $G$ on an arbitrary
Hilbert space, this construction ``brings it home'' to $G$ itself and objects
directly associated with it: the Hilbert space is a closed subspace of
$L^2(d\mu )$, and the representation is induced from the (left) action of $G$
on itself, i.e. $g\mapsto g_1 ^{-1} g$. That is, Klauder's construction exhibits
$U$ as a subrepresentation of the {\sl regular representation  \/}  of $G$. 
(See Mackey [1968] for the definition and discussion of the regular
representation.) 

 \vfill\eject%

 \noindent {\bf  3.3.  Perelomov's Homogeneous G--Frames }
\def\rightheadline{\tenrm\hfil {\sl 3.3. Perelomov's 
Homogeneous G--Frames}\hfil\folio}  

 \skp
\noindent Let us now attempt to apply the group--theoretic method to the
windowed Fourier transform.  As most of our applications will be to the
phase--space formulation of quantum mechanics, we shift gears and replace
the time by a space coordinate, $t\to -x$ (the sign is related to the
Minkowski metric; see section 1.1) and  the frequency by a momentum
coordinate, $2\pi \nu \to p$.  Although it is a trivial matter to extend
everything we do in this section to an arbitrary (finite) number of degrees
of freedom (where $x$ and $p$ belong to $\rl^s$), we restrict ourselves to a
single degree of freedom to keep the notation simple.  The
self--adjoint generators of translations in space and momentum, $P$ and
$X$, are defined by 

\be
 P f(x)=-i{\partial \over{\partial x}}f(x),\qquad (Xf)\hat{\,}(p)=i{\partial
\over{\partial p}}\fhat(p).
 \eno1)
 \ee
Expressed in the space domain,

\be
 Xf(x)=xf(x).
\eno2)
 \ee
Starting with a basic window function $h(x)$, the window centered at $(p,x)$
in phase space is given by

\ba{
 h_{p,x} (x')\a=e^{ipx'}h(x'-x)\cr
\a=e^{ipx'}\left(e^{-ixP}h\right)(x')\cr
\a=\left( e^{ipX}\,e^{-ixP}h \right)(x').
 \cr}
 \eno3)
 \ea
That is, $h_{p,x} $ is obtained from $h$ by a translation in space (by $x$)
followed by a translation in momentum (by $p$).  Defining the
corresponding unitary operators 

\be
 U(p,x)=e^{ipX}\,e^{-ixP},
\eno4)
 \ee
let us see what happens when two such operations are applied in succession:

\ba{
\left( U(p_1,x_1)U(p,x)h  \right)(x')\a=\left( U(p_1,x_1)h_{p,x}  \right)(x')\cr
\a=e^{ip_1x'}h_{p,x} (x'-x_1)\cr
\a=e^{ip_1x'}e^{ip(x'-x_1)}h(x'-x_1-x)\cr
\a=e^{-ipx_1}h_{p_1+p,x_1+x}(x')\cr
\a=e^{-ipx_1}U(p_1+p,x_1+x)h(x').
 \cr}
 \eno5)
 \ea
Hence the operators $U(p,x)$ do not form a group, since two successive
operations give rise to a ``multiplier'' $\exp(-ipx_1)$.  The reason is that
translations in space  do not commute with translations in momentum, as can
also be seen at the infinitesimal level by noting that their respective
generators obey the ``canonical commutation relations'' 

\be
 [X,P]=\left[ x,-i{\partial \over{\partial x }} \right]=iI.
\eno6)
 \ee
The remedy (suggested by Hermann Weyl) is to include the identity
operator as a new generator (it  generates phase factors $e^{-i\phi }$ which
can be used to absorb the multiplier).  To see this in terms of unitary
representations of Lie groups, consider the abstract {\sl real\/} Lie algebra
{\bf w} with three generators $\{-iS, -iT, -iE\}$ and Lie brackets

\be
 [S, T]=iE,\qquad [S, E]=0,\qquad [T,E]=0.
\eno7)
 \ee
The corresponding three--dimensional (simply connected) Lie group $\cal
W$ is known as the {\sl Weyl--\-Heisen\-berg group}.  Topologically, $\cal
W$ is just $\rl^3$.  (If configuration space is $\rl^s$, the corresponding
Weyl--Heisenberg group ${\cal W}_s \approx\rl^{2s+1}$ .)  A general element
in $\cal W$, para\-metrized by $(p,x,\phi )\in \rl^3$, may be expressed as
the product of three factors 

\be
 g(p,x,\phi )=\exp(-i\phi E)\exp(ip S)\exp(-ixT),
\eno8)
 \ee
where ``$\exp$''  denotes the exponential mapping from the Lie algebra
{\bf w} to the Lie group $\cal W$, whose group law is 

\be
 g(p_1,x_1,\phi _1)\,g(p,x,\phi )=g(p_1+p,x_1+x,\phi _1+\phi +px_1).
\eno9)
 \ee
A unitary irreducible representation  of $\cal W$ on $L^2(\rl)$  is obtained
by the correspondence $S\to X,\  T\to P,\  E\to I$.  This is known as the {\sl
Schr\"odinger representation}.  The unitary operator correponding to
$g(x,p,\phi )$ is

\be
 U(p,x,\phi )=e^{-i\phi }\,e^{ipX}\,e^{-ixP}.
\eno10)
 \ee
As expected, these operators are closed under multiplication, with the
composition law

$$ U(p_1,x_1,\phi _1)\,U(p,x,\phi )=U(p_1+p,x_1+x,\phi _1+\phi +px_1).$$

With this unitary irreducible representation  of $\cal W$ on
$L^2(\rl)$, we have all the ingredients needed to attempt the construction of
a group frame for $\cal W$.  The prospective frame vectors are 

\be
 h_{p,x,\phi }=U(p,x,\phi )h=e^{-i\phi }h_{p,x},
\eno11)
 \ee
where $h_{p,x} $ are the vectors defined earlier.  The left--invariant
measure on $\cal W$ (which, in this case, is also right--invariant) is just
Legesgue measure on $\rl^3$, given by the differential form $dp\wedge
dx\wedge d\phi $.  This can be seen by looking at the composition law: For
fixed $(p_1,x_1,\phi _1)$, 

\be
 d(p_1+p)\wedge d(x_1+x)\wedge d(\phi _1+\phi +px_1)
=dp\wedge dx\wedge d\phi 
\eno12)
 \ee
since $dp\wedge dp  =0$.  The method of section 2 then gives the following
candidate for a resolution of unity:

\ba{
J\a=\int_{\cal W}\,dp\wedge dx\wedge d\phi \,|\,h_{p,x,\phi } \,\rangle
\langle\, h_{p,x,\phi }\,|\,\cr
\a=\int_{\cal W}\,dp\wedge dx\wedge d\phi \,|\,h_{p,x} \,\rangle\langle\, 
h_{p,x} \,|\, ,
 \cr}
 \eno13)
 \ea
where the phase factor cancels in the integrand.  This integral clearly
diverges, since the integrand is independent of $\phi $ and the integration
is over all real $\phi $.  Equivalently, the representation  $U$ is not
square--integrable, since for every nonzero $h$,

\be
 c_h\equiv \int dp\,dx\,d\phi \,\,|\,\langle\, h\,|\,U(p,x,\phi )h\,\rangle\,|\,^2
=\infty 
\eno14)
 \ee
due, again, to the constancy of the integrand in $\phi $.  We could get
around the problem by choosing a {\sl multiply connected\/} version 
${\cal W}_1$ of $\cal W$, say with $0\le \phi <2\pi $.  (${\cal W}_1$ has the
same Lie algebra as $\cal W$).  This would indeed give a tight frame, but
this frame is {\sl unnecessarily\/} redundant (as opposed to the beneficial
sort of redundance associated with oversampling) since the vectors $h_{p,x,
\phi } $ are not essentially different for distinct values of $\phi $.  More
significantly, we would miss an important lesson which this example
promises to teach us.  For other important groups, as we will see, the
problem cannot be circumvented by compactifying the troublesome
parameters.  The following solution was  proposed by Perelomov [1972]
(see also Klauder [1963b, p. 1068], where this idea is anticipated).  To get rid
of the $\phi $--dependence, choose a ``slice'' of $\cal W$, $\phi =\alpha
(p,x)$, and define

\be
 h^\alpha _{p,x}=h_{p,x,\alpha (p,x)}=e^{-i\alpha (p,x)}h_{p,x}.
\eno15)
 \ee
Integrating only over this slice with the measure $dp\wedge dx$, we get

\be
 J'\equiv \int dp\wedge dx\,\,|\,h^\alpha _{p,x}\,\rangle\langle\, 
h^\alpha _{p,x}\,|\,=c_hI,
\eno16)
 \ee
where $c_h=2\pi \|h\|^2$, since the integral reduces to the same one we had
for the windowed Fourier transform.

From a computational point of view this is, of course, trivial.  But to extend
the technique to other groups we must understand the group theory behind
it.  Suppose, then, that we return to the general setup we had in the last
section:  Given a unitary irreducible representation  $U$ of a Lie group $G$ on
a Hilbert space $\cal H$,  choose a nonzero vector $h$ in $\cal H$ 
and form its translates $h_g=U(g)h$ under the group action as before. 
Consider now the set $H$ of of all elements $k$ of $G$  for which
the action of $U(k)$ on $h$ reduces to a multiplication by a phase factor 
$\chi (k)=\exp[i\phi (k)]$:
\be
 U(k)h=\chi (k)h,\qquad k\in H.
\eno17)
 \ee
In the case of $\cal W$, $H$ consists of all elements of the form $k=(0,0,\phi
)$ and $U(k)$ is, in fact, nothing but a phase factor.  However, this is
deceptive, since in general $U(k)$ may be a non--trivial operator, acting
trivially only on some vectors $h$.  Hence, in general, $H$ will in fact depend
on the choice of $h$ and should properly be designated $H(h)$.  Now for two
elements $k_1$ and $k_2$ of $H$, we have

\be
 U(k_1 k_2^{-1} )h=U(k_1)U(k_2)^{-1} h
=\chi (k_1)\chi (k_2) ^{-1} h,
\eno18)
 \ee
hence $k_1 k_2^{-1}$ also belongs to $H$ and it follows that $H$ is a {\sl
subgroup\/} of $G$.  Furthermore, the above equation shows  that the
map $k\mapsto \chi (k)$ is a group--homomorphism \/of $H$ into the
unit circle, thus it is a {\sl charac\-ter\/} of $H$, i.e. a unitary representation 
on the one--dimensional Hilbert space $\cx$\/.  $H$ is called the {\sl stability
subgroup\/} for $h$, and its Lie algebra is called the stability subalgebra. 
The reason for this terminology is that $H$ does not affect the
quantum--mechanical {\sl state\/} defined by $h$, since all observable
expectation values in that state are given in terms of sequilinear forms in
$h$.  More precisely, if we choose $\|h\|=1$,  the corresponding state is by
definition the rank--one projection operator 

\be
 P=\,|\,h\,\rangle\langle\, h\,|\,  
\eno19)
 \ee
and the expected value of any observable $A$  in this state is given by

\be
 \langle\, A\,\rangle\equiv \langle\, h\,|\,A\,h
\,\rangle={\rm trace}\,(PA).
\eno20)
 \ee
(This formulation, besides avoiding irrelevant phase facors, also permits
states which are statistical mixtures of pure states as needed, for example,
in statistical quantum mechanics.)  The translate of $P$ under a general
group element $g$ is then

\be
 P_g\equiv \,|\,h_g\,\rangle\langle\, h_g\,|\,=U(g)\,P\, U(g)^*,
\eno21)
 \ee
hence for $g$ in $H$ we have $P_g=P$, i.e., $P$ is {\sl stable\/}
under $H$ as the name implies.  There is, therefore, some advantage to
formulating the theory as much as possible in terms of states rather than
Hilbert space vectors since this automatically eliminates the fictitious degree
of freedom represented by the overall phase. [Incidentally,
a similar situation appears to prevail in communication theory, since an
overall phase shift has no effect whatsoever on the informational content of
the signal.]  However, the Hilbert space
vectors will play an important role in connection with holomorphy (recall
that in the  coherent--state representation, $\ftil (z)$ is analytic, whereas
$\,|\,\ftil (z)\,|\,^2$ is not), hence we work primarily with them. 

If $g\in G$ and $k\in H$, then

\be
 h_{gk}=U(gk)h=U(g)U(k)h=\chi (k)h_g,
\eno22)
 \ee
hence $P_{gk}=P_g$.  That is, $P_g$ is the same for all members
of the {\sl left coset\/}

\be
 gH\equiv \{gk\,|\,k\in H\}.
\eno23)
 \ee
The set of all translates $P_g$ is therefore parametrized by the {\sl left
coset space\/}

\be
 M= G/H\equiv \{gH\,|\,g\in G\}.
\eno24)
 \ee
Members of $M$ will be alternatively denoted by $m$ and by $gH$, and will
play a dual role: as  points in $M$, and  subsets of $G$.  In the case of $\cal
W$, for example, $M$ is parametrized by $(p,x)\in\rl^2$, i.e. it is {\sl phase
space}, precisely the label space for the frame we obtained.  The coset
 $(p,x)H$ is a straight line in ${\cal W}\approx\rl^3$. (We will see
in the next chapter that $\cal W$ can be interpreted as a degenerate
non--relativistic  limit  of phase space$\times $time and $(p,x)H$ then
corresponds to the trajectory of a free classical particle in $\cal W$.)  

To build a frame, we take a ``slice'' of $G$ by choosing a representative from
each coset, i.e. choosing a map

\be
 \sigma {:}\ G/H\to G.
\eno25)
 \ee
{\sl Note:\sp}This is actually a non--trivial process.  The projection $G\to
G/H$ defines a {\sl fiber bundle}, and $\sigma $ is a {\sl section\/} of this
bundle; in general $\sigma $ can be chosen {\sl smoothly\/} only in a
neighborhood of each point of $G/H$, i.e., locally  (see F. Warner [1971],
p.120).  However, this is sufficient for our purposes, since we will ultimately 
deal only with the state $P_m$, which is independent of the choice of $\sigma
$.   \# 
\skp

 Thus $\sigma $ has the form $\sigma (gH)=g\,\alpha (g)$ for some function
$\alpha {:}\ G\to H$.  In the case of $\cal W$, we had $\sigma
(p,x)=(p,x,\alpha (p,x))$.  The frame vectors corresponding to this choice are

\be
 h^\sigma _m\equiv h_{\sigma (m)}=U(\sigma (m))h.
\eno26)
 \ee
To build a frame, we need two more ingredients: an {\sl action\/} of $G$ on
the label space $M$, and a measure on $M$ which is invariant with respect
to this action.  The action is easy, since $G$ acts naturally on $G/H$ by left
translation:

\be
 g_1(gH)=(g_1g)H.
\eno27)
 \ee
If $m=gH\in M$, we will denote $(g_1g)H$ by $g_1m$.  $M$ is called a {\sl
homogeneous space \/} of $G$.  As for an invariant measure, it exists, in
general, only subject to a certain technical condition (Helgasson [1962], p.
369).  It does exist whenever $G$ is a unimodular group (its right-- and
left--invariant measures are proportional by a constant), such as $\cal W$. 
Let us assume that a $G$--invariant measure does exist on $M$ (in which
case it is unique, up to a constant factor) and denote it by $d\mu _{_M} $. 
Once more we consider the formal integral

\be
 J=\int_M\,d\mu _{_M}(m)\,|\,h^\sigma _m\,\rangle\langle\, h^\sigma _m\,|\,
   =\int_M\,d\mu _{_M}(m)\,P_m.
\eno28)
 \ee
If we can show that $J$ commutes with every $U(g_1)$, then irreducibility
again forces $J=cI$.  But

\ba{
 U(g_1)\,h^\sigma _{gH}\a=U(g_1)\,U(\sigma (gH))h\cr
\a=U(g_1\sigma (gH))\,h\cr
\a=U(g_1g\,\alpha (g))\,h\cr
\a=U(g_1g\,\alpha (g_1g)\alpha (g_1g) ^{-1} \alpha (g))\,h\cr
\a=\chi  (\alpha (g_1g) ^{-1} \alpha (g))\,h^\sigma _{g_1gH}.
 \cr}
 \eno29)
 \ea
That is, the vectors $h^\sigma _m$ are ``almost'' covariant under the action
of $G$, with a residual phase factor (multiplier).  This means that the {\sl
states\/} $P_m$ transform covariantly, i.e.,

\be
 U(g_1)\,P_m\,U(g_1)^*=P_{g_1m}.
\eno30)
 \ee
Thus 

\ba{
U(g_1)\,J\,U(g_1)^{-1}\a =\int_M\,d\mu _{_M}(m)\,U(g_1)\,P_m\,U(g_1)^*\cr
\a=\int_M\,d\mu _{_M}(m)\,P_{g_1m}\cr
\a=\int_M\,d\mu _{_M}(m')\,P_{m'}\cr
\a=J,
 \cr}
 \eno31)
 \ea
using the unitarity of $U(g_1)$ and the invariance of $d\mu _{_M}$.  This
proves that $J=cI$, provided the integral converges in some sense. 
\skp
\noindent {\sl Note: \sp} The above proof becomes shorter if one works
directly with the states $P_m$ rather than the frame vectors;  however, it is
important to see the action of $G$ on the frame vectors in terms of the
multipliers since this will play a role in the next section, where we look for
representations of $G$ on spaces of holomorphic functions.  \#
\skp 

 A necessary condition for the {\sl weak\/} convergence of the integral is
that 

\be
 \langle\, h\,|\,J\,h\,\rangle=\int_M\,d\mu _{_M}(m)\,\,|\,\langle\,
h^\sigma  _m\,|\,h\,\rangle\,|\,^2 <\infty,
\eno32)
 \ee
i.e., that $\tilde h(m)\equiv \langle\, h^\sigma _m\,|\,h\,\rangle$ be in
$L^2(d\mu _{_M})$.  As before, this condition is also sufficient, and the same
terminology is used:  $h$ is called admissible, and $U$ is called
square--integrable.  The action of $U$ in the Hilbert space of functions
$\ftil(m)=\langle\, h^\sigma _m\,|\,f\,\rangle$ is somewhat more
complicated than earlier because of the multiplier.  (If we simply dropped
the multipliers we would no longer have a unitary representation of $G$ but
a {\sl projective\/} representation; see Varadarajan [1970].)

The above construction has the advantage that the trivial part of the action
of $G$ is factored out, thereby improving the chances that the integral $J$
converges.  As mentioned above, it depends on the existence of the
invariant measure $d\mu _{_M}$ on $G/H$, which is not guaranteed.  Note
that for fixed $g\in G$, the stability subgroup for the vector $U(g)h$ is $gHg
^{-1} $, i.e., a subgroup of $G$ {\sl conjugate\/} to $H$.  In general, however,
the stability subgroups of two admissible vectors $h_1$ and $h_2$ may not
be conjugates.  It may happen that $G/H_1$ has an invariant measure while
$G/H_2$ does not.  It pays, therefore, to choose the vector $h$ very
carefully.  Intuition suggests that $h$ should be chosen so as to {\sl
maximize\/} the stability subgroup, since this will minimize the
homogeneous space $M$ and improve the chances for convergence. 
Furthermore, maximal use of symmetry would seem to make it more likely
that an invariant measure exists on the quotient.  More will be said about
this in the next section, in connection with the ``weight'' of a representation.

We will refer to the frame $\{\,h^\sigma _m\,\}$ as a {\sl homogeneous
$G$--frame\/} associated with $(U, h)$. The dependence on $\sigma $ will
usually be suppressed, since a change in (the local sections) $\sigma $ gives
an equivalent frame.

\secskp

 \noindent {\bf 3.4.  Onofri's Holomorphic $G$--Frames  }
\def\rightheadline{\tenrm\hfil {\sl 3.4. 
 Onofri's Holomorphic $G$--Frames}\hfil\folio}  

 \skp
\noindent The frames associated with the windowed Fourier transform
and the canonical coherent states are similar in that both provide ``phase
space'' representations of functions in $L^2(\rl^s)$.  The distinguishing
feature is that the representation  determined by the canonical coherent
states is in terms of {\sl holomorphic\/} (analytic) functions.  Neither of the
two general group--theoretic methods covered so far explains the origin of
this analyticity, yet it has been found that all such group--related
phase--space representations have analytic counterparts.  Moreover,
analyticity will play a key role in the coherent--state representations  of
relativistic quantum mechanics  (next chapter) and quantum field theory 
(chapter 5). The method to be described in this section assumes we are
dealing with a compact, semi\-simple group. Since the coherent--state
representations we will develop for relativistic quantum mechanics  are
based on the Poincar\'e group, which is neither compact nor se\-mi\-simple,
the present considerations  do not apply directly to the main body of this
book (chapters 4 and 5).  Nevertheless, we describe them in considerable
detail in the hope that they may shed some light on our later constructions,
which are still not well--understood in general terms.

 Let us  return to the
canonical coherent states in order to isolate the property leading to
analyticicy and find its generalization to other groups.  We follow an
approach first advocated by Onofri [1975].  For related developments, also see
Perelomov [1986].  Again consider the
three--di\-men\-sion\-al Weyl--Heisenberg group $\cal W$, whose (real) Lie
algebra  {\bf w} has a basis $\{-iS, -iT, -iE\}$ which is represented on
$L^2(\rl)$ by $S\to X$, $T\to P$ and $E\to I$. Recall that we arrived at the
canonical coherent states $\chi _z$ as eigenvectors of the non--hermitian
operator $A=X+iP$, which represents the complex combination $S+iT$ of
generators in {\bf w}.  Since {\bf w} is a  real Lie algebra, we must {\sl 
complexify \/} it in order to consider such combinations of its generators. 
This is done in the same way as complexifying a real vector space, namely
by taking the tensor product with the field of complex numbers.  With
obvious notation, ${\bf w}_c={\bf w}\otimes \cx={\bf w}+i{\bf w}$.  As will
be seen below,  complex combinations such as $S\pm iT$ play a very
important role in the theory of {\sl   real\/} Lie groups, exactly for the same
reason that complex eigenvectors and eigenvalues are necessary in order to
study real matrices.

We begin by rederiving the canonical coherent states from an algebraic
point of view which shows their relation to the vectors $h_{p,x} $ associated
with the windowed Fourier transform and pinpoints the property which
makes them analytic.  In the last section we found that 

\be
 h_{p,x} =e^{ipX}\,e^{-ixP}\,h.
\eno1)
 \ee
Now if $B$ and $C$ are operators such that $[B,C]$ commutes with both $B$
and $C$, then the Baker--Campbell--Hausdorff formula (Varadarajan [1974]) 
reduces to

\be
e^B e^C= e^{{\sc {{ 1 }\over{2  }}}[B,C]}\,e^{B+C}.
\eno2)
 \ee
Since $[ipX,-ixP]=ipxI$, we therefore have

\be
 h_{p,x} =\exp(ipx/2)\,\exp\left( ipX-ixP \right)\,h.
\eno3)
 \ee
Substituting 

\be
 X=(A^*+A)/2 \qquad \hbox{and}\qquad -iP=(A^*-A)/2
\eno4)
 \ee
and defining $z\equiv x-ip$, we get

\be
 h_{p,x} =\exp(ipx/2)\,\exp\left[  \zbar A^*/2-zA/2\right]\,h.
\eno5)
 \ee
So far, all our manipulations have been justifiable since we have only
exponentiated skew--adoint operators.  The next one is more delicate since
the operators to be exponentiated are not skew--adjoint; it will be justified
later.  Using the Baker--Campbell--Hausdorff formula again with $B=\zbar
A^*/2$ and $C=-zA/2$, write

\be
 h_{p,x}=\exp\left( ipx/2-\zbar z/4 \right)\,\exp\left( \zbar A^*/2 \right)\,
\exp\left(  -zA/2\right)\,h.
\eno6)
 \ee
If we choose 

\be
 h(x')=N\exp\left( -x'\,^2/2 \right)=\chi _0(x')
\eno7)
 \ee
(where $N=(2\pi )^{ -1/4}$ and $\chi _0$ is just the canonical coherent
state $\chi _z$ with $z=0$), then $Ah=0$, hence $\exp\left( -zA/2
\right)\,h=h$ and

\be
 h_{p,x} =\exp\left( ipx/2-\zbar z/4 \right)\,\exp\left(  \zbar A^*/2\right)\,
\chi _0.
\eno8)
 \ee
We claim that 

\be
 \exp\left(  \zbar A^*/2\right)\, \chi _0=\chi_z.
\eno9)
 \ee
This can be seen by applying $A$ to the left--hand side, then using $A\chi
_0=0$ and $[A,A^*]=2I$:

\ba{
 A\exp\left( \zbar A^*/2 \right)\,\chi _0
\a=\left[A, \exp\left( \zbar A^*/2 \right)\right]\,\chi _0\cr
\a=\zbar\,\exp\left( \zbar A^*/2 \right)\,\chi _0,
 \cr}
 \eno10)
 \ea
which shows that $ \exp\left( \zbar A^*/2 \right)\,\chi _0 $ equals 
$\chi _z$ up to a  constant factor, which is easily shown to be unity.  That is,

\be
 h_{p,x}=\exp\left(ipx-\zbar z/4  \right)\,\chi_z,
\eno11)
 \ee
so the canonical coherent states are a special case (modulo the
$z$--dependent factor in front) of the frame vectors $h_{p,x} $ for the
choice $h=\chi _0$.  Note that the corresponding {\sl states  \/} are related
by

\be
 \,|\,h_{p,x} \,\rangle\langle\, h_{p,x} \,|\,=\exp\left(  -|z|^2/2\right)
\,|\,\chi _z\,\rangle\langle\, \chi _z\,|\,  ,
\eno12)
 \ee
giving the weight function on the right--hand side in the bargain.  The
reason for this is that the $h_{p,x} $'s were obtained by unitarily translating
$h$, whereas the operator  $\exp\left( \zbar A^*/2 \right)$ which
``translates'' $\chi _0$ to $\chi _z$ is not unitary but results in $\|\chi
_z\|=\exp\left( |z|^2/4 \right)\,\|\chi _0\|$; the weight function then corrects
for this.

We can now justify the fine point we glossed over earlier.  The expression

\be
\exp\left( \zbar A^*/2 \right)\,\exp\left(  -zA/2\right)\,\chi _0
\eno13)
 \ee
makes sense because

\item{(a)} 
\ba{
e^{-zA/2}\chi _0=\chi _0
\cr}
\eno14)
\ea
since $A\chi _0=0$, and 
\item{(b)} $\chi  _0$ is an {\sl   analytic vector\/} (Nelson [1959]) for the
operator $A^*$, since 

\ba{
\sum_{n=0}^\infty {{ 1 }\over{ n! }} \a \,\|\left( \zbar A^*/2 \right)^n\,\chi
_0\,\| =\sum_{n=0}^\infty  {{ |z|^n }\over{ 2^n\,n! }} \|{A^*}^n\chi_0\|\cr
\a=\sum_{n=0}^\infty  {{ |z|^n }\over{ 2^n\,n! }} 2^{n/2}\sqrt{n!}\,\|\chi
_0\|<\infty \qquad \forall z\in \cx\ ,
 \cr}
 \eno15)
 \ea
where $\|{A^*}^n\chi_0\|=2^{n/2}\sqrt{n!}$ follows from $[A,A^*]=2I$ and
 $A\chi _0=0$.
 \par \skp
As will be seen below, the key to constructing representations of
real Lie groups on spaces of analytic functions will be to 
\item{(a)} complexify the Lie algebra; 
\item{(b)} find a counterpart to $A\chi _0=0$; 
\item{(c)} define counterparts to $\chi _z=\exp(\zbar A^*/2)\,\chi_0$.\par
\skp
 Before embarking on this task, we must make a brief excursion
into the structure theory of Lie algebras.  For background and
details, see Helgason [1978] or Hermann  [1966].  The Lie groups  and
--algebras we consider below are assumed to be {\sl real  \/} and {\sl
semisimple}. (A Lie algebra is semisimple if it contains no proper abelian
ideals; a Lie group is semisimple if its Lie algebra is semisimple.) Since $\cal
W$ is not semisimple (the subspace spanned by $-iE$ is an ideal of {\bf w}),
these results do not apply to it, strictly speaking. However, as will be shown
in section 3.6, $\cal W$ can be obtained as a (contraction) limit  of a simple
Lie group ($SU(2)$), and this turns out to be sufficient for our purpose.  Thus,
let $G$ be an arbitrary real, semisimple Lie group and {\bf g} its Lie algebra.
The following is known:  \skp
\item{1.} Every element $X$ of {\bf g} defines a linear map ${\rm ad} 
\,X$ on {\bf g} by  \break
${\rm ad} \,X(Y)=[X,Y]$.  Similarly, every $Z\in {\bf
g}_c $ defines  a complex--linear map (denoted by ${\rm ad} \,Z$) on
${\bf g}_c $.  It therefore makes sense to look for eigenvalues and
eigenvectors of ${\rm ad}\,Z$.

\item{2.} {\bf g} has a (Lie) subalgebra {\bf h}  called a {\sl Cartan
subalgebra,  \/} defined as  a maximal abelian subalgebra of
{\bf g} satisfying an additional technical condition.    {\bf h} is analogous to a
``maximal commuting set of observables'' in quantum mechanics.  Its
complexification, denoted by ${\bf h}_c$, is a maximal abelian subalgebra of
${\bf g}_c $. (The technical condition on {\bf h}   essentially
ensures that each of the linear maps ${\rm ad}\,H$, with $H\in {\bf h}_c$,
has a complete set of eigenvectors in ${\bf g}_c$, hence that all the linear
transformations ${\rm \,ad\,} H$ with $H\in {\bf h}_c $ can be diagonalized
simultaneously.)

\item{3.} For every complex--linear form $\alpha {:}\ {\bf h}_c \to \cx$\ ,
let 
\be
 {\bf g} ^\alpha =\lbrace Z\in {\bf g}_c\,|\,\ 
[H,Z]=\alpha (H)Z\quad \forall\,H\in {\bf h}_c     \rbrace.
\eno16)
 \ee
That is, ${\bf g} ^\alpha $ is the set of all vectors in ${\bf g}_c $ which are
common eigenvectors of  ${\rm ad\,}H$ for every $H$  in ${\bf h}_c$, with
corresponding eigenvalue $\alpha (H)$.  For generic $\alpha $, no such
eigenvectors will exist, hence ${\bf g} ^\alpha =\{0\}$.  When ${\bf g} ^\alpha
\ne \{0\}$, $\alpha $ is called a {\sl  root \/} (of ${\bf g}_c $ with respect to
${\bf h}_c $), each $Z$ in ${\bf g}_c $ is called a {\sl root vector  \/} and
${\bf g}^\alpha  $ is called a {\sl  root subspace \/} of ${\bf g}_c $.  Clearly,
the zero--form $\alpha (H)\equiv 0$ is a root, since the fact that ${\bf h}_c$
is abelian means that ${\bf g} ^0$ contains ${\bf h}_c$.  The fact that ${\bf
h}_c $ is maximal--abelian means that actually ${\bf g} ^0={\bf h}_c $, since
every $Z$ in ${\bf g}^0$ commutes with all $H$ in ${\bf h}_c $. 

\item{4.}  Let $Z_\alpha \in {\bf g}^\alpha $ and $Z_\beta \in {\bf g}^\beta
$, where $\alpha $ and $\beta $ are arbitrary linear forms on ${\bf h}_c $.
  Then the Jacobi identity implies that for every $H$ in ${\bf h}_c$,
\ba{
\left[ H, [Z_\alpha ,Z_\beta ] \right]\a=\left[ [H,Z_\alpha ],Z_\beta  \right]
+\left[ Z_\alpha ,[H,Z_\beta ] \right]\cr
\a=\left( \alpha (H)+\beta (H) \right)\,[Z_\alpha ,Z_\beta ],
 \cr}
 \eno17)
 \ea
thus $[Z_\alpha ,Z_\beta ]\in {\bf g} ^{\alpha +\beta }$.  This statement is
abbreviated as 
\be
 \left[ {\bf g} ^\alpha ,{\bf g}^\beta  \right]\,\subset {\bf
g}^{\alpha +\beta }\qquad \forall \alpha ,\beta \in {\bf h}_c ^*.
\eno18)
 \ee

\item{5.}  The set of all {\sl  nonzero \/} roots
is denoted by $\Delta $.  ${\bf g}_c $ has a direct--sum decomposition {\sl 
(root space decomposition)\/}  as
\be
 {\bf g}_c={\bf h}_c +\sum_{\alpha \in\Delta } {\bf g}^\alpha,
\eno19)
 \ee
and each ${\bf g}^\alpha $ (with $\alpha \in \Delta $) is a one--dimensional
subspace of ${\bf g}_c$.

\item{6.}  The {\sl  Killing form \/} of ${\bf g}_c $ is the bilinear symmetric
form defined by 
\be
 B(Z,Z')=\hbox{trace}\left( {\rm ad}\,Z\,{\rm ad}\,Z' \right).
\eno20)
 \ee
It is non--degenerate on ${\bf g}_c $, and its restriction to ${\bf h}_c $ is also
non--degenerate.  Since a non--degenerate bilinear form on a vector space
defines an isomorphism between that space and its dual, the restriction of
$B$ to ${\bf h}_c $ defines an isomorphism between ${\bf
h}_c $ and ${\bf h}_c ^*$.  The vector in ${\bf h}_c $ corresponding to
$\alpha \in {\bf h}_c ^*$ is denoted by $H_\alpha $.  It is defined by 
\be
 B(H_\alpha ,H)=\alpha (H)\qquad \forall H\in {\bf h}_c.
\eno21)
 \ee
Note that the vector corresponding to $\alpha \equiv 0$ is $H_0=0$.

\item{7.} If $\alpha \in \Delta $, then $-\alpha \in \Delta $ and $\alpha
(H_\alpha )\equiv B(H_\alpha ,H_\alpha )\ne 0$.  Furthermore, 
\be
 \left[ {\bf g}^\alpha ,{\bf g} ^{-\alpha } \right]=\cx\,H_\alpha ,
\eno22)
 \ee
i.e. the set of all brackets $[Z_\alpha , Z_{-\alpha }]$ with $Z_\alpha \in {\bf
g}^\alpha $ fills out the one--dimensional subspace spanned by $H_\alpha
$.  ($H_\alpha \ne 0$ since $\alpha \in\Delta $.)

\item{8.} Any two root subspaces ${\bf g}^\alpha $ and ${\bf g}^\beta $
with $\alpha +\beta \ne 0$ are ``orthogonal'' with respect to $B$.

\item{9.} It is possible to choose (non--uniquely) a subset $\Delta ^+$ of
$\Delta $, called a set of {\sl  positive roots, \/} such that 
\itemitem{(a)} If $\alpha $ and $\beta $ belong to $\Delta ^+$  and $\alpha
+\beta $ is a root, then $\alpha +\beta $ belongs to $\Delta ^+$.
\itemitem{(b)} The set $\Delta ^-\equiv -\Delta ^+$ is disjoint from $\Delta
^+$.
\itemitem{(c)}$\Delta =\Delta ^+\cup\Delta ^-$.
\skp
It follows from (4) and (9) that 

\be
 {\bf n} ^+\equiv \sum_{\alpha \in\Delta ^+} {\bf g} ^\alpha 
\qquad \hbox{and}\qquad {\bf n} ^-\equiv \sum_{\alpha \in\Delta ^-} 
{\bf g} ^\alpha 
\eno23)
 \ee
are Lie subalgebras of ${\bf g}_c $, and by (5), 

\be
 {\bf g}_c= {\bf h}_c +{\bf n} ^+ +{\bf n}^-.
\eno24)
 \ee
Since there can only be a finite number of roots, (4) and (9) also imply that
after taking a finite number of brackets of elements in either ${\bf n} ^+$ or
${\bf n} ^-$ we obtain zero; that is,  the subalgebras ${\bf n} ^\pm$ are {\sl  
nilpotent\/}.  We may choose a basis for ${\bf g}_c $ as follows: Pick a
non--zero vector $Z_\alpha $ from each ${\bf g}^\alpha $ with $\alpha \in
\Delta ^+$. Then by (7) there is a unique vector  $Z_{-\alpha }$ in ${\bf
g}^{-\alpha}$ with 

\be
 [Z_\alpha , Z_{-\alpha }]=2H_\alpha/\alpha (H_\alpha ) .
\eno25)
 \ee 
Since $Z_\alpha $ and $Z_{-\alpha }$ are root vectors, they also satisfy

\ba{
 [H_\alpha ,Z_\alpha ]\a=\alpha (H_\alpha ) Z_\alpha  \cr
 [H_\alpha ,Z_{-\alpha} ]\a=-\alpha (H_\alpha)  Z_{-\alpha}  ,
\cr}
 \eno26)
 \ea
and $\alpha (H_\alpha )\ne 0$ (property (7)).  If we define

\be
 C_\alpha =H_\alpha /\alpha (H_\alpha ),
\eno27)
 \ee
then each triplet $(Z_\alpha , C_\alpha , Z_{-\alpha })$ 
spans a Lie subalgebra of ${\bf g}_c $ which is 
isomorphic to ${\bf sl}(2,\cx\,)$:

\be
 [C_\alpha ,Z_{\pm \alpha }]=\pm Z_{\pm \alpha },\qquad 
[Z_\alpha ,Z_{-\alpha }]=2C_\alpha .
\eno28)
 \ee
The root space
decomposition then shows that ${\bf g}_c $ is a direct sum of copies of  ${\bf
sl}(2,\cx\,)$, indexed by $\Delta ^+$.
\skp
The connection with the non--hermitian combinations $X\pm iP$ can now
be explained.  The following argument is heuristic and has no pretense to
rigor.  Its sole purpose  is to motivate the construction of
general holomorphic frames.  It will be made precise in section 3.6 at the
level of unitary  representations, where a clear geometric
interpretation will be given.

Begin with the group $G=SU(2)$, which is a simple, real Lie
group whose Lie algebra ${\bf g} ={\bf su}(2)$ has a basis $\{-iJ_1, -iJ_2,
-iJ_3\}$ satisfying

\ba{
 [J_1, J_2]\a=iJ_3\cr
[J_2, J_3]\a=iJ_1\cr
[J_3, J_1]\a=iJ_2.
 \cr}
 \eno29)
 \ea
We first show that {\bf w} can be obtained as a ``contraction limit''  of {\bf
g}.\footnote*{The idea of group contractions is due to
In\"on\"u and Wigner [1953].}
 Choose a positive number $\kappa $ and define $K_1=\kappa J_1$,
$K_2=\kappa J_2$ and $K_3=\kappa^2 J_3$.  These form a new basis for 
{\bf g}, with 

\ba{
[K_1,K_2]\a=iK_3\cr
[K_2, K_3]\a=i\kappa ^2K_1\cr
[K_3,K_1]\a=i\kappa ^2K_2.
 \cr}
 \eno30)
 \ea
In the limit $\kappa \to 0$, {\bf g} becomes isomorphic to {\bf w}. We will
see in section 3.6 that within a unitary irreducible representation  of
$SU(2)$, the operator $K_3$ can be chosen so that $K_3\to I$ as $\kappa \to
0$, hence we may interpret the limits of $K_1$ and $K_2$ as $X$ and $P$,
respectively.  

Now apply the above structure theory to ${\bf g}_c $, which is just \break
${\bf sl}(2,\cx\,)$.  The direct sum of copies of ${\bf sl}(2,\cx\,)$ therefore
reduces to a single term.  For the Cartan subalgebra of {\bf g} we can choose 
${\bf h} =\rl\,J_3$ (the one--dimensional subspace spanned by $J_3$), so
that    ${\bf h}_c =\cx\,J_3$.  Two linearly independent root vectors are
given by  $J_\pm=J_1\pm iJ_2$, with 

\be
 [J_3, J_+]=J_+,\qquad [J_3, J_-]=-J_-.
\eno31)
 \ee
We choose $\alpha (J_3)=1$ as the single ``positive'' root
and $J_\pm$ for the basis vectors of the two one--dimensional root
subspaces.  Then 

\be
 [J_+,J_-]=2J_3
\eno32)
 \ee
shows that $C_\alpha $ corresponds to $J_3$.
Setting $K_\pm=\kappa J_\pm$ and $K_3=\kappa ^2J_3$, and
taking the limit $\kappa \to 0$, we obtain the correspondence

\ba{
K_+\a \to S+iT\mapsto  X+iP=A\cr
K_-\a \to S-iT \mapsto X-iP=A^*\cr
K_3\a \to E  \mapsto I,
 \cr}
 \eno33)
 \ea
where we have used ``$\mapsto$'' to denote the representation  of {\bf w} on
 $L^2(\rl)$.  [This correspondence is not unique; for example,
$K_+\to A^*$, $K_-\to A$, $K_3\to -E$ is equally good.  As will be shown in
section 3.6, {\sl   both\/}  of these correspondences actually occur as weak
limits, due to the fact that an irreducible representation  of $SU(2)$
contracts to a {\sl   reducible\/}  representation  of {\bf w}.]   Note that the
three roots $\{\kappa ^2, 0, -\kappa ^2\}$  all merge into a single root, zero,
in the contraction limit.  Thus the operators $A$ and $A^*$ are interpreted as
the contraction limits of root vectors.
\skp

\noindent {\sl   Note:\/}  Another way of seeing the
importance and naturality of $A$ and $A^*$ is in the context of the
Kirillov--Kostant--Souriau theory, sometimes called ``Geometric
Quantization,'' applied to the {\sl Oscillator group\/}   (Streater [1967];  see
section 3.6).  For background on Geometric Quantization, see  Kirillov [1976], 
Kostant [1970] and Souriau [1970];  see also Guillemin and Sternberg
[1984], Simms and Woodhouse [1976] and \'Sniatycki [1980]. 

 \skp
We are, at last, ready to generalize the construction of
group--representations on spaces of holomorhic functions to other groups.
We begin with a semisimple, real Lie group $G$ and a unitary irreducible
representation  $U$ of $G$ on a Hilbert space $\cal H$.  To avoid technical
difficulties, we will here assume that $G$ is {\sl compact.}  (The reason for
assuming compactness, as well as ways to get around it, will be discussed
below.)  Then the irreducibility of $U$ implies that ${\cal H}$ be
finite--dimensional, hence the  operators $U(g)$ representing 
group operations are just unitary matrices and the operators $U(X)$
representing elements of {\bf g} are skew--adoint matrices. (Sometimes the
operator representing $X$  is written as $dU(X)$ to emphasize its
``infinitesimal'' nature;  we will write it as $U(X)$ to keep the notation
simple.) At the Lie algebra level, $U$ extends, by complex--linearity, to a
representation $T$ of  ${\bf g}_c $:

\be
 T(X+iY)\equiv U(X)+iU(Y).
\eno34)
 \ee
$T$ is the unique representation of ${\bf g}_c $ that extends $U$; that is,
for all  $ c_1, c_2\in \cx\ $ and all $ Z_1,Z_2\in {\bf g}_c$,

\ba{
T(c_1Z_1+c_2Z_2)\a=c_1T(Z_1)+c_2T(Z_2)  \cr
T\left( [Z_1, Z_2] \right)\a=\left[ T(Z_1), T(Z_2) \right] \cr
T(X)\a=U(X) \qquad \hbox{if}\qquad X\in {\bf g} \subset {\bf g}_c.
 \cr}
 \eno35)
 \ea
The matrices $T(Z)$ are no longer skew--adjoint, but because of the finite
dimensionality of ${\cal H}$, there is no problem in exponentiating them to
give a  representation of $G_c$ on ${\cal H}$, which
we also denote by $T$.  Thus, the representative of the  group element
$\exp Z$ of $G_c$ is defined by\footnote*{Since $G$ is compact, it is
{\sl exponential;\/}  that is, every group element can be written in the form
$\exp Z$ for some $Z$.}

\be
 T(\exp Z)= \exp[T(Z)],\qquad Z\in {\bf g}_c.
\eno36)
 \ee
{\sl Note:\/}  If $G$ is non--compact, $\exp Z$ is still well--defined but
the right--hand side of the above equation is problematic since for any
non--trivial representation $U$, ${\cal H}$ is infinite--dimensional and
$T(Z)$ will, in general,  be a  non--skew--adjoint, unbounded operator.  (Even
the definition  $T(Z)=T(X)+iT(Y)$ becomes troublesome, since the
skew--adjoint operators $T(X)$ and $T(Y)$ may both be unbounded and
their domains may have little in common; additional assumptions must be
made.)  In the case of $\cal W$, this was resolved by restricting $\exp[T(Z)]$
to act on analytic vectors.  A similar approach is used in extending the
present construction to non--compact $G$.  For the present, we continue to
assume that $G$ is compact to avoid this problem. \#
\skp
\noindent \bf Lemma 3.2..   \sl 
\item{(a)} $g\mapsto T(g)$ is an irreducible {\rm (non--unitary)   \/}
representation  of $G_c$. 
\item{(b)} The map $Z\mapsto T(\exp Z)$  is
analytic as a map from the complex vector space ${\bf g}_c $ to the complex
matrices on ${\cal H}$. 
\skp\rm
\noindent {\bf Proof.}  If $A$ is a matrix which commutes with all $T(Z)$ 
for $Z\in {\bf g}_c $, then in particular it commutes with all $U(X)$ for $X\in
{\bf g} $, hence must be a multiple of the identity since $U$ is irreducible. 
Therefore $T$ is irreducible.  To prove (b), note that $Z\mapsto T(\exp Z)$ is,
by definition, the composite of the two analytic maps $Z\mapsto T(Z)$ and
$T(Z)\mapsto \exp [T(Z)]$.
 \qed\skp 
It follows that the map $T$ from $G_c$ (considered as a complex manifold;
see Wells [1980]) to the group $GL ({\cal H})$ of non--singular matrices on
$\cal H$ is also analytic; that is, $T$ is a {\sl   holomorphic representation\/}
of $G_c$, obtained by analytically continuing the representation $U$ of $G$.
Now the point of Onofri's construction is this: We have seen that by choosing
a state which is stable under  $H$,  $U$ can
be reformulated as a representation  of $G$ on a space  of functions $\ftil(m)$
defined over  the homogeneous space $G/H$. In the case $G= \cal W$,
$G/H$ was identified as a {\sl   phase space,} but in general it is not clear that
it can be interpreted as such.  Following Onofri, we will show that:
\skp
\item{(a)} The representation $T$ induces a complex structure on the
homogeneous space $G/H$, making it into a {\sl   complex
manifold\/} on which $G$ acts by {\sl   holomorphic transformations}.
(Such a manifold is called a complex homogeneous
space of $G$, or a {\sl   holomorphic\/}  homogeneous $G$--space.)
\item{(b)} In addition, $G/H$ has the (symplectic) structure of a classical
phase space, and the action of $G$ on $G/H$ is by canonical transformations. 
 Thus it becomes possible to think of $G/H$ as {\sl phase space}.  To
actually identify $G/H$ as {\sl   the\/}  phase space of a classical physical
system, i.e. as the set of dynamical trajectories  followed by that
system, it is necessary for $G$ to include the {\sl   dynamics\/} for the
system, i.e. its evolution group, of which nothing has been said so far.  This
will be discussed in the next chapter.

\skp
The representatives $U(H)$ of the elements $H$ of ${\bf h} $ form a
commuting set of skew--adjoint  matrices, hence can all be diagonalized
simultaneously.  Let $h $ be a common eigenvector:

\be
 U(H)\,h =\lambda (H)\,h,
\eno37)
 \ee
where $\lambda (H)$ is {\sl   imaginary}.  Since $U$ is linear at the Lie
algebra level, it follows that  $\lambda $ is a linear functional on {\bf h},
called the {\sl   weight\/} of $h $.  (Roots are simply
weights in the adjoint representation, where ${\cal H}$ is replaced by ${\bf
g}_c $ and $U(H)$ by ad$H$.)  For any non--zero element $Z_\alpha $
in ${\bf g}^\alpha $ with $\alpha $ in $\Delta ^+$, we have (remembering
that  $U(H)=T(H)$ since $H\in {\bf h} $)

\ba{
T(H)T(Z_\alpha )\,h \a=\left[  T(H), T(Z_\alpha )\right]\,\,h +
T(Z_\alpha)T(H)\,h \cr
\a=T\left( [H, Z_\alpha ] \right)\,\,h +\lambda (H)\,T(Z_\alpha )\,h \cr
\a=\left( \alpha (H)+\lambda (H) \right)\,T(Z_\alpha )\,\,h .
 \cr}
 \eno38)
 \ea
That is, $T(Z_\alpha )$ ``raises'' the weight by $\alpha $. Similarly, for
$\alpha \in\Delta ^-$, $T(Z_\alpha )$ {\sl   lowers\/}  the weight by
$-\alpha $.  Since non--zero
vectors with different weights are linearly independent and ${\cal H}$ is
finite--dimensional, it follows that $\cal H$ must contain a non--zero vector
 with {\sl   lowest weight,\/} i.e. such that 

\be
 T(Z_\alpha )\,h=0\qquad \forall \alpha \in \Delta ^-.
\eno39)
 \ee
Equivalently, 

\be
 T(Z)\,h =0\qquad \forall Z\in {\bf n} ^-.
\eno40)
 \ee
For the group $\cal W$, $\,h $ was the ''ground state''$\,\chi  _0$ and the
above equations correspond to $T(-iE)\,\chi  _0=-i\chi  _0$ (so $\lambda
(-iE)=-i$) and $A\chi _0=0$.

Consider the subalgebra 

\be
 {\bf b}={\bf h}_c +{\bf n} ^+
\eno41)
 \ee
of ${\bf g}_c $, called a {\sl   Borel subalgebra.\/}  If $N\in {\bf n} ^+$ and
we denote by $\bar N$ its  complex--conjugate with respect to the real
subalgebra {\bf g}, then $\bar N\in {\bf n} ^-$.  Hence for arbitrary
$Z=H+N\in {\bf b} $, we have

\be
 T(\bar Z)h=T(\bar H)h+T(\bar N)h=T(\bar H)h,
\eno42)
 \ee
and $\bar H$ belongs to ${\bf h}_c $ since $[\bar H, {\bf h}_c ]=0$.
Extending $\lambda $  by complex--linearity to ${\bf h}_c $, we
therefore have

\be
T(\bar Z)h=\lambda (\bar H)\,h  ,
\eno43)
 \ee
hence 

\be
\exp\left[ T(\bar Z) \right]\,h=
\exp\left[ \lambda (\bar H) \right]\,h.
\eno44)
 \ee
The subgroup $B=\exp({\bf b})$ is called a {\sl   Borel
subgroup\/} of ${\bf g}_c $.  For $Z$ as above, let $b=\exp Z\in B$, $\bar
b=\exp \bar Z$ and $\pi (\bar b)=\exp[\lambda (\bar H)]$.  Then

\be
 T(\bar b)\,h=\pi (\bar b)\,h.
\eno45)
 \ee
Since $\lambda (H)$ is imaginary for $H\in {\bf h} $, complex--linearity
implies that $\lambda (\bar H)=-\br{\lambda (H)}$ for $H\in {\bf h}_c $. 
Hence 

\be
 \pi (\bar b)=\br{\pi (b)}\,^{-1} , \qquad b\in B.
\eno46)
 \ee

The map $\pi{:}\ B\to \cx ^* $ satisfies 
$\pi (b_1b_2)=\pi (b_1)\,\pi (b_2)$, i.e. it  is a   character of $B$. 
Furthermore, since $\pi (b)$ is analytic in the group parameters of $b$, 
Onofri calls it a  {\sl   holomorphic character\/} of $B$.

Notice that  the {\sl   state\/} corresponding to $h$ (i.e.,
the one--dimen\-sion\-al subspace spanned by it) is invariant under $B$.  If
we restrict ourselves to the {\sl   real\/} group $G$, this means that the state
is invariant under the subgroup $H$. If $H$ is the {\sl   maximal\/}  subgroup
of $G$ leaving this state invariant, then the weight $\lambda $ is called {\sl  
non--singular.}  In that case, $H$ plays the same role as it did in the last
section: it is the {\sl   stability subgroup\/}  of the state. 
We will assume this to be the case; if it is not (in which case $\lambda $ is
{\sl   singular}), the present considerations still apply but in modified
form.  Note that in the non--singular case, the stability subgroup is {\sl  
abelian.} 

We now adapt the construction of the last section in a way which respects 
the complex--analytic structure of $G_c$.  Introduce the notation 

\be
 ({T(g)^*}) ^{-1} \equiv T^\#(g),\qquad g\in G_c.
\eno47)
 \ee
Since the representation  $U$ of $G$ is unitary, $U(X)^*=-U(X)$ for $X\in {\bf
g} \,$; hence for $Z\in {\bf g}_c $, we have $T(Z)^*=-T(\bar Z)$.   It follows
that for  group elements $g=\exp(Z)$ of $G_c$, 

\be
 T(g)^\#=T(\bar g),\qquad g\in G_c,
\eno48)
 \ee
where $\bar g=\exp(\bar Z)$ and, in particular, $T(g)^\#=T(g)=U(g)$ for
$g\in G$.  Define the vectors

\be
 h_g=T(g)^\#\,h,\qquad g\in G_c,
\eno49)
 \ee
which, when restricted to $g\in G$, coincide with the earlier frame vectors
but are  anti--holomorphic in the group parameters of $G_c$.  An arbitrary
vector $f\in {\cal H}$ defines a  holomorphic function on $G_c$ by

\be
\ftil(g)\equiv \langle\, h_g\,|\,f\,\rangle.
\eno50)
 \ee
For arbitrary $b\in B$,

\ba{
h_{gb}=T(g)^\#T(b)^\#h=\br{\pi (b)}\, ^{-1} h_g ,
 \cr}
 \eno51)
 \ea
hence 

\be
 \ftil(gb)=\pi (b) ^{-1} \ftil(g).
\eno52)
 \ee
\skp
\noindent {\sl   Note:\/} \quad The reader familiar with fiber bundles
(Kobayashi and Nomizu [1963, 1969]) will recognize the above equation as the
condition defining a holomorphic section of the holomorphic line bundle
associated to the principal bundle $B\to G_c \to G_c/B$ by the character $\pi
{:}\ B\to\cx^*$. 
 We now proceed to construct this section in a naive way, that is,
without assuming any knowledge of bundle theory.  \#
\skp
The above shows that the {\sl   state\/}  determined by $h_g$ depends only
on the left coset $gB$, which we denote by $z$.  Let 

\be
 {\cal Z}=G_c/B
\eno53)
 \ee
be the left coset space.  Now
\item{(a)} $G_c$ is a complex manifold;
\item{(b)} $B$, as a complex subgroup, is a complex submanifold of $G_c$;
\item{(c)} the projection map $G_c \to {\cal Z}$ is holomorphic.
\par
\noindent Hence it follows that ${\cal Z}$ is a  complex manifold.  This
means that  a neighborhood  of each point $z=gB\in {\cal Z}$ can be
parametrized by a local chart, i.e. a set of local complex coordinates $(z_1,
\ldots ,z_n)$  (say, with $(0, \ldots,0)$ corresponding to $z$), and the
transformation from one local chart to another on overlapping
neighborhoods is a local holomorphic function.  In the case of $G=\cal W$, $B$
corresponds  (under the contraction limit) to the complex subalgebra
spanned by $E$ and $A$, and ${\cal Z}$ can be identified with $\cx$, hence
only a single chart is needed to cover all of ${\cal Z}$.  In general, more than
one chart is necessary.  For $G=SU(2)$, we will see that ${\cal Z}$ is the
Riemann sphere $S^2$, hence two charts are needed; however, the north pole
has measure zero, and a single chart will do for $S^2\backslash
\{\infty\}\approx \cx$.

Since $G_c$ acts on itself by holomorphic transformations, its action on
${\cal Z}$ is also by holomorphic transformations.  This means the
following:  For $g_1\in G_c$ and $z=gB\in {\cal Z}$, let $w(z)=g_1z\equiv
(g_1g)B$.  If $(z_1,\ldots,z_n)$ and $(w_1,\ldots,w_n)$ are local charts in
neighborhoods of $z$ and $w$, respectively, then the mapping $\phi {:}\ 
(z_1,\ldots,z_n) \mapsto (w_1,\ldots,w_n)$ is holomorphic in a neighborhood
of $(0,\ldots,0)$.  (It must, of course, be locally invertible with holomorphic
inverse.)

We could proceed as in section 3 and consider the states

\be
 P_z={{ \,|\,h_g\,\rangle\langle\, h_g\,|\, }\over{\langle\,
h_g\,|\,h_g\,\rangle  }}, \qquad z\equiv gB\in {\cal Z},
\eno54)
 \ee
where we must now divide by $\|h_g\|^2$ since the non--unitary operator
$T(g)$ does not preserve norms.  However, this would spoil the holomorphy
which we are attempting to study.   Instead, proceed as follows: Choose an
arbitrary reference point $a$ in $G_c$ and define the {\sl   holomorphic
coherent states \/}

 \be
\chi ^a_z={{ h_g }\over{ \langle\, h_a\,|\,h_g\,\rangle }} .
\eno55)
 \ee
As indicated by the notation, the right--hand side depends only on the coset
$z=g B$.  There is no guarantee that the denominator on the right--hand side
is non--zero, but certainly the open set

\be
 U_\alpha =\lbrace g\in G_c\,|\,\ \langle\, h_a\,|\,h_g\,\rangle\ne 0 \rbrace
 \eno56)
 \ee
(which, as indicated,  depends only on the coset $\alpha =aB$) contains
$a$, and its projection $V_\alpha $ to ${\cal Z}$ is an open set containing
$\alpha $  such that $\chi_z^a$ is defined for all $z$ in $V_\alpha $.  Hence
by choosing more than one  reference point $a$, if necessary, we can 
cover ${\cal Z}$ with patches $V_\alpha $ on which the $\chi ^a_z$'s are
defined.

An arbitrary vector $f\in \cal H$ can now be expressed as a {\sl   local
holomorphic function \/} 

\be
 \ftil^a(z)=\langle\,\chi^a_z\,|\,  f\,\rangle
\eno57)
 \ee
of $z$ in $V_\alpha $, with {\sl   transition functions\/} 

\ba{
 \ftil^b(z)\a={{ \langle\, h_g\,|\,h_a\,\rangle }
\over{ \langle\, h_g\,|\,h_b\,\rangle }}\,\ftil^a(z)  \cr
\a\equiv \tau ^b_a(z)\ftil^a(z),\qquad z\in V_\alpha \cap V_\beta.
 \cr}
 \eno58)
 \ea

The reader may wonder where this is all leading, since we are ultimately
interested in the {\sl   real\/} group $G$ and not in $G_c$. Here is the point:  
It is known (Bott [1957]) that the complex homogeneous space ${\cal
Z}=G_c/B$  actually coincides with the {\sl   real\/} homogeneous space
$M=G/H$ used in Perelomov's construction!  For example, consider the
Weyl--Heisenberg group:  $G/H$ is parametrized by $(x,p)$ while $G_c/B$ is
parametrized by $x-ip$, and they are the same set but with the difference
that the latter has gained a complex structure.  The identification of $M$
with ${\cal Z}$ in general can be obtained by noting that  {\sl    as a 
subgroup of  $G_c$,  $G$ acts on ${\cal Z}$  by holomorphic
transformations\/\/\/}; this action turns out to be transitive, and the
isotropy subgroup at the ``origin'' $z_0=B$ is $H$, hence ${\cal Z}\approx
G/H$. In other words, $M$ {\sl   inherits a complex structure from $G_c$, and
the natural action of  $G$ on $M$ preserves this structure.}  Because of this,
we need not deal directly with $G_c$ to reap the benefits of the complex
structure. Let us therefore restrict ourselves to $G$.  Then

\be
 h_g=\langle\,h_a\,|\,h_g\,\rangle\,\chi ^a_z=U(g)\,h,
\eno59)
 \ee
hence $\|\,h_g\,\|=\|\,h\|\,\equiv 1$  and the state corresponding to
$h_g$ is 

\ba{
 \,|\,h_g\,\rangle\,\langle\, h_g\,|\, \a=\left| \,\langle\, h_a\,|\,h_g\,\rangle \, 
\right| ^2\,|\,\chi ^a_z\,\rangle\,\langle\, \chi ^a_z\,|\,  \cr
\a\equiv e^{-\phi (z,\alpha )}\,|\,\chi^a _z\,\rangle\,\langle\, \chi^a _z\,|\,  ,
 \cr}
 \eno60)
 \ea
where $\phi (z,\alpha )\equiv -2\ln |\langle\,h_a\,|\,h_g\,\rangle| $
depends only on $z=gH$ and $\alpha=aH$ and their complex conjugates (it is
{\sl   not\/} analytic).  Notice that in eq. (60), %
{\sl   the left--hand
side, hence also the right--hand side, is independent of $a$.\/} Only the
three individual factors on the right--hand side depend on $a$.  This
becomes important if several patches are needed to cover ${\cal Z}$, since it
means that we can change the reference point without affecting the
smoothness of the frame. With the above definitions, the
resolution of unity derived in section 3 becomes

\be
 \int _Md\mu_M(z)\ e^{-\phi (z,\alpha)}\,|\,\chi^a _z\,\rangle\langle\, \chi
^a_z\,|\,=I,
 \eno61)
 \ee
where we have assumed for simplicity that a single chart suffices.  If more
than one chart is needed, partition $M$ as a (disjoint) union $\cup_nM_n$,
where each $M_n$ is covered by a single chart.  Since, by the above
remark, the integrand is independent of $a$, the corresponding integrals 
$I_n$ form a {\sl   partition of unity\/}  in the sense that
$\sum_nI_n=I$.  Therefore, 
the holomorphic coherent states form a tight frame which we call the
{\sl   holomorphic $G$--frame\/} associated with the representation  $U$
and the lowest--weight vector $\chi $.  In this connection, note that a choice
of lowest weight actually {\sl   determines \/} the representation  $U$ up
to equivalence, hence a more economical terminology would be to call the
above the holomorphic $G$--frame associated with $\chi $.  The
corresponding  inner product is given by

\be
 \langle\, f_1\,|\,f_2\,\rangle=\int_M d\mu _M(z)\,e^{-\phi (z,\alpha )}\ 
\br{\ftil_1^a(z)}\,\ftil_2^a(z) .
\eno62)
 \ee

In the case $G=\cal W$, since ${\cal Z}=\cx$, everything can be done
globally:  Just one chart is needed,  and the reference point $a$ can be
fixed once for all.  Taking $a=1$ (the identity element of $G$) and $\alpha
=0$, we find  that the $\chi_z $'s reduce to the canonical coherent
states and 

\be
 e^{-\phi (z,0)}=\left|  \,  \langle\, \chi _z\,|\,\chi _0\,\rangle  \, \right|^2
=e^{-\zbar z/2}  .
\eno63)
 \ee
Hence in the general case, $e^{-\phi }$ takes the place of the  Gaussian weight
function:  it corrects for the fact that {\sl   holomorphic translations do not
preserve the norm.}  The action of $G$ on the space of local holomorphic
functions $\ftil^a(z)$ has a ``multiplier'':

\ba{
\langle\, \chi ^a_z\,|\,U(g_1)f\,\rangle\a={{ \langle\,
h_g\,|\,U(g_1)\,f\,\rangle}
\over{\langle\, h_g\,|\,h_a\,\rangle  }}  \cr
\a={  { \langle\, h_{g_1 ^{-1} g}\,|\,h_a\,\rangle }
\over{ \langle\, h_g\,|\,h_a\,\rangle }}
\cdot{{ \langle\, h_{g_1 ^{-1} g}\,|\,f\,\rangle }
\over{ { \langle\, h_{g_1 ^{-1} g}\,|\,h_a\,\rangle } }}\cr
\a\equiv \gamma  (z, g_1, \bar\alpha )\ftil^a({g_1}^{-1} z),
 \cr}
 \eno64)
 \ea
where $z= gH$, $\alpha = aH$ and 

\ba{
 \gamma  (z, g_1, \bar\alpha )\a={   {\langle\,  h_g\,|\,U(g_1)\,h_a\,\rangle }
\over{ \langle\, h_g\,|\,h_a\,\rangle }   }\cr
\a={   {\langle\, h_{g_1 ^{-1} g}\,|\,h_a\,\rangle }
\over{  \langle\, h_g\,|\,h_a\,\rangle  }   }\cr
\a={   { \langle\, h_g\,|\,h_{g_1a}\,\rangle }
\over{ \langle\, h_g\,|\,h_a\,\rangle   }   }
 \cr}
 \eno65)
 \ea
is holomorphic in $z$ and anti--holomorphic in $\alpha $. 
\skp
We have mentioned that $M$ inherits a complex structure from $G_c$. 
Actually, this is only part of the story.  Let $\partial$ and $\bar \partial$
denote the external derivatives with respect to $z$ and $\zbar$,
respectively, i.e., in local coordinates, 

\be
 \partial f={\partial f\over{\partial z_k}}\,dz_k,\quad 
\bar \partial f={\partial f\over{\partial \zbar_k}}\,d\zbar_k
\eno66)
 \ee
(summation over $k$ is implied). Consider the
2--form

\be
 \omega =i\partial \bar \partial \phi =
i{\partial^2  \phi  \over{\partial z_j \partial \zbar _k}}
\,dz^j\wedge d\zbar^k,
 \eno67)
 \ee
where $\phi $ is the function in the exponent of the weight function
above.  
\skp
\noindent  {\bf   Theorem. }{\sl
\item{(a)} $\omega $ is  {\sl   closed,\/} i.e. $\partial\omega =\bar
\partial\omega =0$,
\item{(b)} $\omega $ is independent of the reference point $a$,
\item{(c)} $\omega $ is invariant under the action of $G$, and 
\item{(d)} $\omega $ is non--degenerate, if $\lambda $ is non--singular.
}
\skp
\noindent {\bf Proof.}  
 We prove (a), (b) and (c).  For the proof of (d), see Onofri [1975].

\noindent (a) follows from the fact that $\ \bar \partial+\partial=d$ is the
total exterior derivative, hence the  identity $d^2=0$ implies

\be
 \partial^2=0,\quad  \bar \partial^2=0,\quad \bar \partial\partial
+\partial\bar \partial=0.
\eno68)
 \ee

\noindent (c) follows from the fact that $\gamma (z, g_1, \bar \alpha )$ is
holomorphic in $z$, hence $|\gamma |^2$ is harmonic and  $\partial\bar
\partial |\gamma |^2=0$.  But eq. (65) shows that

\be
|\gamma (z,g_1,\alpha )|^2={{ \exp[-\phi (g_1 ^{-1} z,\alpha )] }
\over{ \exp [-\phi (z,\alpha)] }}, 
\eno69)
 \ee
which implies that the pullback $g_1^*\omega $ of $\omega $ under $g_1$
equals $\omega $, i.e. that $\omega $ is invariant.

\noindent (b) follows from (c) and $\phi (z,g_1\alpha )=\phi (g_1 ^{-1} z,
\alpha )$.
\qed\skp

The property (b) implies  that $\omega $ is defined {\sl   globally\/}  on
${\cal Z}$, whereas (a) and (d) mean that $\omega $ is a {\sl   symplectic
form\/}  (Kobayashi and Nomizu [1969]) on ${\cal Z}$, which makes ${\cal Z}$
a possible classical phase space.  Finally, (c) means that the symplectic
structure defined by $\omega $ is $G$--invariant, hence $G$ acts on ${\cal
Z}$ by {\sl   canonical transformations.\/}   (Actually, the 2--form $\omega $
together with the complex structure define a {\sl   K\"ahler structure\/}  on
${\cal Z}$, i.e. a Hermitian metric such that the complex structure is invariant
under parallel translations.)

\secskp

 \noindent {\bf 3.5. The Rotation Group  }
\def\rightheadline{\tenrm\hfil {\sl 3.5. The Rotation Group}\hfil\folio}  

 \skp
\noindent A simple but important example of the foregoing methods is
provided by their application to the three--dimensional rotation group
$SO(3)$. The resulting frame vectors are known as {\sl   spin coherent
states.\/} 
An excellent and detailed account of this is given in Perelomov [1986]; the
treatment here will be fairly brief.  $SO(3)$ is locally isomorphic to the group
$SU(2)$ of unitary unimodular 2$\times$2 matrices, which we denote by $G$
in this section. This is the set of all matrices 

\be
g=\left(\matrix{\alpha &\beta \cr
-\bar \beta  &\bar \alpha 	\cr}\right),\qquad |\alpha |^2+|\beta |^2=1,
\eno1)
\ee
hence $G\approx S^3$ (the unit sphere in $\rl^4$) as a manifold. The Lie
algebra {\bf g} has a basis  $\{J_1,J_2,J_3\}$ satisfying $[J_1,J_2]=iJ_3$ plus
cyclic permutations (where, as usual, it is actually $iJ_k$ which span the real
algebra {\bf g}) which can be conveniently given as $J_k=(1/2)\sigma _k$ in
terms of the Pauli matrices 

\be
 \sigma _1=\left(\matrix{0&1\cr1&0\cr}\right),\quad 
\sigma _2=\left(\matrix{0&-i\cr i&0\cr}\right),\quad 
\sigma _3=\left(\matrix{1&0\cr 0&-1\cr}\right).
\eno2)
 \ee
Root vectors in ${\bf g}_c$ are given by $J_\pm=J_1\pm iJ_2$, satisfying 

\be
 [J_3,J_\pm]=\pm J_\pm,\qquad [J_+,J_-]=2J_3.
\eno3)
 \ee
The vectors $\{J_3,J_\pm\}$ form a complex basis for ${\bf g}_c$, with 

\be
 J_+=\left(\matrix{ 0&1\cr 0&0\cr}\right),\qquad 
J_-=\left(\matrix{ 0&0\cr 1&0 \cr}\right).
\eno4)
 \ee
Unitary irreducible representations of $G$ are characterized by a single
number (highest weight) $s=0,1/2,1,3/2,\ldots$, with the representation 
space $\cal H$ having dimensionality $2s+1$.  The generators $J_k$ are
represented by hermitian matrices $S_k$ satisfying  the irreducibility
(Casimir) condition 

\be
 {\bf S}^2\equiv S_1^2+S_2^2+S_3^2=(1/2)(S_+S_-+S_-S_+)+S_3^2=s(s+1).
\eno5)
 \ee
A basis for ${\cal H}$ is obtained by starting with a highest--weight vector
$v_s$, i.e.,

\be
 S_+v_s=0,\qquad S_3v_s=sv_s,
\eno6)
 \ee
and applying $S_-$ repeatedly until the commutation relations imply that
the resulting vector vanishes.  This results (after normalization) in an
orthonormal basis $\{v_s,v_{s-1}, \ldots,v_{-s}\}$ satisfying

\ba{
S_3v_m\a=mv_m\cr
S_+v_m\a=\sqrt{(s-m)(s+m+1)}\,v_{m+1}\cr
S_-v_m\a=\sqrt{(s+m)(s-m+1)}\,v_{m-1}.
 \cr}
 \eno7)
 \ea
To build a homogeneous frame as in section 3.3, we use a decomposition of
$G$ in terms of Euler angles,

\be
 g(\phi ,\theta ,\psi )=\exp(-i\phi J_3)\,\exp(-i\theta J_2)\,\exp(-i\psi J_3),
\eno8)
 \ee
with $0\le \phi <2\pi $, $ 0\le \theta \le \pi $ and $0\le \psi <4\pi $, which
gives a corresponding decomposition of $U(g)$ as 

\be
 U(\phi ,\theta ,\psi )=\exp(-i\phi S_3)\,\exp(-i\theta S_2)\,\exp(-i\psi S_3).
\eno9)
 \ee
If $2s$ is odd, $\phi ,\theta $ and $\psi $ have the same ranges as before;  if
$2s$ is even, then $\psi +2\pi $ and $\psi $ give the same operators, hence
$0\le \psi <2\pi $. If we choose one of the basis vectors $v_m$ as our initial
vector $h$, then the stability subgroup is $H=\{g(0,0,\psi )\}\approx S^1$. 
The homogeneous space $G/H\approx S^3/S^1$ is parametrized by $(\phi
,\theta)$, or by the unit vectors ${\bf n} =(\cos \phi \,\sin \theta , \sin \phi
\,\sin \theta , \cos \theta )$, hence $G/H$ can be identified with the unit
sphere $S^2$.  Choosing the section $\sigma {:}\ S^2\to G$ as $\sigma (\phi
,\theta )=(\phi ,\theta ,0)$, we obtain the frame vectors 

\be
 h_{\bf n} =e^{-i\phi S_3}\,e^{-i\theta S_2}\,h.
\eno10)
 \ee
The $G$--invariant measure on $S^2$ is just the area measure

\be
 d{\bf n} =\sin \theta \,d\theta \,d\phi \,  ,
\eno11)
 \ee
and the tight frame is 

\be
 \int_{S^2} d {\bf n} \,|\,h_{\bf n} \,\rangle\langle\, h_{\bf n} \,|\,=cI
\eno12)
 \ee
for some number $c$.  To find $c$, take the trace of both sides and use the
fact that $\,|\,h_{\bf n} \,\rangle\langle\, h_{\bf n} \,|\,$ is a rank--one
projection operator (and hence its trace is unity).  This gives

\be
 4\pi =c\ {\rm Tr}\,I=c(2s+1),
\eno13)
 \ee
hence the resolution of unity is

\be
 {{ 2s+1 }\over{  4\pi }}
 \int_{S^2} d {\bf n} \,|\,h_{\bf n} \,\rangle\langle\, h_{\bf n} \,|\,=I.
\eno14)
 \ee
The overlap between frame vectors can be shown   to be

\be
 \,|\,\langle\, h_{\bf n} \,|\,h_{\bf n'} \,\rangle\,|\,^2=
\left( {{ 1+{\bf n} \cdot {\bf n} ' }\over{ 2 }} \right)^{2s}  ,
\eno15)
 \ee
hence they are orthogonal if and only if ${\bf n} '=-{\bf n} $.

To construct a holomorphic frame, we must consider the complex Lie algebra
${\bf g}_c $ and its Lie group $G_c=SL(2,\cx\ )$ of unimodular $2\times 2$
complex matrices

\be
g= \left(\matrix{ \alpha &\beta \cr \gamma &\delta  \cr}\right),\qquad 
\alpha \delta -\beta \gamma =1.
\eno16)
 \ee
The Lie algebra {\bf h} of the subgroup $H$ of $G$ used above is a Cartan
subalgebra of {\bf g}, and ${\bf h} _c=\cx  J_3$.  The corresponding
root--space decomposition is  

\be
 {\bf g}_c ={\bf h}_c +{\bf n} ^+ +{\bf n} ^-=\cx J_3+\cx J_+ +\cx J_-,
\eno17)
 \ee
and this yields a {\sl   Gaussian\/}  decomposition of (almost all) elements of
$G_c$ as products of lower--triangular, diagonal and upper--triangular
matrices:

\ba{
g(\zeta ,d,\xi )\a=e^{\zeta J_-}\,e^{2dJ_3}\,e^{\xi J_+} \cr
\a=\left(\matrix{ 1&0\cr \zeta &1 \cr}\right)
\left(\matrix{ e^d&0\cr 0&e^{-d} \cr}\right)
\left(\matrix{ 1&\xi \cr 0&1 \cr}\right) \cr
\a\equiv  \zeta _- \ {\bf d} \  \xi _+. 
 \cr}
 \eno18)
 \ea

If we write $N^\pm=\exp({\bf n} ^\pm)$, then the above decomposition is
$G_c\sim N^- H_c N^+$.  Comparison with the original form of $g$ gives

\ba{
\alpha =e^d,\a\qquad \beta =e^d\,\xi , \cr
\gamma =\zeta e^d,\a\qquad \delta =\zeta e^d\,\xi +e^{-d}.
 \cr}
 \eno19)
 \ea
We call $\zeta (g), \ \xi (g)$ and $e^d\equiv \alpha (g)$ the {\sl   Gaussian
parameters\/} of $g$.  Thus 

\be
 \alpha (g)=\alpha ,\quad \zeta (g)=\gamma /\alpha ,\quad 
\xi (g)=\beta /\alpha .
\eno20)
 \ee
\skp
\noindent {\sl   Remarks:\/} 
\item{1.} Matrices with $\alpha =0$, i.e. of the form 
\be
 g=\left(\matrix{ 0&\beta \cr -\beta  ^{-1} &\delta  \cr}\right),
\eno21)
 \ee
clearly do not have this decomposition.  They form a 2--dimen\-sional
complex submanifold of $G_c$, hence have (group--) measure zero.

\item{2.} Elements of $G$ are distinguished by $\delta =\bar \alpha $ and
$\gamma =-\bar \beta $, which implies 
\be
 \bar \alpha  \alpha =(1+\bar \zeta \zeta ) ^{-1} =(1+\bar\xi \xi ) ^{-1} .
\eno22)
 \ee
\skp
The Borel subgroup discussed in section 3.4 is $B=H_cN^+$ and consists of all
matrices of the form 

\be
 b=\left(\matrix{\alpha &\beta  \cr 0&\alpha  ^{-1}  \cr}\right).
\eno23)
 \ee
Its cosets $\zeta _-B$ can therefore be parametrized by $\zeta \in\cx$.  The
unattainable matrices with $\alpha =0$ form a single coset, corresponding to
$\zeta =\infty$.  Hence the homogeneous space $G_c/B$ is the Riemann
sphere:

\be
{\cal Z}\equiv G_c/B\approx\cx \cup\{\infty\}\approx S^2, 
\eno24)
 \ee
in agreement with our earlier $G/H=S^2$ but with an added complex
structure, as claimed in section 3.4.  The action of $G_c$ on ${\cal Z}$ is as
follows:  If $g\zeta '_-B=\zeta ''_-B$, i.e.

\be
 \left(\matrix{\alpha &\beta \cr \gamma &\delta   \cr}\right)
\left(\matrix{ 1&0\cr \zeta '&1 \cr}\right)\,B=
\left(\matrix{ 1&0\cr \zeta ''&1 \cr}\right)\,B  ,
\eno25)
 \ee
then

\be
 \zeta ''=\zeta (g\zeta '_-)={{ \gamma +\delta \zeta ' }\over{\alpha +\beta 
\zeta '  }}.
\eno26)
 \ee
That is, $G_c$ acts on ${\cal Z}$ by {\sl   M\"obius transformations.\/} 

We defined $h_g=T(g)^\#h$, where $T(g)$ was the analytic continuation of
$U(g)$ to $G_c$ and $T(g)^\#={(T(g)^*)}^{-1}$ .  But

\ba{
T(\zeta _-\,{\bf d} \,\xi _+)^\#\a=T(\zeta _+)^\#\,
T({\bf d} )^\#\,T(\xi _-)^\#\cr
\a=\exp(-\bar \zeta S_+)\exp(-2\bar dS_3)\exp(-\bar \xi S_-).
 \cr}
 \eno27)
 \ea
Hence the unique state left invariant by $B$ is that corresponding to the
vector of   lowest weight,  $h=v_{-s}$. For this choice, we
get 

\be
 h_g=e^{2\bar d s}\exp (-\bar \zeta S_+)\,h\equiv e^{2\bar d s}\,h_\zeta .
\eno28)
 \ee
The holomorphic coherent states with respect to the reference point
$g_1\in G_c$ are then

\be
 \chi ^{g_1}_\zeta ={{ h_g }\over{ \langle\, h_{g_1}\,|\,h_g\,\rangle}}
={{ e^{-2sd_1}h_\zeta  }\over{ \langle\, h_{\zeta _1}\,|\,h_\zeta \,\rangle }},
\eno29)
 \ee 
with 

\be
 \langle\, h_{\zeta _1}\,|\,h_\zeta \,\rangle
=\langle\, h\,|\,\exp(-\zeta _1S_-)\,\exp(-\bar \zeta S_+)\,h\,\rangle.
\eno30)
 \ee
To evaluate this, express $\exp(-\zeta _1S_-)\,\exp(-\bar \zeta S_+)$ in the
{\sl   opposite\/}  factorization $N^+H_cN^-$ and use $S_-h=0$.  It suffices to
do the computation at the level of 2$\times$2 matrices since the result
depends only on the commutation relations, which are preserved by the
representation. Thus

\ba{
\left(\matrix{ 1&0\cr -\zeta _1&1 \cr}\right) \a
\left(\matrix{1&-\bar \zeta \cr 0&1 \cr}\right)\cr
\a=\left(\matrix{1&\xi'\cr 0&1 \cr}\right)
\left(\matrix{ e^{d'}&0\cr 0&e^{-d'}\cr}\right)
\left(\matrix{ 1&0\cr \zeta '&1\cr}\right)\equiv\xi'_+{\bf d}' \zeta '_-\ ,
 \cr}
 \eno31)
 \ea
which gives

\be
 e^{-d'}=1+\zeta _1\bar \zeta ,\quad e^{-d'}\zeta '=-\zeta _1,
\quad \xi'e^{-d'}=-\bar \zeta .
\eno32)
 \ee
Hence 

\ba{
 \langle\, h_{\zeta _1}\,|\,h_\zeta \,\rangle\a=
\langle\, h\,|\,T(\zeta '_+{\bf d}' \zeta '_-)\,h\,\rangle\cr
\a=e^{-2sd'}=\left( 1+\zeta _1\bar \zeta  \right)^{2s},
 \cr}
 \eno33)
 \ea
and

\be
 \chi ^{g_1}_\zeta =e^{-2sd_1}\left( 1+\zeta _1\bar \zeta  \right)^{-2s}
\,h_\zeta .
\eno34)
 \ee
Since a single chart covers all of ${\cal Z}\approx S^2$ except for the point at
$\zeta =\infty$, just one reference point $g_1$ is needed in this case.   The
simplest choice is $g_1=1$, the identity of $G$, which gives 

\be
 \chi ^1_\zeta =h_\zeta .
\eno35)
 \ee
With this choice, the weight function is 

\ba{
 e^{-\phi (\zeta )}\a=\,|\,\langle\, h\,|\,h_g\,\rangle\,|\,^2,\quad 
g=\zeta _-{\bf d} \xi _+  \cr
\a=e^{2s(d+\bar d)}\,|\,\langle\, h\,|\,h_\zeta \,\rangle\,|\,^2 \cr
\a=e^{2s(d+\bar d)}.
 \cr}
 \eno36)
 \ea
But for $g\in G$ we have the constraint

\be
 e^{2s(d+\bar d)}=\,|\,\alpha \,|\,^2=(1+\bar \zeta \zeta )^{-1},
\eno37)
 \ee
hence

\be
 e^{-\phi (\zeta )}=(1+\bar \zeta \zeta )^{-2s}.
\eno38)
 \ee
To find the invariant measure on $G_c/B$, recall that the 2--form $\omega
=i\partial\bar \partial\phi $   is invariant under $G_c$ and
non--degenerate.  Hence it defines an invariant measure, once we choose a
positive orientation on $G_c/B$.  An easy computation gives

\ba{
 \omega \a=-2is\partial\bar \partial\ln(1+\bar \zeta \zeta ) \cr
\a={{ 2isd\bar \zeta \wedge d\zeta  }\over{( 1+ \bar  \zeta \zeta )^2}}.
 \cr}
 \eno39)
 \ea
Thus 

\be
2si \int_\cxs {{ d\bar \zeta \wedge d\zeta  }
\over{ ( 1+ \bar  \zeta \zeta )^{2s+2} }}\,|\,h_\zeta \,\rangle\langle\, 
h_\zeta \,|\,=cI
\eno40)
 \ee
for some $c$.  Taking the trace and using

\be
 \langle\, h_\zeta \,|\,h_\zeta \,\rangle=( 1+ \bar  \zeta \zeta )^{2s}
\eno41)
 \ee
gives, with $\zeta =re^{i\theta }$, 

\be
 8\pi s\int_0^\infty{{  rdr  }\over{(1+r^2)^2}}=c(2s+1),
\eno42)
 \ee
from which $c=4\pi s/(2s+1)$.  Thus we have the resolution of unity

\be
 {{2s+1  }\over{\pi  }}\int_\cxs{{ d^2\zeta  }
\over{ ( 1+ \bar  \zeta \zeta )^{2s+2} }}\,|\,h_\zeta \,\rangle\langle\, 
h_\zeta \,|\,=I,
\eno43)
 \ee
where $d^2\zeta $ is now Lebesgue measure on $\cx$.

What do the functions $\ftil(\zeta )=\langle\, h_\zeta \,|\,f\,\rangle$  look
like?   Consider the vectors

\be
 u_n={{ 1 }\over{ n! }}(-S_+)^n\,h,\quad n=0, 1, 2, \cdots .
\eno44)
 \ee
Then $u_0\equiv h, u_1, \ldots ,u_{2s}$ are linearly independent and
$u_{2s+1}=0$, since $S_+^{2s+1}=0$.  Thus 

\ba{
 h_\zeta \a=e^{-\bar \zeta S_+}h\cr
\a=u_0+\bar \zeta u_1+\cdots+(\bar \zeta)^{2s}u_{2s}, 
 \cr}
 \eno45)
 \ea
hence

\be
 \ftil(\zeta )=f_0+\zeta f_1+\cdots +\zeta ^{2s}f_{2s},
\eno46)
 \ee
where $f_n=\langle\, u_n\,|\,f\,\rangle$.  Thus, $\ftil(\zeta )$ is a
polynomial of degree $\le 2s$ in $\zeta $.
\skp
How are the two sets of frame vectors $h_{\bf n} $ and $h_\zeta $ related? 
It turns out that ${\bf n}$ is related to $\zeta $ by stereographic projection of
the 2--sphere onto the complex plane.  The exact relation depends on the
particular factorizations of $G$ used to construct the frames.  In the case of
$h_{\bf n} $, this was the Euler angle decomposition, whereas for $h_\zeta $
it was the Gaussian decomposition.  However, there is an {\sl   intrinsic \/}
way of relating the two sets of vectors, which goes as follows:  Consider the
functions 

\be
 \tilde S_k(\zeta )={{\langle\, h_\zeta \,|\,S_k h_\zeta \,\rangle  }
\over{ \langle\, h_\zeta \,|\,h_\zeta \,\rangle }},
\eno47)
 \ee
which are the {\sl   expectations\/}  of the observables $S_k$ in
the state of $h_\zeta $, and which correspond to the components of the
classical angular momentum of a system  whose only degrees of freedom are
 the rotational motions given by $G$. 
$\tilde S_k(\zeta)$ is a quantum--mechanical version of a statistical
average.  Since $h_\zeta =\exp(-\bar \zeta S_+)h$, we have

\ba{
 \tilde S_+(\zeta )\a=-{\partial \over{\partial \bar \zeta }}
\ln \langle\, h_\zeta \,|\,h_\zeta \,\rangle\cr
\a=-{{ 2s\zeta  }\over{ 1+\bar \zeta \zeta  }}.
 \cr}
 \eno48)
 \ea
To find $\tilde S_3(\zeta )$, note that $[S_3,S_+]=S_+$ implies
$[S_3,S^n_+]=nS^n_+$, hence

\be
 \left[ S_3,e^{-\bar \zeta S_+}  \right]=-\bar \zeta S_+e^{-\bar \zeta S_+} 
\eno49)
 \ee
and

\ba{
 S_3h_\zeta \a=-\bar \zeta S_+h_\zeta +e^{-\bar \zeta S_+} S_3h\cr
\a=-\left(\bar \zeta S_++s  \right)h_\zeta .
 \cr}
 \eno50)
 \ea
Hence

\be
\tilde S_3(\zeta )=-s-\bar \zeta \tilde S_+(\zeta )=s\left[ 
{{\bar \zeta \zeta -1 }\over{ \bar \zeta \zeta +1 }} \right].
\eno51)
 \ee
The equator of $S^2$ corresponds to $S_3=0$, hence to $|\zeta |=1$, and the
south pole ($S_3=-s$) and north pole ($S_3=s$) to $\zeta =0$ and $\zeta
=\infty$, respectively.  The above equations imply that 

\ba{
\tilde {\bf S}^2(\zeta )\a\equiv \tilde S_1(\zeta )^2+\tilde S_2(\zeta )^2
+\tilde S_3(\zeta )^2\cr
\a=|\tilde S_+(\zeta )|^2+\tilde S_3(\zeta )^2\cr
\a=s^2,
 \cr}
 \eno52)
 \ea
thus $\tilde {\bf S}(\zeta )$ belongs to the 2--sphere of radius $s$
centered at the origin.  In fact, the correspondence $\zeta \leftrightarrow
\tilde {\bf S}(\zeta )$ is a bijection if we include $\zeta =\infty$.  The
relation 

\be
 {{ \tilde S_+(\zeta ) }\over{ s-\tilde S_3(\zeta ) }}=-\zeta 
\eno53)
 \ee
shows that {\sl   $-\zeta $ is just the   stereographic projection of
$ \tilde {\bf S}(\zeta )$ from the north pole to the complex plane tangent to
the south pole.\/} We could choose $\bf s$ as a new independent variable
ranging over the 2--sphere of radius $s$ minus the north pole and define
$h_{\bf s}=h_\zeta $, where $\zeta $ is the unique point with $\tilde {\bf
S}(\zeta )=\bf s$. Then the vectors $h_{\bf s}$ are essentially equivalent to
the $h_{\bf n} $'s. Note  that they are eigenvectors of the 
operator ${\bf s}\cdot{\bf S}$ with eigenvalue $s^2$, i.e.,

\be
 {\bf s}\cdot {\bf S} h_{\bf s} =s^2 h_{\bf s} ,
\eno54)
 \ee
since the equation obviously holds for ${\bf s} =(0,0,-s)$, hence for all ${\bf
s}$ by symmetry.

\secskp

 \noindent {\bf 3.6. The Harmonic Oscillator as a Contraction Limit}
\def\rightheadline{\tenrm\hfil {\sl 
3.6. The Harmonic Oscillator as a Contraction Limit}\hfil\folio}  

 \skp

\noindent In section 3.4, we gave a heuristic argument suggesting that the
Weyl--Heisenberg group $\cal W$ is a ``contraction limit'' of
$SU(2)$. This can now be made  precise and given a
geometric interpretation at the representation level.   Furthermore, imitating
the analysis of the non--relativistic  limit of Klein--Gordon theory (next
chapter), we shall gain an understanding of the relation between the
Harmonic Oscillator and the canonical coherent states in the bargain. This
connection between the rotation group and the harmonic oscillator has some
potentially important applications in quantum field theory, which I hope to
explore in the future.

\skp
We will study the limit of the representation  of $G=SU(2)$  with spin $s$ as
$s\to\infty$. Since $s$ is now a variable, we denote the representation 
space by ${\cal H}_s$ and the holomorphic coherent states by $h_\zeta ^s$. 
The matrices representing the generators $J_k$ will be denoted by $S^s_k$. 
By way of motivation, compare the resolution of unity on ${\cal H}_s$,

\be
 {{ 2s+1 }\over{ \pi  }}\int_\cxs d^2\zeta\, (1+\bar \zeta \zeta )^{-2s-2}
\,|\,h^s_\zeta \,\rangle\langle\, h^s_\zeta \,|\,=I_s,
\eno1)
 \ee
with that on the representation  space ${\cal H}$ of $\cal W$ in terms of
the canonical coherent states,

\be
 {{ 1 }\over{ 2\pi  }}\int _\cxs d^2z\,e^{-\bar zz/2}
\,|\,\chi _z\,\rangle\langle\, \chi _z\,|\,=I.
\eno2)
 \ee
Note that if we set 

\be
 \zeta =-{{ z }\over{ 2\sqrt{s+1} }}
\eno3)
 \ee
and define $\chi _z^s\equiv h^s_\zeta $, then the resolution of unity on
${\cal H}_s$  becomes 

\be
   {{2s+1  }\over{4\pi (s+1) }}\int_\cxs\, d^2z \ 
\left( 1+ {{ \bar z z  }\over{ 2(2s+2) }}\right)^{-(2s+2)} \,|\,\chi ^s_z
\,\rangle\langle\,  \chi^s_z \,|\,=I.
\eno4)
 \ee
If we now take the {\sl   formal\/}  limit $s\to\infty$, this coincides with
the resolution of unity on ${\cal H}$, provided we can show that
$\chi ^s_z  \to\chi_z$.  Our task is now (a) to find the {\sl   sense\/}  in
which this limit is to be taken, (b) to show that the generators $S^s_k$ of
{\bf g} go over to the generators  $A,A^*$ and $I$ of {\bf w} and (c) to show
that the coherent states $\chi ^s_z$ go over to the canonical coherent states
$\chi _z$.
\skp
To properly study the limit $s\to\infty$, we will first of all imbed all the
spaces ${\cal H}_s$ into ${\cal H}$, so that the limit may be considered
within ${\cal H}$.  This is done most easily by using the orthonormal bases
obtained by applying $S^s_+$ and $A^*$ to the respective ``ground states''. 
An orthonormal basis for ${\cal H}$ is given by 

\be
 w_n=\left( 2^n n! \right)^{-1/2} {\left( A^* \right)}^n \chi _0,
\qquad n=0,1,2, \ldots,
\eno5)
 \ee
where $A\chi _0=0$ and the normalization is determined by the
commutation relation $[A,A^*]=2I$.  The generators $A$ and $A^*$ act by 

\ba{
Aw_n\a=\sqrt{2n}\,w_{n-1}\cr
A^*w_n\a=\sqrt{2n+2}\,w_{n+1}.
 \cr}
 \eno6)
 \ea
An orthonormal basis for ${\cal H}_s$ is given by 

\be
 w_n^s=\sqrt{{{ (2s-n)! }\over{n!(2s)! } }}\,{\left( S^s_+ \right)}^n h^s,
\qquad n=0,1,2, \ldots, 2s,
\eno7)
 \ee
where $S^s_-h^s=0$.  We imbed ${\cal H}_s$ into ${\cal H}$ by identifying
$w^s_n$ with $w_n$ and defining $S^s_k w_n=0$ for $n>2s$.  Then for $0\le
n\le 2s$, 

\ba{
S^s_-w_n\a=\sqrt{n(2s-n+1)}\,w_{n-1}\cr
S^s_+w_n\a=\sqrt{(n+1)(2s-n)}\,w_{n+1}\cr
S^s_3w_n\a=(n-s)w_n.
 \cr}
 \eno8)
 \ea
To see how the generators $S^s_k$ must be scaled, note that the relation
between $\zeta $ and $z$  implies that 

\be
 h^s_\zeta \equiv \exp(-\bar \zeta S^s_+)h^s=\exp(\bar zK_+/2)w_0
\equiv \chi ^s_z,
\eno9)
 \ee
where 

\be
 K^s_+={{ S^s_+ }\over{ \sqrt{s+1} }}.
\eno10)
 \ee
Define $K^s_-=S^s_-/\sqrt{s+1}$, so that $K^s_-={\left( K^s_+ \right)}^*$, and 

\be
 K^s_3\equiv{1\over 2}[K^s_+,K^s_-]={{S^s_3  }\over{ s+1 }}.
\eno11)
 \ee
Then for $0\le n\le 2s$,

\ba{
K^s_-w_n\a=\sqrt{{{ n(2s-n+1) }\over{ s+1 }}}\,w_{n-1}\cr
K^s_+w_n\a=\sqrt{{{ (n+1)(2s-n) }\over{ s+1 }}}\,w_{n+1}\cr
K^s_3w_n\a={{ n-s }\over{ s+1 }}\,w_n.
\cr}
 \eno12)
 \ea
If we now take the limit $s\to\infty$ while {\sl   keeping $n$ fixed,\/}  we
obtain

\ba{
K^s_-w_n\a\to\sqrt{2n}\,w_{n-1}\cr
K^s_+w_n\a\to\sqrt{2n+2}\,w_{n+1}\cr
K^s_3w_n\a\to -w_n.
 \cr}
 \eno13)
 \ea
Comparing this with the action of $A$ and $A^*$, we see that 

\ba{
K^s_-\a\to A\cr
K^s_+\a\to A^*\cr
K^s_3\a\to-I
 \cr}
 \eno14)
 \ea
as $s\to\infty $ in the {\sl   weak operator topology\/} of ${\cal H}$. (These
limits are not valid in the strong operator topology since it is necessary to
keep $n$ fixed.)  Thus we have shown that in the weak sense, the
representation  of $SU(2)$ goes over to a representation  of $\cal W$ as
$s\to\infty$.  Note  that the operator

\be
 N^s_-=S^s_3+s,
\eno15)
 \ee
which satisfies

\ba{
N^s_- w_n\a=nw_n,\qquad 0\le n\le 2s,\cr
N^s_-w_n\a=sw_n,\qquad n>2s,
 \cr}
 \eno16)
 \ea
has the  weak limit

\be
\hbox{w--lim} \,N^s_-\equiv N={{ 1 }\over{ 2 }}A^*A,
\eno17)
 \ee
which is the Hamiltonian for the harmonic oscillator (minus the \break
ground--state energy).  If we write $A=X+iP$ with $X$ and $P$ self--adjoint,
then 

\be
 N={{1  }\over{ 2 }}\left( X^2+P^2-I \right).
\eno18)
 \ee
\skp
The fact that $K^s_+\to A^*$ means that

\be
 \chi^s_z\equiv \exp\left( \bar zK^s_+/2 \right)w_0\longrightarrow 
\exp(\bar zA^*/2)w_0\equiv \chi _z,
\eno19)
 \ee
where $\chi _z$ are the canonical coherent states and the convergence is in
the weak topology of ${\cal H}$.  
\skp
We now examine the limit $s\to\infty$ from a global geometric point of view,
using the coherent states.  Recall that the expectation vector

\be
 \tilde{\bf S}^s (\zeta )\equiv {{\langle\,  h^s_\zeta \,|\,{\bf S}^s\,h^s_\zeta
\,\rangle  } \over{ \langle\, h^s_\zeta \,|\,h^s_\zeta \,\rangle }}
\eno20)
 \ee
ranges over the  sphere of radius $s$ centered at the origin, with the north
pole corresponding to $\zeta =\infty$.  The transformation from $S^s_k$ to
$K^s_k$ {\sl   deforms \/} this sphere to an ellipsoid,

\be
  {{ |\tilde K^s_+(\zeta )|^2 }\over{ s+1 }}+\tilde K^s_3(\zeta )^2=\left( 
{{ s }\over{ s+1 }} \right)^2.
\eno21)
 \ee
When $s\to\infty$, this ellipsoid splits into the two planes $\tilde
K^s_3\to\pm1$.  Our weak limit $K^s_3\to -I$ only picked out the lower
plane.  We could have picked out the upper plane by imbedding ${\cal H}_s$ 
into ${\cal H}$ differently, namely by identifying $w_{2s-n}$ with $w_n$
for $n=0,1,2,\ldots,2s$.  In that case we would have obtained the weak
limits

\ba{
 K^s_-\a\to A^*\cr
K^s_+\a\to A\cr
K^s_3\a\to I.
 \cr}
 \eno22)
 \ea
In terms of the coherent states, this corresponds to using the
highest--weight vector instead of the lowest--weight vector or,
equivalently, using a chart centered about the north pole rather than the
south pole, for example, by using as reference point the element
$g_1=e^{-i\pi J_2}$.  The corresponding harmonic--oscillator Hamiltonian is
the weak limit of the operator 

\be
 N^s_+\equiv s-S^s_3=2s-N^s_-.
\eno23)
 \ee
The expectation values of $N^s_-$ and $N^s_+$,

\ba{
 \tilde N^s_-(\zeta )\a={{ 2s\bar \zeta \zeta  }\over{ 1+\bar \zeta \zeta  }}\cr
\tilde N^s_+(\zeta )\a={{ 2s}\over{ 1+\bar \zeta \zeta  }},
 \cr}
 \eno24)
 \ea
are related by {\sl   inversion:}

\be
 \tilde N^s_+(\zeta )=\tilde N^s_-(1/\zeta ).
\eno25)
 \ee
\skp
The splitting of the ellipsoid into the two planes can be understood from the
 point of view of representation  theory by writing the irreducibility
condition in terms of the $K$'s:

\be
 {{ 1 }\over{ 2s+2 }}\left( K^s_+K^s_-+K^s_-K^s_+ \right)+{(K^s_3)}^2
={{ s }\over{s+1  }}I_s.
\eno26)
 \ee
When $s\to\infty$, this implies  formally that ${(K^s_3)}^2\to I$.  The
subspaces ${\cal H}_\pm$ on which $K^s_3\to\pm I$ are invariant in the
limit, hence the limiting representation  of $\cal W$ is {\sl   reducible}. 
Evidently, by taking the limits in the weak topology we are able to pick out
just one irreducible component at a time.
\skp
There is an interesting analogy between the above analysis and the
non--relativistic  limit of Klein--Gordon theory, which we now point out for
those readers already familiar with the latter.  (The non--relativistic  limit
will be discussed in the next chapter.)  The relation $S^s_3/(s+1)\to\pm I$
corresponds to $P_0/mc^2\to \pm I$ in the limit $mc^2\to\infty$, where
$P_0$ is the relativistic energy operator.  As will be seen in the next
chapter, the Poincar\'e group contracts, in this approximation, to a semidirect
product of the 7--dimensional Weyl--Heisenberg group and the rotation
group.  A first--order correction is obtained by expanding

\be
 P_0=\pm\sqrt{m^2c^4+{\bf P}^2c^2}\sim 
\pm\left( mc^2+{\bf P}^2/2m  \right)\equiv \pm\left( mc^2+H \right),
\eno27)
 \ee
where $H$ is the non--relativistic  free Schr\"odinger Hamiltonian. 
Similarly, it follows from

\ba{
\a{{ 1 }\over{ 2 }}\left( S^s_+S^s_-+S^s_-S^s_+ \right)+{(S^s_3)}^2=s(s+1)\cr
\a{{ 1 }\over{ 2 }}\left( S^s_+S^s_--S^s_-S^s_+ \right)-S^s_3=0
 \cr}
 \eno28)
 \ea
that

\be
\left( S^s_3-{\sc{1\over 2}} \right)^2=\left( s+{\sc{1\over 2}} \right)^2-
S^s_+S^s_-,
\eno29)
 \ee
hence  formally

\ba{
S^s_3-{\sc{1\over 2}}\a=\pm\sqrt{\left(s+{\sc{1\over 2}} 
              \right)^2-S^s_+S^s_-}\cr
\a\sim\pm\left( s+{\sc{1\over 2}} \right)\mp{{ S^s_+S^s_- }\over{ 2s+1 }},
 \cr}
 \eno30)
 \ea
from which

\ba{
 S^{s(-)}_3\a\sim -s+{1\over 2}K^s_+K^s_-,\cr
 S^{s(+)}_3\a\sim s+1-{1\over 2}K^s_+K^s_-,
 \cr}
 \eno31)
 \ea
which corresponds to $P^{(\pm)}_0\sim \pm (mc^2+H)$.  The analogy
between the large--spin  and the non--relativistic limits can be summarized
as follows: The Poincar\'e group corresponds to $SU(2)$, the energy $P_0$ to
$S_3^s$, the rest energy $mc^2$ to $-s$, and the non--relativistic 
Hamiltonian $H$ to the harmonic oscillator hamiltonian $N$.  The analog of
the central extension of the Galilean group (which is the contraction limit of
the Poincar\'e group, as discussed in the next chapter) is the {\sl   Oscillator
group\/} (Streater [1967]), whose generators are $A,A^*, I$ and $N$,
with the commutation relations

\be
 [N,A]=-A,\qquad [N,A^*]=A^*.
\eno32)
 \ee
\skp
Finally, we note that the above analysis explains a well--known relationship
between the harmonic oscillator and the canonical coherent states, namely
that the latter evolve naturally under the harmonic oscillator dynamics.  We
first derive the corresponding relation within $SU(2)$, i.e. for finite $s$. 
Consider the behavior of $h^s_\zeta $ under the ``evolution operator''
$\exp(-itN^s_-)$.  We wish to express

\be
 e^{-itN^s_-}\chi ^s_z=e^{-its}e^{-itS_3}e^{-\bar \zeta S_+}h^s
\eno33)
 \ee
in terms of the coherent states $\chi ^s_z$, hence we need to display the
operator on the right in the reverse Gaussian form

\be
 \exp(-\bar \zeta'S^s_+) \exp (2d'S_3)\exp(\xi 'S_-).
\eno34)
 \ee
 As explained earlier, it suffices to do the computation on 2$\times$2
matrices:

\ba{
e^{-itJ_3}e^{-\bar \zeta J_+}\a=
\left(\matrix{ e^{-it/2}&0\cr 0&e^{it/2}\cr}\right)
\left(\matrix{ 1&-\bar \zeta \cr 0&1\cr}\right)\cr
\a\equiv \left(\matrix{ 1&-\bar \zeta'\cr 0&1\cr}\right)
\left(\matrix{ e^{d'}&0\cr 0&e^{-d'}\cr}\right)
\left(\matrix{ 1&0\cr \xi '&1\cr}\right).
 \cr}
 \eno35)
 \ea
This gives $d'=-it/2$, $\zeta '=e^{it}\zeta $ and $\xi '=0$.  Hence

\ba{
e^{-itN^s}\chi  ^s_z\a=e^{-its}e^{-\bar \zeta' S^s_+}e^{-itS^s_3}h \cr
\a=e^{-\bar \zeta 'S^s_+}h=h_{\zeta '}\cr
\a=\chi ^s_{z(t)},
 \cr}
 \eno36)
 \ea
where $z(t)=e^{it}z$. (This is intuitively obvious, since $\exp(-itS^s_3)$ rotates
the 1--2 plane clockwise by an angle $t$, hence it rotates  the coordinate $z$
counterclockwise by an angle $t$.) In the limit $s\to \infty$, this gives the
well--known result (Henley and Thirring [1962]) that the {\sl   set\/}  of
canonical coherent states is invariant under the harmonic oscillator time
evolution, with individual coherent states moving along the classical
trajectories $z(t)$ determined by the initial conditions $z=x-ip$ in phase
space.  The above shows that the same is true within $SU(2)$, i.e.  for finite
$s$, where it is a consequence of the fact that $N^s_-$ is essentially the
generator of rotations about the 3--axis.  

\skp

\noindent {\sl   Note:\/}  After this section was written, I learned from
R.~F.~Streater that a related construction was made by Dyson [1956].

\VE

\def\leftheadline{\tenrm\folio\hfil {\sl 4. Complex Spacetime}\hfil} 
\def\rightheadline{\hfill\folio}

\headline={\ifodd\pageno\rightheadline\else\leftheadline\fi}
\def\be{$$}\def\ee{$$}\def\ba{$$\eqalign}\def\a{&}  
\def\eno{\eqno(}\def\ea{$$}

\centerline{\bf Chapter 4}\skp
\centerline{\bf COMPLEX SPACETIME}
\vskip 3 cm
\noindent {\bf  4.1. Introduction}
\skp

\noindent Relativistic quantum mechanics is a synthesis of special relativity
and ordinary (i.e., non-relativistic) quantum mechanics.  The former is based
on the Lorentzian geometry of spacetime, while the latter is usually obtained
from classical mechanics by a somewhat mysterious set of rules known as
``quantization'' in which classical observables, which are functions on phase
space, suddenly become operators on a Hilbert space.  Classical mechanics, in
turn, can be formulated in terms of Newtonian space-time (the Lagrangian
approach) or it can be based on the symplectic geometry of phase space
(see Abraham and Marsden [1978]).  The latter, called the
{\sl Hamiltonian approach,\/}  is usually considered to be deeper and more
powerful, and its study has virtually turned modern classical mechanics into
a branch of differential geometry.  Yet, the standard formalism of
relativistic quantum mechanics rests solely on the geometry of spacetime. 
Symplectic geometry, so prominent in classical mechanics, seems to have
disappeared without a trace.

\skp

 In this chapter we develop a formulation of relativistic
quantum mechanics in which symplectic geometry plays an important role. 
This will be done by studying the role of phase space in relativity and
discovering its counterpart in relation to the Poincar\'e group, which is the
invariance group of Minkowskian spacetime.  It turns out that the
Perelomov construction fails for relativistic particles (the physically relevant
representations  are not square-integrable), and an alternative route must
be taken.  The result is a formalism based on {\sl complex spacetime}
which, we show,  may be regarded as a relativistic extension of classical
phase space.  As a by-product, two long-standing inconsistencies of
relativistic quantum mechanics (in its standard spacetime formulation) are
resolved, namely the problems of {\sl localization} and {\sl covariant
probabilistic interpretation.}  Rather than being {\sl sharply} localizable in
{\sl space} (which leads to conflicts with causality;  see Newton and Wigner
[1949] and Hegerfeldt [1985]), particles in the new formulation are at best 
{\sl softly} localizable in {\sl phase space.}  This is just a covariant version of
the situation in the coherent--state representation.  But whereas for
non-relativistic particles both the Schr\"odinger representation and the 
coherent--state representation give equally consistent theories, the
spacetime formulation of  relativistic quantum mechanics is inconsistent
because it lacks a genuine probabilistic interpretation, a situation remedied
by the phase-space formulation (section 4.5).

\secskp

\noindent {\bf 4.2.  Relativity, Phase Space and Quantization } 
\def\rightheadline{\tenrm\hfil {\sl 
4.2.  Relativity, Phase Space and Quantization}\hfil\folio}  
\skp

\noindent At first it appears that phase space and spacetime are
mutually exclusive: The phase-space coordinates of a
particle are in one-to-one correspondence with the 
initial conditions  which determine its classical motion, i.e. its worldline. 
Hence the  phase space can be identified with either the set of all
initial conditions or the set of worldlines of the classical particle. In either
case, time is treated differently from space: For initial conditions, an
arbitrary ``initial'' time is chosen; for world-lines, the dynamical
``flow'' is factored out.  Furthermore, {\sl locality} in spacetime
is lost  in either case.

 \par In this section we confine ourselves to the physical case $s=3$, i.e.,
three--dimensional space.  Our approach will be to leave spacetime intact
and, instead, consider its group of symmetries, the {\sl Poincar\'e group\/},
which is defined as follows: Let $u$ be a  four-vector.  We denote its time
component (with respect to an arbitrary reference frame) by $u^0$ and its
space components by ${\bf u}=(u^1, u^2, u^3).$  The invariant Lorentzian
 inner product of two such vectors is defined as

\be
(u,v)\equiv uv \equiv c^2u^0v^0-{\bf u}\cdot{\bf v}\equiv
g_{\mu\nu} u^\mu v^\nu. 
  \eno1)																		
\ee
Later we will set $c=1$ (which amounts
to measuring time as the distance traveled by light), but for the present it is
important to include $c,$ since the non-relativistic limit
$c\rightarrow\infty$ will be considered.
\par The {\sl Lorentz group\/} $\cal L$ is the set of all linear transformations
\break $\Lambda:~\rl^4~\to ~\rl^4$ which leave the inner product invariant:

\be
(\Lambda u, \Lambda v)=(u,v) \ \  \forall u,v\in \rl^4. 
  \eno2)
\ee

$\cal L$ includes transformations which reverse the orientation of
time and ones which reverse the orientation of space. Such space-- and time
reflections split ${\cal L}$ into four connected components.  The component
connected to the identity (whose elements reverse neither the orientations
of time nor of space)  is called the {\sl restricted Lorentz group\/} and
denoted by ${\cal L}_0$.     If we let $u^4\equiv cu^0,$ so that $uv=-{\bf
u}\cdot{\bf v}+u^4v^4,$ we can identify ${\cal L}_0$  with  $SO(3,1)_+$ (the
plus sign indicates that the orientation of time, hence also of space, is
preserved separately.)  The {\sl Poincar\'e group\/} $\cal P$  is defined as
the set of all Lorentz transformations combined with spacetime translations: 

\be  
{\cal P} =\{(a,\Lambda )|\Lambda\in {\cal L}, a\in \rl^4\} 
  \eno3)
\ee
where $(a,\Lambda)$ acts on $\rl^4$ by 

\be
(a,\Lambda)u=\Lambda u+a.
  \eno4) 																				
\ee

We will be dealing with the {\sl restricted  
Poincar\'e group\/} ${\cal P}_0$   where $\Lambda\in 
{\cal L}_0$.    The reason for our interest in ${\cal P}_0$    is
that it parametrizes all reference
 frames which can be obtained from a given reference frame  by a
continuous motion.  (By ``motion'' we mean any spacetime translation,
rotation or boost, including physically impossible ``motions'' such as
space--like translations and translations backwards in time.)
 Now to specify a reference frame $(a,\Lambda)$ relative to some
fixed reference
 frame (located, say, at the origin in spacetime), we must give its {\sl
origin\/} (namely, $a$), its {\sl velocity \/} and its {\sl spatial orientation
\/} (all relative to the fixed frame).  Thus if we {\sl  ignore\/} the spatial
orientation by factoring out the rotation subgroup $SO(3)$ of $\cal
L^\uparrow _+$ , the resulting seven-dimensional homogeneous space 

\be
{\cal C} \equiv {\cal P}_0/SO(3)  
  \eno5)  									
\ee
can be identified with the set of positions in spacetime ({\sl 
events})\/ together with all possible velocities at these events.  The set of all
(future-pointing) four-velocities is a hyperboloid

\be
\Omega_c^+\equiv \{u\in \rl^4|\ u^2\equiv (u,u)=c^2,\ u^0>0\}.
  \eno6)
\ee
Thus 

\be
{\cal C}\approx \rl^4\times \Omega_c^+,    	
  \eno7)											
\ee
which is an extension of classical phase space obtained by
including time along with the space coordinates.  Such an object is usually
called a {\sl  state space.} Strictly speaking, ${\cal C}$  is an extended {\sl
velocity\/} phase space rather than an extended momentum phase space; this
appears to be the price for retaining locality in spacetime. What matters for
us is that it will have the required symplectic (more precisely, {\sl
contact})\/ structure.  From its construction, it is clear that ${\cal C}$ is a
relativistically covariant object; it is not {\sl in}variant since the choice of
$SO(3) $ in ${\cal P}_0$   is frame-dependent. We will see that ${\cal C}$
combines the geometries of spacetime and phase space in a natural way.

\par Thus, rather than conflicting with relativity, the concept of phase space
actually {\sl  follows\/} from it:  Appending time to the geometry of
space means appending velocity-changing transformations (which are just
rotations in a space-time plane) to the group of rigid motions, hence the
enlarged group contains velocity coordinates in addition to space coordinates.
In a sense, ${\cal P}_0$     itself is actually a
``super phase space'' since it furthermore contains information on the spatial
orientation, which is needed to include {\sl  spin\/} degrees of freedom along
with the translational degrees of freedom.  ${\cal P}_0$   even has a
natural generalization to the case of {\sl  curved} spacetime, namely the {\sl 
frame bundle\/} of all  ``orthonormal'' frames (with respect to the
given curved metric) at all possible events. (For the definition and study of
frame bundles, see Kobayashi and Nomizu [1963, 1969].)

In our review of canonical coherent states (chapter 1), we saw that the
classical phase space resulted from the Weyl-Heisenberg group $\cal W$,
which, unlike ${\cal P}_0$   was not a symmetry group of the theory but
merely a Lie group generated by the fundamental dynamical observables of
position and momentum at a fixed time.  We will now see that $\cal W $ is
related to the non-relativistic limit of ${\cal P}_0$   in two distinct
ways: as a normal subgroup, and as a homogeneous space. This insight will
play a key role in generalizing the canonical coherent states
 to the relativistic case.  It turns out that the role of $\cal W $ as a
{\sl group\/} has no relativistic counterpart, whereas its role as a
homogeneous space does: its relativistic generalization is $\cal C$.   
\par Consider the Lie algebra $\wp$ of ${\cal P}_0$, which is
spanned by the generators $P_k$ of spatial translations, $P_0$ of time
translations, $J_k$ of spatial rotations and $K_k$ of pure Lorentz
transformations (the ``boosts'').  The Lie brackets of $\wp$ are given by

\be
\matrix{[J_j,J_k]=iJ_l       		& [J_j,K_k]=iK_l \cr  
									[P_0,K_r]=iP_r    &  [J_j,P_k]=iP_l \cr
				[K_j,K_k]=-ic^{-2}J_l			& \quad [P_r,K_s]=ic^{-2}\delta_{rs}P_0}
  \eno8)
\ee
where $(j,k,l)$ is a cyclic permutation of (1,2,3), $r,s=1,2,3$ and
all unspecified brackets vanish.  The physical dimensions of these
generators are as follows:  $P_0$ is a reciprocal time, $P_k$ is a reciprocal
lenght, $J_k$ is dimensionless (reciprocal angle) and $K_k$ is a reciprocal
velocity.

Notice that so far nothing has been said about quantum mechanics.  
$\wp$ is simply the Lie algebra of the infinitesimal motions of Minkowskian
spacetime, a classical concept.  (The unit imaginary $i$ in eq (9)
 can be removed by replacing $P_{\mu}$ with $-iP_{\mu},$  $J_k$ with
$-iJ_k$ and  $K_k$ with $-iK_k$;  $i$ is included because we anticipate that
in quantum mechanics  these generators become self-adjoint operators.) 
Quantum mechanics is now introduced through the following postulate:
\skp
\noindent {\sl (Q). \sp  The formalism of relativistic quantum theory is
based on a unitary (though possibly reducible) representation of }
${\cal P}_0$.    
\skp
\noindent That is, the representation provides the
quantum--mechanical Hilbert \break
 space, and the generators of $\wp$,
which by unitarity are represented by self-adjoint operators, are interpreted
as the basic physical observables:  $P_0$ as the energy, $P_k$ as the
momentum and $J_k$ as the angular momentum.  One may
then consider perturbations by introducing interactions or gauge fields.  In
fact, the assumption {\sl (Q)\/} is very general in scope; it serves as one of
the axioms in axiomatic quantum field theory (Streater and Wightman
[1964]).  Unlike the usual prescription of ``quantization'' in non-relativistic
quantum mechanics, {\sl (Q)\/} is both mathematically and physically
unambiguous.  Yet, {\sl  (Q)\/ implies and, at the same time, supercedes the
canonical commutation relations!\/}  To see this, consider the non-relativistic
limit $c\rightarrow\infty$ of $\wp$.  Letting  

\be
c^{-2}P_0 \equiv M,
  \eno9)
\ee
we see that in the limit $c\rightarrow\infty,$ \ $\wp$ ``contracts''
to a Lie algebra $\bf g_1$ with generators $M, P_k, J_k,K_k$ satisfying

\be
\matrix{[J_j,J_k]=iJ_l       					& [J_j,K_k]=iK_l \cr  
									[M,K_r]=0    					&  [J_j,P_k]=iP_l \cr
									[K_j,K_k]=0			& \quad [P_r,K_s]=i\delta_{rs}M}
  \eno10)
\ee
and all other brackets vanishing.  Note that (a) $M$ is a central 
element of $\bf g_1$ and (b) $M, P_k \ \hbox {and}\  K_k$ span an
invariant subalgebra $\bf w$ of $\bf g_1$ which is isomorphic to the
Weyl--Heisenberg algebra, with $M$ playing the role of the central element
$E$.  Hence if ${\cal G}_1$ denotes the connected, simply connected Lie group
generated by $\bf g_1,$ then the invariant subgroup of ${\cal G}_1$ 
generated by $\bf w$ is isomorphic to the Weyl--Heisenberg group $\cal
W$.  The remaining generators $J_k$ of   $\bf g_1$  span the Lie algebra
$so(3)$ of the spatial rotation group $SO(3),$  so ${\cal G}_1$ is the
semi-direct product of $\cal W$ with $SO(3):$ 

\be
{\cal G}_1= {\cal W}\sdp SO(3) .
  \eno11)																	
\ee

\par Now suppose that the unitary representation of $\cal
P^\uparrow _+$  in assumption \ {\sl (Q)\/ is   irreducible.}  [Assumption
{\sl (Q)\/} means that we are dealing with a general quantum {\sl  system,}
possibly a system of interacting particles or even quantum fields; it is the
additional assumption of irreducibility which makes this system {\sl 
elementary,\/}  roughly a single {\sl  particle.}  Hence the concept of {\sl 
position,} discussed below, is only now admissible.]  Assuming that the
formal limit $c\rightarrow \infty$ of Lie algebras rigorously induces a
corresponding limit at the representation level (and this is indeed the case,
as  will be shown later), assumption \ {\sl (Q)\/} implies that we have a
unitary irreducible representation of  ${\cal G}_1$ in that limit. Then $M,
P_k  \ \hbox {and}\  K_k$ are represented by self-adjoint operators on some
Hilbert space, which we will denote by the same symbols.  Irreducibility
implies that the central element $M$ has the form $mI,$ where $m$ is a real
number and $I$ denotes the identity operator.  Assume $m>0$ (this is
physically necessary since $m$ will be interpreted as a {\sl  mass})\/ and let 

\be
X_k=-(1/m)K_k.		
  \eno12)																											  
\ee
Then eq. (10)
shows that $X_k$ and $P_k $ satisfy the canonical commutation relations:

\be
[X_r,P_s]=i\delta_{rs}I , 	
  \eno13)																								
\ee
thus $X_k$ behaves like a position operator.
This shows that the assumption \ {\sl  (Q)\/}, which is 
conceptually simple, mathematically precise, relativistically invariant and
very general, actually {\sl  implies\/} the much less satisfactory
``quantization'' prescription in the non-relativistic limit, under the additional
assumption of irreducibility. 

\skp 

How does it happen that classical relativistic geometry, as represented by
${\cal P}_0$, when combined with assumption \ {\sl  (Q),\/} yields
the mysterious canonical commutation relations?  To understand this, note
that eq. (10)
 came from the relativistic Lie bracket

\be
[P_r,K_s]=ic^{-2}\delta_{rs}P_0 
  \eno14)  																		
\ee
which states that boosting (accelerating) in any given
spatial direction does not commute with translating in the same direction. 
This, in turn,  is a consequence of the fact that Einsteinian space is not
absolute since  in the accelerated frame  there is a (Lorentz) contraction in
the direction of motion, so first translating and then boosting is not the same
as first boosting and then translating.  In the non-relativistic limit this gives
the canonical commutation relations.  No such easy derivation of these
relations would have been possible without invoking Relativity.    For had we
begun with Newtonian space-time, the appropriate invariance group would
have been not ${\cal P}_0$  but the Galilean group $\cal G.$  Since
Newtonian space is absolute and hence unaffected by boosting to a moving
frame, the Galilean boosts ${K_k}'$ commute with with the Galilean
generators of translations ${P_k}',$ hence yield no canonical commutation 
relations and no associated uncertainty principle.  In the case of ${\cal G}_1,$
the canonical commutation relations are a remnant of relativistic invariance. 
Thus {\sl  the uncertainty principle originates, in some sense, in
``classical'' Relativity theory!}

\skp

Eq.  (12) states that an acceptable set of position operators for a
non-relativistic particle is given in terms of the generators of Galilean
boosts (more precisely,  the boosts of a central extension of the Ga\-li\-lean
group, as explained below).  It is interesting to see how this comes about
from a more intuitive, physical point of view, since position operators are
problematic in relativistic quantum mechanics (as will be discussed later)
but the boosts have natural relativistic counterparts.  To gain insight, we will
now give two additional rough but intuitive arguments for the validity of eq.
(12). \skp

\noindent {\bf 1.}\ \ For a spinless particle,  the generators of the Poincar\'e
group  can be realized as  operators on a space of functions  over spacetime
(namely, the space of solutions of the Klein--Gordon equation) by

\ba{
             P_0 \a=i{\partial \over{\partial x_0}} \cr
              P_k\a=-i{\partial \over{\partial x_k}} \cr
               J_k\a=x_lP_m-x_mP_l               \cr
								K_k\a=x_0P_k-c^{-2}x_kP_0 
  \cr}
\eno15)      
\ea
where $(k,l,m)$ is a cyclic permutation of $(1,2,3)$. In the
non-rela\-tiv\-istic  limit, $P_0\rightarrow mc^2$, so 

\be
 -(1/m)K_k\rightarrow x_k-x_0P_k/m,	
  \eno16)	                 
\ee
which displays $-(1/m)K_k$ as the operator of
multiplication by $x_k$ (the usual non-relativistic  position operator) minus
the distance the particle has traveled in time $t=x_0$ while going at a
velocity ${\bf v}={\bf P}/m$.  This is just the {\sl initial\/} position of the
particle at time $t=0$; that is, the non-relativistic  limit of $-(1/m)K_k$
is just the (non-relativistic) position operator in the {\sl  Schr\"odinger
picture,} where operators are constant while states vary with time.
\skp

\noindent {\bf 2.}\ \ The second explanation of eq. (12) begins with the
relativistic state space  ${\cal C}\approx \rl^4\times \Omega_c^+$ and
considers the non-relativistic  limit.  As $c\rightarrow\infty$, the
hyperboloid $\Omega _c^+$ flattens out to a (three-dimensional) hyperplane
at infinity, the non-relativistic {\sl  velocity space\/} ${\cal V}\approx
\rl^3$.  In that limit, the boosts $K_k$ commute and become the generators
of translations in $\cal V$. If the cartesian coordinates of $\cal V$ are $v_k$,
then

\be
K_k\rightarrow -i{\partial \over{\partial v_k}} .	
  \eno17)			
\ee
Now the velocities $v_k$ are related to the momenta $p_k$ by
$v_k=p_k/m$, where m is the mass.  Thus in the limit 

\be
K_k=-mi{\partial \over{\partial p_k}}.   
  \eno18)                
\ee
But in the {\sl momentum representation\/} of non-relativistic 
quantum mechanics ,  the position operators are represented by

\be
X_k=i{\partial \over{\partial p_k }} , 
  \eno19)                         
\ee
which again gives agreement with eq. (12).
\skp

Now that we have an acceptable relativistic generalization of the classical
phase space, let us return to our goal of extending the
canonical coherent--state representation    to relativistic particles.  We have
seen that the Weyl--Heisenberg group, on which this representation is based,
is isomorphic to an invariant subgroup of the non-relativistic  limit of $\cal
P^\uparrow _+$.  However, this subgroup is {\sl  not\/} the non-relativistic 
limit of any subgroup of ${\cal P}_0$, since the Lie brackets of
$K_k,P_k\ \hbox{ and } \  P_0$ do not close due to 

\be
[K_j,K_k]=-ic^{-2}J_l, \ (j,k,l \ \hbox{cyclic }).  
  \eno20)      
\ee
That is, we cannot simply generalize the canonical coherent states    by
choosing the right subgroup of   ${\cal P}_0$.  However,  eq. (11)
shows  that $\cal W$ also plays another role in the group ${\cal G}_1$,
namely as a {\sl  homogeneous space}:

\be
{\cal W}= {\cal G}_1/SO(3) .
  \eno21)                                           
\ee
In this form, it does have an obvious relativistic counterpart,
namely $\cal C$.  In retrospect, the role of ${\cal W}$ as a group is purely
incidental to quantum mechanics  , since there is no fundamental reason why
the set of all dynamical states should form a group.  On the other hand, this
set should certainly be a homogeneous space under the group of  motions,
since this group must transform dynamical states into one another and this
action can be assumed to be transitive (otherwise we may as well restrict
ourselves to an orbit).  In either of its roles relative to  ${\cal G}_1$, $\cal
W$ acquires a slightly different physical interpretation from the one it had in
relation to the canonical commutation relations:  Since the group--manifold
of  $\cal W$ is generated by the vector fields $P_k, K_k\ \hbox{ and } \ M$,
the coordinates on $\cal W$ are the  positions $x_k$ (generated by $P_k$),
the velocities $v_k$ (generated by $K_k$) and the variable  generated by
$M$, which is a degenerate form of ``time''  inherited form Relativity through
the limit $c^{-2}P_0\rightarrow M$.  By comparison, the coordinates on the
original Weyl--Heisenberg group were $x_k$ (generated by
$P_k$), the momentum $p_k$ (generated by $X_k$) and a dimensionless
``phase angle'' $\phi $ generated by the identity operator.  With this new
interpretation, $\cal W$ is the product of velocity phase space with ``time''
and truly represents the non-relativistic  limit of $\cal C$. 

We will construct a coherent-state representation  of ${\cal P}_0$ 
by first discovering its non-relativistic  limit.   This limit  will be a
 representation   of a {\sl  quantum
mechanical\/} version  ${\cal G}_2$ of the Galilean group $\cal G$ and 
will be seen to be a close relative of the canonical coherent--state
representation     of $\cal  W$.  It is therefore necessary first
to understand just how the group $\cal  G$   is related to
the Poincar\'e group ${\cal P}_0$ and its non--relativistic  limit ${\cal G}_1$.  Again, we
will do everything at the level of Lie algebras. There is no problem with
globalization.  

The universal enveloping algebra of ${\wp}$ contains the mass-squared
operator 

\be
M^2=c^{-4}{P_0}^2-c^{-2}({P_1}^2+ {P_2}^2+{P_3}^2)
																		= c^{-4}{P_0}^2-c^{-2}{\bf P}^2 ,    
  \eno22)   
\ee
which is a Casimir operator, i.e. commutes with all  generators in 
${\wp}$.  Assuming that both $P_0$ and $M$ are represented by positive
operators and that M is invertible (and this will be the case for massive
particles), we have formally for large $c$:

\be
c^{-2}P_0=(M^2+c^{-2}{\bf P}^2)^{1/2}
																			=M+{\bf P}^2/2Mc^2+O(c^{-4}),  
  \eno23)
\ee
where we have used the fact that $M$ commutes with $P_k$. 
The operator 

\be
H={\bf P}^2/2M       
  \eno24)                                                  
\ee
 is just the Hamiltonian for the non-relativistic  free particle,
hence generates its time translations and must be included in the Lie
algebra of the Galilean group.  Let us therefore try to append it to the Lie
algebra ${\bf g}_1$ of ${\cal G}_1$.  Indeed, by  eq. (10),

\be
\matrix{[H,P_k]=0&\quad [H,M]=0\cr
                  [H,J_k]=0&\quad [H,K_k]=iP_k,} 
  \eno25)                 
\ee
 showing that

\be
{\bf g}_2 \equiv \ \hbox{Span}\ \{K_k,P_k,J_k,M,H\}  
  \eno26)
\ee
actually forms a Lie algebra with Lie brackets given by eqs. (10) and (25).
 Clearly $ {\bf g}_2$ contains  $ {\bf g}_1$ as a subalgebra.  We
will refer to the corresponding Lie group ${\cal G}_2$  as the {\sl  quantum
mechanical Galilean group}.  It is the group of  translations, rotations,
boosts, dynamics (i.e., time translations) and multiplications by constant
phase factors (generated by $M$) acting on the wave functions of a
non-relativistic quantum mechanical particle.  The relation of ${\cal G}_2$ to
the {\sl  classical\/} Galilean group $\cal  G$ is as follows:  The subgroup
generated by $M$, which can be identified with the group of real numbers
$\rl$, is central; then $\cal  G$ is obtained from ${\cal G}_2$ by factoring out
$\rl$:

\be
{\cal  G}= {\cal G}_2/\rl .  
  \eno27)                                              
\ee
The action of $\rl$ on quantum mechanical wave functions, which
a\-mounts to a multiplication by a constant phase factor, is a necessary part 
of ${\cal G}_2$ because of  $[P_r,K_s]=i\delta _{rs}M$ which, as we have
seen, is related to the uncertainty principle.  Factoring out this action means
ignoring that phase factor, so it is reasonable that it should give the
classical Galilean group.  On the Lie algebra level, it amounts to setting
$M=0$. Had we included Planck's constant $\hbar$ in eq. (10), this would
have amounted to taking the classical limit $\hbar\to 0$.  The
above relation between $\rl$, ${\cal G}_2$ and $\cal  G$ in an example of a
{\sl  central extension\/}  (Varadarajan [1969]).  One says that ${\cal G}_2$ is
a central extension  of $\rl$ by  $\cal  G$.  
\skp
The fact that $\cal W$ is a subgroup of ${\cal G}_2$ was  noted by
Bargmann [1954]; that representations of ${\cal G}_2$ are contractions of 
representations of the Poincar\'e group was shown by Mackey
[1955].\footnote*{ I thank R.~F.~Streater for these remarks.}

\vfill\eject

 \noindent {\bf 4.3.  Galilean Frames  }
\def\rightheadline{\tenrm\hfil {\sl 4.3.  Galilean Frames }\hfil\folio}  
 \skp

\noindent Our object in this section is to construct coherent states which are
naturally associated with free non--relativistic  particles,  just as the
canonical coherent states are associated with the Weyl--Heisenberg
group or the Harmonic oscillator and the spin coherent states are associated
with with $SU(2)$.  An obvious starting point would be to apply the
Klauder--Perelomov method (chapter 1) to the quantum--mechanical Galilean
group  ${\cal G}_2$, since it is this group which describes such particles. 
However, this method fails, due to the fact that all the representations of
physical interest are not square--integrable.  Therefore we will follow a more
pedestrian route.  Our main guides will be analyticity (which, it turns out,
follows from an important physical condition) and ``physical intuition.''

We return to the general case where the configuration space is $\rl^s$
instead of $\rl^3$.  For simplicity, we restrict ourselves to spinless particles. 
It is not difficult to include spin, as will be shown later.  The states of such
particles are described by complex--valued wave functions  $f({\bf x} ,t)$ of
position ${\bf x} $ and time $t$ which are are square--integrable with
respect to ${\bf x} $ at any time $t$.  Their evolution in time is given by the
Schr\"odinger equation

\be
 i{\partial f\over{\partial t}}=Hf,
\eno1)
 \ee
where 

\be
 H=-{1\over 2m}\Delta 
\eno2)
 \ee
is the Hamiltonian operator, and $\Delta $ is the Laplacian acting on \break
$L^2(\rl^s)$.  Since $H$ is self--adjoint, though unbounded, the solutions are
given through the unitary one--parameter group $U(t)=\exp(-itH)$:

\ba{
f({\bf x} ,t)\a=(U(t)f)({\bf x} )\cr
\a= (2\pi )^{-s}\int_{\rl^s}d^s{\bf p}\,\exp[-it{\bf p}^2/2m
+i{\bf p}\cdot{\bf x} ]\hat f({\bf p}) ,
 \cr}
 \eno3)
 \ea
where it is assumed that $f({\bf x} ,0)$, hence also its Fourier transform 
$\hat f({\bf p})$, is in $L^2(\rl^s)$.

The key to our approach will be to note that $H$ is a non--negative
operator, hence the evolution group $U(t)$ can be analytically continued to
the lower--half complex time plane $\,\cx^-$ as

\be
 U(t-iu)=\exp[-i(t-iu)H]=e^{-itH}e^{-uH}, \quad u>0.
\eno4)
 \ee
(Note also that since $H$ is unbounded, no analytic continuation to the
upper--half time plane is possible.)  The operator $e^{-uH}$ is familiar
from two other contexts: it constitutes the evolution semigroup for the
heat equation (where $u$ is time), and it is also the unnormalized
density matrix for the Gibbs canonical ensemble describing the statistical
equilibrium of a quantum system at temperature $T$ (where $u=1/kT$).  To
get a feel for our use of this operator, let us be heuristic for a moment and
consider what happens when a free {\sl classical\/} free particle of mass
$m$ is evolved in complex time $\tau =t-iu$.  If its initial position and
momentum are ${\bf x} $ and ${\bf p}  $ respectively, then its new position
will be 

\ba{
 {\bf z} (\tau )\a={\bf x} +(t-iu){\bf p} /m\cr
\a=({\bf x} +t {\bf p} /m)-i(u/m){\bf p} \cr
\a={\bf x} (t)-i(u/m){\bf p} .
 \cr}
 \eno5)
 \ea
Since ${\bf x} (t)$ is just the position evolved in real time $t$, we see that 
$ {\bf z} (t)$ is, in fact, a {\sl complex phase space coordinate\/} of the same
type we encountered in the construction of the canonical coherent states! 
Armed with this intuition, let us now return to quantum mechanics  and see
if this idea has a quantum mechanical counterpart.  The operator $e^{-u
H}$, when applied to any function in $L^2(\rl^s)$, gives

\ba{
 f_u({\bf x} )\a \equiv (e^{-uH}f)({\bf x} )\cr
 \a=(2\pi )^{-s}\int_{\rl^s}d^s{\bf p}\,\exp[-u{\bf p}^2/2m
+i{\bf p}\cdot{\bf x} ]\hat f({\bf p}).
 \cr}
 \eno6)
 \ea
If we replace ${\bf x} $ in the integrand by an arbitrary ${\bf z} \in \cx^s$,
the integral still converges absolutely since the quadratic term in the
exponent dominates the linear term for large $|{\bf p} |$.  Clearly the
resulting function is {\sl entire\/} in ${\bf z} $ (differentiating the integrand
with respect to $z_k$ still gives an absolutely convergent integral).  
This shows that the group of Galilean space--time translations,

\be
 U({\bf x} ,t)=\exp\bigl(-itH+i{\bf x} \cdot {\bf P} \bigr) ,
\eno7)
 \ee
extends analytically to a {\sl semigroup\/} of {\sl complex\/} space--time
translations 

\be
 U(\bar{\bf z} ,\bar\tau )=\exp\bigl(-i\bar\tau H+i\bar{\bf z} \cdot {\bf P}
\bigr) 
\eno8)
 \ee
defined over the complex space--time domain

\be
 {\cal D}=\{\,({\bf z}, \tau) \,|\,{\bf z} \in\cx^s, \tau \in\cx^-\,\}.
\eno9)
 \ee
This translation semigroup can be combined with the rotations and boosts to
give an analytic semigroup ${\cal G}_2^c$ extending ${\cal G}_2$.

 Let
${\cal H}_u$ be the vector space of all the entire functions $f_u
({\bf z} )$ as $\hat f({\bf p})$ runs through $L^2(\rl^s)$.  Then 

\be
 f_u({\bf z} )=\langle\, e^u_{\bf z} \,|\,\fhat\,\rangle, 
\eno10)
 \ee
where

\ba{
 e^u_{\bf z}({\bf p} ) \a=(2\pi )^{-s}\exp[-u{\bf p}^2/2m
                -i{\bf p}\cdot{\bar{\bf z} } ]\cr
\a=(2\pi )^{-s}\exp\left [m{\bf y} ^2/2u -{u
\over {2m}} ({\bf p} -m {\bf y} /u)^2-i{\bf p} \cdot {\bf x} \right]
 \cr}
 \eno11)
 \ea
are seen to be Gaussian wave packets in momentum space with expected
position and momentum given in terms of ${\bf z} \equiv {\bf x} -i{\bf y} $
by

\be
 \langle\, X_k\,\rangle=x_k\quad \hbox{and}\quad  \langle\, P_k\,\rangle=
(m/u)y_k.
 \eno12)
 \ee
The $e^u_{\bf z}$'s are easily shown to have minimal uncertainty
products.  The momentum uncertainty can be
read off directly from the exponent and is 

\be
 \Delta _{P_k}=\sqrt{m/2u},
\eno13)
 \ee
hence

\be
 \Delta _{X_k}=\sqrt{u/2m}.
\eno14)
 \ee

We now have our prospective coherent states and their label space
$M=\cx^s$.  To construct a coherent--state representation, we need a
measure on $M$ which will make the $e^u_{\bf z} $'s into a frame. 
Since the $e^u_{\bf z}$'s are Gaussian, the measure in not difficult to
find:

\be
 d\mu _u({\bf z}) =(m/\pi u)^{s/2}\exp\left(  -m {\bf y} ^2/u
       \right) d^s {\bf x} \,d^s {\bf y} .
\eno15)
 \ee
Defining

\be
\langle\, f\,|\,g\,\rangle _{{\cal H}_u}\equiv \langle\, f_u
\,|\,g_u\,\rangle\equiv 
\int d\mu _u({\bf z} ) \br{f_u({\bf z}) }g_u({\bf z} ),
\eno16)
 \ee
we have 
\skp
\noindent { \bf Theorem 4.1.}  {\sl 

\item{(a)}   $\langle\, \cdot\,|\,\cdot\,\rangle_{{\cal H}_u}$ is an
inner product  on ${\cal H}_u$ under which  ${\cal H}_u$ is a
Hilbert space.

\item{(b)}   The map $e^{-uH}$ is unitary from $L^2(\rl^s)$  onto ${\cal H}_u$.

\item{(c)}   The $e^u_{\bf z}$'s define a resolution of unity on
$L^2(\rl^s)$ given by 
\be
 \incs  d\mu _u({\bf z})\,\,|\,e^u_{\bf z}\,\rangle\langle\, 
   e^u_{\bf z}\,|\,=I. 
\eno17)
\ee
}
\skp

\noindent {\bf Proof.}  We  prove   that 
$\|\,f\,\|^2_{{\cal H}_u}\equiv 
\langle\, f\,|\,f\,\rangle _{{\cal H}_u} =\|\fhat\|^2_{L^2}$. The inner
product can be recovered by polarization. To begin with, assume that $\fhat$
is in the Schwartz space ${\cal S}(\rl^s)$ of ra\-pidly decreasing smooth test
functions. Then

\be
 f_u({\bf x} -i {\bf y} )=\left( \exp\bigl[ -u{\bf p}^2/2m+
{\bf y} \cdot {\bf p} \bigr]\,\fhat \right) \check{\,}\,({\bf x} ),
\eno18)
 \ee
hence by Plancherel's theorem,

\ba{
\int d^s {\bf x} \,|\,f_u\a ({\bf x} -i {\bf y} )\,|\,^2\cr
\a=(2\pi )^{-s}\int d^s
     {\bf p}  \exp\left( -u{\bf p} ^2/m+2{\bf y} \cdot {\bf p} 
\right)\,|\,\fhat({\bf p})  \,|\,^2
\cr}
 \eno19)
 \ea
and 

\ba{
\int d\mu_u\a ({\bf z} ) \,|\,f_u({\bf z} )\,|\,^2\cr
   \a=(m/\pi u)^{s/2}(2\pi )^{-s}
        \int d^s {\bf p} \exp\bigl[-u{\bf p} ^2/m\bigr]
         \,|\,\fhat({\bf p} )\,|\,^2\times\cr
\a\qquad\qquad  \int d^s {\bf y} \exp \bigl( 
-m {\bf y} ^2 /u+2 {\bf y} \cdot {\bf p}  \bigr)\cr
\a=(2\pi )^{-s}\int d^s {\bf p} \,|\,\fhat({\bf p} )\,|\,^2=\|\,\fhat\,\|^2,
\cr}
 \eno20)
 \ea
where exchanging the order of integration was justified since the integrals
are absolutely convergent.  This proves (b), hence also (a), for $f\in {\cal
S}(\rl^s)$.  Since the latter space is dense in $L^2(\rl^s)$, the proof extends
to $f\in L^2(\rl^s)$ by continuity.  (c) follows by noting that

\ba{
\langle\, \fhat\,|\,\hat g\,\rangle_{L^2}\a=\langle\, f\,|\,g\,\rangle
          _{{\cal H}_u}\cr
\a=\int d\mu _u({\bf z} ) \br{f_u({\bf z}) }g_u({\bf z} )\cr
\a=\int d\mu _u({\bf z} ) \langle\, \fhat\,|\,e^u_{\bf z}\,\rangle
\langle\, e^u_{\bf z}\,|\,\hat g\,\rangle
 \cr}
 \eno21)
 \ea
and dropping $\langle\, \fhat\,|\,$ and $\,|\,\hat g\,\rangle$.\sp\qed
\skp
Using the map $e^{-uH}$, we can transfer any structure from
$L^2(\rl^s)$ to ${\cal H}_u$.  In particular, time evolution is given by

\ba{
 f_u({\bf z} ,t)\a=\left( e^{-uH}\bigl(e^{-itH}f\bigr)  \right)
           ({\bf z})\cr
\a=(2\pi )^{-s}\int d^s {\bf p} \,\exp\bigl[ -i \tau {\bf p} ^2/2m+i {\bf z} 
 \cdot {\bf p} \bigr]\,\fhat({\bf p} )\cr
\a\equiv \langle\, e_{{\bf z},\tau  }\,|\,\fhat\,\rangle,
 \cr}
 \eno22)
 \ea
where $\tau =t-iu$ and the wave packets 

\be
 e_{{\bf z}, \tau }({\bf p})=(2\pi )^{-s}\exp[-\bar\tau  {\bf p}^2/2m
                -i{\bf p}\cdot{\bar{\bf z} } ]
\eno23)
 \ee
are obtained from the $e^u_{\bf z}$'s by evolving in real time $t$.  They
cannot be of minimal uncertainty since the free-particle Schr\"odinger
equation is neccessarily dissipative.  Instead, they give the following
expectations and uncertainties:

\ba{
\langle\, P_k\,\rangle\a=(m/u)y_k\cr
\langle\, X_k\,\rangle\a=x_k-(t/u)y_k
          =x_k-(t/m)\langle\, P_k\,\rangle\cr
\Delta _{P_k}\a=\sqrt{m/2u}\cr
\Delta _{X_k}\a=\sqrt{{u\over {2m}}\left(1+{t^2\over{u^2}} 
\right)} .
 \cr}
 \eno24)
 \ea
Since 

\be
e_{{\bf z},\tau }=e^{itH}e^u_{\bf z}, 
\eno25)
 \ee
it follows that 

\ba{
\|f\|^2_\tau \a\equiv \incs d\mu _u({\bf z} )\,|\,f({\bf z} ,\tau )\,|\,^2\cr
\a=\|e^{-it{\bf p} ^2/2m}\fhat\|^2=\|\fhat\|^2,
 \cr}
 \eno26)
 \ea
thus we have a frame $\{\,e_{{\bf z},\tau }\,|\,{\bf z} \in \cx^s\,\}$ at each
complex ``instant'' $\tau =t-iu$, with the corresponding resolution of
unity

\be
 \incs d\mu _u({\bf z} )\,|\,e_{{\bf z},\tau }\,\rangle\langle\, 
e_{{\bf z},\tau } \,|\,=I.
\eno27)
 \ee 

The space $L^2(\rl^s)$ carries a  representation  of the
quantum mechanical Galilean group ${\cal G}_2$.  Since the 
$e_{{\bf z},\tau }$'s were obtained from the dynamics associated with this
group, they transform naturally under its action.  A typical element of ${\cal
G}_2$  has the form $g=(R, {\bf v} ,{\bf x}_0, t_0, \theta  )$, where $R$ is a
rotation, ${\bf v} $ is a boost, ${\bf x}_0 $ is a spatial translation, $t_0$ is a
time--translation and $\theta  $ is  the ``phase'' parameter associated with
the central element  $M=m/\hbar\equiv m$ in our
representation (see section 4.2).  $g$ acts on the complex space--time
domain $\cal D$ by sending the point $({\bf z},\tau  )$ to $(\tau' ,{\bf z}' )$,
where

\ba{
{\bf x} '\a=R {\bf x} +t {\bf v} +{\bf x}_0\cr
{\bf y} '\a=R {\bf y} +u{\bf v} \cr
t'\a=t+t_0\cr
u'\a=u.
 \cr}
 \eno28)
 \ea
The parameter $\theta $ has no effect on space--time; it only acts on wave
functions by multiplying them by a phase factor.  The representation of
${\cal G}_2$ is defined by 

\be
\bigl( U_gf  \bigr) ({\bf z} , \tau )=e^{-im\theta }f\bigl( g ^{-1} ({\bf z} , \tau ) 
\bigr) .
\eno29)
 \ee
Thus we have

\be
 U_g\, e_{{\bf z},\tau }=e^{im\theta }e_{g({\bf z} ,\tau )}, 
\eno30)
 \ee
 and the $e_{{\bf z},\tau }$'s are ``projectively covariant'' under the action of
${\cal G}_2$; if we define $e_{{\bf z},\tau ,\phi }\equiv e^{-im\phi }e_{{\bf
z},\tau}$, then this expanded set is invariant under the action of ${\cal G}_2$,
with $\phi '=\phi -\theta $. Since $e_{{\bf z},\tau }$ and $e_{{\bf z},\tau
,\phi} $ represent the same physical state, we won't be fussy and just work
with the $e_{{\bf z},\tau }$'s.  Anyway, this anomaly will disappear when we
construct the corresponding relativistic coherent states.

The above representation  of ${\cal G}_2$ on $L^2(\rl^s)$ can be transfered
to ${\cal H}_u$ using the map $e^{-uH}$.  This map therefore {\sl  
intertwines\/} (see Gelfand, Graev and Vilenkin [1966])  the representations
on ${\cal G}_2$ on $L^2(\rl^s)$ with the one on ${\cal H}_u$.

We conclude with some general remarks.
\skp
\noindent 1.\sp  Since the $e^u_{\bf z}$'s are spherical and therefore
invariant under $SO(n)$ (which is, after all, why they describe spinless
particles), they can be parametrized by the homogeneous space ${\cal
W}={\cal G}_1/SO(n)$ as long as we keep $u$ fixed ($u$ is a
parameter associated with the Hamiltonian, which is a generator of ${\cal
G}_2$ but not of ${\cal G}_1$).  The action of $\cal W$ as a 
{\sl subgroup\/} of  ${\cal G}_1$ on the  $e^u_{\bf z}$'s is
preserved in passing to the homogeneous space, hence $\cal W$ acts to
translate these vectors  in phase space.  This explains the
similarity between the $e^u_{\bf z}$'s and the canonical coherent
states.  On the other hand, {\sl dynamics\/} (in imaginary time) is
responsible for the parameter $u$.  If we write ${\bf k} \equiv
(m/u){\bf y} $, then 
\ba{
e^u_{\bf z}({\bf p} ) \a=(2\pi )^{-s}
\exp\left [{u\over {2m}}{\bf k}^2  -{u
\over {2m}} ({\bf p} -{\bf k})^2 -i{\bf p} \cdot {\bf x} \right]\cr
\a\equiv \exp\bigl[u{\bf k}^2/2m\bigr]e^{-i{\bf p} \cdot {\bf
x}}\,h\left( {\bf p} -{\bf k} \over{\sqrt{2m/u}} \right).
 \cr}
 \eno31)
 \ea
The measure $d\mu _u({\bf z} )$ is now
\be
d\mu _u({\bf x},{\bf k})=(u/\pi m )^{s/2}\exp\bigl( 
-u{\bf k}^2/m \bigr)\,d^s{\bf x} \,d^s {\bf k} .
\eno32)
\ee
Hence the exponential factor $\exp[u{\bf k}^2/2m]$ in $e^u_{\bf z}$,
when squared in
the reconstruction formula, precisely cancels the Gaussian weight factor in
$ d\mu _u({\bf x},{\bf k})$, leaving the measure
\be
 \bigl(u/\pi m\bigr)^{s/2}\,d^s {\bf x} \,d^s {\bf k}
\eno33)
 \ee  
in phase space.  It follows from the above form of $e^u_{\bf z}$  that
$2\Delta _{P_k}=\sqrt{2m/u}$ plays the role of a {\sl scale factor\/}  in
momentum space (as used in the wavelet transforms of   chapter 1), hence
its reciprocal
 $\Delta _{X_k}=\sqrt{u/2m}$ acts as a scale factor in configuration
space. Thus the Gali\-lean coherent states combine the properties of rigid
``windows'' with those of wavelets, due to the fact that their analytic
semigroup ${\cal G}_2^c$  includes both phase--space translations and
scaling, the latter due to the heat operator $e^{-uH}$.  However, note
that $u$ is constant, though arbitrary, in the resolution of unity and the
corresponding reconstruction formula.  Since there is an abundance of
``wavelets'' due to translations in phase space, only a single scale is
needed for reconstruction.  (One could, of course, include a range of scales
by integrating over $u$ with a weight function, but this seems
unnecessary.)  In the treatment of relativistic particles, $u$ becomes
the time component of a four--vector $y=(u,{\bf y} )$, hence will no
longer be constant.  This is because {\sl relativistic windows\/} shrink in the
direction of motion, due to Lorentz contractions, thus automatically adjusting
to the analysis of high--frequency components of the spectrum.
\skp

\noindent 2.\sp Notice that $e^u_{\bf z} $ is essentially the  heat
operator $e^{-uH}$ applied to the $\delta $--function at $x$, then
analytically continued to $\bar{\bf z}={\bf x} +i{\bf y}  $.  The fact that all
the $e^u_{\bf z}$'s have minimal uncertainties shows that the action of
the heat semigroup $\{U(-iu)\}$ is such that  while the position
undergoes the normal diffusion, the momentum undergoes the opposite
process of {\sl refinement,\/} in just such a way that the product of the two
variances remains constant. This is reflected in the fact that the operator
$e^{-uH}$, whose inverse in $L^2(\rl^s)$ is unbounded, becomes unitary when
the functions in its range get analytically continued, and the reconstruction
formula is just a way of inverting $e^{-uH}$. Hence no
information is lost if one looks in phase space rather than
configuration space! It seems to me that this way of ``inverting'' semigroups 
must be an example of a general process.  If such a process exists, I am
unaware of it. In our case, at least, it appears to be possible because of
analyticity.

\skp
\noindent 3.\sp So far, it seems that coherent--state representations are
intimately connected with groups and their representations.  However, there
is a reasonable chance   that coherent--state
representations similar to the  above can be constructed for systems
which, unlike free particles,  do not possess a great deal of
symmetry.  Suppose we are given a system of $s/3$ particles in $\rl^3$
which interact with one  another and/or with an external source
through a potential $V({\bf x} )$.  We assume that $V({\bf x} )$ is
time--independent, so the system is conservative. (This means that we do
have one symmetry, namely under time translations. If, moreover,  the
potential depends only on the differences ${\bf x}_i-{\bf x}_j$ between
individual particles, we also have symmetry with respect to translations of
the center of mass of the entire system; but we do not make this assumption
here.)  This system is then described by a Schr\"odinger equation with the
Hamiltonian operator $H=H_0+V$, where $H_0$ is the free Hamiltonian and
$V$ is the operator of multiplication by $V({\bf x} )$.  We need to assume
that this (unbounded) operator can be extended to a self--adjoint operator on
$L^2(\rl^s)$.  How far can the above construction be carried in this case?
The key to our method was the positivity of the free
Hamiltonian $H_0={\bf P}^2/2m$.  But  a  general  Hamiltonian 
must at least satisfy the {\sl stability condition:}
 \skp 
{\sl (S) \sp The spectrum of $H$ is bounded below.  } 
\skp
\noindent If $H$ fails to meet this condition, then the system it describes is
unstable, and a small perturbation can make it cascade down,
giving off an infinite amount of energy.  For a stable system,
the evolution group $U(t)=e^{-itH}$ can be analytically continued to
an analytic semigroup $U(\tau )=e^{-i\tau H}$ in the
lower--half complex time plane as in the free case.  Depending on the
strength of the potential,  the functions $f_u=U(-iu)f$ may be
continued to some subset of\/ $\cx^s$. Formally, this corresponds to defining
\ba{
f({\bf z},\tau  )\a=\left( e^{i{\bf z}\cdot{\bf P}}e^{-i\tau  H}f \right)({\bf
0}),\quad \tau =t-iu\cr 
\a=\left( e^{{\bf y}\cdot {\bf P}}e^{-i\tau  H}f \right)({\bf x})
 \cr}
 \eno34)
 \ea
for an initial function $f({\bf x} )$ in $L^2(\rl^s)$.  As mentioned, this
expression is formal since the operator  $e^{{\bf y}\cdot {\bf P}}$ is
unbounded and $e^{-i\tau  H}f $ may not be in its domain.  But it
can  make sense operating on the range of $e^{-i\tau H}$, which coincides
with the range of  $e^{-uH}$, provided ${\bf y} $ is not too large.  Let  ${\cal
Y}_u$ be the set of all ${\bf y} $'s for which  $e^{{\bf y}\cdot {\bf P}}$ is
defined on the range of $e^{-uH}$ and, furthermore, the function 
$\exp\bigl[i{\bf z}\cdot{\bf P}\bigr]\*\exp\bigl[-i\tau  H\bigr]f$ is
sufficiently regular to be evaluated at the origin in $\rl^s$, no matter which
initial $f$ was chosen in $L^2(\rl^s)$.  For many potentials, of course, ${\cal
Y}_u$ will consist of the origin alone; in that case there are no coherent
states.  We assume that  ${\cal Y}_u$ contains at least some open
neighborhood of the origin.  Intuitively, we may think of ${\cal Y}_u$ as the
set of all imaginary positions which can be attained by the particle in an
imaginary time--interval $u$, while moving in the potential $V$.
 In the free case, ${\cal Y}_u=\rl^s$ and there is no restriction on ${\bf y} $
provided only that $u>0$. This corresponds to the fact that there is no
``speed limit'' for free non--relativistic   particles, hence a particle can get
to any imaginary position in a given positive imaginary time. For
relativistic free particles, ${\cal Y}_u$ is  the open sphere of radius $uc$,
where $c$ is the speed of light.  
Returning to our system of interacting particles,  define the associated
{\sl complex space--time domain}

\be
 {\cal Z}_H=\{\,({\bf x}-i{\bf y}, t-iu)\in \cx^{s+1}\,|\,u>0,\ {\bf
y}\in{\cal Y}_u\,\}. 
\eno35)
 \ee
This is the set of all complex space--time points which can be reached by
the system in the presence of the potential $V({\bf x} )$, and it is the label
space for our prospective coherent states. These are now defined as {\sl
evaluation maps\/} on the space of analytically continued solutions:

\be
 f({\bf z} ,\tau )=\langle\, e_{{\bf z},\tau }^H\,|\,f\,\rangle, 
\eno36)
 \ee
the inner product being in $L^2(\rl^s)$. Then from the above 
expression, again formally, we have the {\sl dynamical coherent states}

\be
 e_{{\bf z},\tau }^H=e^{i\bar\tau H}e^{-i\bar {\bf z}\cdot {\bf P}} \delta _{\bf
0} 
\eno37)
 \ee
for $({\bf z} ,\tau )$ in ${\cal Z}_H$.

 What is still missing, of course, is the 
measure $d\mu _u ^H $. (Since the potential is $t$--independent, so
will be the measure, if it exists.) Finding the measure promises to be equally
difficult to finding the  propagator for the dynamics.  The latter is closely
related to the reproducing kernel, 

 \be
 K_H({\bf z} ,\tau ;\,\bar{\bf z} ', \bar \tau ')=\langle\, e_{{\bf z},\tau }^H
\,|\,e_{{\bf z'},\tau'}^H\,\rangle.  
\eno38)
 \ee
$K_H$ depends on $\tau $ and $\bar\tau '$ only through the
difference $\tau - \bar\tau '$, and is the analytic continuation of  the
propagator to the domain
 ${\cal Z}_H\times\br{ {\cal Z}_H}$.  It is related to the measure through
the reproducing property,
\ba{
 \int_{\sigma_\tau  } \,d\mu _u ^H({\bf z} )\,
K_H({\bf z'} ,\tau' ;\,\bar{\bf z} , \bar \tau )\,\a 
K_H({\bf z} ,\tau ;\,\bar{\bf z}'' , \bar \tau'' )\cr
\a=K_H({\bf z'} ,\tau' ;\,\bar{\bf z}'' , \bar \tau'' ),
 \cr}
\eno39)
\ea
where the integration is carried out over a ``phase space''  $\sigma_\tau$
 in  ${\cal Z}_H$ with a fixed value of $\tau=t-iu$.  A reasonable candidate
for $d\mu ^H_u$ (see section 4.4) is 

\ba{
d\mu^H_u( {\bf z} )\a=C\,\|e^H_{{\bf z},\tau }\,\|^{-2}
\,d^s {\bf x} \,d^s {\bf y}\cr
\a\equiv C\,e^{-\phi _u({\bf z})}\,d^s {\bf x} \,d^s {\bf y}.
 \cr}
 \eno40)
 \ea

Rather than finding the measure explicitly, a more likely
possibility is that its existence can be proved by functional--analytic
methods for some classes of potentials and approximation techniques may be
used to estimate it or at least derive some of its properties. The theoretical
possibility that such a measure exists raises the prospect of an interesting
analogy between the quantum mechanics  of a {\sl single\/} system
 and a {\sl statistical ensemble} of
corresponding {\sl classical\/} systems at equilibrium with a heat reservoir. 
In the case of a free particle, if we set ${\bf k} =(m/u){\bf y} $ as above
(see remark 1) and define $T$ by  $u=1/2kT$ where $k$ is Boltzmann's
constant, then it so happens that our measure  $d\mu _u$ is identical to
the Gibbs measure for a classical canonical ensemble (see Thirring [1980]) of
$s/3$ free particles of mass $m$ in $\rl^3$, at equilibrium with a heat 
reservoir at absolute temperature $T$.  Thus, integrating with  $d\mu _u$
over phase space is very much like taking the classical thermodynamic
average at equilibrium!  It remains to be seen, of course, whether this is a
mere coincidence or if it has a generalization to interacting systems.

There is also a connection between the expectation values of an operator
$A$ in the coherent states $e^H_{{\bf z} ,\tau }$ and its thermal average in
the Gibbs state,

\be
 \langle\, A\,\rangle_\beta \equiv Z ^{-1}\, {\rm Trace}\,\left(
 e^{- \beta H}\,A\right)=Z ^{-1}\, {\rm Trace}\,\left(  e^{- \beta H/2}\,A\,
 e^{- \beta H/2} \right),
\eno41)
 \ee
where $Z\equiv {\rm Trace}\,\left(  e^{- \beta H}\right)$.  Namely, if we
have the resolution of unity 

\be
 \int_{\sigma_\tau  } \,d\mu _u ^H({\bf z} )\,\,|\,e^H_{{\bf z},\tau }\,\rangle
\langle\, e^H_{{\bf z},\tau }\,|\,=I,
\eno42)
 \ee
then

\ba{
\langle\, A\,\rangle_\beta \a=Z ^{-1}  \int_{\sigma_\tau  } 
\,d\mu _u ^H({\bf z} )\, \langle\, e^H_{{\bf z},\tau }\,|e^{-\beta H/2}\,A
e^{-\beta H/2}\,e^H_{{\bf z},\tau }\,\rangle\cr
\a=Z ^{-1}  \int_{\sigma_\tau  } 
\,d\mu _u ^H({\bf z} )\, \langle\, e^H_{{\bf z},\tau-i\beta /2 }\,A
\,e^H_{{\bf z},\tau-i\beta /2 }\,\rangle\cr
\a\equiv Z ^{-1}  \int_{\sigma_\tau  } \,d\mu _u ^H({\bf z} )\,
\tilde A({\bf z}, \tau -i\beta /2),
 \cr}
 \eno43)
 \ea
where we have used the formula

\be
 {\rm Trace}\, A=\int_{\sigma_\tau  } \,d\mu _u ^H({\bf z} )\,\langle\, 
e^H_{{\bf z},\tau }\,|\,A\,e^H_{{\bf z},\tau }\,\rangle,
\eno44)
 \ee
which follows easily from eq. (42).
Thus  taking the thermal average means shifting the imaginary part 
$u$ by $\beta /2$ in the integral. 
\secskp

 \noindent {\bf 4.4. Relativistic Frames  }
\def\rightheadline{\tenrm\hfil {\sl 
4.4. Relativistic Frames}\hfil\folio}  
 \skp
\noindent We are at last ready to embark on the main theme of this 
book:  A new synthesis of Relativity and quantum mechanics  through the
geometry of complex spacetime.  The main tool for this synthesis will be 
the physically necessary condition that the energy operator of the total
system be non--negative, also known in quantum field theory  as the
{\sl spectral condition.\/} The (unique) relativistically covariant statement
of this condition gives rise to a canonical complexification of spacetime
which embodies in its geometry the structure of quantum mechanics  as
well as that of Special Relativity. The complex spacetime also has the
structure of a classical phase space underlying the quantum system under
consideration. Quantum physics is developed through the
construction of frames labeled by the complex spacetime manifold,  
which thus forms a natural bridge between the  classical
and quantum aspects of the system.  It is hoped that this marriage, once fully
developed, will survive the transition from Special to General Relativity.  

As mentioned at the beginning of this chapter, the Perelomov--type
constructions of chapter 3 do not apply directly to the Poincar\'e group
since its time evolution (dynamics) is non--trivial.   Pending a generalization
of these methods to  dynamical groups, we merely use the ideas
of chapter 3 for inspiration rather than substance.  In fact, it may well be 
that a closer examination of the construction to be developed here may
suggest such a generalization.

We begin with the most basic object of relativistic quantum mechanics, the
Klein--Gordon equation, which describes a simple relativistic particle in the
same way that the Schr\"odinger equation describes a non--relativistic
particle.  The spectral condition will enable us to analytically continue 
the solutions of this equation to complex spacetime, and the evaluation maps
on the space of these analytic solutions will be bounded linear functionals,
giving rise to a reproducing kernel as in section 1.4.  Physically, the
evaluation maps are optimal wave packets, or coherent states, and it is this
interpretation which establishes the underlying complex manifold as an
extension of classical phase space.  The next step is to build {\sl frames \/} of
such coherent states. (Recall from section 1.4 that a frame determines a
reproducing kernel, but not vice versa.)

The coherent states we are about to construct are covariant under the
restricted Poincar\'e group, hence they represent  {\sl relativistic wave
packets\/} .  As we have seen, such a covariant family  is closely
related to a unitary irreducible representation  of the appropriate group,  in
this case ${\cal P}_0$. Such representations are called {\sl elementary
systems,\/}  and correspond roughly to the classical notion of {\sl
particles,\/} though with a definite quantum flavor. (For example, physical
considerations prohibit them from being localized at a point in space, as will
be discussed later.) We will focus on representations corresponding to {\sl
massive particles.\/}  (A phase--space formalism for massless particles
would be of great interest, but to my knowledge, no satisfactory formulation
exists as yet.)  Such representations are characterized by two parameters, the
mass $m> 0$ and the spin $j=0,1/2, 1, 3/2, \ldots$ of the corresponding
particle.  We will specialize to spinless particles ($j=0$) for simplicity.  The
extension of our construction to particles with spin is not difficult and will be
taken up later.  Thus we are interested in the (unique, up to equivalence)
representation  of ${\cal P}_0$ with $m>0$ and $j=0$.  A natural way to
construct this representation  is to consider the space of solutions of  the
Klein--Gordon equation 

\be
 (\del +m^2c^2)f(x)=0,
\eno1)
 \ee
where 

\ba{
 \del\a=c^{-2}{\partial^2 \over{\partial t^2}}-\Delta \cr
\a=\partial^{\,\mu} \partial_\mu 
 \cr}
 \eno2)
 \ea
is the Del'Ambertian, or wave operator, $\Delta $ is the usual spatial
Laplacian and $\partial_\mu =\partial/\partial x^\mu $.  The function $f$ is
to be complex--valued (for spin $j$, $f$ is valued in $\cx^{2j+1}$).  We set
$c=1$ except as needed for future reference.  If we write $f(x)$ as a Fourier
transform,

\be
 f(x)=(2\pi )^{-s-1}\int_{\, {\rm I\!R}^{s+1}} d^{s+1}p\ e^{-ixp}\,\ftil(p),
\eno3)
 \ee
then the Klein--Gordon equation requires that $\ftil(p)$ be a distribution
supported on the {\sl mass shell}

\be
 \Omega _m=\lbrace p=(p_0, {\bf p} )\in \rl^{s+1}\,|\,p^2\equiv p_0^2- {\bf
p}^2 =m^2\rbrace.
\eno4)
 \ee
$\Omega _m$ is a two--sheeted hyperboloid,

\be
 \Omega _m=\Omega _m^+\cup \Omega _m^-,
\eno5)
 \ee
where 

\be
 p_0=\pm\sqrt{m^2+ {\bf p} ^2}\equiv \pm\omega ( {\bf p} )
\ \hbox{on}\    \Omega _m^\pm.
\eno6)
 \ee
Taking  

\be
 \ftil(p)=2\pi \,\delta (p^2-m^2)\,a(p)
\eno7)
 \ee
for some function $a(p)$ on $\Omega _m$, and using 

\ba{
 \delta (p^2-m^2)\a=\delta \left((p_0-\omega )(p_0+\omega )  \right)\cr
\a={ 1\over 2\omega }\left[ \delta (p_0-\omega )+\delta (p_0+\omega )
\right],
 \cr}
 \eno8)
 \ea
we get

\ba{
f(x)\a=\ino e^{-ixp} \,a(p)\cr
\a=\inop \left[ e^{-ixp} \,a(p)+e^{ixp} \,a(-p) \right],
 \cr}
 \eno9)
 \ea
where 

\be
 d\tilde p=(2\pi )^{-s}(2\omega ) ^{-1} d^s {\bf p} 
\eno10)
 \ee
is the unique (up to a constant factor) Lorentz--invariant measure on
$\Omega _m$. (The factor $\omega ^{-1} $ corrects for Lorentz contraction in
frames at momentum $p$.)   For physical particles, we must require that the
energy be positive, i.e. that $a(p)=0$ on $\Omega_m ^-$.  Hence the physical
states are given as {\sl positive--energy solutions,}

\be
 f(x)=\inop e^{-ixp} \,a(p).
\eno11)
 \ee
The function $a(p)$ can now be related to the initial data by setting
$x^0\equiv t=0$, which shows that 

\ba{
 f_0({\bf x} )\a\equiv f({\bf x},0)=\inop e^{i{\bf x} \cdot {\bf p} } a(p)\cr
\a=\left( (2\omega ) ^{-1} a \right)\check{ }\ ( {\bf x} ),
 \cr}
 \eno12)
 \ea
so 

\be
 a(p)\equiv a(\omega ,{\bf p})=2\omega \fhat_0 ({\bf p} ),
\eno13)
 \ee
where $\hat{ }$ denotes the spatial Fourier transform.  In particular, $f(x)$ is
determined by its values on the Cauchy surface $t=0$.  For general solutions
of the Klein--Gordon equation, we would also need to specify $\partial f
/\partial t$ on that surface, but restricting ourselves to
positive--energy solutions means that $f(x)$ actually satisfies the
first--order pseudo--differential {\sl non--local\/} equation

\be
 i{\partial f\over{\partial t}}=\sqrt{m^2-\Delta }\,f(x)
\eno14)
 \ee
(which implies the Klein--Gordon equation), hence only $f({\bf x},0)$ is
necessary to determine $f$.  (We will see that when analytically continued
to complex spacetime, positive--energy solutions have a local
characterization.) The inner product on the space of positive--energy
solutions is defined using the Poincar\'e--invariant norm in momentum
space,

\be
 \|f\|^2\equiv \inop \,|\,a(p)\,|\,^2.
\eno15)
 \ee
We will refer to the Hilbert space 

\be
 L^2_+(d\tilde p)\equiv \lbrace a\in L^2(d\tilde p)\,|\,a(p)=0 \ {\rm on }\ 
\Omega _m^- \rbrace
\eno16)
 \ee
as the space of positive--energy solutions in the {\sl momentum
representation.}  It carries a unitary irreducible
representation  of ${\cal P}_0$ defined as follows.  The natural action of
${\cal P}_0$ on spacetime is 

\be
 (b, \Lambda)x=\Lambda x+b,
\eno17)
 \ee
where $\Lambda $ is a resticted Lorentz transformation ($\Lambda \in
{\cal L}_0$ )  and $b$ is a spacetime translation.  Since the Klein--Gordon
equation is invariant under ${\cal P}_0$, the induced action on functions
over spacetime transforms solutions to solutions.  Since the positivity of the
energy is also invariant under ${\cal P}_0$, the  subspace of
positive--energy solutions is also left invariant.  ${\cal P}_0$ acts on
solutions  by

\be
 \left( U(b,\Lambda)f \right)(x)\equiv f\left( \Lambda ^{-1} (x-b) \right).
\eno18)
 \ee
The invariance of the inner product on $L^2_+(d\tilde p)$ then implies that
the induced action on that space  (which we denote by the same operator)
is 

\be
 \left( U(b,\Lambda)\,a \right)(p)=e^{ibp}\,a\left( \Lambda ^{-1} p \right).
\eno19)
 \ee
The invariance of the measure $d\tilde p$ then shows that $U(b,\Lambda)$
is unitary, thus $(b,\Lambda)\mapsto U(b,\Lambda)$ is a unitary
representation  of ${\cal P}_0$.  It can be shown that it is, furthermore,
irreducible.  
\skp

Neither of  the ``function'' spaces $\{f(x)\}$ and $L^2_+(d\tilde p)$ are 
re\-pro\-ducing--kernel Hilbert spaces, since the evaluation maps $f\mapsto
f(x)$ and $a\mapsto a(p)$ are unbounded.  To obtain a space with bounded
evaluation maps, we proceed as in the last section.  Due to the positivity of
the energy, solutions can be continued analytically to the lower--half time
plane:

\be
 f({\bf x},t-iu)=\inop \exp\left( -it\omega -u\omega +i{\bf x} \cdot
{\bf p}  \right)\,a(p),
\eno20)
 \ee
where $u>0$.  As in the non--relativistic  case, the factor $\exp(-u\omega )$
decays rapidly as $\,|\,{\bf p} \,|\,\to \infty$, which permits an analytic
continuation of the solution to complex {\sl spatial\/} coordinates ${\bf z}
={\bf x}-i {\bf y} $.  But since $\omega ({\bf p})\equiv \sqrt{m^2+{\bf p} ^2}$
is no longer quadratic in $\,|\,{\bf p}\,|\,$, ${\bf y}$ cannot be arbitrarily
large.  Rather, we must require that the four--vector $(u,{\bf y})$ satisfy
the condition 

\be
 u\omega -{\bf y} \cdot {\bf p} >0\qquad \forall (\omega ,{\bf p})\in
\Omega _m^+.
 \eno21)
 \ee
In covariant notation, setting $y^0\equiv u$, we must have 

\be
 yp>0\qquad \forall p\in \Omega _m^+,
\eno22)
 \ee
so that the complex exponential $\exp\left[  -i(x-iy)p\right]$ remains
bounded as $p$ varies over $\Omega_m^+$.  This implies that $yp>0$ for all
$p\in V_+$, where 

\be
 V_+\equiv \lbrace p=(p_0,{\bf p})\in \rl^{s+1}\,|\,\ |{\bf p}|<p_0/c\rbrace
\eno23)
 \ee
is the  {\sl open forward light cone\/}   in momentum space.  In general, we
need to consider  the closure of $V_+$, i.e. the cone 

\be
\overline{  V}_+=\lbrace p\in \rl^{s+1}\,|\,\,|{\bf p}|\le p_0/c\rbrace,
\eno24)
 \ee
which contains the light cone $\{p^2=0\,\,|\,\,p_0>0\}$ (corresponding to
massless particles) and the point $\{p=0\}$ (corresponding in quantum field
theory  to the vacuum state). The set of all  $y$'s  with $yp>0$ is called
the {\sl dual cone\/}  of $\overline{V}_+$, i.e.,

\be
 V_+'\equiv \lbrace y\in \rl^{s+1}\,|\,\ yp>0\ \forall p\in
\overline{V}_+\rbrace. 
\eno25)
 \ee
It is easily seen that $y$ belongs to $V_+'$ if and only if  $\,|\,{\bf y} \,|\,<
cy^0$.  Note that as $c\to \infty$, $\overline {V}_+$ contracts to the
non--negative $p_0$--axis while $V_+'$ expands to the half--space $\lbrace
(u,{\bf y}) \,|\,\, u>0,\ {\bf y} \in\rl^s\rbrace$ which we have encountered in
the last section.  $V_+'$ coincides with $V_+$ when $c=1$, but it is
important to distinguish between them since they ``live'' in different spaces
(see section 1.1).  

Thus for $y\in V_+'$, setting $z=x-iy$, we define 

\be
 f(z)=\inop e^{-izp}\,a(p).
\eno26)
 \ee
The integral converges absolutely for any $a\in L^2_+(d\tilde p)$ and
defines a function on the {\sl forward tube\/} 

\be
 {\cal T}_+\equiv \lbrace x-iy\in \cx^{s+1}\,|\,\, y\in V_+'\rbrace,
\eno27)
 \ee
also known as the {\sl future tube\/}  and,  in the mathematical literature,
as the {\sl tube over\/}  $V_+'$. Differentiation with respect to $z^\mu $
under the integral sign leaves the integral absolutely convergent, hence the
function $f(z)$ is {\sl holomorphic,\/} or analytic, in ${\cal T}_+$.  As $y\to
0$ in $V_+'$, $f(z)\to f(x)$ in the sense of $L^2_+(d\tilde p)$. Thus $f(x)$ is 
 a {\sl boundary value\/}  of $f(z)$.  Clearly $f(z)$ is
a solution of the Klein--Gordon equation in either of the variables $z$ or $x$.
Let $\cal K$ be the space of all such holomorphic solutions:

\be
 {\cal K}=\lbrace f(z)\,|\,\,a\in L^2_+(d\tilde p) \rbrace.
\eno28)
 \ee
Then the map $a(p)\mapsto f(z)$ is one--to--one from $L^2_+(d\tilde p)$
onto $\cal K$.  Hence we can make $\cal K$ into a Hilbert space by defining 

\be
 \langle\, f_1\,|\,f_2\,\rangle_{\cal K}\equiv \langle\, a_1\,|\,a_2\,\rangle,
\eno29)
 \ee
where the inner product on the right--hand side  is understood to be  that of
$L^2_+(d\tilde p)$. We now show that $\cal K$ is a reproducing--kernel
Hilbert space.  Its evaluation maps are given by

\be
 E_z(f)\equiv f(z)=\inop e^{-izp}\,a(p)\equiv \langle\, e_z\,|\,a\,\rangle,
 \eno30)
 \ee 
where 

\be
 e_z(p)=e^{i\bar z p}.
\eno31)
 \ee

\vfill\eject

\noindent {\bf Lemma 4.2.} {\sl
\item{1.}  For each $z\in {\cal T}_+$, $e_z$ belongs to $L^2_+(d\tilde p)$,
with  \be
 \|e_z\|^2=(2\pi )^{-1} \left( {mc \over 4\pi \lambda } \right)^\nu 
K_\nu (2\lambda mc),
\eno32)
 \ee
where $\nu =(s-1)/2$,
\be
 \lambda \equiv \sqrt{y^2}=\sqrt{c^2(y^0)^2-{\bf y} ^2}>0
\eno33)
 \ee
and $K_\nu $ is a modified Bessel function {\rm (Abramowitz and Stegun
[1964];  the speed of light has been inserted for future reference.)}
\item{2.} In particular, the evaluation maps on $\cal K$ are bounded, with 
\be
\,|\,f(z)\,|\,\le \|e_z\|\,\|f\|.
\eno34)
 \ee

}
\skp 
\noindent {\bf Proof.}  
Set $c=1$. (To recover $c$, replace $m$ by $mc$ in the end.) Then

\be
\|e_z\|^2=\inop e^{-2yp}\equiv G(y).
\eno35)
 \ee
Since $G(y)$ is Lorentz--invariant and $y\in V_+'$, we can evaluate the
integral in a Lorentz frame in which $y=(\lambda ,{\bf 0} )$:

\ba{
 G(y)\a=G(\lambda ,{\bf 0} )=\inop e^{-2\lambda \omega }\cr
\a=(2\pi )^{-s}\int  { d^s {\bf p} \over 2\sqrt{m^2+{\bf p} ^2}}
\exp\left[ -2\lambda \sqrt{m^2+{\bf p} ^2} \right].
 \cr}
 \eno36)
 \ea
Set ${\bf p}=m {\bf q} $.  Then

\ba{
G(y)\a=(2\pi )^{-s} m^{s-1} 
\int  { d^s {\bf q} \over 2\sqrt{1+{\bf q} ^2}}
\exp\left[ -2\lambda m\sqrt{1+{\bf q} ^2} \right]\cr
\a=(2\pi )^{-s} m^{s-1} { \pi ^{s/2}\over  \Gamma (s/2) }
\int _0^\infty { r^{s-1} dr \over \sqrt{1+r^2}}
\exp\left[ -2\lambda m\sqrt{1+r ^2} \right]\cr
\a={ m^{s-1}\over (4\pi) ^{s/2}\Gamma(s/2)}\int_0^\infty dt\,\sinh^{s-1}t\,\,
\exp\left[ -2\lambda m \cosh t \right]\cr
\a=(2\pi )^{-1} \left( { m\over 4\pi \lambda } \right)^\nu K_\nu (2\lambda
m). \sp\qed
 \cr}
 \eno37)
 \ea

The reproducing kernel can be obtained by analytic continuation from
$\|e_z\|^2$:

\ba{
K(z',\bar z)\a\equiv \langle\, e_{z'}\,|\,e_z\,\rangle\cr
\a=\inop \exp\left[ -i(z'-\bar z) p \right]\cr
\a=(2\pi )^{-1} \left( { mc\over 2\pi \eta } \right)^\nu K_\nu (\eta mc),
 \cr}
 \eno38)
 \ea
where 

\ba{
\eta \a\equiv \sqrt{-(z'-\bar z)^2}\cr
\a=\left[ (y'+y)^2-(x'-x)^2+2i(y'+y)(x'-x) \right]^{1/2}
 \cr}
 \eno39)
 \ea
is defined by analytic continuation from $z'=z$ (when $\eta =2\lambda $) as
follows:  The square--root function is defined on the complex plane cut along
the negative real axis.  Since $y$ and $y'$ both belong to $V_+'$, so does
$y'+y$.  Now the argument of the square root is real if and only if 
$ (y'+y)(x'-x) =0$, and this can happen only when $(x'-x)^2\le 0$. (Otherwise,
either $x'-x$  or $x-x'$ belongs  to $V_+'$, hence $ (y'+y)(x'-x)$ is positive or
negative, respectively.)  But then, 

\be
 -(z'-\bar z)^2=(y'+y)^2-(x'-x)^2\ge (y'+y)^2>0.
\eno40)
 \ee
Thus for $z', z \in {\cal T}_+$, the quantity $-(z'-z)^2$ cannot belong to the
negative real axis, so $\eta $ is well--defined.
\skp
The reproducing kernel is closely related to the analytically continued
(Wightman) {\sl 2--point function  \/}  for the scalar quantum field of mass
$m$ (Streater and Wightman [1964]):

\be
 K(z', \bar z)=-i\Delta ^+(z'-\bar z).
\eno41)
 \ee
We will encounter this and other 2--point functions again in the next
chapter, in connection with quantum field theory.
\skp
\noindent {\sl Note:\/} We will be interested in the behavior of
$\,\|\,e_z\,\|\,$ near the boundary of ${\cal T}_+$, i.e. when $\lambda \sim
0$.  From the properties of $K_\nu $ it follows that 

\be
 \|e_z\|^2\sim { \Gamma(\nu )\over (4\pi )^{\nu +1}}\,\lambda ^{-2\nu }
\ \hbox{when}\  \lambda \sim 0.
\eno42)
 \ee
In particular, the evaluation maps are no longer bounded when $\lambda
=0$.\skp

${\cal P}_0$ acts on  ${\cal T}_+$ by a complex
extension of its action on real spacetime, i.e., 

\be
z'\equiv  (b,\Lambda)\,z=\Lambda z+b.
\eno43)
 \ee
This means that $x'=\Lambda x+b $ as before, and $y'=\Lambda y$. (This is
consistent with the phase--space interpretation of ${\cal T}_+$ to be
established below.)  The induced action on $\cal K$ is therefore 

\be
 \left( U(b,\Lambda)f \right)(z)=f\left( \Lambda  ^{-1} (z-b) \right).
\eno44)
 \ee
This implies that the wave packets $e_z$  transform covariantly under ${\cal
P}_0$, i.e.

\be
 U(b,\Lambda)\,e_z=e_{\Lambda z+b}.
\eno45)
 \ee
\skp
We have now established that the space $\cal K$ of holomorphic 
po\-si\-tive--energy solutions is a reproducing--kernel  Hilbert space.  Recall
that picking out the positive--energy part of $f(x)$ was a non--local
operation in {\sl real\/} spacetime, involving the pseudodifferential operator 
$\sqrt{m^2-\Delta }$.   However, when extended to ${\cal T}_+$, such
functions may be characterized locally, as simultaneaous solutions of the
Klein--Gordon equation and the Cauchy--Riemann equations, since the
negative--energy part of $f(x)$ does not have an analytic continuation to
${\cal T}_+$.
\skp
We now show that ${\cal T}_+$ may, in fact, be interpreted as an extended
phase space  for the underlying classical relativistic particles.  Clearly,
$x^\mu \equiv \Re \,z^\mu $ are the spacetime coordinates.  Their relation to
the expectation values of the relativistic (Newton--Wigner) position
operators will be discussed below.   We now wish to investigate the relation
of the imaginary coordinates $y^\mu \equiv -\Im \,z^\mu $ to the
energy--momentum vector. The bridge between the ``classical'' coordinates
$y^\mu $ and the quantum--mechanical observables $P_\mu $ will be, as
usual,  the (future) coherent states $e_z $. Before getting involved in
computations, let us take a closer look at these wave packets in order to get
a qualitative picture.  Since $yp$ is Lorentz--invariant, it can be evaluated
in a reference  frame where $p=(mc^2,{\bf 0} )$.  Thus

\be
 yp=y^0mc^2=\sqrt{\lambda ^2+ {\bf y} ^2}\,mc \ge \lambda mc,
\eno46)
 \ee
with equality if and only if ${\bf y}={\bf 0}$, i.e. when $y$ and $p$ are {\sl
parallel.\/}  This is a kind of {\sl reverse \/}  Schwarz inequality which
holds in $V_+'\times \overline {V}_+$ under the pairing provided by the 
Lorentzian scalar product.  Thus for {\sl fixed \/} $y\in V_+'$ and variable
$p\in \Omega_m^+$, we have

\be
 \,|\,e_z(p)\,|\,\le e^{-\lambda mc},
\eno47)
 \ee
the maximum being attained when and only when
 $p=(mc/\lambda )y\equiv p_y$.  Hence we expect, roughly, that

\ba{
\langle\, P_\mu \,\rangle\equiv { \langle\, e_z\,|\,P_\mu e_z\,\rangle
\over  \langle\, e_z\,|\, e_z\,\rangle }
\sim(mc/\lambda)y_\mu .
 \cr}
 \eno48)
 \ea
Therefore the vector $y$, while itself {\sl not\/}  an energy--momentum,
 acts as a {\sl control vector\/} for the energy--momentum by {\sl filtering
out\/} $p$'s which are ``far'' from $p_y$. The larger we take the parameter
$\lambda $, the {\sl finer\/}  the filter.   The expected
energy--momentum can be computed exactly by noting that

\ba{
 \langle\, e_z\,|\,P_\mu e_z\,\rangle\a=\inop p_\mu \,e^{-2yp}\cr
\a=-{ 1\over 2 }{\partial G(y)\over{\partial y^\mu }}
 \cr}
 \eno49)
 \ea
where $G(y)=\|e_z\|^2$ as before.  Since $G$ depends on $y$ only through the
invariant quantity $\lambda $, we have

\ba{
 \langle\, P_\mu \,\rangle\a=-{ 1\over 2 }G ^{-1}\,{\partial \lambda 
\over{\partial y^\mu }} {\partial  G\over{\partial \lambda }}\cr
\a=\left[ { K_{\nu +1}(2\lambda mc)
\over K_\nu (2\lambda mc)} \cdot{ mc\over \lambda }\right]\,y_\mu , 
 \cr}
 \eno50)
 \ea
where we have used the recurrence relation (Abramowitz and Stegun
[1964])

\be
- {\partial \over{\partial \lambda  }}\left( \lambda  ^{-\nu }\,K_\nu
(2\lambda m ) \right)= 2m \lambda ^{-\nu }K_{\nu +1}(2\lambda m ).
\eno51)
 \ee
This verifies and corrects the above qualitative estimate.  In view of the
above relation, the hyperboloid

\be
 \Omega _\lambda ^+\equiv \{y\in V_+'\,|\,y^2=\lambda ^2\}
\eno52)
 \ee
corresponds to the mass shell. Hence the submanifold

\be
 {\cal C}_\lambda =\{x-iy\in {\cal T}_+\,|\,\,y^2=\lambda ^2\}
\eno53)
 \ee
corresponds to the extended phase space  ${\cal C}$ defined in section 4.2
which, we recall, was a homogeneous space of ${\cal P}_0$ that was
interpreted as spacetime$\times$velocity space.
 Let us define the {\sl effective mass\/}
$m_\lambda $ of the particle on $\Omega _\lambda ^+$ by 

\be
 (m_\lambda c)^2\equiv \langle\, P_\mu \,\rangle\langle\, P^\mu \,\rangle
\equiv \langle\, P\,\rangle^2.
\eno54)
 \ee
Then 

\be
 m_\lambda =m \cdot{ K_{\nu +1}(2\lambda mc)\over K_\nu (2\lambda
mc) } ,
\eno55)
 \ee
and 

\be
 \langle\, P_\mu \,\rangle={ m_\lambda c\over \lambda } y_\mu .
\eno56)
 \ee
We claim that $m_\lambda >m$, which can be seen as follows.  For all
$p,p'\in\Omega _m^+$ we have the ``reverse Schwarz inequality'' $pp'\ge
m^2$, with equality if and only if $p=p'$.  Hence

\ba{
m_\lambda ^2\a= \langle\, P\,\rangle^2\cr
\a=G^{-2}\int \!\!\int d\tilde p\,d\tilde p'\,\,pp'\,
\exp(-2yp-2yp')\cr
\a> m^2.
 \cr}
 \eno57)
 \ea
This is a kind of ``renormalization effect'' due to the uncertainty, or
fluctuation, of the energy--momentum in the state $e_z$.  It appears to go
in the ``wrong direction'' (i.e., $\langle\, P\,\rangle^2>\langle\, P^2\,\rangle$)
for the same reason as does the inequality $pp'\ge m^2$, namely because of
the Lorentz metric.

Thus $\langle\, P_\mu \,\rangle$ is proportional,
by a $y$--dependent but ${\cal P}_0$--invariant factor, to $y_\mu $.  We
may therefore consider the
 $y_\mu $'s as {\sl homogeneous coodinates\/}  for the direction of motion
of the classical particle in (real) spacetime.  Alternatively, the expectation of
the velocity operator ${\bf P} /P_0$ can be shown to be ${\bf y}/y_0$. 
Thus of the $s+1$ coordinates $y_\mu $, only $s$ have a ``classical''
interpretation. It is important to understand that the parameter $\lambda $
{\sl has  no relation  to the mass;\/}  it can be chosen to be an arbitrary
positive number and has the physical dimensions of length.  It is the
relativistic counterpart of the parameter $u$ encountered in connection
with the non--relativistic  coherent states, and its significance will be
studied later. At this point we simply note that $\lambda $ {\sl  measures 
the invariant  distance of $z$ from the boundary of \/} ${\cal T}_+$. The
larger $\lambda $, the more smeared out are the spatial features and the
more refined are the features in momentum space.  (Recall that the
imaginary part $u$ of the time played a similar role in the non--relativistic 
theory.)

\skp
Because the vector $y$ is so fundamental to our approach, it deserves a
name of its own.  We will call it the {\sl temper
vector.\/}  The name is motivated in part by the  smoothing effect which
$y$ has on spacetime quantities, and also by the fact that $y$ plays a role 
similar to that played by the inverse
teperature $\beta =1/kT$ in statistical mechanics; the latter controls the
energy.
\skp
 From the asymptotic properties of the $K_\nu $'s it follows that 

\be
 \langle\, P_\mu \,\rangle\sim \cases{ (mc/\lambda )y_\mu  , &when 
$\lambda mc \to \infty$;\cr 
(\nu /\lambda ^2)y_\mu , &when $\lambda mc \to
0$.\cr} 
\eno58)
 \ee
This can be understood as follows:  When $\lambda mc\to
\infty$, e.g. $c\to \infty$ for fixed $\lambda m$,  we recover the
non--relativistic  results.  For example, the expectations of the spatial
momenta  approach those in the non--relativistic  coherent states:

\be
 \langle\, P_k\,\rangle \sim (mc/\lambda )y_k \equiv (m/u)y_k.
\eno59)
 \ee
When $\lambda mc\to 0$, say 
$\lambda \to 0$ for fixed $mc$, then $z$ approaches the  boundary  of ${\cal
T}_+$.  In that case,  fluctuations take over and the expectations become
independent of the mass $m$.
\skp
The relation of the  spacetime parameters $x_\mu \equiv \Re z_\mu $ to
the\break 
 Newton--Wigner position operators  is as follows.  Since a fixed
$f\in {\cal T}_+$  describes the entire history of the particle, the associated
state  does not change with time (i.e., the dynamics is already built in).  This
means that we are in the so--called {\sl Heisenberg picture,  \/} and
time--behavior must be described by evolving the observables $A$:

\be
 A(t)=e^{itP_0}\,A(0)\,e^{-itP_0}.
\eno60)
 \ee
The Newton--Wigner position operators are uniquely determined by a set of
seemingly reasonable assumptions concerning the localizability of the
particle (Newton and Wigner [1949], Wightman [1962]), and are given in the
momentum representation  {\sl at time $x_0=0$\/} by

\ba{
 X_k(0)\a=\omega ^{1/2}\,i{\partial \over{\partial p_k}}\,\omega ^{-1/2}\cr
\a=i\left( {\partial \over{\partial p_k}} -{ p_k\over \omega ^2}\right),
\quad k=1,2,\ldots,s.
 \cr}
 \eno61)
 \ea
Now choose $z=x-iy\in {\cal T}_+$ with $x_0=0$.  Then 

\ba{
\langle\, e_z\,|\,X_k(0)\,e_z\,\rangle\a=\inop \omega ^{1/2}\,e^{-izp}\,
i{\partial \over{\partial p_k}}\left( \omega ^{-1/2}\,e^{i\zbar p} \right)\cr
\a=\Re\,\inop  \omega ^{1/2}\,e^{-izp}\,
i{\partial \over{\partial p_k}}\left( \omega ^{-1/2}\,e^{i\zbar p} \right)\cr
\a=x_k\,\|e_z\|^2,
 \cr}
 \eno62)
 \ea
hence

\be
 \langle\, X_k(0)\,\rangle=x_k.
\eno63)
 \ee
It must be noted, however, that the concept of position for relativistic
particles is highly unsatisfactory.  Not only are the position operators
non--covariant (this would seem to require a time operator on equal footing
with them, which would exclude the possibility of dynamics);  but the very
concept of localizability for such particles is fraught with difficulties.  For
example, eigenvectors of the Newton--Wigner position operators, known as
``localized states,'' spread out from a single point at time $x_0$ to fill the
entire universe an arbitrarily small time later, violating relativistic 
causality.   (The same phenomenon in the non--relativistic  theory presents
no conceptual problem, since propagation velocities are unrestricted there.)
Even much weaker notions of localization give rise to problems with
causality (Hegerfeldt [1985]).  In my opinion, it is best to admit that  position
is simply a non--relativistic concept, and in the relativistic theory events $x$
should be regarded as mere {\sl parameters\/} of the spacetime manifold. 
As such, our formalism extends them to $z=x-iy$, with the new parameters
$y$ playing the role of a control vector for the energy--momentum
observables.  Thus, in place of a set of  pairs  of canonically conjugate
observables $X_k, P_k$, we have a set of observables $P_\mu $ and a dual
set of complex {\sl parameters\/}  $z^\mu$.  The symmetry between
position-- and momentum operators in the non--relativistic  theory was
based on the Weyl--Heisenberg group, and we have seen that this symmetry
is ``accidental,'' being broken in the transition to relativity.
\skp
\noindent {\sl Note:\/} This further reinforces the idea discussed in section
4.2, that  ``group--theoretic quantization,'' i.e. the quantization of classical
systems obtained by requiring states and observables to transform under
unitary representations of the associated dynamical groups or central
extensions thereof, is superior to ``canonical quantization''
(sending the classical phase--space observables to operators, with Poisson
brackets going to commutators).\sp\#

\skp
Let us now consider  the uncertainties in the energy-- and
momentum operators.  More generally, we compute the correlation matrix

\be
 C_{\mu \nu }\equiv \langle\, P_\mu P_\nu \,\rangle-\langle\, P_\mu
\,\rangle\langle\, P_\nu \,\rangle,
 \eno64)
 \ee
of which the uncertainties $\Delta _{P_\mu }^2$ are the diagonal elements. 
From

\be
 \langle\, e_z\,|\,P_\mu P_\nu  e_z\,\rangle={1 \over 4}
{\partial ^2G\over{\partial y^\mu  \partial y^\nu }}
\eno65)
 \ee
it follows that 

\ba{
4C_{\mu \nu }\a=G ^{-1} {\partial ^2G\over{\partial y^\mu  \partial y^\nu }}
-G^{-2}{\partial G\over{\partial y^\mu }}
\,{\partial G\over{\partial y^\nu }}\cr
\a={\partial ^2 \ln G\over{\partial y^\mu  \partial y^\nu }}.
 \cr}
 \eno66)
 \ea
(Incidentally, this shows that the function $\ln G$ is of some interest in
itself:  its first partials are the expected momenta, while its second partials
are the correlations.)  A computation
similar to that for $\langle\, P_\mu \,\rangle$ gives 

\be
 C_{\mu \nu }=y_\mu y_\nu {m^2 \over \lambda ^2}\left( { K_{\nu
+2}(2\lambda m)\over K_\nu (2\lambda m)} -{ m_\lambda ^2\over
m^2}\right) -g_{\mu \nu }{ m_\lambda \over 2\lambda } .
\eno67)
 \ee
Although this expression does not appear to be enlightening in any obvious
way, the uncertainties $\Delta _{P_\mu }$ can be estimated from it in
various limits such as $\lambda m\to \infty$ and $\lambda m\to 0$.
\skp
The reproducing kernel by itself is of limited use.  Although it makes it
possible to establish the interpretation of ${\cal T}_+$ as an extended
classical phase space, it does not provide us with a direct physical
interpretation of the function values $f(z)$.  The inner product in $\cal K$ is
borrowed from $L^2_+(d\tilde p)$, hence a probability interpretation exists,
so far,  only in momentum space.  In the standard formulation of
Klein--Gordon theory, it is possible to define the inner product in
configuration space, but the corresponding density turns out not to be
positive, thus precluding a probabilistic interpretation.  This is one of the
well--known difficulties with the first--quantized Klein--Gordon theory, 
and is one of the reasons cited for the necessity to go to quantum field
theory  (second quantization). We will see that the phase--space approach
{\sl does\/} admit a covariant probabilistic picture of relativistic quantum
mechanics, thus making the theory more complete even before second
quantization.  These topics will be discussed further in the next section and
the next chapter.  At this point we wish only to define an ``autonomous''
inner product on $\cal K$ as an integral over a ``phase space'' lying in ${\cal
T}_+$. This will provide us with a normal frame of evaluation maps (chapter
1).  

 Recall that in the non--relativistic  theory of the last section,  the norm in the
space ${\cal H}_u$ of holomorphic solutions was obtained by integrating
$\,|\,f({\bf z},\tau  )\,|\,^2$ with respect to the Gaussian measure $d\mu _u
({\bf z})$ over the phase space $\tau \equiv t-iu=\,$constant.  But now, for
given $y^0=u$, only those ${\bf y}$'s with $\,|\,{\bf y} \,|\,<cu$ belong to
$V_+'$.  That is, the particle can only travel a finite imaginary distance in a
finite imaginary time.  In view of the relation $\langle\, P_\mu \,\rangle
\propto y_\mu $, the
obvious candidate for a phase space is the set defined for given $t\in \rl$ and
$\lambda >0$ by

\be
 \sigma \equiv \sigma _{t,\lambda }=\lbrace x-iy\in {\cal T}_+\,|\,\ 
 x^0=t,\ y^2=\lambda ^2\rbrace.
\eno68)
 \ee
Such sets are not covariant, but a covariant extension will be found in the
next section. As for the measure, a Gaussian weight function (such as
$\exp(-m {\bf y} ^2/u)$, which occured in $d\mu _u({\bf y} )$)  is no longer
satisfactory since it cannot be covariant.  It turns out that we do not need a
weight function at all!  This can be seen as follows:  In the non--relativistic 
case, the shift to complex time was performed once and for all by the
operator $e^{-uH}$.  For fixed $u>0$, the weight function served to correct for
the non--unitary translation from the real point ${\bf x} $ in space to the
complex point ${\bf z} = {\bf x} -i {\bf y} $.    However, if we restrict
ourselves to the subset $\sigma $, then a translation to imaginary space is
necessarily accompanied by a translation in imaginary time.  The analog of
the above translation is $(t-i\lambda, {\bf x})\mapsto (t-i \sqrt{\lambda ^2+
{\bf y} ^2}, {\bf x} -i {\bf y} )$.  The increase in $y^0$, it turns out, {\sl
precisely compensates\/}  for the shift to complex space! This follows from
the fact that the operator $e^{-yP}$, which affects the  total shift to complex
spacetime,  is relativistically invariant, hence the point ${\bf y} =
{\bf 0} $ no longer plays a special role.  We will show
later that in the non--relativistic  limit, we recover the weight function
naturally.  Hence the Gaussian weight function associated with the Galilean
coherent states (which, as we have seen, is closely related to that associated
with the canonical coherent states) has its origin in the {\sl geometry\/} of
the relativistic phase space, i.e. in the curvature of the hyperboloid $\{
y^2=\lambda ^2\}$.

For $\sigma \equiv \sigma _{t,\lambda }$ as above and $f\in \cal K$, define 

\be
 \|f\|^2_\sigma =\int _\sigma  d\sigma \, \,|\,f(z)\,|\,^2,
\eno69)
 \ee
where we parametrize $\sigma $ by $({\bf x}, {\bf y} )\in
\rl^{2s}$ and the measure $d\sigma $ is given by

\be
 d\sigma (z)=A_\lambda  ^{-1} d^s {\bf x} \,d^s {\bf y} 
\eno70)
 \ee
with 

\ba{
 A_\lambda \a=\pi ^{-1} \left( { \pi \lambda \over mc } \right)^{\nu +1}
K_{\nu +1}(2\lambda mc)\cr
\a=\left( {2\pi \lambda  \over mc} \right)^s\cdot { m_\lambda \over m}\,
G(\lambda ).
 \cr}
 \eno71)
 \ea
Then we have the folowing result.
\skp

\noindent {\bf Theorem 4.3. } {\sl For all $f\in \cal K$, 

\be
 \|f\|_\sigma = \|f\|_{\cal K} . 
\eno72)
 \ee
}

\skp 
\noindent {\bf Proof.}  Assume, to begin with, that $a( {\bf p} )\equiv
a(\omega , {\bf p} )$ belongs to the Schwartz space ${\cal S}(\rl^s)$ of
rapidly decreasing test functions.  Then 

\ba{
f(z)\a=(2\pi )^{-s}\int d^s {\bf p} \,(2\omega )^{-1} \exp(-ixp -yp)\,a({\bf p} )
\cr
\a=\left[ (2\omega )^{-1} \exp(-it\omega  -yp)\,a \right]\check {\ } ( {\bf
x}) .
 \cr}
 \eno73)
 \ea
Hence, by Plancherel's theorem, 

\be
 \int_{x^0=t}d^s {\bf x} \,\,|\,f(x-iy)\,|\,^2=(2\pi )^{-s}\int_{\rl^s} d^s {\bf p} \,
(4\omega ^2) ^{-1} e^{-2yp}\,\,|\,a(p)\,|\,^2.
\eno74)
 \ee
Exchanging the order of integration in the integral representing  \break
$ \|f\|^2_\sigma $, we obtain 

\be
  \|f\|^2_\sigma =A_\lambda  ^{-1} (2\pi)^{-s} \int d^s {\bf p} \,(4\omega ^2)
^{-1}  \,|\,a(p)\,|\,^2\,\int d^s {\bf y} e^{-2yp} .
\eno75)
 \ee 
We now evaluate 

\be
 J(p)\equiv \int d^s {\bf y} \,e^{-2yp} 
\eno76)
 \ee
as follows:  Consider all $s+1$ components of $p$ as independent and {\sl
define\/}  $m(p)\equiv \sqrt{p^2}$. From the integral computed earlier, i.e. 

\ba{
G(y)\a\equiv (2\pi)^{-s} \int d^s {\bf p} \,(2\omega ) ^{-1} e^{-2yp} \cr
\a=(2\pi)^{-1} \left( { m\over 4\pi \lambda } \right)^\nu K_\nu (2\lambda 
m),
 \cr}
 \eno77)
 \ea
we obtain by exchanging $p$ and $y$ (as well as $m$ and $\lambda $):

\be
\int d^s {\bf y} \,(2y_0) ^{-1} e^{-2yp} =
\left( { \pi \lambda \over m } \right)^\nu K_\nu
(2\lambda  m).
\eno78)
 \ee
Taking the partial derivative with respect to $p^0$ on both sides gives

\ba{
J(p)\a=-{\partial \over{\partial p^0}}\,\left[  
\left( { \pi \lambda \over m } \right)^\nu K_\nu (2\lambda  m)\right]\cr
\a=-(2\pi \lambda ^2)^\nu {\partial \over{\partial p^0}}\left[ 
\xi ^{-\nu }K_\nu (\xi ) \right]\cr
\a=-(2\pi \lambda ^2)^\nu {\partial \xi \over{\partial p^0}}\,
{\partial \over{\partial \xi }}\left[ \xi ^{-\nu }K_\nu (\xi ) \right],
 \cr}
 \eno79)
 \ea
where $\xi (p)\equiv 2\lambda m(p)$.  Using again the recurrence relation
for the $K_\nu $'s, we get

\be
 J(p)=2p_0\,A_\lambda .
\eno80)
 \ee
Thus

\be
 \|f\|^2_\sigma =(2\pi)^{-s} \int d^s {\bf p} \,(2\omega ) ^{-1} 
\,|\,a(p)\,|\,^2=\|f\|^2_{\cal K}
\eno81)
 \ee
for $a({\bf p} )\in {\cal S}(\rl^s)$.  By continuity, this extends to all
$a\in L^2_+(d\tilde p)$ since ${\cal S}(\rl^s)$ is dense in $L^2_+(d\tilde p)$.
 \qed\skp
If we define 

\be
 \langle\, f_1\,|\,f_2\,\rangle_\sigma \equiv \int_\sigma  d\sigma
(z)\,\overline{f_1(z) }f_2(z),
 \eno82)
 \ee
then by polarization we have the following immediate consequence of the
theorem.

\skp

\noindent {\bf Corollary 4.4. } {\sl  For all $f_1, f_2 \in \cal K$, 
\be
 \langle\, f_1\,|\,f_2\,\rangle_\sigma = \langle\, f_1\,|\,f_2\,\rangle_{\cal K}.
\eno83)
 \ee
In particular,  $\langle\, \cdot \,|\,\cdot \,\rangle_\sigma $ defines a ${\cal
P}_0$--invariant inner product on $\cal K$, and we have the resolution of
unity 
\be
\int_\sigma d\sigma_\lambda (z)\,\,|\,e_z\,\rangle\langle\, e_z\,|\,=I,
\eno84)
\ee
making the vectors $e_z$ with $z\in \sigma $ a normal frame.  \sp \qed
} %
\skp
\noindent {\sl Note:\/}  The above results extend to the case $\lambda =0$
by continuity.  The normalization constant in the measure $d\sigma $ is
then given by

\be
 A_0\equiv \lim_{\lambda \downarrow 0} A_\lambda 
={\pi ^\nu \Gamma(\nu +1)\over 2(mc)^{s +1} }.
\eno85)
 \ee
The same formula applies for fixed $\lambda >0$ and $m\sim 0$, and shows
that $A_\lambda $ becomes unbounded as $m\to 0$. \sp\#
\skp 
The vectors $e_z$ belong to $L^2_+(d\tilde p)$, but correspond to vectors
$\tilde e_z$  in ${\cal K}$ defined by 

\be
 \tilde e_z(z')=\langle\, e_{z'}\,|\,e_z\,\rangle=K(z'-\bar z).  
\eno86)
 \ee

\skp

The norm $\|f\|^2_\sigma \equiv \langle\, f\,|\,f\,\rangle_\sigma $ on ${\cal
K}$ provides us with an interpretation of $\,|\,f(z)\,|\,^2$ as a {\sl probability
density\/}  with respect to the measure $d\sigma _\lambda $ on the phase
space $\sigma $.  Within this interpretation, the wave packets $e_z $ have
the following optimality property:  For fixed $z\in {\cal T}_+$ let 

\be
 \hat e_z(z')={ \langle\, e_{z'}\,|\,e_z\,\rangle\over \|e_z\| }.
\eno87)
 \ee
\skp

\noindent {\bf Proposition 4.5. } {\sl Up to a constant phase factor, the
function $\hat e_z$ is the unique solution to the following variational
problem: Find $f\in {\cal K}$ such that $\|f\|=1$ and $\,|\,f(z)\,|\,$ is a
maximum.
}

\skp 
\noindent {\bf Proof.}  This follows at once from the Schwarz inequality and
theorem 4.3, since by eq. (26),

\ba{
 \,|\,f(z)\,|\,\a=|\,\langle\, e_z\,|\,a\,\rangle\,|=|\,\langle\, \tilde
e_z\,|\,f\,\rangle\,|\cr
\a\le \|\tilde e_z\|\,\|f\|=\|e_z\|\,\|f\|,
 \cr}
 \eno88)
 \ea
with equality if and only if $f$ is a constant multiple of $\tilde e_z$.\sp\qed
\skp
\noindent According to our probability interpretation of $\,|\,f(z)\,|\,^2$, this
means that the normalized wave packet $\hat e_z$ maximizes the
probability of finding the particle at $z$. 
\skp
\noindent {\sl Note:\/} Unlike the non--relativistic  coherent states of the
last section, the $e_z$'s do not have minimum uncertainty products.  In
fact, since the uncertainty product is not a Lorentz--invariant  notion, it is
{\sl a priori\/}  impossible to have relativistic coherent states with
minimum uncertainty products.  The above optimality, which {\sl is\/} 
invariant, may be regarded as a reasonable substitute.  Actually, there are
better ways to measure uncertainty than the standard one used in quantum
mechanics, which is just the variance.   From a statistical point of view, the
variance is just the  second moment of the probability distribution. Perhaps
the {\sl best\/}  definition of uncertainty, which includes all moments, is in
terms of {\sl entropy\/}  (Bialynicki--Birula and Mycielski [1975], Zakai
[1960]).  Being necessarily non--linear, however, makes this definition less
tractable.

\secskp

\noindent {\bf 4.5.  Geometry and Probability  }
\def\rightheadline{\tenrm\hfil {\sl 
4.5.  Geometry and Probability }\hfil\folio}  
 \skp
\noindent  The formalism of the last section was based on the phase space
$\sigma _{t, \lambda }\equiv \sigma $ and the measure
$d\sigma$, neither of which is invariant under the action of 
${\cal P}_0$ on ${\cal T}_+$.  Yet, the resulting inner product $\langle\, \cdot
\,|\,\cdot \,\rangle_\sigma $ is clearly invariant.  It is therefore reasonable to
expect that $\sigma $ and $d\sigma $ merely represent one choice out of
many.  Our purpose here is to construct a large  natural class of such  phase
spaces and associated measures  to which our previous results can be
extended.  This class will include $\sigma $ and will be invariant under
${\cal P}_0$.  In this way our formalism is freed from its dependence on
$\sigma $ and becomes manifestly covariant.  As a byproduct, we find that
positive--energy solutions of the Klein--Gordon equation give rise to a {\sl
conserved probability current,\/}  so the probabilistic
interpretation becomes entirely compatible with the spacetime geometry.  
As is well--known, no such compatibility is possible in the usual approach to
Klein--Gordon theory.  
\skp
We begin by regarding ${\cal T}_+$ as an extended phase space  (symplectic
manifold) on which ${\cal P}_0$ acts by canonical transformations. 
Candidates for phase space  are $2s$--dimensional symplectic  submanifolds
$\sigma \subset {\cal T}_+$, and ${\cal P}_0$ maps different $\sigma
$'s into one another by canonical transformations.  A submanifold of the
``product'' form $\sigma =S-i\Omega _\lambda ^+$, where $S$ (interpreted
as a generalized configuration space) is an $s$--submanifold of (real)
spacetime $\rl^{s+1}$, turns out to be symplectic if and only if $S$ is given
by $x^0=t\,( {\bf x} )$ with $|\nabla t|\le 1$, that is, if and only if $S$ {\sl is
nowhere timelike.\/} (This is slightly larger than the class of all {\sl
spacelike\/}  configuration spaces admissible in the standard theory.) The
original $\sigma_{t,\lambda }$ corresponds to $t\,( {\bf x} )\equiv $constant. 
The results of the last section are extended to all such phase spaces of
product form.  
\skp
The action of ${\cal P}_0$ on ${\cal T}_+$ is not transitive but leaves each of
the  $(2s+1)$--dimensional submanifolds 

\be
 {\cal T}^+_\lambda \equiv\{x-iy\in {\cal T}_+ \,|\,\,y^2=\lambda ^2\}
\eno1)
 \ee
invariant.  Each ${\cal
T}^+_\lambda$ is a homogeneous space of ${\cal P}_0$, with isotropy
subgroup $SO(s)$, hence

\be
 {\cal T}^+_\lambda\approx {\cal P}_0/SO(s) .
\eno2)
 \ee
Thus ${\cal T}^+_\lambda$ corresponds to the homogeneous space ${\cal
C}$ of section 4.2 (where we had  specialized to $s=3$). In view of the
considerations in sections 4.2 and 4.4, each ${\cal T}^+_\lambda $ can be
interpreted  as the product of spacetime with ``momentum space''.  Phase
spaces $\sigma $ will be obtained by taking slices to eliminate the time
variable.

On the other hand,  we also need a covariantly assigned measure for each
$\sigma $.  The most natural way this can be accomplished is to
begin with a single ${\cal P}_0$--{\sl invariant  symplectic form\/}  on
${\cal T}_+$ and require that its restriction to each $\sigma $ be  symplectic. 
This will make each $\sigma $ a symplectic manifold (which, in any case, it
must be to be interpreted as a classical phase space) and thus provide it
with a canonical (Liouville) measure.  Thus we look for the most general
2--form $\alpha $ on ${\cal T}_+$ such that 

\item{(a)} $\alpha $ is closed, i.e., $d\alpha =0$; 
\item{(b)}  $\alpha $ is non--degenerate, i.e., the $(2s+2)$--form $\alpha
^{s+1}\equiv \alpha \w\alpha \w\cdots\w\alpha $ vanishes nowhere;
\item{(c)} for every $g=(a,\Lambda )\in {\cal P}_0$, $g^*\alpha =\alpha $,
where $g^*\alpha $ denotes the pull--back of $\alpha $ under $g$ (see
Abraham and Marsden [1978]).
Since every ${\cal P}_0$--invariant function on ${\cal T}_+$ depends on $z$
only through $y^2$, the most general  invariant 2--form is given by 

\be
 \alpha =\phi (y^2)\,dy_\mu \w dx^\mu +\psi (y^2)\,y_\mu y_\nu dy^\mu
\w dx^\nu .
 \eno3)
 \ee
Now the restriction (pullback) of the second term to ${\cal T}^+_\lambda $
vanishes, since it contains the factor $y_\mu dy^\mu =d(y^2)/2$. 
Furthermore, the coefficient $\phi (y^2)$ of the first term is constant on
${\cal T}_\lambda^+$.  Hence we may confine our attention to  the form

\be
 \alpha =dy_\mu \w dx^\mu
\eno4)
 \ee
without any essential  loss of generality.  This form is symplectic as well as
invariant, hence it fulfills all of the above conditions. ${\cal T}_+$, together
with $\alpha $, is a symplectic manifold, and invariance means that each
$g\in {\cal P}_0$ maps ${\cal T}_+$ into itself by a canonical transformation.
\skp
A general $2s$--dimensional submanifold $\sigma $ of ${\cal T}_+$ will be a
potential phase space only if the restriction, or pullback, of $\alpha $ to
$\sigma $ is a symplectic form.  We denote this restriction by $\alpha
_\sigma $.  Let $\sigma $ be given by

\be
 \sigma =\{z\in {\cal T}_+ \,|\,\,s(z)=h(z)=0\},
\eno5)
 \ee
where $s(z)$ and $h(z)$ are two real--valued, $C^\infty$  (or at least $C^1$)
functions on ${\cal T}_+$ such that $ds\w dh\ne 0$ on $\sigma $.  For
example, $\sigma _{t,\lambda }$ can be obtained from $s(z)=x^0-t$ and
$h(z)=y^2-\lambda ^2$.  The pullback $\alpha _\sigma $ depends only on the
submanifold $\sigma $, not on the particular choice of $s$ and $h$. 
\skp

\noindent {\bf Proposition 4.6. } {\sl The form $\alpha _\sigma $ is
symplectic if and only if the Poisson bracket
\be
 \{s,h\}\equiv {\partial s\over{\partial x^\mu }}{\partial h\over{\partial
y_\mu }}-{\partial h\over{\partial x^\mu }}{\partial s\over{\partial
y_\mu}}  \ne 0
\eno6)
 \ee
everywhere on $\sigma $.
}
\skp 
\noindent {\bf Proof.}  $\alpha _\sigma $ is closed since $\alpha $ is closed
and $d(\alpha _\sigma )=(d\alpha)_\sigma $.  Hence $\alpha_\sigma $ is
symplectic if and only if it is non--degenerate, i.e. if and only if its $s$-th
exterior power $\alpha ^s$ vanishes nowhere on $\sigma $.  Now $(\alpha
_\sigma )^s$ equals the pullback of $\alpha ^s$ to $\sigma $, and a
straightforward computation gives 

\be
 \alpha ^s=s!\,\,\dy^\mu \w \dx_\mu ,
\eno7)
 \ee
where 

\ba{
\dy^\mu \a=(-1)^s\,dy_0\w dy_1\w \cdots \w dy_{\mu -1}\w dy_{\mu
+1}\w \cdots \w dy_s\cr
\dx_\mu \a=(-1)^s\,dx^s\w dx^{s-1}\w \cdots \w dx^{\mu +1}\w dx^{\mu
-1}\w \cdots \w dx^0.
\cr}
 \eno8)
 \ea
($\dy^\mu $ and $\dx_\mu $ are essentially the Hodge duals (Warner
[1971]) of $dy_\mu $ and $dx^\mu $, respectively, with respect to the
Minkowski metric.) 

\noindent Let $\{u_1, \ldots, u_{2s}, v_1, v_2\}$ be a basis for the tangent space  of
${\cal T}_+$ at $z\in \sigma $, with $\{u_1, \ldots, u_{2s} \}$ a basis for the
tangent space $\sigma _z$ of $\sigma $.  Since $ds$ and $dh$ vanish on the
vectors $u_j$, 

\ba{
(\alpha ^s\w ds\w dh)\a(u_1, \ldots, u_{2s}, v_1, v_2)\cr
\a=\alpha ^s(u_1, \ldots, u_{2s})\,(ds \w dh )(v_1, v_2)\cr
\a=\alpha ^s_\sigma (u_1, \ldots, u_{2s})\,(ds \w dh )(v_1, v_2).
 \cr}
 \eno9)
 \ea
By assumption,  $(ds \w dh )(v_1, v_2)\ne 0$. 
Therefore $\alpha _\sigma $ is non--degenerate at $z$ if and only if 
$\alpha ^s\w ds\w dh\ne0 $ at $z$.  But by eq. (7), 

\be
 \alpha ^s\w ds\w dh=s!\,\{s,h\}\,dy \w dx,
\eno10)
 \ee
where 

\ba{
 dy\a=dy_0\w\cdots \w dy_s\cr
dx\a=dx^s\w \cdots \w dx^0.
\cr}
 \eno11)
 \ea
Hence $\alpha _\sigma ^s\ne 0$ at $z$ if and only if $\{s, h\}\ne 0$ at $z$.
\qed\skp 
Let us denote the family of all such symplectic submanifolds $\sigma $ by
$\Sigma _0$.
\skp

\noindent {\bf Proposition 4.7. } {\sl Let $\sigma \in \Sigma _0$ and $g\in
{\cal P}_0$.  Then $g\sigma \in \Sigma _0$ and the restriction $g{:}\ \sigma
\to g\sigma $ is a canonical transformation from $(\sigma , \alpha _\sigma
)$ to $(g\sigma , \alpha _{g\sigma })$.

}
\skp  
\noindent {\bf Proof.} Let $g^*$ denote the pullback map defined by $g$,
taking forms on $g\sigma $ to forms on $\sigma $.  Then the invariance of
$\alpha $ implies that

\be
 g^*\alpha _{g\sigma }=\alpha _\sigma .
\eno12)
 \ee 
Hence $\alpha _{g\sigma }$ is non--degenerate.  It is automatically closed
since $\alpha $ is closed.  Thus $g\sigma \in \Sigma _0$.  To say that 
$g{:}\ \sigma \to g\sigma $ is a canonical transformation means precisely
that $\alpha _\sigma $ and $\alpha _{g\sigma }$ are related as above.
\qed\skp 
We will be interested mainly in the special  case where $h(z)=y^2-\lambda
^2$ for some $\lambda >0$ and $s(z)$ depends only on $x$.  Then the
$s$--dimensional manifold

\be
 S\equiv \{x\in \rl^{s+1}\,|\,\,s(x)=0\}
\eno13)
 \ee
is a potential generalized configuration space, and $\sigma $ has the
``product'' form

\be
 \sigma =S-i \Omega_\lambda ^+ \equiv \{x-iy\in {\cal T}_+\,|\,\,
x\in S, y\in  \Omega_\lambda ^+\}.
\eno14)
 \ee
The following result is physically significant in that it relates the
pseudo--Euclidean geometry of spacetime and the symplectic geometry of
classical phase space. It says that $\sigma $ is a phase space if and
only if $S$ is a (generalized) configuration space. \skp

\noindent {\bf Theorem 4.8. } {\sl Let $\sigma =S-i \Omega_\lambda ^+ $ be
as above.  Then $(\sigma , \alpha _\sigma )$ is symplectic if and only if

\be
 {\partial s\over{\partial x^\mu }}\,{\partial s\over{\partial x\mu }}\ge 0,
\eno15)
 \ee
that is, if and only if $S$ is {\rm nowhere timelike.}
}
\skp 
\noindent {\bf Proof.}  On $\sigma $, we have 

\be
 \{s, h\}=2{\partial s\over{\partial x^\mu }}\,y^\mu \ne 0,
\eno16)
 \ee
and we may assume $\{s, h\}>0$ without loss.  For fixed $x\in S$, the above
inequality must hold for all $y\in \Omega_\lambda ^+$, hence for all $y\in
V_+'$.  This implies that the vector $\partial s/\partial x^\mu $ is in the dual 
$\overline{ V}_+$ of $V_+'$.
 \qed\skp 
We denote the family of all $\sigma $'s as above (i.e., with $S$ nowhere
timelike) by $\Sigma $. It is a subfamily of $\Sigma _0$ and is clearly
invariant under ${\cal P}_0$.  Note that $\Sigma $ admits
{\sl lightlike\/}  as well as {\sl spacelike\/} configuration spaces, whereas the
standard theory only allows spacelike ones.  
\skp
 We will now  generalize the results  of the last section to all $\sigma \in
\Sigma $.  The $2s$--form $\alpha _\sigma ^s$ defines a positive measure on
$\sigma $, once we choose an orientation (Warner [1971]) for $\sigma $. (This
can be done, for example,  by choosing an {\sl ordered\/}  set of vector fields
on $\sigma $ which span the tangent space at each point; the order of such a
basis is a generalization of the idea of a ``right--handed'' coordinate system
in three dimensions.) The appropriate measure generalizing $d\sigma $ of
the last section is now defined as

\be
 d\sigma =(s!\,A_\lambda )^{-1} \alpha _\sigma ^s.
\eno17)
 \ee
$d\sigma $ is the restriction to $\sigma $ of a $2s$--form defined on all of
${\cal T}_+$, which we also denote by $d\sigma $. (This is a mild abuse of
notation; in particular, the ``$d$'' here must {\sl not\/}  be confused with
exterior differentiation!)  By eq. (7), we have

\be
 d\sigma =A_\lambda ^{-1} \,\dy^\mu \w \dx_\mu .
\eno18)
 \ee
We now derive a concrete expression for $d\sigma $.  Since $s$ obeys eq.
(15) and $ds\ne 0$ 
 on $\sigma $, we can solve $ds=0$ (satisfied by the {\sl
restriction\/}  of $ds$ to $\sigma $) for $dx^0$ and substitute this into
$\dx_k$.   This (and a similar procedure for $y$) gives 

\ba{
\dx_\mu \a=\left( {\partial s\over{\partial x^0}} \right)^{-1} \,
{\partial s\over{\partial x^\mu }}\,\dx_0\cr
\dy\,^\mu \a=\left( {\partial h\over{\partial y_0}} \right)^{-1} \,
{\partial h\over{\partial y_\mu }}\,\dy\,^0=(y^\mu /y^0)\,\dy\,^0
 \cr}
 \eno19)
 \ea
on $\sigma $.  Hence

\be
 d\sigma =A_\lambda  ^{-1}\, \left( y^0\,{\partial s\over{\partial x^0}} \right)
^{-1}  \left( y^\mu \,{\partial s\over{\partial x^\mu }}
\right)\,\dy\,^0\w\dx_0.
 \eno20)
 \ee
We identify $\sigma $ with $\rl^{2s}$ by solving $s(x)=0\ $ for $x^0=t\,({\bf
x})$ and mapping 

\be
 (t\,({\bf x})-i\sqrt{\lambda ^2+ {\bf y}^2}, \,{\bf x}-i{\bf y} )\mapsto 
({\bf x}, {\bf y} ).
\eno21)
 \ee
We further identify $\dy\,^0\w\dx_0$ with the Lebesgue measure
$d^s{\bf y} \,d^s{\bf x} $ on $\rl^{2s}$ (this amounts to choosing a
non--standard orientation of $\rl^{2s}$).  Thus we obtain an expression for
$d\sigma $ as a measure on $\rl^{2s}$.  Now $s(x)=0$ on $\sigma $ implies
that

\ba{
 0\a={\partial \over{\partial x^k}}\,s\,(t\,({\bf x}), {\bf x} )\cr
\a={\partial s\over{\partial x^0}}\,{\partial t\over{\partial x^k}}
+{\partial s\over{\partial x^k}},
 \cr}
 \eno22)
 \ea
which can be substituted into the above expression to give

\ba{
d\sigma \a=A_\lambda ^{-1} \,\left( 1-{\partial t\over{\partial x^k}} \,
{ y^k\over y^0}\right)\,d^s{\bf y} \,d^s{\bf x} \cr
\a=A_\lambda ^{-1} \bigl( 1-\nabla t\cdot ({\bf y} /y_0) \bigr)\,d^s{\bf y}
\,d^s{\bf x} .
 \cr}
 \eno23)
 \ea
But eq. (15) implies that $\,|\,\nabla t\,( {\bf x} )\,|\,\le 1$, hence for  $y\in
V_+'$,

\be
 \bigr| \nabla t \cdot \left(  {\bf y} /y_0\right) \bigl|<1
\eno24)
 \ee
and $d\sigma $ is a positive measure as claimed.  The above also shows that
if $\,|\,\nabla t\,( {\bf x} )\,|\,= 1$ for some ${\bf x} $, then $d\sigma $
becomes ``asymptotically'' degenerate as $\,|\,{\bf y} \,|\,\to \infty$ in the
direction of $\nabla t\,( {\bf x}) $.  That is, if $\sigma $ is {\sl lightlike \/} at
$(t\,({\bf x}), {\bf x} )$, then $d\sigma $ becomes small as the velocity ${\bf
y}/y_0$ approaches the speed of light in the direction of $\nabla t\,( {\bf
x})$.  This means that functions in $L^2(d\sigma )$ \/(and, in particular, as
we shall see, in ${\cal K}$) are allowed high velocities in the direction $\nabla
t\,({\bf x})$ at $(t\,({\bf x}), {\bf x} )\in S$. This argument is an example of
the kind of {\sl microlocal analysis\/} which is possible in the phase--space
 formalism.  (In the usual spacetime framework, one cannot say anything
about the velocity distribution of a function at a given point in spacetime,
since this would require taking the Fourier transform and hence losing the
spatial information.)
\skp
For $\sigma \in\Sigma $,  denote by $L^2(d\sigma )$  the Hilbert space of all
complex--valued, measurable functions on $\sigma $ with 

\be
 \|f\|^2_\sigma \equiv \ins \,|\,f\,|\,^2<\infty.
\eno25)
 \ee
If $f$ is a $C^\infty$ function on ${\cal T}_+$, we restrict it to $\sigma $ and
define $\|f\|_\sigma $ as above.  Our goal is to show that $\|f\|_\sigma
=\|f\|_{\cal K}$ for every $f\in {\cal K}$.  To do this, we first prove that each 
$f\in {\cal K}$ defines a conserved current in spacetime, which, by Stokes'
theorem,  makes it possible to deform the phase space $\sigma _{t,\lambda }$
of the last section to an arbitrary $\sigma=S-i\olp \in \Sigma $ without
changing the norm.  For $f\in {\cal K}$, define 

\be
 J^\mu (x)=A_\lambda ^{-1} \int_\olp\dy\,^\mu  \,|\,f(x-iy)\,|\,^2,
\eno26)
 \ee
where $\Omega _\lambda ^+$ has the orientation defined by $\dy\,^0$, so
that $J^0(x)$ is positive.  Then

\be
 \|f\|^2_\sigma =\int_S\dx_\mu \,J^\mu (x),
\eno27)
 \ee
where $S$ has the orientation defined by $\dx_0$.  (The restriction of
$\dx_0$ to $S$ does not vanish since $\,|\,\nabla t\,({\bf x})\,|\,\le 1$.)
\skp

\noindent {\bf Theorem 4.9. } {\sl Let $\fhat({\bf p} )$ be $C^\infty$ with
compact support.  Then $J^\mu (x)$ \break is $C^\infty$ and satisfies the
continuity equation

\be
 {\partial J^\mu \over{\partial x^\mu }}=0.
\eno28)
 \ee

}
\noindent {\bf Proof.}  By eq. (19), 

\be
 J^\mu (x)=A_\lambda ^{-1} \int_\olp d\tilde y\ y^\mu \,|\,f(x-iy)\,|\,^2,
\eno29)
 \ee
where $d\tilde y\equiv \dy\,^0/y^0$.  The function 

\be
 F_x^\mu (y,p,q)\equiv y^\mu \exp\left[ ix(p-q)-y(p+q) \right]\,
\overline{\fhat(p) }\,\fhat(q)
\eno30)
 \ee
is in $L^1(d\tilde y\times d\tilde p\times d\tilde q)$, hence by Fubini's
theorem,

\ba{
J^\mu (x)\a=A_\lambda ^{-1} \int_\olp d\tilde y\,\int_{\omp\times
\omp}d\tilde p \,d\tilde q\,F_x^\mu (y,p,q)\cr
\a=A_\lambda ^{-1} \int_{\omp\times \omp}d\tilde p \,d\tilde q\,\exp\left[
ix(p-q) \right]\, \overline{\fhat(p) }\,\fhat(q)\,H^\mu (p+q),
 \cr}
 \eno31)
 \ea
where, setting $k\equiv p+q$, $\eta \equiv \sqrt{k^2}$ and using the
recurrence relation for the $K_\nu $'s given by eq. (51) in section 4.4, we
compute

\ba{
H^\mu (k)\a\equiv  \int_\olp d\tilde y\ y^\mu \,e^{-yk}\cr
\a=-{\partial \over{\partial k_\mu }}\, \int d\tilde y \,e^{-yk}\cr
\a=-{\partial \over{\partial k_\mu }}\,\left[ 2\left( 2\pi \lambda \over \eta 
\right)^\nu K_\nu (\lambda \eta ) \right]\cr
\a=\left(k^\mu /\pi   \right)\,\left( { 2\pi \lambda \over \eta } \right)^{\nu
+1}K_{\nu +1}(\lambda \eta )\cr
\a\equiv k^\mu H(\eta ).
 \cr}
 \eno32)
 \ea
$H(\eta )$ is a bounded, continuous function of $\eta$ for $\eta \ge
2m$, and 

\be
 J^\mu (x)=A_\lambda ^{-1} \int_{\omp\times \omp}d\tilde p \,d\tilde q\,
\exp\left[ ix(p-q) \right]\, \overline{\fhat(p) }\,\fhat(q)\,\left( 
p^\mu +q^\mu  \right)\,H(\eta ).
\eno33)
 \ee
Since $\fhat({\bf p})$ has compact support, differentiation under the
integral sign to any order in $x$ gives an absolutely convergent integral,
proving that $J^\mu $ is $C^\infty$.  Differentiation with respect to $x^\mu $
brings down the factor $i(p^\mu -q^\mu) $ from the exponent, hence the
continuity equation follows from $p^2=q^2=m^2$.\sp\qed\skp 
\noindent {\sl Remark.\/} The continuity equation  also follows from a more
intuitive,  geometric argument.  Let 

\be
\blp=\{y\in V_+'\,|\,\,y^0>\sqrt{\lambda ^2+{\bf y} ^2}\},
\eno34)
 \ee
oriented such that 

\be
 \olp=-\partial \blp
\eno35)
 \ee
(the outward normal on $\partial \blp$ points ``down,'' whereas $\olp$ is
oriented ``up'').  Then by Stokes' theorem,

\ba{
 J^\mu (x)\a=A_\lambda ^{-1} \int_{\olp}\dy\,^\mu
\,|\,f(x-iy)\,|\,^2\cr 
\a=-A_\lambda ^{-1} \int_\blp d\left( \dy\,^\mu\,|\,f(x-iy)\,|\,^2 \right).
 \cr}
 \eno36)
 \ea
Here, $d$ represents exterior differentiation with respect to $y$, and since
the $s$--form $\dy\,^\mu $ contains all the $dy_\nu $'s except for $dy_\mu
$, we have

\be
 J^\mu (x)=-A_\lambda ^{-1} \int_\blp dy\ {\partial 
\over{\partial y_\mu }}\,\,|\,f(x-iy)\,|\,^2,
\eno37)
 \ee
where $dy$ is Lebesgue measure on $\blp$.
To justify the use of Stokes' theorem, it must be shown that the contribution
from $\,|\,{\bf y} \,|\,\to \infty$ to the first integral vanishes. This depends
on the behavior of $f(z)$, which is why we have given the previous
analytic proof using the Fourier transform.  Then the continuity equation is
obtained by differentiating under the integral sign (which must also be
justified) and using

\ba{
{\partial ^2 \,|\,f\,|\,^2\over{\partial x^\mu \partial y_\mu }}=0,
 \cr}
 \eno38)
 \ea
which follows from the Klein--Gordon equation {\sl combined with
analyticity,\/} since

\ba{
{\partial ^2\over{\partial x^\mu \partial y_\mu }}\a=
\left( {\partial \over{\partial {{\zbar}}^\mu  }}+{\partial \over{\partial 
z^\mu }} \right)\cdot i\left( {\partial \over{\partial {\zbar} _\mu }} 
-{\partial \over{\partial z_\mu }}\right)\cr
\a=i{\partial ^2\over{\partial {\zbar} ^\mu \partial {\zbar} _\mu }}
-i{\partial ^2\over{\partial z ^\mu \partial z _\mu }}\cr
\a\equiv i\left( \del_{\!\!\zbar}-\del_{\!\!z}\right).
 \cr}
 \eno39)
 \ea
Incidentally, this shows that 

\ba{
 j^\mu (z)\a\equiv -{\partial \over{\partial y_\mu }}\,|\,f(z)\,|\,^2\cr
\a=i\left[ \overline{ f\z} \,\partial_\mu f\z-\overline{ \partial_\mu
f\z}\cdot f\z \right]
 \cr}
 \eno40)
 \ea
is a ``microlocal,''  spacetime--conserved probability current for each fixed
$y\in V_+'$, so the scalar function $\,|\,f(z)\,|\,^2$ is a {\sl potential\/}  for
the probability current.  We shall see that this is a general trend in the
holomorphic formalism:  many vector and  tensor quantities can
be derived from scalar potentials. 

\skp

 Eqs. (37) and (40) also show that our
probability current is a {\sl regularized version\/}  of the usual current
 associated with solutions in {\sl real\/}  spacetime.  The latter
 (Itzykson and Zuber [1980]) is given by 

$$J^\mu _{\rm usual}\x=
i\left[ \overline{ f\x} \,\partial_\mu f\x-\overline{ \partial_\mu
f\x}\cdot f\x \right],$$
which leads to a conceptual problem since the time component, which
should serve as a probability density, can become negative even for
positive--energy solutions (Gerlach, Gomes and Petzold [1967], Barut and
Malin [1968]).  By contrast, eq. (36) shows that $J^0\x$ is stricly
non--negative. The tendency of quantities in complex spacetime to give
regularizations of their counterparts in real spacetime is further discussed
in chapter 5.

 \skp
We can now prove the  main result of this section.
\skp

\noindent {\bf Theorem  4.10. } {\sl Let $\sigma =S-i\olp \in \Sigma $ and
$f\in {\cal K}$ .  Then $\|f\|_\sigma =\|f\|_{\cal K}$.

}
\skp
\noindent {\bf Proof.}  We will  prove the theorem for $\fhat({\bf p} )$
in the space ${\cal D}(\rl^s)$ of $C^\infty$ functions with compact support,
which implies it for arbitrary $\fhat \in L^2_+(d\tilde p)$ by continuity.  Let
$S$ be given by $x^0=t\,({\bf x})$, and for $R>0$  let

\ba{
D_R\a=\{x\in\rl^{s+1}\,\bigm|\ |{\bf x}|<R, x^0\in [0, t\,({\bf x})] \,\},\cr
E_R\a=\{x\in\rl^{s+1}\,\bigm|\ |{\bf x}|=R, x^0\in [0, t\,({\bf x})] \,\},\cr
S_{0R}\a=\{x\in\rl^{s+1}\,\bigm|\ |{\bf x}|<R, x^0=0\},\cr
S_R\a=\{x\in\rl^{s+1}\,\bigm|\ |{\bf x}|<R, x^0= t\,({\bf x}) \},
 \cr}
 \eno41)
 \ea
where $[0, t\,({\bf x})]$ means $[t\,({\bf x}), 0]$ if $t\,({\bf x})<0$.  We orient
$S_{0R}$ and $S_R$ by $\dx_0$, $E_R$ by the ``outward normal''

\be
 \hat {\bf r}=R ^{-1}\sum_{k=1}^s x^k\dx_k,
\eno42)
 \ee
and $D_R$ so that $\partial D_R=S_R-S_{0R}+E_R$.  Now let $\fhat ({\bf
p})\in {\cal D}(\rl^s)$.  Then $J^\mu (x)$ is $C^\infty$, hence by Stokes'
theorem,

\ba{
\int_{S_R-S_{0R}+E_R}J^\mu (x)\,\dx_\mu \a=\int_{D_R} d\left( J^\mu \,
\dx_\mu  \right)\cr
\a=(-1)^s\int_{D_R}dx\,{\partial J^\mu \over{\partial x^\mu }}=0.
 \cr}
 \eno43)
 \ea
We will show that 

\be
 \Delta (R)\equiv \int_{E_R} J^\mu \dx_\mu \to 0 \ \hbox{as}\  R\to \infty
\eno44)
 \ee
(i.e., there is no leakage to $| {\bf x} | \to \infty$), which implies that 

\ba{
\|f\|^2_\sigma \a\equiv \lim_{R\to\infty}\int_{S_R} J^\mu\, \dx_\mu \cr
\a=\lim_{R\to\infty}\int_{S_{0R}} J^\mu \,\dx_\mu \cr
\a=\|f\|^2_{\sigma_{0\lambda }}=\|f\|^2_ {\cal K}  
 \cr}
 \eno45)
 \ea
by theorem 1 of section 4.4.  To prove that $\Delta (R)\to 0$, note that on
$E_R$, $\dx_0=0$ and 

\be
 \dx_k=x_k{\dx_1 \over x_1}=x_k{\dx_2 \over x_2}=\cdots=
x_k{\dx_s \over x_s},
\eno46)
 \ee
each form being defined except on a set of measure zero; hence

\be
 \hat {\bf r}=R\,{\dx_1 \over x_1}.
\eno47)
 \ee
By eq. (29),

\be
 |J^k (x)|\le J^0(x),
\eno48)
 \ee
hence

\ba{
\bigr| \Delta (R) \bigl|\a=\bigr|  \sum_{k=1}^s\int_{E_R}J^k\,\dx_k\bigl|\cr
\a=\bigr|  \sum_{k=1}^s\int_{E_R}J^k\,x_k{\dx_1 \over x_1}\bigl|\cr
\a\le s\,\int_{E_R}J^0\,R\,{\dx_1 \over x_1}\cr
\a=s\,\int_{E_R}J^0\,\hat {\bf r}\equiv a(R).
 \cr}
 \eno49)
 \ea
Now by eqs. (31) and (32),

\be
 J^0(x)=\int_{\rl^{2s}}d^s {\bf p}\, d^s {\bf q} \,\,e^{ix(p-q)}\,\phi ( {\bf p},
{\bf q}), 
\eno50)
 \ee
where

\ba{
\phi ({\bf p} ,{\bf q} )=\overline{\fhat(p) }\,\fhat(q)\,\psi ({\bf p} ,{\bf q} ),
 \cr}
 \eno51)
 \ea
and $\psi \in C^\infty(\rl^{2s})$.  Hence $\phi \in {\cal D}(\rl^{2s})$.  Let

\be
 D=\hat {\bf x}\cdot \nabla_{\bf p},
\eno52)
 \ee
where $\hat {\bf x}= {\bf x} /R$, and observe that for ${\bf x}\in
E_R$, 

\ba{
De^{ixp}\a=-iR\left( 1-{ x^0\over R} \hat {\bf x}\cdot {\bf v} \right)e^{ixp}\cr
\a\equiv -iR\,\xi (x, p)\,e^{ixp},
 \cr}
 \eno53)
 \ea
where ${\bf v} ={\bf p}/p_0$.  Since $\phi $ has compact support, there
exists a constant $\alpha <1$ such that $|{\bf v} |\le \alpha $ and
 $|{\bf v'} |\le \alpha $ for all $( {\bf p},{\bf p'})$ in the support of $\phi $. 
Furthermore, since $|\nabla t\,(x)|\le 1$, given any $\epsilon >0$ we have 
$|x^0|<R(1+\epsilon )$ for $\xi \in E_R$ for $R$ sufficiently large; hence

\ba{
|\xi (x,p)|\ge 1-\alpha (1+\epsilon )\ \hbox{for}\  x\in E_R\
\hbox{and}\   {\bf p} \in \ \hbox{supp}\,\,\phi .
 \cr}
 \eno54)
 \ea 
Choose $0<\epsilon <\alpha ^{-1} -1$, substitute

\be
 e^{ixp}={i \over R\,\xi (x,p)}\,D\,e^{ixp},\qquad x\in E_R
\eno55)
 \ee
into the expression for $J^0(x)$ and integrate by parts:

\ba{
J^0(x)\a=(iR)^{-1} \int_{\rl^{2s}}d^s {\bf p} d^s {\bf q} \,e^{ix(p-q)}\,
D\left( { \phi ( {\bf p}, {\bf q} )\over \xi (x,p)} \right)\cr
\a\equiv (iR)^{-1} \int_{\rl^{2s}}d^s {\bf p} d^s {\bf q} \,e^{ix(p-q)}\,
\phi '_x({\bf p}, {\bf q} ).
 \cr}
 \eno56)
 \ea
This procedure can be continued, giving (for $x\in E_R$)

\be
 J^0(x)=(iR)^{-n}\int_{\rl^{2s}}d^s {\bf p} d^s {\bf q} \,e^{ix(p-q)}\,
\phi^{(n)}_x({\bf p}, {\bf q} ), \qquad n=1,2, \cdots,
\eno57)
 \ee
where 

\ba{
 \phi^{(n)}_x({\bf p}, {\bf q} )\a=\left( D\circ \xi ^{-1}  \right)^n
 \phi ({\bf p}, {\bf q} )\cr
\a=\left[ \hat {\bf x}\cdot \nabla_{\bf p} 
\left( 1-{x^0 \over R}\,\hat {\bf x}\cdot {\bf v}  \right)^{-1} \right]^n\, \phi
({\bf p}, {\bf q} ).
 \cr}
 \eno58)
 \ea
Now $\left( D\circ \xi ^{-1}  \right)^n$ is a partial differential operator in
${\bf p} $ whose coefficients are polynomials in $D^k(\xi ^{-1})$ with $k=0,1,
\cdots, n$.  We will show that for $x\in E_R$ with $R$ sufficiently large,
there are constants $b_k$ such that 

\be
 \bigr| D^k(\xi ^{-1}) \bigl|<b_k,\qquad k=0,1, \cdots,
\eno59)
 \ee
which implies that 

\be
 \|\phi ^{(n)}_x\|_{L^1(\rl^{2s})}<c_n,\qquad x\in E_R, \,n=1,2,\cdots
\eno60)
 \ee
for some constants $c_n$, so that by eqs (49) and (57),

\ba{
a(R)\a=s\,\int_{E_R}J^0\hat {\bf r} \cr
\a=\le sR^{-n}\int_{E_R} \|\phi ^{(n)}_x\|_{L^1(\rl^{2s})}\,\hat {\bf r}(x)\cr
\a\le sR^{-n}\,c_n{ 2\pi ^{s/2}\over \Gamma (s/2)}\,R^{s-1}
\int_0^{R(1+\epsilon )}\,dx^0\cr
\a={ 2s\pi ^{s/2}\over \Gamma (s/2)}\,c_n R^{s-n}\,(1+\epsilon )\cr
\a\to 0 \ \hbox{as}\  R\to\infty
 \cr}
 \eno61)
 \ea
if we choose $n>s$.  To prove eq. (59), note that it holds for $k=0$ by eq. (54)
and let $u=\hat {\bf x} \cdot {\bf v} $.  Then 

\be
 Du\equiv \left( \hat {\bf x} \cdot \nabla_ {\bf p}  \right)\,
\left(  \hat {\bf x} \cdot {\bf p} /p_0 \right)={1-u^2 \over p_0},
\eno62)
 \ee
and if for some $k$

\be
 D^ku={ P_k(u)\over p_0^k}
\eno63)
 \ee
where $P_k$ is a constant--coefficient polynomial, then 

\ba{
D^{k+1}u\a={P_k'(u)Du \over p_0^k}-{kP_k(u) \over p_0^{k+1}}\cr
\a\equiv { P_{k+1}(u)\over p_0^{k+1}},
 \cr}
 \eno64)
 \ea
hence eq. (63) holds for $k=1,2, \cdots  $ by induction.  Thus 

\ba{
D^k\xi =-{x^0 \over R}\,D^k u=-{x^0 \over R}\,{ P_k(u)\over p_0^k},
\qquad k=1,2,\cdots,
 \cr}
 \eno65)
 \ea
which implies 

\be
 \bigr| D^k\xi  \bigl|\le { 1+\epsilon \over m^k}\,\max_{|u|\le 1}
\bigr| P_k(u) \bigl|.
\eno66)
 \ee
But $D^k(\xi ^{-1} )$ is a polynomial in   $\xi ^{-1} $ and $D\xi ,
D^2\xi , \cdots,  D^k\xi $; hence eq. (59) follows from eqs. (54)
and (66).\sp\qed\skp

The following is an immediate consequence of the above theorem.

\skp

\noindent {\bf Corollary 4.11. } {\sl 
\item{ (a)} For every $\sigma \in \Sigma $, the form
\be
 \langle\, f_1\,|\,f_2\,\rangle_\sigma =\ins \overline{f_1(z) }f_2(z)
\eno67)
 \ee
defines a ${\cal P}_0$--invariant inner product on ${\cal K}$, under which
${\cal K}$ is a Hilbert space.
\item{(b) } The transformations $(U_gf)(z)\equiv f(g ^{-1} z), \ g\in {\cal
P}_0$,  form a unitary irreducible representation  of ${\cal P}_0$ under the
above inner product, and the map $\fhat\mapsto f$ from $L^2_+(d\tilde p)$
to ${\cal K}$ intertwines this representation  with the usual one on
$L^2_+(d\tilde p)$.
\item{(c) } For each $\sigma \in\Sigma $, we have the resolution of unity 
\be
\ins \,|\,e_z\,\rangle\langle\, e_z\,|\,=I
\eno68)
\ee  
\sp \sp  on $L^2_+(d\tilde p)$ (or, equivalently, on ${\cal K}$ if $e_z$ is
replaced by $\tilde e_z.\ $) \qed
}
\skp
\noindent {\sl Note:\/} As in section 4.4, all the above results extend by
continuity to the case $\lambda =0$.\sp\#

\secskp

\noindent {\bf 4.6. The Non--Relativistic Limit }
\def\rightheadline{\tenrm\hfil {\sl 
4.6. The Non--Relativistic Limit}\hfil\folio}  
 \skp
\noindent  We now show that in the non--relativistic  limit
$c\to\infty$, the foregoing coherent--state representation  of ${\cal P}_0$
reduces to the representation  of ${\cal G}_2$ derived in section 4.3, in a
certain sense to be made precise.  As a by--product, we discover that the
Gaussian weight function associated with the latter representation (hence
also the closely related weight function associated with the canonical
coherent states)   has its origin in the geometry of the relativistic (dual)
``momentum space'' $\olp$.  That is, for large $|{\bf y}| $ the solutions in
${\cal K}$ are dampened by the factor $\exp[-\sqrt{\lambda ^2+ {\bf y}
^2}\omega] $ in momentum space, which in the non--relativistic  limit
amounts to having a Gaussian weight function in phase space.

 \skp
In considering the non--relativistic  limit, we make all dependence on
$c$ explicit but set $\hbar=1$.   Also, it is convenient to choose a coordinate
system in which the spacetime metric is $g={\rm diag}(1,-1, \ldots, -1)$, so
that $y^0=y_0=\sqrt{\lambda ^2+ {\bf y} ^2}$ and  $p^0=p_0=\sqrt{m ^2c^2+
{\bf p} ^2}$.  Fix $u>0$ and let $\lambda =uc$.  Then

\ba{
y_0\omega \a=\sqrt{u^2c^2+{\bf y} ^2}\sqrt{m^2c^2+{\bf p} ^2}\cr
\a=umc^2+{ u{\bf p} ^2\over 2m}+ { m{\bf y} ^2\over 2u} +O(c^{-2}).
 \cr}
 \eno1)
 \ea  
Working heuristically at first, we expect that for large $c$, holomorphic 
solutions of the  Klein--Gordon equation can be approximated by

\ba{
f(x-iy)\a\equiv \inop \exp\left[ -it\omega +i {\bf x} \cdot {\bf p} 
-y_0\omega + {\bf y} \cdot {\bf p}  \right] \fhat( {\bf p} )\cr
\a\sim\int {d^s {\bf p}  \over (2\pi )^2\cdot 2mc }\,\exp\left[  
-it\left( mc^2+ { {\bf p} ^2\over 2m} \right)+i {\bf x} \cdot {\bf p} \right]
\times \cr
\a\qquad \exp\left[ -umc^2-{ u {\bf p} ^2\over 2m} -
{ m {\bf y} ^2\over 2u}+{\bf y}\cdot {\bf p} \right]\fhat ({\bf p})\cr
\a\sim (2mc)^{-1} \exp\left[ -i\tau mc^2-m{\bf y} ^2/2u
\right]\,f_{\sc{\rm N R}}  ({\bf x} -i {\bf y} , \tau ),
 \cr}
 \eno2)
 \ea
where $\tau =t-iu$ and  $f_{\sc{\rm N R}} $ is the corresponding holomorphic
solution of the Schr\"o\-din\-ger equation defined in section 4.3.  Note that
 the Gaussian factor $\exp[-m {\bf y} ^2/2u]$
is the square root of the weight function for the Galilean coherent states,
hence if we choose  $\fhat( {\bf p} )\in L^2(\rl^s)\subset L^2_+(d\tilde p)$,
then

\be
  \|e^{-m {\bf y} ^2/2u}f_{\sc{\rm N R}} \|^2_{L^2(\,\cxs^s)}=\left( \pi u/m
\right) ^s\|f_{\sc{\rm N R}} \|^2_ {{\cal H}_u}<\infty. 
\eno3)
 \ee

We now rigorously justify the above heuristic argument.  Let  $f(z)$ be
the function in ${\cal K}$ corresponding to $\fhat({\bf p} )$ and denote by
$f_c$  its restriction to $x^0=t$ and $y^2=u^2c^2$, for {\sl fixed\/}  $u>0$. 
 \skp

\noindent {\bf Theorem 4.12. } {\sl Let $u>0$ and $\fhat( {\bf p} )\in
L^2(\rl^s)$.  Then

\ba{
 J(c)\a\equiv \|2mc e^{i\tau mc^2}\,f_c-e^{-m {\bf y} ^2/2u}
f_{\sc{\rm N R}} \|^2_{L^2(\,\cxs^s) }\cr
 \a\to 0 \ \hbox{as}\  c\to\infty.
 \cr}
 \eno4)
 \ea
}
\skp 
\noindent {\bf Proof.}  
Without loss of generality, we set $u=m=1$ and $t=0$ to simplify the
notation.  Note first of all that 
\ba{
\|2ce^{c^2}f_c\|^2_{L^2(\,\cxs^s)}\a=4c^2e^{2c^2}A_c\|f\|^2_{\cal K}\cr
\a= 4c^2e^{2c^2}A_c\|\fhat\|^2_{L^2_+(d\tilde p)}, 
 \cr}
 \eno5)
 \ea
where $A_c\equiv A_\lambda $ ($\lambda \equiv uc=c$). But

\ba{
 A_c\a=\pi ^\nu K_{\nu +1}(2c^2)\cr
\a=(2c) ^{-1} \pi ^{s/2}e^{-2c^2}\left[ 1+O(c^{-2}) \right]
\cr}
 \eno6)
 \ea
and

\be
\ |\fhat\|^2_{L^2_+(d\tilde p)}\le (2c) ^{-1} (2\pi )^{-s}|\fhat\|^2_{L^2(\rl^s)}.
\eno7)
 \ee
Thus 

\ba{
\|2ce^{c^2}f_c\|^2_{L^2(\,\cxs^s)}\a\le (4\pi ) ^{-s/2} 
\|\fhat\|^2_{L^2(\rl^s)} \left[ 1+O(c^{-2}) \right],
 \cr}
 \eno8)
 \ea
showing that $2ce^{c^2}\!f_c$ approaches a limit in $L^2(\cx^s)$ as
$c\to\infty$.   Now 

\ba{
J(c)\a=\int\!\!\!\int d^s {\bf x} \,d^s {\bf y} \left| \left[ \left(  
{ c\over \omega } e^{c^2-yp}-e^{-( {\bf p} -{\bf y} )^2/2}\right)\fhat
\right]\,\check{\ }( {\bf x} ) \right|^2\cr
\a=\int d^s {\bf p} |\fhat( {\bf p} )|^2\int d^s {\bf y}  \left(  
{ c\over \omega } e^{c^2-yp}-e^{-( {\bf p} -{\bf y} )^2/2}\right)^2.
 \cr}
 \eno9)
 \ea
Choose $\alpha , \gamma $ such that $1/2<\gamma <\alpha <1$.  Then 

\ba{
 J_1\a\equiv \int_{| {\bf p} |>c^{1-\alpha }} d^s {\bf p} |\fhat( {\bf p} )|^2\int
_{\rl ^s} d^s {\bf y}  \left(  { c\over \omega } e^{c^2-yp}-e^{-( {\bf p} -{\bf y}
)^2/2}\right)^2 \cr
\a\le 4\pi ^{s/2}\|\chi _c\fhat\|^2_{L^2(\rl^s)}\cr
\a\to0 \ \hbox{as}\   c\to\infty,
 \cr}
 \eno10)
 \ea
where $\chi _c$ is the indicator function of the set $\{ {\bf p} \bigm|\, | {\bf
p} |>c^{1-\alpha }\}$.  Define $\theta $ and $\phi $ by $|{\bf y} |=c\sinh
\theta $ and $|{\bf p}|=c\sinh \phi $.  Then $y^0=\cosh \theta $ and
$\omega =c^2\cosh \phi $, hence 

\be
 yp\ge c^2\cosh(\theta -\phi )\ge c^2\left[ 1+(\theta -\phi )^2/2 \right].
\eno11)
 \ee
Thus for arbitrary $a\ge 0$,

\ba{
G_a({\bf p} )\a\equiv \int_{|{\bf y}|>c\cosh a}d^s {\bf y} \,e^{2c^2-2yp}\cr
\a\le { 2c^s\pi ^{s/2}\over \Gamma(s/2)}\int_a^\infty d\theta \,
\sinh ^{s-1}\theta \cosh \theta e^{-c^2(\theta -\phi )^2}\cr
\a\le {2^{1-s}c^s\pi ^{s/.2} \over \Gamma(s/2)}\int_a^\infty d\theta \,
e^{(s-1)\theta }\left( e^\theta +e^{-\theta } \right)e^{-c^2(\theta -\phi )^2}\cr
\a\le {2^{1-s}c^s\pi ^{s/.2} \over \Gamma(s/2)}\int_a^\infty d\theta \,
e^{s\theta -c^2(\theta -\phi )^2}\cr
\a={2^{1-s}c^s\pi ^{s/.2} \over \Gamma(s/2)}e^{s\phi
+s^2/4c^2}\int_{c(a-\phi )-s/2c}^\infty du\, e^{-u^2} .
 \cr}
 \eno12)
 \ea
Let $a=\sinh ^{-1} (c^{-\gamma })$.  Then for $|{\bf p}|<c^{1-\alpha }$, 

\be
 c(a-\phi )-s/2c\ge c\left[ \sinh ^{-1} (c^{-\gamma })- 
\sinh ^{-1} (c^{-\alpha  })\right]-s/2c\equiv g(c).
\eno13)
 \ee
$g(c)$ is independent of ${\bf p} $ and $g(c)\sim c^{1-\gamma }$ as
$c\to\infty$.  Also, $\phi <c^{-\alpha }$ when $|{\bf p}|<c^{1-\alpha }$.  Hence

\ba{
J_2\a\equiv \int_{|{\bf p}|<c^{1-\alpha }}d^s {\bf p} |\fhat({\bf p} )|^2\int_
{|{\bf y}|>c^{1-\gamma }} d^s {\bf y} \,e^{2c^2-2yp}\cr
\a\le { 2^{2-s}c^{s-1}\pi ^{s/2}\over \Gamma(s/2)}e^{sc^{-\alpha }+s^2/4c^2}
\int_{g(c)}^\infty e^{-u^2}du\,\|\fhat\|^2_{L^2(\rl^s)}\cr
\a\to 0 \ \hbox{as}\   c\to\infty.
 \cr}
 \eno14)
 \ea
Now

\ba{
2c^2-2yp\a=y^2+p^2-2yp=(y-p)^2\cr
\a=(y_0-\omega )^2/c^2-( {\bf y}-{\bf p} )^2\ge -( {\bf y}-{\bf p} )^2.
 \cr}
 \eno15)
 \ea
Hence 

\be
 \int_{|{\bf p}|<c^{1-\alpha }}d^s {\bf p} \,|\fhat( {\bf p} )|^2
\int_{|{\bf y}|>c^{1-\gamma }}d^s{\bf y}\,e^{-( {\bf y}-{\bf p} )^2}
\le J_2,
\eno16)
 \ee
and 

\ba{
\int_{|{\bf p}|<c^{1-\alpha }}\a d^s{\bf p} \,|\fhat( {\bf p} )|^2
\int_{|{\bf y}|>c^{1-\gamma }} d^s{\bf y}\,\left(  
{ c\over \omega } e^{c^2-yp}-e^{-( {\bf p} -{\bf y} )^2/2}\right)^2\cr
\a\qquad \le 4J_2\to 0 \ \hbox{as}\  c\to\infty.
 \cr}
 \eno17)
 \ea
Finally, 

\ba{
J_3\a\equiv \int_{|{\bf p}|<c^{1-\alpha }} d^s{\bf p} \,|\fhat( {\bf p} )|^2
\int_{|{\bf y}|<c^{1-\gamma }} d^s{\bf y}\left(  
{ c\over \omega } e^{c^2-yp}-e^{-( {\bf y} -{\bf p} )^2/2}\right)^2\cr
\a=\int_{|{\bf p}|<c^{1-\alpha }} d^s{\bf p} \,|\fhat( {\bf p} )|^2
\int_{|{\bf y}|<c^{1-\gamma }}d^s{\bf y}\,e^{-({\bf y}-{\bf p})^2}
\left( {c \over \omega }  e^{c^2\delta ^2/2}-1\right)^2,
\cr}
 \eno18)
 \ea
where 

\ba{
\delta \a=\bigr| \sqrt{1+{\bf y} ^2/c^2} -\sqrt{1+{\bf p} ^2/c^2}\bigl|\cr
\a\le {1\over 2c^2}\bigr| {\bf y} ^2-{\bf p}^2 \bigl|\cr
\a\le {1\over 2}\left( c^{-2\gamma }+  c^{-2\alpha  }\right)\cr
\a\le  c^{-2\gamma }.
 \cr}
 \eno19)
 \ea
We have used the estimate

\ba{
\left| \sqrt{1+u^2}- \sqrt{1+v^2}\right|\a=\left| \int_u^v {xdx
 \over \sqrt{1+x^2}} \right|\cr
\a\le \left|\int_u^v xdx\right|={1\over 2}\left|  v^2-u^2\right|.
 \cr}
 \eno20)
 \ea
Hence for sufficiently large $c$ and $|{\bf p}|<c^{1-\alpha }$,

\ba{
\left({ c\over \omega }e^{c^2\delta ^2/2}-1  \right)^2
\a\le e^{c^2\delta ^2}+1-{ 2c\over \omega }e^{c^2\delta ^2/2}\cr
\a \le \left( 1+2c^2\delta ^2 \right)+1-2\left(1-{\bf p}^2/2c^2 \right)
e^{c^2\delta ^2/2}\cr
\a \le 2\left(1-e^{c^2\delta ^2/2}  \right)+2c^2\delta^2+c^{-2\alpha}
e^{c^2\delta ^2/2}\cr 
\a \le 2c^2\delta ^2+c^{-2\alpha }\left( 1+c^2\delta ^2 \right)\cr 
\a \equiv h(c)\to 0 \ \hbox{as}\  c\to\infty.
 \cr}
 \eno21)
 \ea
Thus 

\ba{
J_3\a\le h(c)\int_{|{\bf p}|<c^{1-\alpha }} d^s{\bf p} \,|\fhat( {\bf p} )|^2
\int_{|{\bf y}|<c^{1-\gamma }} d^s{\bf y}\,e^{-({\bf y} -{\bf p})^2}\cr
\a\le h(c)\pi ^{s/2}\|\fhat\|^2_{L^2(\rl^s)}\cr
\a \to 0 \ \hbox{as}\   c\to\infty,
 \cr}
 \eno22)
 \ea
which proves that $J(c)\to 0$ as
$c\to\infty.$ \sp\qed\skp 

\secskp
\def\rightheadline{\tenrm\hfil {\sl  Notes}\hfil\folio}

\noindent {\bf  Notes}
\skp
\noindent This chapter represents the main body of the author's mathematics
thesis at the University of Toronto (Kaiser [1977c]). All the theorems,
corollaries, lemmas and propositions (labeled 4.1-4.12) have appeared in
the literature (Kaiser [1977b, 1978a]).   In 1966, when the idea of complex
spacetime as a unification of spacetime and phase space first occurred to me,
I had found a kind of frame  in which {\sl both\/}  the bras
and the kets were holomorphic in $\bar z$  and the resolution of unity was
obtained by a contour integral, using Cauchy's theorem.  During a seminar I
gave in 1971 at Carleton University in Ottawa (where I was then a
post--doctoral fellow in physics), L.~Resnick pointed out to me that this
``wave--packet representation'' appeared to be related to the
coherent--state representation,  which was at that time unknown to me.
 The kets were identical to the   canonical coherent states,
 but the bras were not their Riesz duals;  in the language of chapter 1, they
belonged to a (generalized) frame {\sl reciprocal\/}  to that of the kets, and
the resolution of unity was of the type given by eq. (24) in section 1.3, which
may be called a continuous version of {\sl biorthogonality.\/}  A version of
this result was reported at a conference in Marseille (Kaiser [1974]).  I was
later  informed  by J.~R.~Klauder that a similar representation  had
 been developed by Dirac in connection with quantum electrodynamics (Dirac
[1943, 1946]).  

The original idea of complex spacetime as phase space was to consider a
complex combination of the (symmetric) Lorentzian metric with the
(antisymmetric) symplectic structure of phase space,
obtaining a hermitian metric on the complex spacetime parametrized by
local coordinates of the type $x+ibp$.  (I have since learned that this
structure, augmented by some technical conditions,  is known as a {\sl
K\"ahler metric;\/}  see Wells [1980].)  The above ``wave--packet
representation'' indicated that this combination may in fact be interesting, 
but so far it was {\sl ad hoc\/}  and  lacked a physical basis.  Also, the
representation  was non--relativistic, and it was not at all clear how to
extend it to the relativistic domain, as pointed out to me by V.~Bargmann in
1975.  The standard method of arriving at canonical coherent states is to use
an integral transform with a Gaussian kernel in the configuration--space
representation, and there is no obvious relativistic candidate for such a
kernel.  The more general methods  described in
chapter 3 do not work, since the representations of interest are not
square--integrable (section 4.3). An important clue came in 1974 from the
study of axiomatic quantum field theory,  where   I was fascinated by the
appearance of  tube domains.  These domains occur in connection with the
analytic continuation of vacuum expectation values of products of fields, and
are therefore  extensions of such products to complex spcetime.  However,
the complexified spacetimes themselves are not taken seriously as possible
arenas for physics.  They are merely used to justify the application of
powerful  methods from the theory of several complex variables, in order to
obtain results concerning the restrictions of vacuum expectaion values to {\sl
real\/} spacetime.  (However, the restrictions to {\sl Euclidean\/}  spacetime
do have important consequences for statistical mechanics;  see Glimm and
Jaffe [1981].)  I felt that if these tube domains could somehow be given a 
physical interpretation as extended classical phase spaces, this would give
the phase--space formulation of relativistic quantum mechanics  a firm
physical foundation, since in quantum field theory  the extension to complex
spacetime  is based on solid physical principles such as the spectral
condition.  This idea was first worked out at the level of non--relativistic 
quantum mechanics, leading to the representation  of the Galilean group
given in section 4.3.  That amounted to a reformulation of the canonical
coherent--state representation  in which the Gaussian kernel appears
naturally in the {\sl momentum representation,\/} as a result of the analytic
continuation of solutions of the Schr\"odinger equation.  This ``explained'' the
combination $x+ibp$ (section 4.3, eq. (5)) and gave the coherent--state
representation  a {\sl dynamical\/}  significance.   It also cleared the way to
the construction of relativistic coherent states, since now the Gaussian kernel
merely had to be replaced with the analytic Fourier kernel $e^{-izp}$ on the
mass shell.  An important tool was the use of groups to compute certain
invariant integrals, which I learned from a lecture by E.~Stein on Hardy
spaces in 1975.  The  construction of the relativistic coherent states  given in
sections 4.4 and 4.5  was  carried out in 1975--76, culminating in the 1977 
thesis.  Related results were announced  at a conference in 1976 (Kaiser
[1977a]) and at two conferences in 1977 (Kaiser [1977d, 1978b]).  To my
knowledge, this was the first successful formulation of  relativistic coherent
states, which have since then gained some popularity (see De Bi\'evre [1989],
Ali and Antoine [1989]).  An earlier attempt  to formulate such states was
made by Prugove${\rm\check c}$ki [1976], but this was shown  to be
inadequate since the proposed states were merely the Gaussian canonical
coherent states in disguise, hence not covariant under the Poincar\'e group
(Kaiser [1977c], remark 4 in sec. II.5 and addendum, p. 133.)   After the
results of the thesis appeared in the literature,   Prugove${\rm\check c}$ki
[1978; see also 1984] discovered that they can be generalized by replacing
the invariant functions $e^{-yp}$ with arbitrary (sufficiently regular)
invariant functions.  The price of this generalization is that solutions of the
Klein--Gordon equation are no longer represented by holomorphic functions 
and  the close connection with quantum field theory  (chapter 5) appears to
be lost.  The relation between the two formalisms and their history was
discussed at a conference in Boulder in 1983 (Kaiser [1984b]), where an
inconsistency in Prugove${\rm\check c}$ki's formalism was also pointed out.

 The classical limit  of solutions of the  Klein--Gordon equation
in the coherent--state representation  was studied in Kaiser [1979].  In an
effort to understand interactions, the notion of {\sl holomorphic gauge
theory\/}  was introduced (Kaiser [1980a, 1981]).  This is reviewed in section
6.1.  An early attempt was also made to extend the theory to the framework
of interacting quantum fields (Kaiser [1980b]), but that  was soon
abandoned as unsatisfactory.  A  more promising approach was developed
later (Kaiser [1987a]) and  is presented in the next chapter. 

\skp

 Note that our phase spaces $\sigma $ are not unique, since the configuration
space $S$ can be chosen arbitrarily as long as it is nowhere timelike and
$\lambda>0 $ can be chosen arbitrarily.  The freedom in $S$ is, in fact, 
related to the probability--current conservation, while the freedom in
$\lambda $, combined with holomorphy, allowed us to express the
probability current as a regularization of the usual current by the use of
Stokes' theorem (section 4.5, eqs. (37) and (40)).  By contrast, the phase
spaces obtained by De~Bi\'evre [1989] are unique. They are ``coadjoint
orbits'' of the Poincar\'e group, related to  ``geometric quantization''
theory  (Kirillov [1976], Kostant [1970], Souriau [1970]).   Although this
uniqueness seems attractive, it involves a high cost:  the dynamics must be
factored out.  This means that the ensuing theory is no longer ``local in
time.''  Since one of the attractions of coherent--state representations is their
``pseudo--locality'' in both space and momentum, and since in a relativistic
theory time ought to be treated like space, it seems to me an advantage to
retain time in the theory.  Perhaps a
 more persuasive argument for this comes from holomorphic gauge theory
(section 6.1), where a  theory describing a free particle can, in principle, be
``perturbed'' by introducing a non--trivial fiber metric to obtain a theory
describing a particle in an electromagnetic (or Yang--Mills) field.  This
cannot  be done in a natural way once time has been factored out.

\skp
Some very interesting work done recently  by Unterberger [1988] uses
coherent states which are essentially equivalent to ours to develop a
pseudo\-differential calculus based upon the Poincar\'e group as an
alternative to the usual Weyl calculus, which is based on the
Weyl--Heisenberg group.  Since the Poincar\'e group contracts to a group
containing the Weyl--Heisenberg group in the non--relativistic  limit
(section 4.2), Unterberger's ``Klein--Gordon calculus'' similarly contracts to
the Weyl calculus.

\VE

\def\leftheadline{\tenrm\folio\hfil {\sl 5. Quantized Fields}\hfil} 
\def\rightheadline{\hfill\folio}

\headline={\ifodd\pageno\rightheadline\else\leftheadline\fi}
\def\be{$$}\def\ee{$$}\def\ba{$$\eqalign}\def\a{&}  
\def\eno{\eqno(}\def\ea{$$}

\centerline{\bf Chapter 5}\skp
\centerline{\bf QUANTIZED FIELDS}
\vskip 3 cm
\noindent {\bf  5.1. Introduction}
\skp

\noindent We have regarded solutions of the Klein--Gordon equation as the
quantum states of  a  relativistic particle.  But such solutions also
possess another interpretation:  they can be viewed as {\sl classical fields,\/}
something like the electromagnetic field (whose components, in fact,  satisfy
the wave equation, which is the Klein--Gordon equation with zero mass). 
This interpretation is the basis for quantum field theory.  The general idea is
that just as the finite number of degrees of freedom of a system of classical
particles was quantized to give ordinary (``point'')  quantum mechanics, a
similar prescription can be used to quantize the infinite number of degrees of
freedom of a classical field.  It turns out that the resulting theory {\sl
implies\/}  the existence of particles.  In fact, the asymptotic free in-- and
out--fields are represented by operators which create and destroy particles
and antiparticles, in agreement with the fact that such creation and
destruction processes occur in nature.  These particles and antiparticles are
represented by positive--energy solutions of the asymptotic {\sl free\/} 
wave equation, e.g. the  Klein--Gordon or Dirac equation.  Thus the formalism
of relativistic quantum mechanics  appears to be, at least partially,  {\sl
absorbed\/} into quantum field theory.  
\skp
 In regarding solutions of the Klein--Gordon equation as the
physical states of a relativistic particle, it was appropriate to restrict our
attention to functions having only positive--frequency Fourier components,
since the energy of the particle must be positive.  Even a small negative
energy can be made arbitrarily large and negative by a Lorentz
transformation, leading to instability.  When the solutions are  regarded as
classical fields, however, no such restriction on the frequency is necessary or
even justifiable.  For example, in the case of a {\sl neutral \/}  field (i.e., one
not carrying any electric  charge), the solutions  must be
real--valued, hence their Fourier transforms {\sl must\/}  contain negative--
as well as positive--frequency components.  On the other hand, the analytic
extension of the solutions to complex spacetime appeared to  depend
crucially on the positivity of the energy.  We must therefore ask  whether
an  extension   is still possible for fields, or if it is even desirable from a
physical standpoint, since the connection between solutions and particles is
not as immediate as it was earlier.  In this chapter we find an affirmative
answer to both of these questions.  A natural method, which we call the
Analytic--Signal transform, will be developed to extend {\sl arbitrary\/} 
functions  from $\rl^{s+1} $ to $\cx^{s+1}$,  and when the functions
represent  physical fields, the double tube $\tb=\tp\cup\tm$ in $\cx^{s+1}$
will be shown to have a direct physical significance as an extended classical
phase space, not for the fields themselves but for certain  ``particle''-- and
``antiparticle'' coherent states $e_z^\pm$  associated with them.  These states
are  related directly to the dynamical (interpolating)  fields, {\sl not \/} their
asymptotic free in-- and out--fields.  To be precise, they should be called
{\sl charge\/}  coherent states rather than particle coherent states, since
they have a well--defined charge whereas, in general, the concept of
individual  particles does not make sense while interactions are present.  
 If the given fields satisfy some (possibly non--linear) equations, the 
coherent states satisfy a Klein--Gordon equation with a source term.  Hence
they represent {\sl dynamical\/} rather than ``bare''  
particles.  For free fields,  $e_z^+$ reduces  to the state $e_z$
defined in the last chapter and $e_z^-$ to its complex conjugate, which is a
negative--energy solution of the Klein--Gordon equation holomorphic in
$\tm$.

\def\rightheadline{\tenrm\hfil {\sl 5.1. Introduction}\hfil\folio}  

 Complex tube domains also   appear in the
contexts of axiomatic and constructive quantum field theory, and our results
suggest that those domains, too,  may have  interpretations related to
classical phase space,   a point of view which, to my knowledge, has not been
explored heretofore.\footnote*{R.~F.~Streater has recently told
me that G.~K\"all\'en was informally advocating the interpretation of the
holomorphic  Wightman two--point function as a  correlation function
in phase  space  around 1957.  Nothing appears to have been published
on this, however.}

 \skp

While our extended fields are not analytic in general, they are
``analyticity--friendly,'' i.e.  have certain features which yield various
analytic objects under different circumstances.  For example, their two--point
functions are piecewise analytic, and the pieces agree with the analytic
Wightman functions.     In the special case when the given fields are  free, 
the extended fields  themselves are analytic in $\tb$.  Furthermore, the
fields in general  possess a {\sl directional\/}  analyticity which looks like a
covariant version of analyticity in time.  Since the latter forms the basis for 
the continuation of the theory (in the form of vacuum expectaion
values)  from Lorentzian to Euclidean spacetime (see Nelson [1973a,b]
and Glimm and Jaffe [1981]),  it may be that our extended fields, when
restricted to the Euclidean region,  bear some  relation to the corresponding
Euclidean fields.

 \skp 
The formalism we are about to develop for fields is a
natural extension of the one constructed  for particles in the last chapter. 
Like its predecessor, it posesses a degree of regularity not found in the usual
spacetime formalism.   Some examples of this regularity are:  
\skp
\item{(a)} The extended fields $\phi \z$ are, under reasonable
assumptions, oper\-ator--valued {\sl functions\/}  (rather than
distributions, as usual) when restricted to $\tb$.

\item{(b)}  The theory contains a natural, covariant ultraviolet damping,
which is a  permanent feature of the theory.  This
comes from the possibility of working directly in phase space, away from
real spacetime.  From the point of view of the usual ({\sl real\/}  spacetime)
theory, our formalism looks like a ``regularization''.   From our point of view,
however, no regularization is necessary since, it is suggested, {\sl reality
takes place in complex spacetime!\/}   In other words, this ``regularization'' is
permanent and is not to be regarded as a kind of trick, used to obtain finite
quantities, which must later be removed from the theory.

 \item{(c)}   In the  case of free fields, the formalism
automatically  avoids  zero--point energies without normal ordering, due to a
{\sl polarization\/}  of the positive-- and negative frequency components into
the forward and backward tubes, respectively.     Observables such
as charge, energy--momentum and angular momentum are obtained as
conserved integrals of bilinear expressions in the fields over phase spaces
$\sigma \subset \tb$.  These expressions, which are densities for the
corresponding observables, look like {\sl regularizations\/}  of the
corresponding expressions in the usual spacetime theory.  The analytic
(Wightman) two--point function acts as a reproducing kernel for the fields,
much as it did for the wave functions in chapter 4.

\item{(d)}  The particles and antiparticles associated with the free 
Dirac field do not undergo the random motion known as Zitterbewegung
(Messiah [1963]), again because of the aforementioned polarization.

\secskp

 \noindent {\bf  5.2. The Multivariate Analytic--Signal Transform }
\def\rightheadline{\tenrm\hfil {\sl 
5.2. The Multivariate Analytic--Signal Transform}\hfil\folio}  

\skp
 \noindent As mentioned above, in dealing with physical {\sl fields\/} such
as the  electromagnetic field,  rather than quantum {\sl states,\/} we can no
longer justify the restriction that frequencies must be positive.  For one
thing, as we shall see,  in the presence of interactions there is no longer a
covariant way to eliminate negative frequencies.  Hence the method used in
chapter 4 to analytically continue solutions of the Klein--Gordon equation  to
complex spacetime will no longer work directly.  In this section we devise a
method for extending {\sl arbitrary\/}  functions from $\rl^{s+1}$ to
$\cx^{s+1}$.  When the given functions are  positive--energy solutions of the 
Klein--Gordon or the Wave equation, this method reduces to the analytic
extension used in chapter 4. But it is much more general, and will enable us
to extend {\sl quantized\/}  fields, whether Bose or Fermi, interacting or
free, to complex spacetime.
 We begin by formulating the method for functions of {\sl one\/}  variable,
where it is closely related to the concept of {\sl analytic signals.}  For
motivational purposes, we think of the variable as time ($s=0$).  In this
chapter, Fourier transforms will usually be with respect to {\sl spacetime\/}
($\rl^{s+1}$) rather than just space ($\rl^s$).  Hence we will denote them by
$\ftil$, reserving $\fhat$ for the spatial Fourier
transform, as done so far. 
 \skp

Suppose we are given a ``time--signal,'' i.e. a real-- or
complex--valued function of a single real variable $x$. To begin with, assume
that $f$ is a Schwartz test function, although most of our considerations will
extend to   certain kinds of  distributions.  Consider the positive-- and
negative-- frequency parts of $f$, defined by

\ba{
f_+(x)\a\equiv (2\pi)^{-1} \int_0^\infty dp\,e^{-ixp}\,\ftil(p)\cr
f_-(x)\a\equiv (2\pi)^{-1} \int_{-\infty}^0 dp\,e^{-ixp}\,\ftil(p).
\cr}
 \eno1)
 \ea
Then $f_+$ and $f_-$ extend analytically to the lower--half and upper--half
complex planes, respectively, i.e.

\ba{
f_+(x-iy)\a= (2\pi)^{-1} \int_0^\infty dp\,e^{-i(x-iy)p}\,\ftil(p),\quad 
y>0\cr
f_-(x-iy)\a= (2\pi)^{-1} \int_{-\infty}^0
dp\,e^{-i(x-iy)p}\,\ftil(p),\quad y<0.
 \cr}
 \eno2)
 \ea
$f_+$ and $f_-$ are just the {\sl Fourier--Laplace transforms\/}  of the
restrictions of $\ftil$ to the positive and negative frequencies.

If $f$ is complex--valued, then $f_+$ and $f_-$ are independent
and the original signal  can be recovered from them as

\be
 f(x)=\lim_{y \downarrow 0}\left[ f_+(x-iy)+ f_-(x+iy ) \right].
\eno3)
 \ee 
If $f$ is real--valued, then 

\be
 \ftil(p)=\overline{ \ftil(-p)},
\eno4)
 \ee
hence $f_+$ and $f_-$ are related by reflection,
\be
 f_+(z)=\overline{f_-(\zbar) }, \quad z\in\cx^-,
\eno5)
 \ee
and

\be
 f(x)=\lim_{y \downarrow 0} 2\Re f_+(x-iy )
=\lim_{y \downarrow 0} 2\Re f_-(x+iy ).
\eno6)
 \ee
When $f$ is real, the function $f_+(z)$ is known as the {\sl analytic signal\/} 
associated with $f(x)$.  A complex--valued signal would
have {\sl two\/}  independent associated analytic signals $f_+$ and $f_-$. 
What significance do $f_\pm$ have?  For one thing, they are {\sl
regularizations\/}  of $f$.  The above equation states that $f$ is jointly  a
boundary--value of the pair $f_+$ and $f_-$.  As such, $f$ may actually be
quite singular while remaining the boundary--value of analytic functions. 
Also,  $f_\pm$ provide a kind of 
 ``envelope'' description of  $f$ (see  Klauder and Sudarshan [1968], section
1.2).  For example, if $f(x)=\cos ax$ ($a>0$), then 
 $f_\pm(z)={\sc 1\over 2}\exp(\mp iaz)$, so
 the boundary values are $f_\pm(x)={\sc 1\over 2}\exp(\mp iax)$.  

In order to extend the concept of analytic signals to more than one
dimension, let us first of all unify the definitions of $f_+$ and $f_-$ by
defining 

\be
 f(x-iy)\equiv (2\pi)^{-1} \int_{-\infty}^\infty dp\,\theta
(yp)\,e^{-i(x-iy)p}\,\ftil(p)
 \eno7)
 \ee
for {\sl arbitrary\/}  $x-iy\in\cx$, where $\theta $ is the 
{\sl unit step function,\/} 
defined by

\be
 \theta (u)=\cases{0,\quad &$u<0$\cr 1/2,\quad &$u=0$\cr 1,\quad &
$u>0$.\cr}
 \eno8)
 \ee
Then we have

\ba{
f(z)=\cases{f_+(z),\quad &$y>0$\cr {\sc 1\over 2}f(x),\quad &$y=0$\cr
 f_-(z),\quad &$y<0.$\cr}
 \cr}
 \eno9)
 \ea
[The apparent inconsistency $f(x)={\sc 1\over 2}f(x)$ for $y=0$  is due to a
mild abuse of notation.  It could be removed by redefining $f(z)$ by  a factor
of 2 or, more correctly but laboriously, rewriting it as $(Sf)(z)$.  We prefer
the above notation, since the boundary--values $f(x)$ will not actually be
used in the phase--space formalism.]

Let us define the {\sl exponential step function\/} by

\be
 \theta ^\zeta \equiv \theta (-\Re \zeta )\,e^\zeta ,\qquad \zeta \in\cx,
\eno10)
 \ee
so that our extension is given by

\be
 f(z)=(2\pi)^{-1} \int_{-\infty}^\infty dp\,\theta ^{-izp}\,\ftil(p).
\eno11)
 \ee
The identity

\be
 \theta (u)\,\theta (u')=\theta (uu')\,\theta (u+u')
\eno12)
 \ee
shows that $\theta ^\zeta $ has the ``pseudo--exponential'' property

\be
 \theta ^\zeta \,\theta ^{\zeta'}=\theta (\Re \zeta \, \Re \zeta ')\,\theta ^{\zeta
+\zeta '\,},
\eno13)
 \ee
which will be useful later. 

\skp

Although this unification of $f_+$ and $f_-$ may at first appear to be
somewhat artificial, we shall now see that it is actually very natural. Note
first of all that for any real $u$, we have 

\be
 \theta (u)\,e^{-u}=\inds e^{i\tau u},
\eno14)
 \ee
since the contour on the right--hand side  may be closed in the lower
half--plane when $u<0$ and in the upper half--plane when $u>0$.  For
$u=0$, the equation states that 

\ba{
\theta (0)\a={ 1\over 2\pi i} \int_{-\infty}^\infty {(\tau +i)\,d\tau 
 \over \tau ^2+1}\cr
\a={1 \over 2\pi } \int_{-\infty}^\infty { d\tau \over \tau ^2+1}={1\over 2},
 \cr}
 \eno15)
 \ea
 in agreement with our definition, if we interpret the integral as 
the limit as $L\to\infty$ of the integral from $-L$ to $L$. The exponential
step function therefore has the integral representation  

\be
 \theta ^{-i(x-iy)p}=\inds e^{-i(x-\tau y)p}.
\eno16)
 \ee
If this is substituted into our expression for $f(z)$ and the order of
integrations on $\tau $ and $p$ is exchanged, we obtain

\be
 f(x-iy)=\inds f(x-\tau y)
\eno17)
 \ee
for arbitrary $x-iy\in\cx$.   We shall refer to the right--hand side   as the
{\sl Analytic--Signal transform\/} of $f(x)$.  It bears a close relation to the
{\sl Hilbert transform,\/} which is defined by

\be
 ( Hf)(x)={ 1\over \pi }{\rm PV}\int_{-\infty}^\infty
 {du \over u}\,f (x-u), 
\eno18)
 \ee
where PV denotes the principal value of the integral.  Consider the complex
combination 

\ba{
 f (x)-i(Hf )(x)\a={1\over \pi i}\int_{-\infty}^\infty du
\left[ \pi i\delta (u) +{\rm PV} {1 \over u}\right]\,f(x-u)\cr
\a={1\over \pi i}\lim_{\epsilon \downarrow 0}\,\int_{-\infty}^\infty 
{ du\over u-i\epsilon }\,f (x-u)\cr
\a={1\over \pi i}\lim_{\epsilon \downarrow 0}\,\int_{-\infty}^\infty 
{ d\tau \over {\tau -i}}\,f (x-\tau \epsilon )\cr
\a=\lim_{\epsilon \downarrow 0}\,2f (x-i\epsilon ).
 \cr}
 \eno19)
 \ea
Similarly, 

\be
  f (x)+i(Hf )(x)=\lim_{\epsilon \downarrow 0}\,2f (x+i\epsilon ).
\eno20)
 \ee
Hence

\be
( Hf)(x)=-i\lim_{\epsilon \downarrow 0}\,[f(x+i\epsilon )-
f(x-i\epsilon )],
\eno21)
 \ee
which, for real--valued $f$, reduces to 

\be
 (Hf)(x)=\lim_{\epsilon \downarrow 0}\,2\Im f(x+i\epsilon )
=-\lim_{\epsilon \downarrow 0}\,2\Im f(x-i\epsilon ).
\eno22)
 \ee
\skp
We are now ready to generalize the idea of analytic signals to an arbitrary
number of dimensions.
 \skp

\noindent {\bf Definition. }\sp Let $f\in {\cal S} (\rl^{s+1})$.  The {\sl
Analytic--Signal transform\/}  of $f$ is defined by

\be
 f(x-iy)=\inds f(x-\tau y).
\eno23)
 \ee
\skp 
\noindent The same argument as above shows that 

\ba{
 f(z)\a=\inp \theta ^{-izp}\,\ftil(p)\cr
\a=\inp \theta (yp)\,e^{-ixp-yp}\,\ftil(p).
 \cr}
 \eno24)
 \ea
We shall refer to the right--hand side  of this equation as the (inverse)  {\sl
Fourier--Laplace transform   of $\ftil$ in the half--space\/} 

\ba{
M_y\equiv \{p\in \rl^{s+1} \,|\,yp\ge 0\}.
 \cr}
 \eno25)
 \ea
 The integral converges absolutely whenever
$\ftil \in L^1(\rl^{s+1})$, since $|\theta ^{-izp}|\le 1$.  Hence $f(z)$ can
actually be defined for some distributions, not only for test functions.  The
extension of the  transform to distributions is  complicated by the fact that
$\theta ^{-izp}$ is not a test function in the variable $p$, hence for a
tempered distribution $T$, 

\be
 T(z)\equiv \tilde T(\theta ^{-izp})
\eno26)
 \ee
(defined through the Fourier transform $\tilde T$ of $T$) may not make
sense as a function on $\cx^{s+1} $.  It would be intersting to find a natural
class of distributions for which $T(z)$ does make sense as a function.  In
general, however, it may be necessary to consider distributions $T$ such
that $T(z)$ is some kind of  {\sl distribution\/}  on $\cx^{s+1} $.  The solutions
to both of these problems are unknown to me, so a certain amount of
vagueness will be necessary on this point.  In the next section, where $T$ is
a quantized field, it will be assumed to satisfy some physically reasonable
conditions which imply that $T\z$ is well--defined in an important {\sl
subset\/} $\tb$ of $\cx^{s+1}$.

 \skp
Recall that for $s=0$, $f(z)$ was analytic in the upper-- and lower--
half--planes.  In more than one dimension, $f(z)$ need not be analytic, even
though, for brevity, we still write it as a function of $z$ rather than $z$ and
$\zbar$.  However, $f(z)$ does in general possess a {\sl partial\/}
analyticity which reduces to the above when $s=0$.  Consider the partial
derivative of $f(x-iy)$ with respect to $\zbar^\mu $, defined by

\ba{
2\bar \partial_\mu f(z)\a\equiv  2{\partial f\over{\partial \zbar^\mu}}\cr
\a\equiv {\partial f\over{\partial x^\mu }}-i{\partial f\over{\partial
y^\mu}}.  
 \cr}
 \eno27)
 \ea
Then $f$ is analytic at $z$ if and only if $\bar\partial_\mu f=0$ for all $\mu
$.  But using our definition of $f(z)$, we find that 

\ba{
2\bar \partial_\mu f(x-iy)={ 1\over 2\pi }\int_{-\infty}^\infty d\tau \,
{\partial f\over{\partial x^\mu }}(x-\tau y).
 \cr}
 \eno28)
 \ea
The right--hand side  is known as the {\sl X--Ray transform\/} (Helgason
[1984]) of the function $\partial f/\partial x^\mu $, given in terms of the
parameters $x$ and $y$ defining the line $x(\tau )=x-\tau y$. It follows that
the complex  {\sl $\bar \partial$--derivative  in the direction of $y$\/}
vanishes, i.e.

\ba{
4\pi y^\mu \bar \partial_\mu f(z)\a=\int_{-\infty}^\infty
d\tau \, y^\mu {\partial f\over{\partial x^\mu }}(x-\tau y)\cr
\a=-\int_{-\infty}^\infty d\tau \,{\partial \over{\partial \tau }}
f(x-\tau y)\cr
\a=0,
 \cr}
 \eno29)
 \ea
if $f$ decays for large $x$ (e.g., if $f$ is a test function).  Equivalently,
using

\ba{
2\bar\partial_\mu \,\left[ \theta (yp)\,e^{-izp} \right]
\a=2\bar\partial_\mu \,\left[ \theta (yp)\right]\,e^{-izp}\cr
\a=-i{\partial \theta (yp)\over{\partial y^\mu }}\,\,e^{-izp}\cr
\a=-ip_\mu \,\delta (yp)\,e^{-izp}\cr
\a=-ip_\mu \,\delta (yp)\,e^{-ixp},
 \cr}
 \eno30)
 \ea
we have

\ba{
2i\bar \partial_\mu f(z)\a= \inp p_\mu  \,\delta (yp)\,e^{-ixp}\,\ftil(p),
\cr}
 \eno31)
 \ea
so $4\pi i\bar \partial_\mu f$ is the inverse Fourier transform of  $p_\mu
\ftil(p)$ in the hyperplane 
\be
 N_y\equiv \{p\in\rl^{s+1} \,|\, yp=0\}=\partial M_y.
\eno32)
 \ee
Hence, if the intersection of the support of $\ftil$ with $N_y$
has positive Lebesgue measure in $N_y$, then $f$ will
not be analytic at $x-iy$ in general. However, in any case,

\be
2iy^\mu \bar \partial_\mu f(z)=
\inp yp \,\delta (yp)\,e^{-ixp}\,\ftil(p)=0,
\eno33)
 \ee
in agreement with the above conclusion.  In the one--dimensional case
$s=0$, this reduces to

\be
{\partial f(z)\over{\partial \zbar}}=0\quad \forall y\ne 0,
\eno34)
 \ee
which states that $f(z)$ is analytic in the upper-- and lower-- half--planes. 
The point is that in one dimension, there are only {\sl two\/}  imaginary
directions (up or down), whereas in $s+1$ dimensions, every  $y\ne 0$
defines a direction. This motivates the following.
\skp

\noindent {\bf Definition. } Let $Y(z)$ be a  vector field of type $(0, 1)$ on
$\cx^{s+1}$, i.e. $Y=  Y^\mu (z)\,\bar\partial_\mu$. Then
a function $f$ on $\cx^{s+1}$ is {\sl holomorphic along\/} 
$Y$ if

\be
Yf(z)\equiv Y^\mu (z)\,\bar\partial_\mu f(z)\equiv 0.
\eno35)
 \ee
 \skp 
Thus, our Analytic--Signal transform $f(z)$ is holomorphic  along \break
$Y(x-iy)=y^\mu\,\bar\partial_\mu$.  
 \skp
\noindent {\sl Note:\/}  The ``functorial'' way to look at $f(z)$ is as an
extension of $f(x)$ to the {\sl tangent bundle\/} $T(\rl^{s+1}). $ 
Then the above states that $f(z)$ is holomorphic  {\sl along the fibers\/} of
this bundle. It makes sense that $f(z)$ ought to satisfy some constraints,
since it is determined by a function $f(x)$ depending on half the number of
variables.     If $\rl^{s+1}$ is replaced by a differentiable
manifold, the line $x(\tau )=x-\tau y$ would have to be replaced by a
geodesic.  This gives a generalized   transform which  can be used to extend
functions on an arbitrary {\sl Riemannian or Pseudo--Riemannian\/}
manifold, such as a curved spacetime, to its tangent bundle.  \sp\#
\skp

The multivariate Analytic--Signal transform is related to the {\sl Hil\-bert
transform in the direction\/}  $y$    (Stein [1970], p. 49), defined as

\ba{
 ( H_yf)(x)={ 1\over \pi }{\rm PV}\int_{-\infty}^\infty 
 {du \over u}\,f(x-uy), \quad x,y\in\rl^{s+1},\ y\ne 0.
 \cr}
 \eno36)
 \ea
Namely, an argument similar to the above shows that 

\be
f(x)+i(H_yf)(x)=\lim_{\epsilon \downarrow 0}\,2f (x+i\epsilon y),
\eno37)
 \ee
hence

\be
( H_yf)(x)=-i\lim_{\epsilon \downarrow 0}\,[f(x+i\epsilon y)-
f(x-i\epsilon y)],
\eno38)
 \ee
which, for $s=0$, reduces to the previous relation with the ordinary Hilbert
transform.  As in the one--dimensional case,  $f(x)$ is the 
boundary--value of $f(z)$ in the sense that 

\be
f(x)=\lim_{\epsilon \to 0}\,[f(x+i\epsilon y)+
f(x-i\epsilon y)].
\eno39)
 \ee
For real--valued $f$, eqs. (38) and (39) reduce to

\ba{
f(x)=\lim_{\epsilon \to 0}\,2\Re\, f(x+i\epsilon y)\cr
(H_yf)(x)=\lim_{\epsilon \downarrow 0}\,2\Im\,f(x+i\epsilon y).
 \cr}
 \eno40)
 \ea
\skp
\noindent {\sl Note:\/}  The  Analytic--Signal transform is remarkable in
that it combines in a single entity elements of  the Hilbert, Fourier--Laplace
and X--Ray transforms.  In fact, it is an example of a much more general
transform which, furthermore, includes the Radon transform and an
$n$--dimensional version of the wavelet transform as special cases! (Kaiser
[1990b,c]).

\secskp

 \noindent {\bf  5.3. Axiomatic Field Theory and Particle Phase Spaces }
\def\rightheadline{\tenrm\hfil {\sl 
5.3. Axiomatic Field Theory and Particle Phase Spaces }\hfil\folio}  

 \skp
\noindent We begin with a study of {\sl general \/} fields, i.e. fields not
necessarily satisfying any differential equation or governed by any
particular model of interactions. The  Analytic--Signal transform
 defines (at least {\sl formally\/})  a canonical extension of such fields to
complex spacetime.   In this section we show that for general {\sl
quantized\/}  fields satisfying the Wightman axioms (Streater and Wightman
[1964]), the proposed extension  is natural and interesting.   In particular,
the {\sl double tube domain\/}  ${\cal T}\equiv {\cal T}_+\cup {\cal T}_-$
(where $y\in V_\pm'$ in $\tb_\pm$)
can be interpreted as an extended classical   phase space  for certain 
``particle''-- and ``antiparticle'' coherent states naturally associated with the
quantized fields.  Although the extended fields are in general not analytic
(they need not even be {\sl functions,\/}  only ditributions), they are
``analyticity--friendly''  in the sense that various  objects associated with
them {\sl are\/}  analytic functions.  Consequently,  they are more regular
than the  original fields, which are their boundary values.  In
particular, we will see that under some reasonable assumtions, the
extended fields are operator--valued {\sl functions\/} (rather than
distributions) when restricted to $\tb$.  The extensions of {\sl free\/}  fields  and
generalized free fields are, moreover, weakly holomorphic in $\tb$.

 \skp
 Recall (chapter 4)  that a system with a finite number of degrees of freedom,
say a free classical particle, can be quantized in at least two ways:  by
replacing the position--and momentum variables with operators satisfying
the canonical commutation relations  (this is called {\sl canonical
quantization\/}), or by considering unitary irreducible representations of the
underlying dynamical symmetry group (${\cal G}_2$ for non--relativistic 
particles and ${\cal P}_0$ for relativistic particles).  The second scheme
seems to be  more natural, especially in the relativistic case, where position
variables are not a covariant concept.  But the first scheme has the
advantage that interactions (potentials) can be introduced more easily, since
it generates a kind of functional calculus for operators. 

 Both schemes have counterparts in field
quantization.  In canonical quantization, the particle position is replaced with
the initial configuration of the field $\phi $, i.e. with the values of $\phi (x)$
at all space points ${\bf x}$ at some initial  time $x^0=t$; its momentum,
according to Lagrangian field theory,  then corresponds to the 
time--derivative of the field at time $t$.  Thus the ``phase space''  of the field
is simply the set of all possible initial data on the Cauchy surface $x_0=t$ in
real spacetime.  Quantization is implemented by requiring the initial field
and its conjugate momentum to satisfy an infinite--dimensional
generalization of the canonical commutation relations.  Although this  is the
standard approach to field quantization in the physics literature, its physical
and mathematical soundness is open to question.  Whereas the two schemes
are equivalent when applied to non--relativistic  quantum
mechanics, it is not clear that they remain equivalent when applied to field
quantization, in general.  However, they {\sl are \/} equivalent for the
quantization of free fields.  We will apply canonical
quantization to the free Klein--Gordon  and Dirac fields in sections 5.4 and
5.5.

 The second approach to field quantization  is subsumed in the so--called
axiomatic approach   to quantum field theory.  The unitary representations
of ${\cal P}_0$  under which the fields transform are no longer irreducible.
This is a consequence of the fact that while ${\cal P}_0$ is transitive on the
phase space  of a classical particle (i.e., any two locations, orientations and
states of motion can be transformed into one another), it is no longer
transitive on the phase space  of a classical field (two sets of initial
conditions need not be related by ${\cal P}_0$). 
The reducibility of the representation   corresponds to the presence of an
infinite set of degrees of freedom or, at the particle level, to an
indefinite number of particles. (Even {\sl two\/} particles would result in
reducibility, since the Lorentz norm of  their total
energy--momentum can be arbitrarily large.)  Consequently, the
representation is now characterized by an infinite number of parameters
and  no longer uniquely determines the theory, as it did (for a given mass
and spin) when it was irreducible.   On the other hand,  the label space for
the configuration observables, which for $N$ particles in $\rl^s$ was $\{1, 2,
3, \cdots , sN\}$, is now $\rl^s$, and Lorentz invariance means that the
``labels'' ${\bf x}$ mix with the time variable.  This gives an additional
structure to quantized fields not shared by ordinary quantum mechanics,
and some of this structure is codified in terms of the Wightman axioms,
partly making up for the indeterminacy due to the reducibility of the
representation.

\skp
For simplicity of notation we confine our attention to a  single {\sl scalar\/} 
field.  The results of this section  extend to an arbitrary {\sl system\/} 
of scalar, spinor or tensor fields. Thus, let $\phi (x)$ be an arbitrary  scalar
{\sl quantized\/}  field.  ``Quantized'' means that rather than being a real-- or
complex--valued function on spacetime (like the components of the classical
electromagnetic field), $\phi (x)$ is an {\sl operator\/}  on some Hilbert
space ${\cal H}$.  Actually, as noted by Bohr and Rosenfeld [1950], quantized
fields are too singular to be measured at a single point in spacetime.  
 This led Wightman to postulate that  $\phi $ is an
operator--valued  distribution, i.e., when smeared over a
  test function $f(x)$ it gives an operator $\phi (f)$ (unbounded,
in general) on ${\cal H}$. The axiomatic approach turns out to provide a
surprisingly rich mathematical framework common to all quantum field
theories; therefore, it is model--independent.   It was followed by {\sl
constructive quantum field theory,\/} in which model field theories are
built and shown to satisfy the Wightman axioms.
 For simplicity, we state the axioms in terms of the  formal expression $\phi
(x)$ rather than its smeared form $\phi (f)$.    Also,  we
assume, to begin with, that the field $\phi $ is {\sl neutral,\/} which means
that the operators $\phi (x)$ (or, rather, their smeared forms $\phi (f)$ with
$f$ real--valued) are essentially self--adjoint.  Charged fields, which are
described by non--Hermitian operators, and spinor (Dirac) fields, will be
considered later.  The axioms are:
 \skp
\item{ 1. } {\sl Relativistic Invariance:\/} Poincar\'e 
transformations are im\-ple\-men\-ted by a continuous unitary
representation $U$  of ${\cal P}_0$ acting on the Hilbert space ${\cal H}$ 
of the theory.   (For particles of half--integral spin, such as electrons, ${\cal
P}_0$ must be replaced by its universal cover; see section 5.5.)  Thus, 
\be
 \phi (\Lambda x+a)=U(a,\Lambda )\,\phi (x)\,U(a,\Lambda )^*.
\eno1)
 \ee
Let $P_\mu $ and $M_{\mu \nu }$ be the self--adjoint generators of
spacetime  translations and Lorentz trnsformations, respectively. 
They are interpreted physically as the total  energy--momentum and angular
momentum operators of the field ---or, more generally, of the {\sl system\/} 
of (possibly coupled) fields. They satisfy the commutation relations of the Lie
algebra $\wp$ of ${\cal P}_0$.  In particular, the  $P_\mu $'s  must commute
with one another, thus have a joint spectrum $\Sigma \subset \rl^{s+1}$.
(Actually, $\Sigma $ is a subset of the {\sl dual\/} ${(\rl^{s+1})}^*$ of
$\rl^{s+1} $, which we may identify with $\rl^{s+1}$ using the Minkowski
metric;  see section 1.1.)

 \skp
 \item{2. } {\sl Vacuum:\/}   The Hilbert space
contains a unit vector $\Psi _0$, called the vacuum vector, which is invariant
under the representation  $U$, i.e. $U(a,\Lambda )\Psi _0=\Psi _0$
for all $(a,\Lambda )\in {\cal P}_0$.  $\Psi _0$ is unique up to a
 constant phase factor, and it is {\sl cyclic,\/} meaning that the set of all
vectors of the form
\be
 \phi (f_1)\,\phi (f_2)\,\cdots \phi (f_n)\Psi _0
\eno2)
 \ee
spans ${\cal H}$.  Invariance under ${\cal P}_0$ means that $\Psi _0$ is a
common eigenvector of the generators $P_\mu $ and $M_{\mu \nu }$ with
eigenvalue zero:

\ba{
 P_\mu \Psi _0\a=0\cr
M_{\mu \nu }\Psi _0\a=0.
 \cr}
 \eno3)
 \ea
\skp
\item{3. } {\sl Spectral Condition:\/} The joint spectrum $\Sigma $
of the energy--mo\-men\-tum operators $P_\mu $ is contained in the closed
forward light cone:
\be
 \Sigma \subset \overline{ V}_+.
\eno4)
 \ee
This axiom follows from the physical requirement of {\sl
stability,\/}  which merely states that the energy is bounded below.  To see
this, note first of all that $\Sigma $ must be invariant under ${\cal L}_0$,
by Axiom 1.  Hence, if any physical state had a spectral component with $p
\notin \overline{V }_+$, that component could be made to have an
arbitrarily large negative energy by a Lorentz transformation.   Note that the
existence and uniqueness of the vacuum means that $\Sigma $ contains the 
origin in its point spectrum  with  multiplicity one.
 \skp
\item{4. } {\sl Locality:}  The field operators $\phi (x)$ and $\phi
(x')$ at points with spacelike separation commute, i.e.,
\be
 [\phi (x), \phi (x')]=0 \qquad \hbox{if}\  (x-x')^2<0.
\eno5)
 \ee
This corresponds to the physical requirement that measurements of the
field at  points with spacelike separations must be independent, since no
signal can travel faster than light.  (For fields of half--integral spin, such as
the Dirac field treated in section 5.5, commutators must be  replaced with
anticommutators.)
\item{ 5. }  {\sl Asymptotic Condition:\/}   As the
time $x_0\to\pm\infty$, the field $\phi (x)$ has {\sl weak\/} asymptotic
limits 
\ba{
\phi (x)\a\to\phi _{\rm in}(x)\quad \hbox{(weakly) as}\  x_0\to-\infty\cr
\phi (x)\a\to\phi _{\rm out}(x)\quad \hbox{(weakly) as}\  x_0\to\infty.
 \cr}
 \eno6)
 \ea
 $\phi _{\rm in}$ and $\phi _{\rm out}$ are {\sl free\/}  fields of mass
$m>0$, i.e. they satisfy the Klein--Gordon equation:
\be
 (\del+m^2)\,\phi _{\rm in}=(\del+m^2)\,\phi _{\rm out}=0.
\eno7)
 \ee
Physically, this means that in the far past and future, all particles are
sufficiently far apart to be decoupled, and that $\phi $ {\sl interpolates\/} 
the in-- and out-- fields.  Furthermore,  the Hilbert spaces  on which $\phi
_{\rm in}, \phi _{\rm out}$ and $\phi $ operate all coincide: \be
 {\cal H}_{\rm in}={\cal H}_{\rm out}={\cal H}.
\eno8)
 \ee
 \skp

\noindent  In addition to the above, there are axioms
concerning the regularity of the distributions $\phi (x)$ (they are assumed
to be {\sl tempered\/}, i.e. the test functions belong to ${\cal S}(\rl^{s+1})$)
and the  domains of the smeared operators $\phi (f)$, which we use
implicitly, and  clustering,  which we will not use  here.
 \skp

Let us now draw some conclusions from these axioms. Since the
energy--momentum operators generate spacetime translations, it follows
from eq. (1) that 

\be
 i\partial_\mu \phi (x)=[\phi (x),P_\mu ].
\eno9)
 \ee  
For the Fourier transform $\ftd (p)$, this means that 

\be
 [\ftd(p), P_\mu ]=p_\mu \,\ftd(p).
\eno10)
 \ee
Consequently for any $p\in \rl^{s+1}$, the ``vector''

\be
 \Phi _p\equiv \ftd (-p)\,\Psi _0
\eno11)
 \ee 
(which is in general non--normalizable) satisfies

\be
 P_\mu \Phi _p=[P_\mu, \ftd (-p)]\Psi _0=p_\mu \Phi_p.
\eno12)
 \ee
That is, $\Phi _p$ is a {\sl generalized eigenvector\/} of energy--momentum
with eigenvalue $p$, unless it vanishes.  The spectral condition therefore
requires that 

\be
 \ftd (-p)\Psi _0=0\qquad \forall p\notin \Sigma _1,
\eno13)
 \ee
where $\Sigma _1$ is the intersection of the spectrum $\Sigma $ with the
support of the distribution $\ftd  $.
More generally, if $\Psi  _p$ is any
generalized eigenvector of energy--momentum with eigenvalue $p\in \Sigma
$, then the above commutation relations show that 

\be
P_\mu  \ftd (p')\,\Psi  _p =(p_\mu -p'_\mu )\,\ftd (p')\Psi  _p,
\eno14)
 \ee
thus either $ \ftd (p')\,\Psi  _p$ vanishes or it is a generalized eigenvector of
energy--momentum with eigenvalue $p-p'$, a necessary condition for which
is that $p-p'\in \Sigma $.  Thus we conclude that {\sl the operator $\ftd(p)$,
when it does not vanish,  removes an energy--momentum p from the
field.\/} This  establishes the physical significance of the field operators, as
well as the  connection between the (mathematical)  Fourier variable $p$ and
the (physical)  energy--momentum operators $P_\mu $. From the Jacobi
identity, it follows that 

\be
[\, [\ftd(p), \ftd(p')], P_\mu ]=(p_\mu +p'_\mu )\,[\ftd(p), \ftd(p')],
\eno15)
 \ee
showing that $[\ftd(p), \ftd(p')]$ (if not zero) removes a total
energy--mo\-men\-tum $p+p'$ from the field.  The cyclic property of the
vacuum, furthermore, implies that the entire spectrum $\Sigma $ is
generated by repeated applications of $\ftd$ to the vacuum.  Note that in
general we {\sl cannot\/}  draw any conclusions about the support of
$\ftd(p)$.  For example, $\ftd(p)$ need not vanish for spacelike $p$, since
$\Sigma$ may contain points $p'$ such that $p'-p\in \Sigma $.  In fact, very
strong conclusions  can be drawn from the nature of  the support of $\ftd$. 
From Lorentz invariance and  $\ftd (p)^*=\ftd (-p)$  it follows that $\ftd
(p)=0$ if and only if $\ftd(p')=0$, where $p'=\pm\Lambda p$ for some
$\Lambda \in {\cal L}_0$.   We conclude that the support of $\ftd$ must be
a union of sets  of the form

\ba{
\Omega _m\a=\{p\,|\,p^2=m^2\},\qquad m>0\cr
\Omega _0\a=\{p\,|\,p^2=0, \ p\ne 0\}\cr
\Omega _{00}\a=\{0\}\cr
\Omega _{im}\a=\{p\,|\,p^2=-m^2\},\qquad m>0,\cr
\cr}
\eno16)
\ea
which are, in fact, the various orbits of the {\sl full\/}  Lorentz group ${\cal
L}$.   Greenberg [1962] has shown that $\phi $ is a {\sl generalized free
field\/} (i.e., a sum or integral of free fields of varying masses $m\ge 0$) if
$\ftd$ vanishes on any of the following types of sets:

\ba{
A\a=\Omega _{im},\qquad m>0\cr
B\a=\Omega _{00}\cup \bigcup_{0\le m< M}\Omega _m,\qquad M>0\cr
C\a=\bigcup_{m>M} \Omega _m,\qquad M>0.
 \cr}
 \eno17)
 \ea
He has also shown, by giving counter--examples, that this conclusion cannot
be drawn if $\ftd$ vanishes on sets of the type

\ba{
D\a=\bigcup_{M_1\le m<M_2} \Omega _m,\qquad 0\le M_1<M_2\cr
E\a=\Omega _{00}.
 \cr}
 \eno18)
 \ea
(See also Dell'Antonio [1961] and Robinson [1962].)
\skp
\noindent {\sl Note:\/}  Up to now, {\sl we have not assumed that the field
satisfies the canonical commutation relations\/} (section 5.4),  hence our 
conclusions  are quite general and should hold for an arbitrary (system of
mutually) interacting  field(s).  The ``Lie algebra'' generated by the fields
(obtained by including, along with the fields, their commutators \break
$[\ftd(p), \ftd(p')]$ as well as higher--order commutators) has a very interesting 
formal structure, although it is  not a Lie algebra in the usual sense. (For one
thing, it is uncountably infinite--dimensional rather than
finite--dimensional.)   Namely, the above relations suggest that the operators
$P_\mu $ be regarded as belonging to a  Cartan subalgebra and that
$\ftd(p)$ (with $p$ in the support of $\ftd$)  is a {\sl root vector\/}  with
associated root $-p$.   The spectrum $\Sigma $ is therefore reminiscent of a
set $\Delta ^+$ of positive roots.  In general, the Cartan subalgebra consists of
a maximal set of commuting observables.  When considering charged fields,
the charge will also  belong to the Cartan subalgebra, with root values $0,
\pm 	\ve, \pm 2	\ve, \ldots$, where $	\ve$ is a fundamental unit of
charge.  The vacuum is a vector (not in the Lie algebra but in an associated
representation     space) of ``highest'' (or lowest) weight which, as  in the
finite--dimensional case, generates a representation  of the algebra because
of its cyclic property.  To my knowledge, this important analogy between the
structures of {\sl general\/}  quantized fields  (i.e., apart from the canonical
commutation relations or any particular models of interactions)  and Lie
algebras has not been explored, although the methods of Lie--algebra theory
could add a powerful new tool to the study of quantized fields.  (In a
somewhat different context, the structures of quantized fields and
infinite--dimensional Lie algebras are united in string theory;  see Green,
Schwarz and Witten [1987].)  \sp\#  \skp

Let us now formally extend the quantized field $\phi (x)$ to $\cx^{s+1}$,
using the Analytic--Signal transform developed in section 5.2. Recall that
this transform was originally defined for Schwartz test functions.  In
principle, we would like to define $\phi (z)$ by using its distributional
Fourier transform $\ftd(p)$:

\ba{
\phi (z)\a=\inp \theta ^{-izp}\,\ftd(p)\cr
\a\equiv \ftd\left( \theta ^{-izp} \right).
 \cr}
 \eno19)
 \ea
This presents us with a technical problem,  as already noted in the last
section, since $\theta ^{-izp}$ is not a  Schwartz test function in $p$. One way
out is to smear $\phi \z$ with a test function $f\z$ over $\cx^{s+1}$. 
Although this is the safest solution, it is not very interesting since not much
appears to have been gained by extending the field to complex spacetime:
the new field is still an operator--valued distribution.  However, we shall
see that there are reasons to expect  $\phi \z$ to be more regular than a
``generic'' Analytic--Signal transform, due in part to the fact that $\phi \x$
satisfies the Wightman axioms.  When $\phi $ is a  (generalized) {\sl
free\/} field, the restriction of $\phi \z$  to the double tube $\tb$ turns out
to be   a  {\sl holomorphic \/}  operator--valued function. We will see that
even  for  general Wightman fields, $\tb$ is an important
subset of $\cx^{s+1}$. In the presence of
interactions, holomorphy is lost but some regularity  in $\tb$  is expected to
remain.  We now proceed to find conditions  which  do {\sl not\/}
force $\phi$ to be  a generalized free field but still allow $\phi \z$ to be an
operator--valued {\sl function\/}  on $\tb$.  The arguments given below
have no pretense to rigor; they are only meant to serve as a possible
framework for a more precise analysis in the future.  All  statements and
conditions  concerning convergence, integrability and decay of
operator--valued expressions are meant to hold in the {\sl weak\/}  sense,
i.e. for matrix elements between  fixed vectors.  Since the operators involved
are  unbounded, we must furthermore assume that the vectors
used to form the matrix elements are in their (form) domains.

 \skp
For a  fixed timelike ``temper'' vector  $y$,   $\theta ^{-izp}$ fails to be a
\break
Schwartz test function in two distinct ways:  (a) It has a discontinuity
on the spacelike hyperplane $N_y=\{p\,|\,yp=0\}$, and (b) it has a constant
modulus on hyperplanes parallel to $N_y$, hence cannot decay there. On
the other hand, by relativistic covariance, the support of $\ftd$
must be smeared over the orbits of $\cal L$, given by eq. (16).  This gives a 
``stratification'' of $\ftd$ as a sum of tempered distributions

\be
 \ftd=\ftd_+ +\ftd_0 +\ftd_{00}+\ftd_-
\eno20)
 \ee
with support properties

\ba{
{\rm supp}\,\,\ftd_+\a\subseteq\overline{\bigcup_{m>0}\Omega_m}
                 \equiv\Omega_+\cr
 {\rm supp}\, \,\ftd_0\a \subseteq \Omega _0  \cr
{\rm supp}\, \,\ftd_{00}\a \subseteq\Omega _{00}  \cr
{\rm supp}\,\,\ftd_-\a\subseteq\overline{\bigcup_{m>0}\Omega_{im}}
                  \equiv\Omega_-.
 \cr}
 \eno21)
 \ea

\skp

Although $\Omega _+$ and $\Omega _-$ contain $\Omega _0$, the 
distributions $\ftd_+$ and $\ftd_-$ have no contributions from $p^2=0$. (For
example, the distribution $1+\delta (p)$ on $\rl$ has a  decomposition
$T_1+T_0$ where $T_1$ has support $\rl$ and $T_0$ has
support $\{0\}\subset\rl$, but the $p=0$ contribution to $T_1$ vanishes.) 
Similarly, although $\Omega _0$ contains $\Omega _{00}$, $\ftd_0$
has no contribution from $p=0$.  Corresponding to the above
decomposition of $\ftd$, we have, formally, 

\be
 \phi \z=\phi _+\z+\phi _0\z+\phi _{00}\z+\phi _-\z.
\eno22)
 \ee
We will  show that 
\skp
\noindent 1.  $\phi _+\z$ and $\phi _0\z$ are
holomorphic  operator--valued functions in $\tb$;    

\noindent 2. $\phi _{00}\z$ must be a constant field, to be
physically reasonable; and

\noindent 3. under certain (hopefully not
too restrictive) conditions, $\phi _-\z$, though
not holomorphic, is an operator--valued function on $\tb$.
\skp
 First of all, we claim that each of the fields $\phi _\alpha$,  
$(\alpha =\pm, 0, 00)$ is still covariant under ${\cal
P}_0$.\footnote*{However, $\phi _\alpha $ need not satisfy  other
Wightman axioms such as locality;  for example, $\phi _+$ need not be a
generalized free field.}  To see this, take the Fourier transform of eq. (1) and
use the invariance of Lebesgue measure $dp$ under ${\cal L}_0$.  This
leads to 

\be
 \ftd\p=e^{iap}\,U(a,\Lambda )\,\ftd(\Lambda' p)\,U(a, \Lambda )^*,
\eno23)
 \ee
where $\Lambda ' $ is the transpose of $\Lambda \in {\cal L}_0$.  Since
the different components are ``essentially'' supported on disjoint subsets
 and these subsets are invariant under ${\cal L}_0$,
we conclude that eq. (23) holds for each $\ftd_\alpha $.   To show that
$\phi _+\z$ and $\phi _0\z$ are holomorphic in $\tp$, let $z\in\tp$. Since
$\theta ^{-izp}$ vanishes for $p\in \overline{V }_-\backslash \{0\}$, and
since $\ftd_+$ and $\ftd_0$ have no  contribution from $p=0$, we have 

\be
 \phi _\alpha \z=\int_{\overline{ V}_+}dp\,e^{-izp}\,\ftd_\alpha \p
\quad \alpha =+, 0.
\eno24)
 \ee
Now $ e^{-izp}$ may be regarded as the restriction to $\overline{V}_+$ of
 a Schwartz test function $f_z\p$ which is of compact support in ${\bf p} $
for each fixed $p_0$ and vanishes when $p_0<-E$, for some $E>0$.  Thus
$\phi _+\z$ and $\phi _0\z$ make sense as operator--valued functions on
$\tp$, and they are cleary holomorphic there.  (This will be shown
explicitly below.)  A similar analysis shows that the same can be done for 
$z\in\tm$.

Next, we consider $\phi _{00}\z$.  Since $\ftd_{00}$ is supported on
$\{p=0\}$, it must have the form $P(\partial)\,\delta \p$, where
$P(\partial)$ is a partial differential operator.    In the $x$--domain (i.e., in
real spacetime), this corresponds to a polynomial $P(-ix)$, for which the
Analytic--Signal transform is not well--defined,  although it is possible that
a regularization procedure would cure this. But in any case, non--constant
polynomials in  $x$ do not appear to be of physical interest since they
correspond to unbounded fields even in the classical sense (as functions
of $x$).  Hence we assume that  

\skp

{\sl 
\item{(a)} $\ftd_{00}\p=2A\,\delta \p$, where $A$ is a constant
operator. \/} 
\skp

\noindent This corresponds to a constant field $\phi \x\equiv 2A$ and,
correspondingly,  $\phi _{00}\z\equiv A$ (see eq. (15) in section 5.2).  In
order that $\phi \z$ be an operator--valued function on $\tb$, it therefore
remains  only for $\phi _-\z$ to be one.  Note that so far, the  only
assumption we needed to make, in addition to the Wightman axioms, was
{\sl (a).\/}   To make $\phi _-\z$ a function, we now make our second
assumption:
\skp

{\sl 

\item{(b)} $\ftd_-(p)$ is  integrable on all spacelike hyperplanes. 
Furthermore, the integral of $\ftd$ over the hyperplane $H_{y,\nu}\equiv
\{p\,|\,yp=\nu \}$  $(y\in V')$ grows at most polynomially in $\nu $. 

} 
\skp

\noindent It is not clear what specific {\sl minimal\/}  conditions on
$\ftd_-$ produce this property. The integral  occurring in {\sl (b)\/\/}  
 is known as the {\sl Radon transform\/}  $\,(R\ftd)(y,\nu )$ of
$\ftd$ when $y$ is a (Euclidean) unit vector, and will be further discussed
in  section 6.2.  (See also Helgason [1984].) Unlike the Fourier
transform, the Radon transform does not readily generalize to tempered
distributions (which were, after all, designed specifically for the Fourier
transform).  However, it does extend to distributions of compact support
and can be further generalized to distributions with only mild decay.  Also,
the relation of assumptions {\sl (a), (b)\/\/} (or their future replacements,
if any) to the Wightman axioms needs to be investigated.

 In order to compute $\phi _-\z$ for $z\in\tb$ it
suffices, by covariance, to do so for $x=0$ and $y=(u, {\bf 0} )$, for all $u\ne
0$.  The analyses for $u>0$ and $u<0$ are similar, so we restrict ourselves to
$u>0$.  Eq. (19) then gives

\ba{
\phi_-(-iu, {\bf 0})=(2\pi )^{-s-1}\int_0^\infty dp_0\,e^{-up_0}
\inrs d^s {\bf p}  \,\ftd_-(p_0, {\bf p}).
 \cr}
 \eno25)
 \ea
 For fixed $p_0\ge 0$, condition  {\sl (b)\/}  implies that the integral
over ${\bf p}$ converges, giving an operator--valued function $F(p_0)$
which is of at most  polynomial growth in $p_0$.   $\phi_-(-iu, {\bf 0})$ is
then  the Laplace transform of $F(p_0)$, which is indeed
well--defined.
\skp

\noindent {\sl Note:\/}  The behaviors of $\phi \z$ and $\ftd(p)$ exhibit a
certain duality which reflects the dual nature of $y\in\rl^{s+1}$ and 
$p\in (\rl^{s+1})^*$  (section 1.1). We have just seen that when $\ftd(p)$
behaves reasonably for spacelike $p$, then $\phi \z$  behaves
reasonably for timelike $y$.  In the trivial case when $\ftd(p)\equiv 0$ for
$p^2<  0$ and {\sl (a)\/\/} holds, $\phi \z$ is holomorphic for $y^2>0$.  In
fact, $\phi $ is then a generalized free field, hence may  be said to be
``trivial.''  This dual behavior also extends to $p^2\ge 0$ and $y^2<0$: For {\sl
any\/} non--constant field,  $\ftd(p)$ is non--trivial for $p^2\ge 0$;  for
spacelike $y$, the hyperplane $H_{y,\nu}$ (which contains timelike as well
as spacelike directions)  therefore intersects the support of
$\ftd$ in a non--compact set and we do not expect $\phi\z$ to make sense
as an operator--valued function outside of $\tb$.  \#

\skp

 No  claims of analyticity can be made for $\phi (z)$ in general. In fact,
Greenberg's results show that $\phi (z)$ may not  be analytic {\sl
anywhere\/} in $\cx^{s+1}$ unless $\phi $ is a generalized free field.  For as
in the classical case, formal differentiation  with respect to $\zbar^\mu $
gives 

\ba{
2i\bar\partial_\mu \,\phi (z)=\inp p_\mu \,\delta (yp)\,e^{-ixp}\,\ftd(p),
\cr}
 \eno26)
 \ea
hence $4\pi i\bar\partial_\mu \,\phi $  is the inverse Fourier transform of 
$p_\mu \ftd$ in the hyperplane $N_y$. If $\phi $ is not a
generalized free field, then, according to Greenberg,  the support of $\ftd$
contains sets of timelike as well as sets of spacelike $p$'s with positive
Lebesgue measure.  Hence, for any nonzero $y\in \rl^{s+1}$, the
intersection of the support of $\ftd$ with
$N_y$ has positive measure in $N_y$, so $\phi $ will not be
holomorphic at $x-iy$ in general. As in the classical case, however,  the
above equation for $\bar \partial_\mu \phi $ implies that $\phi (z)$ is 
holomorphic along the vector field $y$, i.e.,

\be
 y^\mu \bar\partial_\mu \phi =0.
\eno27)
 \ee
This is a covariant condition which, when specialized to $y=(y_0, {\bf 0} )$,
states that $\phi (z)$ is holomorphic in the complex time--direction.   As we
have seen, this result simply follows from the nature of the Analytic--Signal
transform.  A similar situation forms the basis of {\sl Euclidean quantum field
theory.\/}  However, there one is dealing not directly with the field but with
its vacuum expectation values, and the mathematical reason for the
analyticity is the spectral condition, which would appear to have little in
common with  the Analytic--Signal transform.  

\skp
Incidentally, eq. (26)  provides a simple formal proof that $\phi _+\z$ is
holomorphic in $\tb$.  The support of $\ftd_+\p$  is contained in
$\overline{ V}$, hence for any $y$ in $V'$, its intersection with $N_y$ is
either empty or equal to $\Omega _{00}$. But the contribution from $p=0$
vanishes, hence eq. (26) shows that $\bar\partial_\mu  \phi _+\z=0$ in
$\tb$.   The same argument also shows that free fields and generalized free
fields are holomorphic in $\tb$.
In the next two sections   we shall study  free  Klein--Gordon  and
Dirac fields in more detail.  
 \skp

Although $\phi (z)$ is not holomorphic in general, we will be able to
establish for it one essential ingredient of the foregoing phase--space
formalism, namely the interpretation of the double tube $\tb$  as an
extended classical phase space for certain  ``particles'' and ``antiparticles'' 
associated with the quantized field $\phi $.

 First, let us expand the above
considerations to include {\sl charged\/}  fields  by allowing $\phi (x)$ to be
non--Hermitian (i.e., a non--Hermitian operator--valued distribution).   
Then the extended field $\phi (z)$ need not  satisfy the reflection
condition $\phi (z)^*=\phi (\zbar )$.  The {\sl charge\/} $Q$ is defined as a
self--adjoint  operator which generates overall {\sl phase translations\/}  of
the field, i.e.,

\be
 e^{-i	\alpha Q}\,\phi (x)\,e^{i\alpha Q}=e^{i
\alpha \ve }\,\phi (x) 
\eno28)
 \ee
for real $\alpha $, where $	\ve$ is a fundamental unit of charge. 
This implies

\ba{
[\phi (x), Q]\a=\ve\, \phi (x)\cr
[\ftd(p), Q]\a=	\ve\,\ftd(p),
 \cr}
 \eno29)
 \ea
showing that $\phi (x)$ and $\ftd(p)$ {\sl remove \/} a unit  $	\ve$
of charge from the field, while their adjoints {\sl add\/}  a unit of charge.  
 We assume that phase translations commute with Poincar\'e
transformations, and in particular with spacetime translations.   Thus

\be
 [Q,P_\mu ]=0,
\eno30)
 \ee
so charge is {\sl conserved.\/} $Q$ can be included in the ``Cartan
subalgebra'' containing the $P_\mu $'s, and the above commutation relations
show that $\ftd(p)$ and $\ftd(p)^*$ are still ``root vectors,'' with $Q$--root
values $-\ve $ and $\ve $, respectively.  We also assume that the vacuum is
neutral, i.e. $Q\Psi _0=0$.   Repeated applications of $\ftd$ and $\ftd ^*$ to
$\Psi _0$  show that the spectrum of $Q$ is $\{0, \pm \ve , \pm 2\ve ,
\ldots\}$.

\skp

 Recall that the commutation relation between $\ftd(p)$ and $P_\mu $
implied that $\ftd(p)$ removes an energy--mometum $p$ from the field. 
Similarly, its adjoint relation

\be
 [P_\mu , \ftd(p)^*]=p_\mu \,\ftd(p)^*
\eno31)
 \ee
shows that $\ftd(p)^*$ {\sl adds\/} an energy--momentum $p$ to the
field.  In place of the
generalized eigenvectors $\Phi _p$ of energy--momentum which we had for
the Hermitian field, we can now define {\sl two\/} eigenvectors,

\ba{
\Phi _p^+\a\equiv \ftd(p)^*\,\Psi _0\cr
\Phi _p^-\a\equiv \ftd(-p)\,\Psi _0,
 \cr}
 \eno32)
 \ea
for each $p\in \Sigma _1$.  For a non--Hermitian field, these vectors are
independent. They  are  states of  charge $ \ve $
and $-\ve $, respectively. We may think of them as particles and
antiparticles, although they do not have a well--defined mass since $p^2$
will be variable on $\Sigma _1$, unless $\phi $ is a free field.  Each $p\ne 0$
in $\Sigma _1$ belongs to the continuous spectrum of the $P_\mu $'s, since
it can be changed continuously by Lorentz transformations.  Hence the
``vectors'' $\Phi _p^\pm$ are non--normalizable.  Since the $P_\mu $'s are
self--adjoint, and since $\Phi _p^+$ and $\Phi _p^-$  belong to
different eigenvalues of the charge operator (which is also self--adjoint), we
have (with the usual abuse of Dirac notation, where ``inner products'' of
distributions are taken)

\ba{
\langle\, \Phi _p^+\,|\,\Phi _q^-\,\rangle\a=0\cr
\langle\, \Phi _p^\pm\,|\,\Phi _q^\pm\,\rangle\a=\sigma (p^2)\,(2\pi)^{s+1}
\delta (p-q),
 \cr}
 \eno33)
 \ea
where $\sigma $, a distribution with support in $\Sigma _1$, depends
only on $p^2$ by Lorentz  invariance.  (Charge symmetry requires that
$\sigma $ be the same for particles as for antiparticles.)  If $\phi $ is the free
field of mass $m>0$, $\sigma (p^2)=\theta (p_0)\,2\pi \delta (p^2-m^2)=
2\pi  \delta (p^2-m^2)$ in $\vp\backslash \{0\}$. 

Now define the {\sl  particle coherent states\/}   by

\ba{
e_z^+\a\equiv \phi (z)^*\Psi _0\cr
\a=\inp \theta ^{i\zbar p}\,\ftd(p)^*\,\Psi _0\cr
\a=\inpv \theta^{i\zbar p}\,\Phi _p^+.
 \cr}
 \eno34)
 \ea
Like the $\Phi _p^+$'s, these do not have a well--defined mass; in addition,
they are {\sl wave packets,\/}  i.e.  have a smeared energy--momentum,
but they  still have a definite charge $\ve$ .  Their spectral components are
given by 

\be
 \langle\, \Phi _p^+\,|\,e_z^+\,\rangle=\sigma (p^2)\,\theta ^{i\zbar
p}=\sigma (p^2)\theta(yp)\,e^{i\zbar p} . 
\eno35)
 \ee
If $z$ belongs to the backward tube $\tm$, then $yp<0$ on
$\vp\backslash \{0\}$,   hence $e_z^+=0$.  If $z$ belongs to the forward tube
$\tp$, then $yp>0$ and the vector $e_z^+$ is weakly holomorphic in $\zbar$. 
For  the free field, it reduces  to the coherent
state $e_z$ defined in chapter 4.

Similarly,  define the   {\sl coherent antiparticle states\/} by

\ba{
e_z^-\a\equiv \phi (z)\Psi _0\cr
\a=\inp \theta^{-izp}\ftd(p)\,\Psi _0\cr
\a=\inpv \theta^{izp}\,\Phi _p^-.
 \cr}
 \eno36)
 \ea
These are wave packets of charge $-\ve $ for which

\be
 \langle\, \Phi _p^-\,|\,e_z^-\,\rangle=\sigma (p^2)\,\theta^{iz p}=
\sigma (p^2)\,\theta (-yp)\,e^{iz p}.
\eno37)
 \ee
Thus $e_z^-$ vanishes in the forward tube and is weakly holomorphic in the
backward tube.
\skp
In the usual formulation of quantum field theory, particles are associated
not directly with the interacting, or interpolating, field $\phi $ but with its
asymptotic fields $\phi _{\rm in}$ and $\phi _{\rm out}$, which are free. 
(We will construct such free--particle coherent states in the next two
sections.)  However, the coherent states $e_z^\pm$ {\sl are \/} directly
associated with the interpolating field. We shall refer to them as {\sl
interpolating particle coherent states\/}  (section 5.6).

\skp
	 We are now ready to establish the phase--space interpretation of $\tb$ in
the general case.   We will show that $\tp$ and $\tm$ are extended
phase spaces associated with the  particle-- and antiparticle coherent states
$e_z^+$ and $e_z^-$, respectively, in the sense that they parametrize the
classical states of these particles.  

We first discuss $x$ as a ``position'' coordinate.  In the case of interacting
fields there is no hope of finding even a ``bad'' version of position operators. 
Recall that position operators were in trouble even in the case of a
one--particle theory {\sl without\/}  interactions!  In the general case of
interacting fields, this problem becomes even more serious, since one is
dealing with an indefinite number of particles which may be dynamically
created and destroyed.   (As argued in section 4.2, the generators $M_{0k}$ of
Lorentz boosts qualify as a natural, albeit non--commutative, set of 
center--of--mass operators;  although I believe this idea has merit, it will
not be discussed here.)  Since no position operators are expected to exist,  we
must not think of $x$ as  eigenvalues or even expectation values of anything, 
but rather simply as spacetime  parameters or labels.

  On the other hand, $y$ will now be shown to be related to the expectations
of the energy--momentum operators (which do survive the transition to
quantum field theory, as we have seen).   For $z, z'\in\tp$, we
have

\ba{
\langle\, e_{z'}^+\,|\,e_z^+\,\rangle
\a=\langle\, \Psi _0\,|\,\phi (z')\,\phi (z)^*\Psi _0\,\rangle\cr
\a=\inpv e^{-i(z'-\zbar)p}\,\sigma (p^2)\cr
\a={ 1\over 2\pi } \int_0^\infty dm^2\,\sigma  (m^2)\inop e^{-i(z'-\zbar)p}\cr
\a={ 1\over 2\pi i}\int_0^\infty dm^2\,\sigma  (m^2)\,\Delta ^+(z'-\zbar; m),
 \cr}
 \eno38)
 \ea
where we have set $m^2\equiv p^2$ and used

\be
 (2\pi)^{-s-1} \,dp=(2\pi)^{-s-1}\,dp_0\,d^s {\bf p} =(2\pi)^{-1} \,dm^2\,
d\tilde p,
\eno39)
 \ee
with 

\be
 d\tilde p\equiv \left[ 2(2\pi )^s\, \sqrt{m^2+{\bf p} ^2}\,
 \right] ^{-1} \,d^s {\bf p} 
\eno40)
 \ee
the Lorentz--invariant measure on $\omp$.  $\Delta ^+(w; m)$ is the
two--point function for the {\sl free\/} Klein--Gordon field of mass $m$,
{\sl analytically continued\/}  to $w\equiv z'-\zbar\in\tp$.  In the limit $y,
y'\to 0$, this gives the {\sl K\"all\'en--Lehmann representation  \/} 
(Itzykson and Zuber [1980]) for the usual two--point function,

\be
 \langle\, \Psi _0\,|\,\phi (x')\,\phi (x)^*\Psi _0\,\rangle=
{ 1\over 2\pi i}\int_0^\infty dm^2\,\sigma  (m^2)\,\Delta ^+(x'-x; m),
\eno41)
 \ee
which is a distribution.  In Wightman field theory, such vacuum expectation
values are analytically continued using the spectral condition, and conclusions
are drawn from these analytic functions about the field in {\sl real\/} 
spacetime.  In our case, we have {\sl first\/}  extended the field (albeit
non--analytically), {\sl then\/} taken its vacuum expectation
values (which, due to the spectral condition,  are seen to be analytic
functions, not mere distributions).  The fact that we arrived at the same
result (i.e., that ``the diagram commutes'') indicates that our approach is not
unrelated to Wightman's.  However, there is a fundamental difference:  The
thesis underlying our work is that the ``real'' physics actually takes place in
{\sl complex\/}  spacetime, and that there is no need to work with the
singular limits $y\to 0$.
 \skp
 The norm of $e_z^+$ is given by

\ba{
\|e_z^+\|^2\a=\inpv \sigma (p^2)\,e^{-2yp}\cr
\a={1 \over 2\pi } \int_0^\infty dm^2\,\sigma (m^2)\,G(y; m),
 \cr}
 \eno42)
 \ea
where $G(y; m)$, computed in section 4.4, is given by

\be
 G(y; m)=(2\pi)^{-1} \left( {m \over 4\pi \lambda } \right)^\nu \,K_\nu 
(2\lambda m).
\eno43)
 \ee
Recall that $\lambda \equiv \sqrt{y^2}$, $\nu \equiv (s-1)/2$ and $K_\nu $
is a modified Bessel function.  We assume that $e_z^+$ is normalizable, 
which means that  the spectral density
function $\sigma (m^2)$ satisfies the {\sl regularity condition\/}

\ba{
 \|e_z^+\|^2\a=(2\pi)^{-2}  \int_0^\infty
dm^2\,\sigma (m^2)\,\left( {m \over 4\pi \lambda} \right)^\nu 
\,K_\nu (2\lambda m)\cr
\a\equiv F(\lambda )<\infty.
 \cr}
 \eno44)
 \ea
(This condition is automatically satisfied for Wightman fields, where it
follows from the assumption that $\phi $ is a tempered distribution; 
however, it is also satisfied by more singular fields since $K_\nu $ decays
exponentially.) It follows that 

\ba{
\langle\, e_z^+ \,|\,P_\mu \,e_z^+ \,\rangle\a=\inpv \sigma (p^2)\,p_\mu
\,e^{-2yp} \cr
\a=-{1\over 2}\,{\partial F(\lambda )\over{\partial y^\mu }}\cr
\a=-y_\mu \, F'(\lambda )/ 2\lambda.
 \cr}
 \eno45)
 \ea
Using the recursion relation (Abramowitz and Stegun [1964])

\be
- {\partial \over{\partial \lambda  }}\left( \lambda  ^{-\nu }\,K_\nu
(2\lambda m ) \right)= 2m\lambda  ^{-\nu }K_{\nu +1}(2\lambda m ),
\eno46)
 \ee
we find that the state $e_z^+ $ has an expected energy--momentum 

\be
 \langle\, P_\mu \,\rangle=\left( { m_\lambda\over \lambda }
\right)\,y_\mu , \qquad y\in V_+',
\eno47)
 \ee
where 

\ba{
m_\lambda 
\equiv -{F'(\lambda ) \over 2 F(\lambda )}
={{\int_0^\infty dm^2\,\sigma (m^2)\,m^{\nu +1}\,K_{\nu +1}
(2\lambda m) }\over 
{\int_0^\infty dm^2\,\sigma (m^2)\,m^\nu \,K_\nu 
(2\lambda m) }}.
\cr}
 \eno48)
 \ea
We call $m_\lambda $ the  {\sl effective mass \/} of the  particle coherent
states;  it  generalizes  the corresponding quantity for  Klein--Gordon particles
(section 4.4). The name derives from the relation

\be
 \langle\, P\,\rangle^2\equiv \langle\, P_\mu \,\rangle\langle\, P^\mu
\,\rangle=m_\lambda^2.
 \eno49)
 \ee

It is important to keep in mind that in quantum field theory, the natural
picture is the {\sl Heisenberg picture,\/}  where operators evolve in
spacetime and states are fixed.  
Recall that for a free Klein--Gordon particle (chapter 4), we interpreted
$e_z$ as a wave packet {\sl focused\/}  about the event $x\equiv \Re z$ and
moving with an expected energy--momentum $(m_\lambda /\lambda )\,y$. 
This suggests that the above states $e^\pm_z$ be given a similar
interpretation. Thus $z$ becomes simply a set of {\sl labels\/} parametrizing
the classical states of the particles.

  \skp
    This establishes the interpretation of $\tp$ as an extended
classical phase space  associated with the ``particle'' states $e_z^+ $.  A similar
computation shows that $\tm$ acts as an extended classical phase space  for
the ``antiparticle'' states $e_z^- $, whose expected energy--momentum is

\be
 \langle\, P_\mu \,\rangle=\left( { m_\lambda \over \lambda }
 \right)\,(-y_\mu),\qquad y\in V'_-. 
\eno50)
 \ee
\skp
The expected angular momentum in the states $e_z^\pm$ can  be
computed similarly.  The angular momentum operator $M_{\mu \nu }$ is the
generator of rotations in the $\mu$--$\nu $ plane, hence

\be
 i\left(x_\mu {\partial \over{\partial x_\nu }} -
x_\nu {\partial \over{\partial x_\mu }} \right)\,\phi (x)=
[\phi (x), M_{\mu \nu }].
\eno51)
 \ee
This implies for the Fourier transform 

\be
 [\ftd(p), M_{\mu \nu }]=i\left(p_\mu {\partial \over{\partial p^\nu }}-
p_\nu  {\partial \over{\partial p^\mu  }}\right)\ftd(p).
\eno52)
 \ee
Since the vacuum is Lorentz--invariant, we have $M_{\mu \nu }\Psi _0=0$,
hence for $z\in\tp$,

\ba{
M\a_{\mu \nu }\,e_z^+=\inpv e^{i\zbar p}\,[M_{\mu \nu }, \ftd(p)^*]\,\Psi
_0\cr
\a=i\inpv e ^{i\zbar p}\,\left(  p_\nu  {\partial \over{\partial p^\mu  }}-
p_\mu   {\partial \over{\partial p^\nu   }}\right)\,\ftd(p)^*\Psi _0\cr
\a=\inpv \left(\zbar_\mu p_\nu  - \zbar_\nu p_\mu  \right)\,e ^{i\zbar
p}\,\Phi _p^+,
 \cr}
 \eno53)
 \ea
provided that $\tilde \phi (p)$ vanishes on the boundary of 
$\overline{V}_+$ (this excludes massless fields). The expectation of $M_{\mu
\nu }$ is therefore related to that of $P_\mu $ by

\ba{
\langle\, M_{\mu \nu }\,\rangle\a=\zbar_\mu \langle\, P_\nu \,\rangle-
\zbar_\nu  \langle\, P_\mu  \,\rangle\cr
\a=x_\mu \langle\, P_\nu \,\rangle-x_\nu  \langle\, P_\mu  \,\rangle\cr
\a=\left( {m_\lambda  \over \lambda } \right)(x_\mu y_\nu -x_\nu y_\mu ).
\cr}
 \eno54)
 \ea
Similarly, in $e_z^-$ with $z\in\tm$, 

\be
 \langle\, M_{\mu \nu }\,\rangle=-\left( {m_\lambda  \over \lambda }
\right)(x_\mu y_\nu -x_\nu y_\mu ).
\eno55)
 \ee

\skp
This section can be summarized by saying that the  vector $y$ 
plays a similar role for general quantized fields as it did for
positive--energy solutions of the Klein--Gordon equation, namely it acts as
a {\sl control vector\/} for the energy--momentum.   In other words, the
function $\theta (yp)\,e^{-yp}$ {\sl acts as a window in momentum space,
filtering out from each mass shell $\Omega _m$ momenta which are not
approximately parallel to\/}  $y$. The step function $\theta (yp)$ makes
certain that only parallel components of $p$ pass through this filter by
eliminating the antiparallel ones (which would make the integrals
diverge).  Thus we may think of $\theta (yp)\,e^{-yp} $ as a kind of ``ray
filter'' in $\overline{V }$, when $y\in V'$.  We  continue to refer to $y$ as
a {\sl temper vector\/} (section 4.4).
 \skp
\noindent {\sl Note:\/}  The regularity condition given by eq. (44) for
$\sigma (m^2)$, i.e.  the requirement that $e_z^+ $ be normalizable, shows
that $\lambda $ acts as an effective {\sl ultraviolet cutoff,\/} since $K_\nu
(2\lambda m)$ decays exponentially as $m\to\infty$, giving finite   values
to $m_\lambda $ and other quantities associated with the field.  \sp\#

\secskp

 \noindent {\bf 5.4.  Free Klein--Gordon Fields  }
\def\rightheadline{\tenrm\hfil {\sl 
5.4.  Free Klein--Gordon Fields}\hfil\folio}

 \skp
\noindent In the context of general quantum field theory, we were able to
show that $\tb$ plays the role of an extended phase space for certain
``particle'' states of the fields.  The question arises whether the
phase--space  formalism of chapter 4 can be generalized to quantized
fields.  There,  we saw that all free--particle
states in the Hilbert space could be reconstructed from the values of their
wave function on any phase space  $\sigma \subset \tp$.  There are two
possible ways in which this result might extend to quantized fields: (a) The
vectors $e_z^\pm$ belong the subspaces ${\cal H}_{\pm1}$ with charge
$\pm \ve$, hence we may try to get continuous resolutions, not of the
identity on ${\cal H}$  but of the orthogonal  projection operators  $\Pi
_{\pm 1}$  to ${\cal H}_{\pm1}$, in terms of these vectors. (This can then
be generalized to the resolution of the  projection operator $\Pi _n$ to the
subspace ${\cal H}_n$ with charge $n\ve,\, n\in\zz$.)   (b) The global
observables of the theory, such as the energy--momentum, the angular
momentum and the charge operators, are usually expressed as conserved
integrals of the field operators and their derivatives over an arbitrary
configuration space $S$ in spacetime (i.e., an $s$--dimensional spacelike
submanifold of $\rl^{s+1}$);  our approach would be to express them as
integrals of the extended fields over $2s$--dimensional phase spaces
$\sigma $ in $\tb$, much as the inner products of positive--energy solutions
were expressed as such integrals.  In this section we do both of these things
for the {\sl free Klein--Gordon field of mass\/}  $m>0$,  which is a quantized 
solution of 

\be
 (\del+m^2)\,\phi (x)=0.
\eno1)
 \ee
We consider classical solutions at first. The  Fourier transform $\ftd(p)$  has
the form

\be
\ftd (p)=2\pi \delta (p^2-m^2)\,a(p)
\eno2)
 \ee
for some complex--valued function $a(p)$  defined on the two--sheeted
mass hyperboloid $\om=\omp\cup\omm$.  Write

\be
 b(p)\equiv\overline{ a(-p) },\qquad p\in\omp. 
 \eno3)
 \ee
 If the field  is neutral, then $\phi (x)$ is real--valued and $b(p)\equiv a(p)$. 
For charged fields, $a(p)$ and $b(p)$ are independent.  At this point, we
keep both options open.  Then

\ba{
\phi (x)\a=(2\pi)^{-s} \int_{\rl^{s+1}} dp\,\,\delta(p^2-m^2)\,e^{-ixp}\,a(p)\cr
\a=\ino e^{-ixp}\,a(p)\cr
\a=\inop \left[e^{-ixp}\,a(p)+ e^{ixp}\,\overline{ b(p) }\,\right].
 \cr}
 \eno4)
 \ea
The extension of $\phi (x)$ to complex spacetime  given by the
Analytic--Signal transform is

\ba{
\phi (x-iy)\a=\inds \phi (x-\tau y)\cr
\a=\inp \theta ^{-izp}\,\ftd(p)\cr
\a=\ino \theta ^{-izp}\,a(p)\cr
\a=\inop \left[ \theta (yp)\,e^{-izp}\,a(p)+\theta (-yp)\,e^{izp}\,\,
\overline{ b(p)}\,\right].
 \cr}
 \eno5)
 \ea
If $y\in V_+'$, then $yp>0$ for all $p\in\omp$,
hence 

\be
 \phi (z)=\inop e^{-izp}\,a(p)
\eno6)
 \ee
is  analytic in $\tp$, containing only the
positive--frequency part of the field.  Similarly, when $y\in V_-'$, then
$yp<0$ for all $p\in\omp$ and

\be
 \phi (z)=\inop e^{izp}\,\overline{ b(p)},\quad z\in\tm.
\eno7)
 \ee
Thus $\phi (z)$ is also analytic in ${\cal T}_-$, where it contains only the 
negative--frequency part of the field.  However, note that the two domains
of analyticity $\tp$ and $\tm$ do not intersect, hence the corresponding 
restrictions of $\phi (z)$   need not be  analytic continuations of
one another.  
 \skp

We are now ready to quantize $\phi (z)$.  This will be done by first 
quantizing $\phi (x)$ and then using the Analytic--Signal transform 
 to extend it to  complex spacetime.   We assume, to
begin with, that $\phi$ is a {\sl neutral\/}  field, so $b(p)\equiv a(p)$. 
According to the standard rules (Itzykson and Zuber [1980]) of field
quantization, $\phi (x)$ becomes an operator on a Hilbert space ${\cal H}$ 
such that at any fixed time $x_0$,  the field ``configuration'' operators $\phi
(x_0, {\bf x})$ and their conjugate ``momenta'' $\partial_0 \phi (x_0, {\bf
x})$ obey the {\sl equal--time  commutation relations\/} 

\ba{
[\phi (x), \phi (x')]_{x_0'=x_0}\a=0\cr
 [\phi (x), \partial_0\phi (x')]_{x_0'=x_0}\a=i\delta ( {\bf x} -{\bf x} ').
 \cr}
 \eno8)
 \ea
This is an extension  to infinite degrees of freedom of the canonical
commutation relations  obeyed by the quantum--mechanical position-- and
momentum operators.  Note that since time evolution is to be implemented
by a unitary operator, the same commutation relations will then hold  at any
other time as well.   For the non--Hermitian operators $a(p)$,  the
corresponding commutation relations are

\ba{
[a(p), a(p')]=2p_0\,\theta (-p_0p_0')\,(2\pi )^s\,\delta ({\bf p} +{\bf p}')
 \cr}
 \eno9)
 \ea
for $p,p'\in\om$.  Using the neutrality condition $a(-p)=a(p)^*$, these can
be rewritten in their conventional form

\ba{
[a(p), a(p')]\a=0\cr
 [a(p), a(p')^*]\a=2\omega \,(2\pi )^s\,\delta ( {\bf p}-{\bf p} ')
 \cr}
 \eno10)
 \ea
where now  $p,p'\in\omp$. 
\skp

A  charged  field can be built up from a pair of neutral fields as 

\be
 \phi (x)={ \phi _1(x)+i\phi _2(x)\over \sqrt{2}},
\eno11)
 \ee
where the two fields $\phi _1, \phi _2$  each obey the equal--time 
commutation relations and commute with one another.  Equivalently, the
operators $a(p)$ and $b(p)$ become independent and satisfy

\ba{
[a(p), a(p')]=[b(p), b(p')]=[a(p), b(p')]=0, \cr
[a(p), a^*(p')]=[b(p), b^*(p')]=2\omega \,(2\pi )^s\,\delta ( {\bf p} -{\bf p} ')
 \cr}
 \eno12)
 \ea
for $p, p'\in\omp$.  
The canonical commutation relations  for both neutral and charged fields
can be put in the manifestly covariant form 

\ba{
[\ftd(p), \ftd(p')^*]\a=4\pi ^2\,\delta (p^2-m^2)\,\delta ({p'}^2-m^2)\,[a(p),
a(p')^*]\cr
\a=(2\pi)^{s+2}  2p_0\,\theta (p_0 p_0')\,\delta (p^2-m^2)\,\delta
({p'}^2-p^2)\,\delta ({\bf p} -{\bf p} ')\cr 
\a=(2\pi)^{s+2}  2p_0\,\theta (p_0 p_0')\,\delta(p^2-m^2)\,\delta
({p_0'}^2-p_0^2)\, \delta ({\bf p} -{\bf p} ')\cr
 \a={\rm sign}(p_0) \,(2\pi)^{s+2} \,\delta (p^2-m^2)\,\delta (p-p')
 \cr}
 \eno13)
 \ea
for arbitrary $p, p'\in\rl^{s+1}$.  For charged fields, this must be
supplemented by

\be
 [\ftd(p), \ftd(p')]=0,\qquad p,p'\in\rl^{s+1}.
\eno14)
 \ee
Recall that in the general case we had

\be
 \langle\, \Phi _p^+\,|\,\Phi _{p'}^+\,\rangle=\sigma (p^2)\,(2\pi )^{s+1}\,\delta
(p-p')
 \eno15)
 \ee
for $p,p'\in\vp$, where $\sigma (p^2)$ is the spectral density  for
the two--point function (sec. 5.3,  eq. (33)).  For the free field now
under consideration  we have

\ba{
 \langle\, \Phi _p^+\,|\,\Phi _{p'}^+\,\rangle\a=\langle\, \Psi _0\,|\,\ftd(p)\,
\ftd(p')^*\,\Psi _0\,\rangle\cr
\a=\langle\, \Psi _0\,|\,[\ftd(p)\, ,\ftd(p')^*]\,\Psi _0\,\rangle\cr
\a=(2\pi )^{s+2} \,\delta (p^2-m^2)\,\delta (p-p'),
 \cr}
 \eno16)
 \ea
which shows that the spectral density for the free field is

\ba{
\sigma  (p^2)=2\pi \,\delta (p^2-m^2).
 \cr}
 \eno17)
 \ea

 \skp
The spectral condition implies that 

\ba{
a(p)\Psi _0=0,\qquad b(p)\Psi _0=0\qquad \forall p\in\omp,
 \cr}
 \eno18)
 \ea
since otherwise these would be states of energy--momentum $-p$. \break
Hence the particle coherent  states  defined in the last section are now given
by

\ba{
 e_z^+\a\equiv \phi (z)^*\,\Psi _0=\inop e^{i\zbar p}\,a(p)^*\,\Psi _0\cr
\a\equiv\inop e^{i\zbar p}\,\tilde\Phi_p^+,  \qquad z\in\tp,
\cr}
 \eno19)
 \ea
where the vectors $\tilde\Phi_p^+$ are generalized eigenvectors of
energy--mo\-men\-tum $p\in\omp$ with the normalization

\ba{
 \langle\, \tilde\Phi_p^+\,|\,\tilde\Phi_{p'}^+\,\rangle\a=
\langle\, \Psi _0\,|\,a(p)\,a(p')^*\Psi_0\,\rangle\cr
\a=\langle\, \Psi _0\,|\,[a(p),\,a(p')^*]\Psi _0\,\rangle\cr
\a=2\omega (2\pi )^s\,\delta ({\bf p} -{\bf p}').
 \cr}
 \eno20)
 \ea
The wave packets  $e_z^+$ span the one--particle subspace ${\cal H}_1$ of 
${\cal H}$ and  have the momentum representation  

\be
 \langle\, \tilde\Phi _p^+\,|\,e_z^+\,\rangle=e^{i\zbar p}.
\eno21)
 \ee
A dense subspace of ${\cal H}_1$ is obtained by applying the smeared
operators

\be
 \phi ^*(f)\equiv \phi (\bar f)^*\equiv \int dx \,\phi (x)^*\,f(x)=
\inp\,\ftd (p)^*\,\ftil(p)
\eno22)
 \ee
to the vacuum, where $f$ is a test function.  This gives

\ba{
 \Psi _f^+\a\equiv \phi^*(f)\,\Psi _0=\inp \ftil(p)\,\ftd ^*(p)\,\Psi _0\cr
\a=\inp \ftil(p)\,\Phi ^+_p=\inop \fhat(p) \,\tilde \Phi ^+_p,
 \cr}
 \eno23)
 \ea
 where $\fhat$ is the restriction of $\tilde f$ to $\omp$.
Hence the  functions

\be
 \langle\, e_z^+\,|\,\Psi _f^+\,\rangle=\inop e^{-izp}\,\fhat(p)
\eno24)
 \ee
are exactly  the holomorphic  positive--energy solutions  of the
Klein--Gordon equation discussed in section 4.4, with $e_z^+$ corresponding
to the evaluation maps $e_z$. The space ${\cal K}$ of these solutions  can thus
be identified with ${\cal H}_1$, and the orthogonal projection from ${\cal H}$ 
to ${\cal H}_1$ is given by

\be
 (\Pi_1\Psi )(z)=\langle\, e^+_z\,|\,\Psi \,\rangle.
\eno25)
 \ee
 Consequently, the resolution of unity
developed in chapter 4 can now be restated as a resolution of $\Pi_1$:

\be
 \Pi_1=\insp \,|\, e_z^+\,\rangle\langle\, e_z^+\,|\, ,
\eno26)
 \ee
where $\sigma _+$, earlier denoted by $\sigma $, is a {\sl particle phase
space,\/} i.e. has the form

\be
 \sigma _+=\{x-iy\,|\,x\in S,\ y\in \olp\}
\eno27)
 \ee
for some $\lambda >0$ and some spacelike or, more generally, nowhere
timelike  (see section 4.5) submanifold $S$ of real spacetime.  As in section
4.5, the measure $d\sigma $ is given in terms of the Poincar\'e--invariant
symplectic form $\alpha =dy_\mu \w dx^\mu $ by restricting $\alpha
^s\equiv \alpha \w\cdots \w\alpha $ to $\sigma _+$ and choosing an
orientation:

\ba{
d\sigma =(s! A_\lambda )^{-1} \,\alpha ^s =A_\lambda  ^{-1}\,
\dy^\mu \w\dx_\mu .
 \cr}
 \eno28)
 \ea
\skp
Similarly, the  antiparticle coherent states for the free field are  given by

\ba{
  e_z^-\a\equiv \phi (z)\,\Psi _0=\inop e^{iz p}\,b(p)^*\,\Psi _0\cr
\a\equiv\inop e^{iz p}\,\tilde\Phi_p^-,  \qquad z\in\tm.
\cr}
 \eno29)
 \ea
Since for $p\in\omp$ and $z\in\tm$ we have

\be
 \langle\, \tilde \Phi _p^-\,|\,e_z^-\,\rangle=e^{izp}=
\langle\, \tilde \Phi _p^+\,|\,e_{\zbar}^+\,\rangle,
\eno30)
 \ee
it follows that $e_z^-$ has exactly the same spacetime behavior as
$e_{\zbar}^+$,  confirming the interpretation of an antiparticle as
a particle moving backward in time.  An {\sl antiparticle phase
space\/}  is defined as a submanifold of $\tm$ given by

\be
 \sigma _-=\{x-iy\,|\,x\in S,\ y\in\olm\},
\eno31)
 \ee
where $S$ is as above.   The resolution of $\Pi_{-1}$ is then given by

\be
 \Pi_{-1}=\int_{\sigma _-} d\sigma \,\,|\,e_z^-\,\rangle\langle\,  e_z^-\,|\, .
\eno32)
 \ee
\skp
Many--particle or --antiparticle coherent states and their corresponding
phase spaces    can be defined similarly, and the commutation relations
imply that such states are {\sl symmetric\/} with respect to permutations
of the particles' complex coordinates.  For example, 

\be
e^+_{z_1z_2}\equiv  \phi (z_1)^*\phi (z_2)^*\Psi _0= e^+_{z_2z_1},
 \ee
since $\phi (z_1)$ and $\phi(z_2)$ commute. In this way, a
phase--space formalism can be buit for an indefinite number of particles (or
charges),  analogous to the grand--canonical ensemble in classical statistical
mechanics.  This idea will not be further pursued here. 
Instead, we now embark on option (b) above, i.e. the construction of global,
conserved field observables as integrals over particle  and antiparticle phase
spaces.  \skp

The {\sl particle number\/}  and {\sl antiparticle number\/} operators
 are given by 

\ba{
N_+\a=\inop a(p)^*\,a(p)\cr
N_-\a=\inop b(p)^*\,b(p),
 \cr}
 \eno33)
 \ea
$N_+$ and $N_-$ generalize the harmonic--oscillator Hamiltonian $A^*A$ to
the infinite number of degrees of freedom possessed by the field, where
normal modes of vibration are labeled by $p\in\omp$ for particles and
$p\in\omm$ for antiparticles.
The total charge operator is  $Q=\ve \,(N_+-N_-)$, as can be seen from its
commutation relations with $a(p)$ and $b(p)$.  But the resolution of unity
derived in chapter 4 can now be restated as 

\ba{
\insp\exp(i\zbar p-izp')\a=(2\pi)^s\,2\omega 
({\bf p} )\,\delta ({\bf p}-{\bf p}')=\langle\, \tilde \Phi ^+_p\,|\,
\tilde \Phi ^+_{p'}\,\rangle\cr
\insm\exp(izp-i\zbar  p')\a=(2\pi)^s\,2\omega 
({\bf p} )\,\delta ({\bf p}-{\bf p}')=\langle\, \tilde \Phi ^-_p\,|\,
\tilde \Phi ^-_{p'}\,\rangle
\cr}
 \eno34)
 \ea
for $p, p'\in\omp$, where the second identity follows from the first by
replacing $z$ with $\zbar$ and $\sigma_+$ with $\sigma _-$.   It follows
that $N_\pm$ can be expressed as phase--space integrals of the extended
field $\phi (z)$:

\ba{
N_+\a=\insp \phi (z)^*\,\phi (z)\cr
N_-\a=\insm \phi (z)\,\phi (z)^*.
 \cr}
 \eno35)
 \ea
Hence the charge is given by
\be
 Q=\ve \insp \phi (z)^*\,\phi (z)-\ve \insm \phi (z)\,\phi (z)^*.
\eno36)
 \ee
The two integrals can be combined into one as follows:  Define the {\sl
total\/} phase space as $ \sigma =\sigma _+-\sigma _-$, where the minus
sign means that $\sigma _-$ enters with the {\sl opposite\/}  (``negative'')
orientation to that of $\sigma _+$, in the sense of {\sl chains\/}  (Warner
[1971]).  The reason for this choice of orientation  is that $\blp$ and $\blm$
are both open sets of $\rl^{s+1}$, hence must have the same orientation, and
we orient $\olp$ and $\olm$ so that 

\be
  \Omega _\lambda ^\pm=-\partial\,B_\lambda ^\pm.
\eno37)
 \ee
Since the  outward normal of $\blp$ points ``down''
and that of $\blm$ points ``up,''  this means that $\olm$ must have the
opposite orientation to that of $\olp$.  Thus, setting $B_\lambda \equiv
\blp+\blm$   and $\Omega_\lambda  \equiv \olp-\olm$, we have

\be
 \Omega_\lambda  =-\partial B_\lambda .
\eno38)
 \ee
This gives $\sigma _-$ the orientation  opposite to that of $\sigma _+$, and
we have 

\be
 \sigma \equiv S\times \Omega _\lambda =\sigma _+-\sigma _-.
\eno39)
 \ee
 Next, define the {\sl Wick--ordered product\/}  (or normal
product) by

\be
 :\phi (z)^*\,\phi (z):\equiv \cases{\phi (z)^*\,\phi (z),\quad &$z\in\tp$\cr
\phi (z)\,\phi (z)^*,\quad &$z\in\tm$.}
\eno40)
 \ee
This coincides with the usual definition, since in $\tp$,  $\phi ^*$ is a creation
operator and $\phi $ is an annihilation operator, while in $\tm$ these roles
are reversed.    The charge can now be written in the compact form

\be
 Q=\ve \ins :\phi (z)^*\,\phi (z):
\eno41)
 \ee
We may therefore interpret the operator 

\be
 \rho (z)\equiv \ve:\phi (z)^*\,\phi (z):
\eno42)
 \ee
as a {\sl scalar phase--space charge density\/}  with respect to the measure
$d\sigma $. 

 The Wick ordering  can  be viewed as a special case of {\sl 
imaginary--time ordering,\/} if we define $\phi ^*(z)\equiv \phi (\zbar)^*$:

\be
 :\phi (z')^*\,\phi (z):\equiv :\phi ^*(\bar z')\,\phi (z):=
T_i\left[ \phi ^*(\bar z')\,\phi (z)\right],
\eno43)
 \ee
where
\ba{
 T_i\a\left[\phi^*(z')\,\phi (z)\right]\cr
\a\equiv \theta\left(\Im(z_0'-z_0)  \right)\,\phi ^*(z')\,\phi (z)+
\theta \left(\Im(z_0-z_0')  \right)\,\phi (z)\,\phi ^*(z')
 \cr}
 \eno44)
 \ea
for $z,z'\in\tb$.  This definition is Lorentz--invariant, since 

\ba{
T_i[\phi(z')\,\phi (z)]=\phi (z')\,\phi (z)=\phi (z)\,\phi (z')\quad 
\forall z,z'\in\tb
\cr}
 \eno45)
 \ea
and

\ba{
T_i[\phi^*(z')\,\phi (z)]=\phi ^*(z')\,\phi (z)=\phi (z)\,\phi ^*(z')
 \cr}
 \eno46)
 \ea
when $z$ and $z'$ are in the same half of $\tb$, whereas if they are in
opposite halves of $\tb$, the sign of $\Im (z_0'-z_0)$ is invariant.

\skp
\noindent {\sl Note:\/} For the extended fields, the Wick ordering is not a
necessity but a mere convenience, allowing us to combine the integrals over
$\sigma _+$ and $\sigma _-$ into a single integral.  Each of these integrals is
{\sl already\/}  in normal order, since the extension to complex spacetime
{\sl polarizes\/}  the free field into its positive--and negative--frequency
parts.  Also, the extended fields are operator--valued {\sl functions\/} 
rather than distributions, hence products such as $\phi (z)^*\phi (z)$ are
well--defined, which is {\sl not\/}  the case in the usual formalism.  A similar
situation will occur in the expressions for the other observables
(energy--momentum, angular momentum, etc.) as phase--space integrals. 
Hence {\sl the phase--space formalism resolves the problem of zero--point
energies\/}  without the need to subtract infinite terms ``by hand''!  In this
connection, see the remarks on p. 21 of Henley and Thirring [1962].\sp\#
\skp The above expression for the charge can be related to the usual one in
the spacetime formalism, which is

\ba{
 Q_{\rm usual}\a=i\ve \int_S \dx_\mu \,:\phi^*
{\partial \phi \over{\partial x_\mu }}-
{\partial \phi^*\over{\partial x_\mu }} \cdot\phi (x):\cr
\a\equiv \int_S \dx_\mu \,J^\mu (x),
 \cr}
 \eno47)
 \ea
by using $\ol =-\partial \bl $ and applying Stokes' theorem:

\ba{
Q\a=\ve \,A_\lambda  ^{-1}\,\int_S \dx_\mu \int_\ol \dy^\mu :\phi
^*\,\phi :\cr
\a=-\ve \,\,A_\lambda  ^{-1}\,\int_S \dx_\mu \int_\bl dy\,
{\partial \over{\partial y_\mu }}:\phi ^*\,\phi :\cr
\a\equiv A_\lambda  ^{-1} \int_S\dx_\mu \int_\bl dy\,j^\mu (x-iy),
 \cr}
 \eno48)
 \ea
where 

\be
 j^\mu (z)\equiv -{\partial \over{\partial y_\mu }}\,\rho (z)
\eno49)
 \ee
 is the {\sl phase--space  current density.\/} Using the
notation

\ba{
\partial^\mu \equiv {\partial \over{\partial z_\mu }}\equiv 
{1\over 2}\left({\partial \over{\partial x_\mu }}  +
i{\partial \over{\partial y_\mu }}\right),
 \cr}
 \eno50)
 \ea
we have 

\be
 -{\partial \over{\partial y_\mu }}=i(\partial^\mu -\bar \partial^\mu ). 
\eno51)
 \ee
Hence, by the holomorphy of $\phi $, 

\ba{
j^\mu (z)\a\equiv -\ve \, {\partial \over{\partial y_\mu }}:\phi ^*\,\phi :\cr
\a=i\ve \,\left( \partial^\mu
-\bar\partial^\mu  \right):\phi ^*\,\phi : \cr
\a=i\ve :\phi ^*\,\partial^\mu \phi  -
\bar\partial^\mu  \phi ^*\cdot \phi : \cr
\a=i\ve :\phi ^*{\partial \phi \over{\partial x^\mu }}-
{\partial \phi ^*\over{\partial x^\mu }}\,\phi :.
\cr}
 \eno52)
 \ea
Our expression for the charge is therefore

\be
Q= \int_S\dx_\mu \,J^\mu_{(\lambda)}  (x),
\eno53)
 \ee
where 

\ba{
 J^\mu_{(\lambda)}  (x)\a\equiv A_\lambda ^{-1} \int_\bl dy\,j^\mu (x-iy)\cr
\a=i\ve \,A_\lambda  ^{-1}\,\int_\bl dy 
:\phi ^*{\partial \phi \over{\partial x^\mu }}-
{\partial \phi ^*\over{\partial x^\mu }}\,\phi :
 \cr}
 \eno54)
 \ea
is seen to be a {\sl regularized version\/}  of the usual current density
$J^\mu (x)$ obtained by first extending it to $\tb$ and then integrating it
over $\bl$. 

The vector field $j^\mu (z)$ is conserved in {\sl real\/}  spacetime for each
fixed $y\in V'$, since

\ba{
{\partial j^\mu \over{\partial x^\mu }}\a=-\ve \,
{\partial ^2  \over{\partial x^\mu \partial y_\mu }}:\phi ^*\,\phi :\cr
\a=i\ve \, (\partial_\mu +\bar\partial _\mu )\,
(\partial^\mu -\bar\partial ^\mu ):\phi ^*\,\phi :\cr
\a=i\ve \, (\del\!\!_z-\del\!\!_\zbar):\phi ^*\,\phi :\cr 
\a=i\ve :\left(  \phi ^*\,\del\!\!_z \phi -\del\!\!_\zbar \phi ^*\cdot \phi 
\right):\cr \a=0,
 \cr}
 \eno55)
 \ea
by virtue of  the Klein--Gordon equation combined with the
holomorphy of $\phi $ in $\tb$.  This implies that $J^\mu _{(\lambda)} (x)$ is
also conserved, hence the charge does not depend on the choice of $S$ or
$\sigma $. \skp
\noindent {\sl Note:\/} In using Stokes' theorem above, we have assumed
that the contribution from $|y_0|\to\infty$ vanishes. (This was implicit in
writing the non--compact manifold $\ol$ as $-\partial \bl$.) This is  indeed
the case, as has been shown rigorously in the context of the one--particle
theory in chapter 4 (theorem 4.10). Also, we see another example of the
pattern, mentioned before, that in the phase--space formalism vector-- and
tensor fields can often be derived from {\sl scalar potentials.\/}   Here,
$\rho (z)$  acts as a potential for  $j^\mu (z)$.  Note also that the
Klein--Gordon equation can be written in the form

\be
 (\del\!\!_z+m^2)\rho (z)=\ve:\phi ^*(\del\!\!_z+m^2)\phi :=0,
\eno56)
 \ee
which is manifestly gauge--invariant.   \sp\# 
\skp 
Recall now that $\phi (z)$ is a ``root vector'' of the charge with root
value $-\ve $:

\be
 [\phi (z'), Q]=\ve \,\phi (z')\qquad \forall z'\in\cx^{s+1}.
\eno57)
 \ee
Substituting our expression for $Q$, we obtain the identity

\ba{
\phi (z')\a=\int_\sigma  d\sigma (z)\,[\phi (z'), :\phi (z)^*\,\phi (z):\,]\cr
\a=\int_\sigma  d\sigma (z)\,[\phi (z'), \phi (z)^*]\,\phi (z)\cr
\a\equiv \int_\sigma d\sigma (z)\,K(z',\zbar)\,\phi (z),
 \cr}
 \eno58)
 \ea
where, by the canonical commutation relations,

\ba{
K\a(z',\zbar)\equiv [\phi (z'), \phi (z)^*]=[\phi (z'), \phi^* (\zbar)]\cr
\a=(2\pi)^{-2s-2} \int dp'\int dp\,\theta ^{-iz'p'}\,\theta ^{i\zbar p}\,
[\ftd(p'), \ftd(p)^*]\cr
\a=(2\pi)^{-s} \int dp'\int dp\,\theta ^{-iz'p'}\,\theta ^{i\zbar p}\,{\rm
sign}(p_0)\,\delta (p^2-m^2)\,\delta (p'-p)\cr
\a=\ino\,{\rm sign}(p_0)\,\theta (yp)\,\theta (y'p)\,\exp[-i(z'-\zbar )p].
 \cr}
 \eno59)
 \ea
$K$ is a distribution on $\cx^{s+1}\times\cx^{s+1}$ which is piecewise
analytic in $\tb\times\tb$, with

\be
 K(z', \zbar)=\cases{-i\Delta ^+(z'-\zbar; m), &$\quad z', z\in\tp$\cr
i\Delta ^-(z'-\zbar ; m), &$\quad z', z\in\tm$\cr
0, &$\quad z'\in\tp, z\in\tm$\cr
0, &$\quad z'\in \tm, z\in\tp$.\cr}
\eno60)
 \ee
The two--point functions $-i\Delta ^+$ and $i\Delta ^-$ are analytic in $\tp$
and $\tm$, respectively, and act as reproducing kernels for the subspaces
with charge $\ve $ and $-\ve $.  Because of the above property, it
is reasonable to call $K(z',\zbar) $ a reproducing kernel for the field $\phi
(z)$,  though this differs somewhat from the standard usage of the term as
applied to Hilbert spaces (see chapter 1).  Note that $K$ propagates
positive--frequency components of the field into the forward (``future'')
tube and negative--frequency components into the backward (``past'') tube. 
This is somewhat reminiscent of the Feynman propagator, but $K$ is a
solution of the {\sl homogeneous\/} Klein--Gordon equation in the real
spacetime variables rather than a Green function.
\skp
The energy--momentum and angular momentum operators may be likewise
expressed as conserved phase--space integrals of the extended field:

\ba{
P_\mu \a=i\ins :\phi ^*\partial_\mu \phi :\cr
M_{\mu \nu }\a=i\ins :\phi ^*(x_\mu \partial_\nu -
x_\nu\partial_\mu)\phi :.
 \cr}
 \eno61)
 \ea
Like $Q$, these may be displayed as regularizations of the usual, more
complicated expressions in real spacetime.  Note first that  $P_\mu $ can be
re--written as

\ba{
P_\mu \a= i\,A_\lambda  ^{-1}\,\int_S \dx\,^\tau  \int_\ol \dy\!_\tau  :\phi
^*\,\partial_\mu \phi :\cr
\a={i\over 2}\,A_\lambda  ^{-1}\,\int_S \dx\,^\tau  \int_\ol \dy\!
_\tau  :\phi ^*\,\partial_\mu \phi -\bar \partial_\mu  \phi ^*\cdot \phi :\cr
\a=-{1 \over 2}\,A_\lambda  ^{-1}\,\int_S \dx\,^\tau  \int_\ol \dy\!_\tau  \,
{\partial \over{\partial y^\mu }}\,:\phi ^*\,\phi :.
 \cr}
 \eno62)
 \ea
The angular momentum can be recast similarly as

\ba{
M_{\mu \nu }\a=i\,A_\lambda  ^{-1}\,\int_S \dx\,^\tau  \int_\ol \dy\!_\tau 
:\phi^*\,\left( x_\mu \partial_\nu -x_\nu \partial_\nu   \right)\phi :\cr
\a={ i\over2 }\,A_\lambda  ^{-1}\,\int_S \dx\,^\tau  \int_\ol \dy\!_\tau 
\,\left[ x_\mu (\partial_\nu-\bar\partial_\nu ) -x_\nu (\partial_\mu  
-\bar\partial_\mu ) \right]:\phi^*\phi :\cr
\a=-{ 1\over2 }\,A_\lambda  ^{-1}\,\int_S \dx\,^\tau  \int_\ol \dy\!_\tau 
\,\left[ x_\mu {\partial \over{\partial y^\nu }} -x_\nu {\partial
\over{\partial y^\mu }} \right]:\phi^*\phi : .
 \cr}
 \eno63)
 \ea
Using $\ol=-\partial\bl$ and applying Stokes' theorem, we therefore have

\ba{
P_\mu \a={1 \over 2}\,A_\lambda  ^{-1}\,\int_S \dx\,^\tau  \int_\bl
dy \, {\partial ^2\over{\partial y^\mu \partial y^\tau }}\,:\phi ^*\,\phi :\cr
\a= \int_S\dx\,^\tau \,T^{(\lambda )}_{\mu \tau }(x),
 \cr}
\eno64)
 \ea 
where 

\be
 T^{(\lambda )}_{\mu \tau }(x)\equiv {1 \over 2}\,A_\lambda  ^{-1}\,\int_\bl
dy \, {\partial ^2\over{\partial y^\mu \partial y^\tau }}\,:\phi ^*\,\phi :
\eno65)
 \ee
is a {\sl regularized energy--momentum density tensor\/}  which,
incidentally, is automatically symmetric.  Similarly,  

\ba{
M_{\mu \nu }\a=
{ 1\over2 }\,A_\lambda  ^{-1}\,\int_S \dx\,^\tau  \int_\bl dy
\,\left[ x_\mu {\partial ^2\over{\partial y^\nu \partial y^\tau }} -x_\nu
{\partial ^2\over{\partial y^\mu \partial y^\tau }} \right]:\phi^*\phi :\cr
\a=\int_S\dx\,^\tau \,\Theta ^{(\lambda )}_{\mu \nu \tau }(x),
 \cr}
 \eno66)
 \ea
where

\ba{
\Theta ^{(\lambda )}_{\mu \nu \tau }(x)\a\equiv 
{ 1\over2 }\,A_\lambda  ^{-1}\,
 \int_\bl dy
\,\left[ x_\mu {\partial ^2\over{\partial y^\nu \partial y^\tau }} -x_\nu
{\partial ^2\over{\partial y^\mu \partial y^\tau }} \right]:\phi^*\phi :\cr
\a=x_\mu  T^{(\lambda )}_{\nu  \tau }(x)-
x_\nu   T^{(\lambda )}_{\mu   \tau }(x)
\cr}
 \eno67)
 \ea
is a regularized angular momentum density tensor.

\secskp
 \noindent {\bf  5.5.   Free Dirac Fields} 
\def\rightheadline{\tenrm\hfil {\sl
5.5.   Free Dirac Fields}\hfil\folio}  

 \skp
\noindent For simplicity, we specialize in this section (only) to the physical
case of three spatial dimensions, $s=3$.  The  proper Lorentz group ${\cal
L}_0$   is then $SO(3,1)_+$, where the plus sign indicates that  $\Lambda
_0^0>0$, so that $\Lambda $ preserves the orientations of space and time
separately. Its universal covering group can be identified with
$SL(2,\cx)$ as follows (Streater and Wightman [1964]):  An event $x\in\rl^4$
is identified with the Hermitian $2\times 2$ matrix 

\be
 X=x^\mu \sigma _\mu =\left(\matrix{x^0+x^3&x^1-ix^2\cr
	x^1+ix^2&x^0-x^3	\cr}\right)
\eno1)
 \ee
where $\sigma _0=I$ ($2\times 2$ identity) and $\sigma _k$ ($k=1,2,3$)
are the Pauli spin matrices.  Note that   $\det  X=x^2\equiv x\cdot x$. 
The action of  $SL(2,\cx)$  on Hermitian  $2\times 2$ matrices given by

\be
 X'=AXA^*,\qquad A\in SL(2,\cx),
\eno2)
 \ee
induces a linear transformation on $\rl^4$ which we denote by $\pi (A)$:

\be
 x'=\pi (A)\,x
\eno3)
 \ee
From 

\be
{x'}^2=\det X'=|\det A |^2\,\det X=\det X=x^2
\eno4)
 \ee
it follows  that $\pi (A)$ is a Lorentz transformation, and it can easily be
seen to be  proper.  Hence $\pi $ defines a map

\be
 \pi {:}\ SL(2,\cx)\to {\cal L}_0, 
\eno5)
 \ee
which is readily  seen to be a group homomorphism.  Clearly, $\pi
(-A)=\pi (A)$, and it can be shown that if $\pi (A)=\pi (B)$, then $A=\pm
B$.  Since $SL(2,\cx)$ is simply connected, it follows that $SL(2,\cx)$ is the
{\sl universal covering group\/}  of ${\cal L}_0$, the correspondence being
two--to--one.

  The relativistic transformation law as stated in section 5.3,

\be
 \phi (\Lambda x+a)=U(a,\Lambda )\,\phi (x)\,U(a, \Lambda )^*,
\eno6)
 \ee
applies to scalar fields, i.e. fields without any intrinsic orientation or {\sl
spin.\/}    To generalize it to fields with spin, note first of all   that since
the representing operator $U(a, \Lambda )$ occurs quadratically, the law is
invariant under $U\to -U$.  This means that $U$ could, in fact, be a
representation,  not of ${\cal P}_0$, but of the inhomogeneous version of
$SL(2,\cx)$, 

\be
 {\cal P}_2\equiv \rl^4 \sdp\,SL(2,\cx),
\eno7)
 \ee
which acts on $\rl^4$ by

\be
 (a,A )\,x=\pi (A)\,x+a.
\eno8)
 \ee
${\cal P}_2$  is the two--fold universal covering group of ${\cal P}_0$.  A
field $\psi(x) $ of arbitrary spin  is a distribution taking its values in the
tensor product $L({\cal H})\otimes {\cal V}$   of the operator algebra of the
quantum Hilbert space ${\cal H}$  with some finite--dimensional
representation space ${\cal V}$  of $SL(2,\cx)$.    The  transformation law is

\be
 U(a,A)\,\psi (x)\,U(a,A)^*=S(A^{-1} )\,\psi (\pi (A)x+a),
\eno9)
 \ee
where  $S$ is a given 
representation  of $SL(2,\cx)$ in ${\cal V}$. $S$ determines the spin of the
field, which can take the values $j=0, 1/2, 1, 3/2, 2, \ldots$.
\skp
The locality condition for the scalar field (axiom 4) can be
extended to non--scalar fields as  

\ba{
 [\psi _\alpha(x) , \psi _\beta(x') ]=0\quad \ \hbox{if}\   (x-x')^2< 0
 \cr}
 \eno10)
 \ea
where $\psi _\alpha $ are the components of $\psi $.  
Now it follows from the axioms that if the  field has half--integral spin
($j=1/2, 3/2, \ldots$), then the above locality condition implies that it is
trivial, i.e. that $\psi (x)\equiv 0$.  A non--trivial field of half--integral spin
can be obtained, however, if we modify the locality condition by replacing
commutators with anticommutators:

\ba{
 \{\psi _\alpha(x) , \psi _\beta(x') \}\a\equiv \psi _\alpha(x)  \,\psi
_\beta(x')+\psi _\beta(x')\,\psi _\alpha(x)\cr
\a=0\quad \ \hbox{if}\   (x-x')^2< 0.
 \cr}
 \eno11)
 \ea
Replacing the commutators with 
anticommutators means that changing the order in which $\psi (x)$ and
$\psi (x')$ are applied to a state vector in Hilbert space  merely changes the
sign of the vector, which has no observable effect.   Hence
the physical interpretation that events at spacelike separations cannot
influence one another is still valid.  

Similarly,  for fields of integral spin ($j=0,1, \ldots$), the locality condition
with anticommutators gives a trivial theory, whereas  a
non--trivial theory can exist using commutators. 

The choice of commutators or anticommutators in the locality condition
does, however, have an important physical consequence.  For  we have seen
that the free asymptotic fields can be written as sums of creation and
destruction operators for particles and antiparticles.  If $x_1, x_2, \cdots x_n$
are $n$ distinct  points in the hyperplane $x^0=0$ and $\psi_+(x)$ denotes
the positive--frequency part of the field (which can be obtained from $\psi
(x-iy)$ by taking $y\to 0$ in $V_+'$), then

\be
 \psi_+(x_1)^*\psi_+(x_2)^*\cdots \psi_+(x_n)^*\,\Psi _0
\eno12)
 \ee
is a state with $n$ particles  located at these points.  Since
 any two of these points are  separated by a spacelike distance, the
locality condition  implies that this state is {\sl symmetric\/}  with
respect to the exchange of any two particles if commutators are used 
 and {\sl antisymmetric\/} with respect to the exchange if
anticommutators are used.  Particles whose states
are symmetric under exchange are called {\sl Bosons,\/} and ones
antisymmetric under exchange are called {\sl Fermions.\/}   The choice of
symmetry or antisymmetry crucially affects the large--scale  
statistical behavior of the particles.  For example,  no two Fermi\-ons can
occupy the same state due to the antisymmetry under exchange;  this is the
{\sl Pauli exclusion principle.\/}   Hence the choice of commutators or
anticommutators is known  as 
the choice of {\sl statistics,\/} and the above theorem correlating this choice
with the spin is known as the {\sl Spin--Statistics theorem\/}  of quantum
field theory (Streater and Wightman [1964]).  This theorem is fully supported
by experiment, and represents one of the successes of the theory.
\skp

 The Dirac field is a quantized field with spin 1/2 whose
associated particles and antiparticles are typically taken to be electrons and
positrons, though it is also used (albeit less accurately)  to model neutrons
and protons.   Our treatment follows the notation used in Itzykson and Zuber
[1980], with minor modifications.   The free Dirac field is a solution of the
Dirac equation

\be
( i\dir-m)\psi(x) =0,
\eno13)
 \ee
where 

\be
 \dir\equiv \gamma ^\mu {\partial \over{\partial x^\mu }}
\eno14)
 \ee
is the Dirac operator 
and the $\gamma $'s are a set of $4\times 4$ Dirac matrices, meaning they
 satisfy the Clifford condition with respect to the Minkowski metric:

\ba{
\{\gamma ^\mu , \gamma ^\nu \}\equiv \gamma ^\mu \gamma ^\nu
+\gamma ^\nu \gamma ^\mu =2g^{\mu \nu }.
 \cr}
 \eno15)
 \ea
The components of $\psi $ satisfy the Klein--Gordon equation, since \break
$\dir^2=\del,$ and the solutions can be written as 

\ba{
\psi (x)=\inop \left[e^{-ixp}\,u^\alpha(p) \,b_\alpha(p) +
e^{ixp}\,v^\alpha(p)\, d_\alpha^*(p)\right],
 \cr}
 \eno16)
 \ea  
where $u^\alpha $ and $v^\alpha $ are positive-- and negative--frequency
four--com\-po\-nent spinors and summation over the polarization index
$\alpha =1,2$ is implied.  $b_\alpha $ and $d_\alpha $ are operators
satisfying the ``canonical anticommutation relations'' 

\ba{
\{b_\alpha (p), b_\beta (q)\}\a=\{d_\alpha (p), d_\beta (q)\}=
\{b_\alpha (p), d_\beta (q)\}=0,\cr
\{b_\alpha (p), b_\beta^* (q)\}\a=\{d_\alpha (p), d_\beta^* (q)\}=
2\omega (2\pi )^3\,\delta ({\bf p} -{\bf q} )
 \cr}
 \eno17)
 \ea
for all $p, q\in\omp$.
The spinors satisfy the orthogonality and completeness relations

\ba{
\bar u^\alpha (p)\,u^\beta (p)\a=-\bar v^\alpha (p)\,v^\beta (p)=\delta
^{\alpha \beta }\cr
 u^\alpha (p)\otimes\bar u^\alpha (p)\a={ \pslt+m\over
2m}\cr 
 v^\alpha (p)\otimes\bar v^\alpha (p)\a={ \pslt-m\over
2m},
 \cr}
 \eno18)
 \ea
where a summation on $\alpha $ is implied in the last two equations and the
adjoint spinors are defined by

\be
 \bar u^\alpha (p)=u^\alpha (p)^*\gamma ^0,\quad \bar v^\alpha (p)=v^\alpha
(p)^*\gamma ^0. 
\eno19)
 \ee
In addition, 

\ba{
\bar u^\alpha (p)\,\gamma _\mu \,u^\beta (p)=
\bar v^\alpha (p)\,\gamma _\mu \,v^\beta (p)={ p_\mu \over m}\,\delta ^
{\alpha \beta }.
 \cr}
 \eno20)
 \ea
$b_\alpha (p)$ and $d_\alpha (p)$ are interpreted as
annihilation operators for particles  and antiparticles, respectively, while
their adjoints are creation operators.  The adjoint field is defined by

\be
 \overline{\psi (x) } =\psi (x)^*\,\gamma ^0
\eno21)
 \ee
and satisfies 

\be
 i{\partial \bar \psi \over{\partial x^\mu }}\,\gamma ^\mu =m\bar \psi .
\eno22)
 \ee
\skp
The particle-- and antiparticle number operators are now

\ba{
N_+\a=\inop b_\alpha ^*(p)\,b_\alpha (p)\cr
N_-\a=\inop d_\alpha ^*(p)\,d_\alpha (p),
 \cr}
 \eno23)
 \ea
and the charge operator is 

\be
 Q=\ve(N_+-N_-).
\eno24)
 \ee
As for the Klein--Gordon field, we wish to give a phase--space
representation  of $Q$.  The first step is to extend  $\psi (x)$ to $\cx^4$ using
the Analytic--Signal transform,  which gives

\ba{
\psi (z)=\inop \left[\theta^{-izp}\,u^\alpha(p) \,b_\alpha(p) +
\theta^{izp}\,v^\alpha(p)\, d_\alpha^*(p)\right].
 \cr}
 \eno25)
 \ea
Again, the extended field is analytic in $\tb$, with the parts in $\tp$ and
$\tm$ containing only positive  and negative frequncies, respectively.  
Using the above orthogonality relations, as well as 

\ba{
\insp e^{i\zbar p-izq}=\insm e^{iz p-i\zbar q}=
2\omega ({\bf p})\,(2\pi )^3\,\delta ({\bf p} -{\bf q} )
 \cr}
 \eno26)
 \ea
for $p, q\in\omp$, we obtain the following expressions for the particle-- and
antiparticle number operators as phase--space integrals:

\ba{
N_+\a=\insp \overline{ \psi (z)} \,\psi (z)\,\equiv \insp :\overline{ \psi (z)} \,
\psi (z):\cr
N_-\a=\insm  :\overline{ \psi (z)} \,\psi (z):,
 \cr}
 \eno27)
 \ea
where the fields in the first integral are already in normal order and the
second integral involves two changes of sign: one due to the normal ordering,
and another due to the orthogonality relation for the $v^\alpha $'s.  The 
charge operator can therefore be given the following compact expression as a
phase--space  integral over the oriented phase space  $\sigma =\sigma
_+-\sigma _-$:

\be
 Q=\ve\ins :\bar \psi \,\psi :=\ve\ins \rho (z),
\eno28)
 \ee
where $\rho \equiv \,:\bar\psi \psi :$ is the {\sl scalar phase--space charge
density.\/}  The usual expression for the charge as an integral over a
configuration space $S$ is

\be
 Q_{\rm usual}=\ve\int_S\dx_\mu :\overline{\psi (x) }\,\gamma ^\mu \,\psi
(x): \,\equiv \int_S\dx_\mu J^\mu (x).
\eno29)
 \ee
To compare these two expressions, we again use $\ol=-\partial\bl$ and
invoke Stokes' theorem:

\ba{
Q\a=\ve A_\lambda  ^{-1} \int_S\dx_\mu \int_\ol \dy^\mu :\bar \psi \,\psi
:\cr
\a=-\ve A_\lambda  ^{-1}
\int_S\dx_\mu \int_\bl dy\,{\partial \over{\partial y_\mu }}
:\bar \psi \,\psi:.
 \cr}
 \eno30)
 \ea
Define the {\sl phase--space current density\/} 

\be
 j^\mu (z)\equiv 2m\ve:\overline{\psi(z) } \gamma ^\mu  \,\psi(z):,
\eno31)
 \ee
where the factor $2m$ is included to give $j^\mu $ the correct physical
dimensions, given our normalization. Note that $j^\mu (z)$ is conserved in
spacetime, i.e.

\ba{
 {\partial j^\mu \over{\partial x^\mu }}\a=(\bar\partial_\mu +\partial_\mu
)j^\mu \cr
\a=2m\ve:{\partial \bar \psi \over{\partial \zbar ^\mu }}\,\gamma ^\mu
\,\psi + \bar\psi \gamma ^\mu {\partial \psi \over{\partial z^\mu }}:\cr
\a=0
 \cr}
 \eno32)
 \ea
by the Dirac equation combined with the analyticity of $\psi $ in $\tb$.
The same combination also implies 
\ba{
{1\over 2}\,j^\mu (z)\a=\ve:\bar \psi \gamma ^\mu i\gamma ^\nu
\partial_\nu  \psi :\cr 
\a=i\ve\partial_\nu :\bar \psi \gamma ^\mu \gamma ^\nu \psi :\cr
\a=i\ve\partial_\nu  :\bar \psi (g^{\mu \nu }-i\sigma  ^{\mu \nu })\psi :\cr
\a=i\ve\partial^\mu :\bar \psi \psi :+\ve\partial_\nu :\bar \psi \sigma 
^{\mu \nu }\psi :,
 \cr}
 \eno33)
 \ea
where 

\be
 \sigma ^{\mu \nu }={i\over 2}\,[\gamma ^\mu ,\gamma ^\nu ]
\eno34)
 \ee
are the spin matrices.  The real part of this equation gives a phase--space
version of the {\sl Gordon identity\/} 

\ba{
 j^\mu (z)\a=i\ve(\partial^\mu -\bar \partial^\mu ):\bar \psi \psi :+
\ve(\partial_\nu +\bar \partial_\nu ):\bar \psi \sigma ^{\mu \nu }\psi :\cr
\a=-\ve{\partial \rho \over{\partial y_\mu }}+
\ve{\partial \over{\partial x^\nu }}:\bar \psi \sigma ^{\mu \nu }\psi :.
 \cr}
 \eno35)
 \ea
The two terms are conserved separately, since

\ba{
{\partial^2\rho  \over{\partial x^\mu \partial y_\mu}}
=i\ve(\del\!\!_\zbar- \del\!\!_z):\a\bar \psi \psi :\,=0\cr
{\partial^2 \over{\partial x^\mu \partial x^\nu  }}
:\bar \psi \sigma ^{\mu \nu }\psi :\,\a=0,
\cr}
 \eno36)
 \ea
and the second term, which is due to spin, does not contribute to the total
charge since it is a pure divergence with respect to $x$.  Thus 

\be
 Q= A_\lambda ^{-1} \int_S\dx_\mu \int _\bl dy\, j^\mu (z)=
\int_S\dx_\mu \,J^\mu _{(\lambda )}(x),
\eno37)
 \ee
where 

\be
 J^\mu _{(\lambda )}(x)\equiv  A_\lambda ^{-1} \int_\bl dy\,\,j^\mu
(x-iy) 
\eno38)
 \ee
is a ``regularized'' spacetime current.  
\skp
{\sl Note:\/} The Dirac equation can also be written in the manifestly
gauge--invariant form

\be
 i\partial_\mu j^\mu (z)=2m\ve:\bar \psi \gamma ^\mu \partial_\mu \psi
:\,= 2m^2\rho(z).  \sp \#
\eno39)
 \ee

\skp
 Again, $\psi $ is a
``root vector'' of the charge operator, since it removes a charge $\ve$ from
any state to which it is applied:

\be
[\psi (z'), Q]=\ve\psi (z')\qquad \forall z'\in\cx\,^4.
\eno40)
 \ee
Substituting for $Q$ the above phase--space integral and using the
commutator identity

\ba{
[A,BC]=\{A,B\}C-B\{A,C\}
 \cr}
 \eno41)
 \ea
and the canonical anticommutation relations,  we obtain
\ba{
\psi (z')=\ins \{\psi (z'),\overline{\psi (z) }\}\,\psi (z)\equiv 
\ins  K_D(z', \zbar)\psi (z),
 \cr}
 \eno42)
 \ea
where the ``reproducing kernel'' for the Dirac field is a matrix--valued
distribution on $\cx^4\times\cx^4$  given by

\ba{
K_D(z',\zbar)\a=\{\psi (z'),\overline{\psi (z) }\}\cr
\a=\inop \Bigl[  \theta (y'p)\,\theta (yp)\,e^{-i(z'-\zbar)p}\,u^\alpha \otimes
\bar u^\alpha \cr
\a\qquad +\theta (-y'p)\,\theta (-yp)\,e^{i(z'-\zbar)p}\,v^\alpha
\otimes \bar v^\alpha\Bigr]\cr
\a=\left( { i\dir'+m\over 2m} \right)K(z',\zbar).
 \cr}
 \eno43)
 \ea
Here,  $K$ is the reproducing kernel for the Klein--Gordon field and $\dir'$
is the Dirac operator with respect to the real part $x'$ of $z'$.  Like $K$,
$K_D$ is piecewise  holomorphic in $z'-\zbar$ for $z', z\in\tb$.  Another form
of the reproducing relation can be obtained by substituting the more
complicated expression for $Q$ given by eq. (37)  into eq. (40):

\ba{
\psi (z')=2mA_\lambda ^{-1} \int_S \dx_\mu \int_\bl
dy\,\,K_D(z',\zbar)\,\gamma ^\mu \,\psi (z).
 \cr}
 \eno44)
 \ea
This form is closer to the usual relation.  
\skp
The energy--momentum and angular momentum operators for the Dirac
field can likewise be represented by phase--space integrals as

\ba{
P_\mu \a=\ins :\bar \psi\, i\partial_\mu \psi :\cr
M_{\mu \nu }\a=\ins :\bar \psi\left( 
ix_\mu \partial_\nu -ix_\nu \partial_\mu
+{\sst{1\over 2}}\sigma _{\mu \nu } \right)\psi :.
 \cr}
 \eno45)
 \ea
More generally, let $\psi (z)$ represent either a Klein--Gordon field (in
which case $\bar \psi $ will mean $\psi ^*$) or a Dirac field, and let $T_a$
be the local generators of an arbitrary {\sl internal or external\/} 
symmetry group, so that the infinitesimal change in $\psi (z)$ is given by

\be
 \delta \psi (z)=-i\epsilon ^aT_a\psi (z).
\eno46)
 \ee  
For example, $T_a$ is multiplication by $\ve$ for $U(1)$ gauge symmetry,
$T_\mu =i\partial_\mu $ for spacetime translations (where the
derivative is with respect to $x^\mu$), etc.  (In case the theory has an
internal symmetry higher than $U(1)$, of course, $\psi $ must have extra
indices since it must be valued in a representation  space of the
corresponding Lie algebra.)  The generators satisfy the Lie relations

\be
 [T_a, T_b]=C_{ab}^c \,T_c,
\eno47)
 \ee
where $C_{ab}^c$ are the structure constants.  Then we claim that the
conserved  {\sl global\/}  field observable corresponding to $T_a$ is

\be
 Q_a=\ins :\bar \psi \,T_a\psi :.
\eno48)
 \ee
For this implies

\be
 [\psi (z'), Q_a]=\ins K_D(z',\zbar)\,T_a\psi (z),
\eno49)
 \ee
where $K_D$ is replaced by $K$ if $\psi $ is a Klein--Gordon field.  Since
$T_a$ generates a symmetry,  it follows that $T_a\psi (z)$ is also a solution of
the appropriate wave equation,   hence it is reproduced by $K_D$:

\be
 \ins K_D(z', \zbar)T_a\psi (z)=T_a\psi (z').
\eno50)
 \ee
Therefore $Q_a$ has the required property

\be
 [\psi (z'), Q_a]=T_a\psi (z').
\eno51)
 \ee
It can furthermore be checked that 

\be
 [Q_a,Q_b]=\ins :\bar \psi\, [T_a, T_b]\,\psi :=C^c_{ab}\,Q_c,
\eno52)
 \ee
hence the mapping $T_a\mapsto Q_a$ is a Lie algebra
homomorphism.  
\skp
Finally, we show that due to the separation of positive and negative
frequencies in $\tb$, the interference effect known as Zitterbewegung does
not occur for Fermions  in the phase--space formalism.  Let $S_t$ be the
configuration space defined by $x^0=t$.  Then the components of the 
``regularized''  three--current at time $t$ are 

\be
 J^k_{(\lambda )}(t)=2mA_\lambda ^{-1} \int_{S_t}d^3 {\bf x} \int_\bl dy
:\bar \psi  \gamma ^k\psi :,
\eno53)
 \ee
and a straightforward computation gives

\be
J^k_{(\lambda )}(t)=\inop \left( { p^k\over m} \right)\left[ b_\alpha ^*(p)
\,b_\alpha (p)-d_\alpha ^*(p)\,d_\alpha (p) \right] .
\eno54)
 \ee
The right--hand side  is independent of $t$, hence no Zitterbewegung
occurs.  In {\sl real\/}  spacetime, Zitterbewegung is the result of the
inevitable interference between the positive-- and negative--frequency
components of $\psi $. Its absence in complex spacetime is due to the
{\sl polarization\/}  of the positive and negative frequencies of $\psi $ into
$\tp$ and $\tm$, respectively.

 In the usual theory, Zitterbewegung is shown to occur in the
single--particle theory;  the above computation can be repeated for the
classical (i.e., ``first--quantized'') Dirac field, with an identical result except
for a change in sign in the second term due to the commutation of $d_\alpha
^*$ and $d_\alpha $.  Alternatively, the above argument also implies the
absence of Zitterbewegung for the one--particle and one--antiparticle states
of the Dirac field. 

\secskp

 \noindent {\bf  5.6.   Interpolating Particle Coherent States  }
\def\rightheadline{\tenrm\hfil {\sl 
 5.6.   Interpolating Particle Coherent States}\hfil\folio}  

 \skp

\noindent We now return to the interpolating charged scalar field $\phi $.
The asymptotic fields satisfy the Klein--Gordon equation,

\be
 (\del+m^2)\,\fin(x)=0,\quad (\del+m^2)\,\fout(x)=0
\eno1)
 \ee
and have the same vacuum expectation values as the free Klein--Gordon
field discussed in section 5.4.  Hence, by Wightman's reconstruction
theorem (Streater and Wightman [1964]), these three fields are unitarily
related.  We identify the free field of section 5.4 with $\fin$.  Then there is a
unitary operator $S$ such that 

\be
 \fout(x)=S\,\fin(x)\,S^*.
\eno2)
 \ee
$S$ is known as the {\sl scattering operator.\/} 
\skp

Define the {\sl source field\/} $j(x)$ by

\be
 j(x)\equiv (\del+m^2)\,\phi (x).
\eno3)
 \ee
It is a measure of the extent of the interaction at $x$, and by axiom 5,

\be
 j(x)\to 0 \quad \hbox{(weakly)  as}\  x^0\to\pm\infty.
\eno4)
 \ee
Note that we are not making any additional assumptions about $j$.  If $j$ is
a known {\sl function\/} (i.e., if it is a multiple of the identity on ${\cal H}$
for each $x$), then it acts as an {\sl external source\/} for $\phi $. 
If, on the other hand, $j$ is a  local   function of $\phi $ such as $:\phi
^3\!:$, it represents a self--interaction of $\phi $.  In any case, the above
equations can be ``solved'' using the  Green functions of the Klein--Gordon
operator, which satisfy

\be
 (\del\!\!_x+m^2)\,G(x)=\delta (x).
\eno5)
 \ee 
In general, we have formally

\be
 \phi (x)=\phi _0(x)+\int dx'\,G(x- x')\,j(x'),
\eno6)
 \ee
where $\phi _0$ is a free field determined by the initial or boundary
conditions at infinity used to determine $G$.  The {\sl retarded \/}  Green
function (we are back to $s$ spatial dimensions) is defined as

\be
 \gret (x)=\inp {e^{-ixp} \over (p_+^2-m^2) },
\eno7)
 \ee
where 

\be
 p_+\equiv (p_0+i\epsilon , {\bf p})
\eno8)
 \ee
with $\epsilon >0$ and the limit $\epsilon \downarrow 0$ is taken after the
integral is evaluated.  $\gret$ propagates both positive and negative
frequencies {\sl forward\/}  in time, which means that it is {\sl causal,\/}
i.e. vanishes when $x_0<0$.  Since it is also Lorentz--invariant, it follows
that 

\be
 \gret(x- x')=0\quad \ \hbox{unless}\  x-x'\in \overline{ V_+'} .
\eno9)
 \ee 
 $\gret(x-x')$ is interpreted as  the causal effect at $x$ due to a unit
disturbance at $x'$. The corresponding choice of free field $\phi _0$   is
$\fin$, hence

\be
 \phi (x)=\fin(x)+\int dx'\,\gret(x- x')\,j(x').
\eno10)
 \ee
If $j$ is a known external source, this gives a complete solution for $\phi
(x)$.  If $j$ is a known function of $\phi $,  it merely gives an integral
equation which $\phi $ must satisfy.

Similarly, the {\sl advanced\/}  Green function is defined by

\be
  \gadv (x- x')=\inp {e^{-i(x-x')p} \over (p_-^2-m^2) },
\eno11)
 \ee
with $p_-\equiv (p_0-i\epsilon , {\bf p})$  and $\epsilon \downarrow 0$,
and propagates both positive and negative frequencies {\sl backward \/}  in
time, which means it is {\sl anticausal.\/}   The corresponding free field is
$\fout$, hence

\be
 \phi (x)=\fout(x)+\int dx'\,\gadv(x- x')\,j(x').
\eno12)
 \ee

Let us now apply the Analytic--Signal transform to both of these equations:

\ba{
\phi (z)\a=\fin(z)+\int dx'\,\gret(z- x')\,j(x')\cr
\phi (z)\a=\fout(z)+\int dx'\,\gadv(z- x')\,j(x'),
 \cr}
 \eno13)
 \ea
where (with $z=x-iy$)

\ba{
 \gret (z- x')\a\equiv \inds \gret (x-\tau y- x')\cr
\a=\inp {\theta (yp)\,e^{-i(z-x')p} \over (p_+^2-m^2) }
\cr}
 \eno14)
 \ea
and 

\ba{
 \gadv (z- x')\a\equiv \inds \gadv(x-\tau y- x')\cr
\a=\inp {\theta (yp)\,e^{-i(z-x')p} \over (p_-^2-m^2) } .
\cr}
 \eno15)
 \ea
 Since the Analytic--Signal transform involves an integration over the entire
line $x(\tau )=x-\tau y$, the effect of $\gret (z-x')$ is no longer causal when
regarded  as a function of $z$ and $x'$.  Rather, it  might be interpreted as  
the causal effect of   a unit  disturbance at $x'$ on the line parametrized by
$z$. (Note that only those values of $\tau $  for which $x-\tau
y-z'\in \overline{V_+' }$ contribute to the integral.)    A similar statement goes
for $\gadv(z-x')$.

Whereas $\fin(z)$ and $\fout(z)$ are holomorphic in $\tb$, $\phi
(z)$ is {\sl not\/} (unless $j(x)\equiv 0$), since $\gret(z-x')$ and
$\gadv(z-x')$ are not holomorphic.  This breakdown of holomorphy in the
presence of interactions is by now expected.   Of course $\phi , \ \gret$ and
$\gadv$ are all holomorphic along the vector field $y$, as are all
Analytic--Signal transforms. 
\skp
In Wightman field theory, the vacua $\Psi _0^{\rm in}$, 
$\Psi _0^{\rm out}$ and $\Psi _0$ of the in--, out-- and interpolating fields
all coincide (the theory is ``already--renormalized'').
Let us define the {\sl asymptotic particle coherent states \/} by

\ba{
\einp\a=\fin\z^*\Psi _0 \cr
\einm\a=\fin\z\Psi _0 \cr
\eoutp\a=\fout\z^*\Psi _0 \cr
\eoutm\a=\fout\z\Psi _0 .
 \cr}
 \eno16)
 \ea
We will refer to 

\be
 e^+_z=\phi \z^*\Psi _0,\quad e^-_z=\phi \z\Psi _0
\eno17)
 \ee
as the {\sl interpolating particle coherent states.\/} By eq. (13), 

\ba{
e_z^+\a=\einp+\int dx'\, \overline{\gret (z-x') }\,j(x')^*\,\Psi _0 \cr
\a=\eoutp+\int dx'\, \overline{\gadv (z-x') }\,j(x')^*\,\Psi _0 
 \cr}
 \eno18)
 \ea
and

\ba{
e_z^-\a=\einm+\int dx'\, \gret (z-x')\,j(x')\,\Psi _0 \cr
\a=\eoutm+\int dx'\, \gadv (z-x')\,j(x')\,\Psi _0 .
 \cr}
 \eno19)
 \ea
From the definitions it follows that 

\ba{
\overline{\gret(z-x') }\a=\gadv(x'-\zbar) \cr
\overline{\gadv(z-x') }\a=\gret(x'-\zbar),
 \cr}
 \eno20)
 \ea
hence eq. (18) can be rewritten as

\ba{
e^+_z\a=\einp+\int dx'\,\gadv(x'-\zbar)\,j(x')^*\,\Psi _0  \cr
       \a=\eoutp+\int dx'\,\gret(x'-\zbar)\,j(x')^*\,\Psi _0 .
 \cr}
 \eno21)
 \ea
Eqs. (19) and (21) display the interpolating character of $e_z^\pm$.
Note that when $j(x)$ is an external source, then the interpolating particle
coherent states differ from the asymptotic ones by a  multiple of the
vacuum.
\skp

 \skp
 As in the case of the free theory,  a general state with
a single positive charge $\ve$ can be written in the form

\be
 \Psi _f^+=\phi ^*(f)\,\Psi _0.
\eno22)
 \ee
For interacting fields, this may, in general,  no longer be interpreted as a
one--particle state, since no particle--number operator exists.\footnote*{If
the spectrum $\Sigma $ contains an isolated mass shell $\omp$ and
$\ftil(p)$ is concentrated around $\omp$, then $\Psi _f^+$ is, in fact, a
one--particle state.  This is the starting point of the Haag--Ruelle scattering
theory (Jost [1965]).  I thank R.~F.~Streater for this remark.} 
 But the charge operator does exist since charge (unlike particle--number) is
conserved in general, due to gauge invariance; hence $\Psi _f^+$ makes sense
as an eigenvector of charge with eigenvalue $\ve$.   $\Psi _f^+$ can be
expressed in terms of particle coherent states  as

\ba{
\ftil(z)\a\equiv  \l\, e_z^+\,|\,\Psi _f^+\,\r\cr
\a=\l\,\einp\,|\,\, \Psi _f^+\,\r  +
\int dx'\,\gret(z-x')\,\l\,\Psi _0\,|\,j(x')\,\Psi _f^+\r  \cr
 \a=\l\,\eoutp\,|\,\, \Psi _f^+\,\r  +
\int dx'\,\gadv(z-x')\,\l\,\Psi _0\,|\,j(x')\,\Psi _f^+\r . \cr
\cr}
 \eno23)
 \ea
$\ftil(z)$ satisfies the inhomogeneous equations

\ba{
(\del\!\!_x+m^2)\,\ftil(z)\a=\int dx'\,(\del\!\!_x+m^2)\,\gret(z-x')\,
\langle\, \Psi _0\,|\,j(x')\,\Psi _f^+\,\rangle\cr
\a=\int dx'\,(\del\!\!_x+m^2)\,\gadv(z-x')\,
\langle\, \Psi _0\,|\,j(x')\,\Psi _f^+\,\rangle.
 \cr}
 \eno24)
 \ea
But from the definitions it follows that 

\ba{
(\del\!\!_x+m^2)\,\gret(z-x')\a=(\del\!\!_x+m^2)\,\gadv(z-x')\cr
\a=\inp \theta (yp)\,e^{-i(z-x')p}\cr
\a\equiv \delta (z-x'),
 \cr}
 \eno25)
 \ea
where the last equation is a {\sl definition\/}  of $\delta (z-x')$ as the
Analytic--Signal transform with respect to $x$ of $\delta (x-x')$.  
 The above is easily seen to reduce to

\be
 (\del\!\!_x+m^2)\,\ftil(z)=\langle\, \Psi _0\,|\,j(z)\,\Psi _f^+\,\rangle,
\eno26)
 \ee
where $j(z)$ is the Analytic--Signal transform of $j(x)$. 
Equivalently, eq. (3) can be extended to $\cx^{s+1}$ by applying the
Analytic--Signal transform, giving

\be
 (\del\!\!_x+m^2)\,\phi (z)=j(z),
\eno27)
 \ee
hence

\be
 (\del\!\!_x+m^2)\,\ftil(z)=\langle\, \Psi _0\,|\,(\del\!\!_x+m^2)\,\phi (z)\,
\,|\,\Psi _f^+\,\rangle=\langle\, \Psi _0\,|\,j(z)\,\Psi _f^+\,\rangle.
\eno28)
 \ee
For a known external source, this is a ``perturbed'' Klein--Gordon equation
for $\ftil(z)$;  if $j$ depends on $\phi $, it appears to be of little value.

\secskp

 \noindent {\bf 5.7.  Field Coherent States and Functional Integrals }
\def\rightheadline{\tenrm\hfil {\sl 
5.7.  Field Coherent States and Functional Integrals}\hfil\folio}  

 \skp
\noindent So far, all our coherent states have been states with a single
particle or antiparticle.  In this section, we construct coherent states in
which the entire field participates, involving an indefinite number of
particles.   We do so first for a neutral free  Klein--Gordon field
(or a generalized free field; see section 5.3), then for a free charged scalar 
field.  A similar construction works for  Dirac fields, but the  ``functions'' 
labeling the coherent states must then  anticommute instead of being
``classical'' functions and a generalized  type of functional integral 
 must be used (Berezin [1966], Segal [1956b, 1965]). We also indulge in some
speculation on generalizing the construction to interpolating fields.

    An extended neutral free  Klein--Gordon field  satisfies
the canonical commutation relations  

\ba{
[\phi (z), \phi (z')]\a=0\cr
[\phi (z), \phi (z')^*]\a=K(z, \zbar')=-i\Delta ^+(z-\zbar')
 \cr}
 \eno1)
 \ea
for all $z, z'\in\tp$, as well as the reality condition $\phi (z)^*=\phi
(\zbar)$.  
 The basic idea is that since all the operators
$\phi (z)$ ($z\in\tp$) commute, it {\sl may\/}  be possible to
find a total set of simultaneous eigenvectors for them.  This is not guaranteed,
since $\phi (z)$ is not self--adjoint (it is not even {\sl normal,\/} by eq. (1))
and, in any case, it is unbounded and thus may present us with domain
problems.  However, this hope is realized by  explicitly constructing such
eigenvectors. This construction mimics that of the canonical coherent states 
in section 3.4, which used the  lowering and raising operators $A$ and
$A^*$.  As in the case of finitely many degrees of freedom, the canonical
commutation relations  mean that $\phi ^*$ acts as a {\sl generator of
translations\/}  in the space in which $\phi $ is ``diagonal.''  The
construction  proceeds as follows: 
 Let $\fhat({\bf p})$ be a function on
$\rl^s$, which will also be regarded as a function on $\omp$. To
simplify the analysis, we assume to begin with that $\fhat$ is a
(complex--valued) Schwartz test function, although this will be relaxed later.
$\fhat$ determines a   holomorphic  positive--energy solution of the
Klein--Gordon equation,

\be
 f(z)=\inop e^{-izp}\,\fhat({\bf p}).
\eno2)
 \ee
Define

\be
 \phi ^*(f)\equiv \inop a^*(p)\,\fhat({\bf p})=\insp \phi (z)^*\,f(z),
\eno3)
 \ee
where $\sigma _+$ is any particle phase space  and  the second equality
follows from theorem 4.10 and its corollary.  ({\sl Note:\/}  this is {\sl not\/} 
the same as the smeared field in real spacetime, since the latter would
involve an integration over time, which diverges when $f$ is itself a
solution rather than a test function in spacetime.)  The canonical commutation
relations imply  that for $z\in\tp$, 

\be
 [\phi (z), \phi ^*(f)]=\insp\!(z') K(z, \zbar')\,f(z')=f(z),
\eno4)
 \ee
and for $n\ge 1$,

\be
 [\phi (z), \phi ^*(f)^n]=f(z)\cdot n\phi ^*(f)^{n-1}.
\eno5)
 \ee
We now define the {\sl field coherent states \/}  of $\phi $ by the formal
expression

\ba{
E^f=e^{\phi ^*(f)}\,\Psi _0.
 \cr}
 \eno6)
 \ea
Then if $z\in\tp$, so that $\phi (z)\,\Psi _0=0$, eq. (5) implies that

\ba{
\phi (z)\,E^f\a=[\phi (z), e^{\phi ^*(f)}]\,\Psi _0\cr
\a=f(z)\,E^f.
 \cr}
 \eno7)
 \ea
Hence $E^f$ is a common eigenvector of all the
operators  $\phi (z),\ z\in{\tp}$. This eigenvalue equation
implies that  the {\sl state\/} corresponding to $E^f$ is left unchanged by the
removal of a single particle, which requires that $E^f$ be a superposition of
states with  $0, 1,2,  \cdots$ particles.   Indeed, 

\be
 E^f=\sum_{n=0}^\infty {1 \over n!}\,\phi ^*(f)^n\,\Psi _0.
\eno8)
 \ee
The projection of $E^f$ to  the one--particle subspace can be obtained by
using the particle coherent states $e_z$:

\ba{
 \langle\, e_z\,|\,E^f\,\rangle\a=\langle\, \Psi _0\,|\,\phi (z)\,E^f\,\rangle\cr
\a=f(z)\langle\, \Psi _0\,|\,E^f\,\rangle=  f(z),
 \cr}
 \eno9)
 \ea
where the last equality follows from $\phi (\bar f)\,\Psi _0\equiv 
(\phi ^*(f))^*\,\Psi _0=0$.   More generally, the $n$--particle component of
$E^f$ is given by projecting to the $n$--particle coherent state

\be
 e_{z_1z_2\cdots z_n}\equiv \phi (z_1)^*\phi (z_2)^*\cdots\phi (z_n)^*\,\Psi
_0,
 \eno10)
 \ee
which gives 

\be
 \langle\, e_{z_1z_2\cdots z_n}\,|\,E^f\,\rangle=f(z_1)f(z_2)\cdots  f(z_n),
\eno11)
 \ee
 so all particles are in the same state $f$ and  the entire system of
particles is {\sl coherent!\/}     Similar  states have been found to be very
useful in the analysis of the phenomenon of coherence in quantum optics
(Glauber [1963], Klauder and Sudarshan [1968]),  where the name ``coherent
states'' in fact originated.  In the usual treatment, the positive--frequency
components have to be separated out ``by hand'' using their Fourier
representation, since one is dealing with the fields in {\sl real\/}  spacetime. 
For us, this separation occured automatically though the use of the 
Analytic--Signal transform, i.e. $\phi ^*(f)$ can be defined directly as an
integral of $f(z)$ over $\sigma _+$. (This would remain true even if $f$ had a
negative--frequency component, since  the integration over $\sigma _+$
would still restrict $f$ to  positive frequencies.)

 \skp
The inner product of two field coherent states can be computed as
follows.  Note first that if $g(z)$ is another positive--energy  solution, then

\ba{
\phi (\bar f)\,E^g\a\equiv \insp \overline{f(z) }\,\phi (z)\,E^g\cr
\a=\insp\overline{f(z) }\,g(z)\,E^g\cr
\a=\langle\, f\,|\,g\,\rangle\,E^g,
 \cr}
 \eno12)
 \ea
where, by theorem 4.10,

\be
 \langle\, f\,|\,g\,\rangle\equiv \insp \overline{ f(z)}\,g(z)=\inop
\overline{ \fhat({\bf p})}\,\ghat({\bf p}).
\eno13)
 \ee
 Hence

\ba{
\langle\, E^f\,|\,E^g\,\rangle\a=\langle\, \Psi _0\,|\,e^{\phi (\bar
f)}\,E^g\,\rangle\cr
\a=e^{\langle\, f\,|\,g\,\rangle}.
 \cr}
 \eno14)
 \ea
Thus $E^f$ belongs to ${\cal H}$ (i.e., is normalizable) if and only if
$\fhat({\bf p})$ belongs to $L^2_+(d\tilde p)$ or, equivalently, $f(z)$ belongs
to the one--particle space ${\cal K}$ of holomorphic positive--energy
solutions. If we   suppose  this to be the case for the time being, 
then the field coherent states $E^f$ are parametrized by the vectors
$\fhat\in L^2_+(d\tilde p)$ or   $f\in \cal K$.  Next, we look for  a resolution
of unity in ${\cal H}$ in terms of the $E^f$'s. The standard procedure
(section 1.3) would be to look for an appropriate measure $d\mu(f) $ on
$\kc$.    Actually, it turns out that due to the infinite dimensionality of 
$\kc$,  a larger space $\kcp\supset\kc$ will be needed to support $d\mu $.
Thus, for the time being, we leave the domain of integration unspecified and
write  formally

\be
\intf\,|\,E^f\,\rangle\langle\,E^f\,|\,=I_{\cal H},
\eno15)
 \ee
where $d\mu $ is to be found.
Taking the matrix element of this equation
between the states $E^h$ and $E^g$, we obtain

\ba{
\intf\, e^{ \langle\,  h\,|\,f\,\rangle+
\langle\,  f\,|\,g\,\rangle }=e^{\langle\,  h\,|\, g\,\rangle}.
 \cr}
 \eno16)
 \ea
With $h=-g$ this gives

\ba{
\intf\, e^{ \langle\,  f\,|\,g\,\rangle-
\langle\,  g\,|\,f\,\rangle }=e^{-\langle\,  g\,|\, g\,\rangle}\equiv S[g].
 \cr}
 \eno17)
\ea
The left--hand side  is an infinite--dimensional version of the Fourier
transform of  $d\mu $, as becomes apparent if we decompose 
$f$ and $g$ into their real and imaginary parts.  The Fourier transform of a
measure is called its {\sl characteristic function.\/}   Hence we conclude that
a {\sl necessary\/}  condition for the existence of $d\mu $ is that its
characteristic function  be $S[g]$.  In turn, a function must
satisfy certain conditions in order to be the characteistic function  of a
measure.  In the finite--dimensional case, Bochner's theorem (Yosida [1971])
guarantees the existence of the measure if these conditions are satisfied. 
If the infinite--dimensional space of $f$'s is replaced by $\cx^n$, the above
relation would uniquely determine $d\mu $ as a Gaussian measure.  For the
identity

\be
 \int_{\,\cxs^n}\det A^{-1} d^{2n}\zeta \,\exp[-\pi (\zeta -A\xi )^*\,A
^{-1}  (\zeta -A\xi )]=1,
 \eno18)
 \ee
where  $A$ is a positive--definite matrix,  implies

\be
 \int_{\,\cxs^n}d\mu (\zeta )\,e^{\pi (\zeta ^*\xi +\xi ^*\zeta) }
=e^{\pi \xi ^*A\xi },
\eno19)
 \ee
with 

\be
 d\mu (\zeta )=\det A^{-1}\, \exp[-\pi \zeta^* A ^{-1} \zeta ]\,d^{2n}\zeta .
\eno20)
 \ee
The integral in eq. (19) is entire in the variables $\xi $ and $\xi ^*$ {\sl
separately,\/}  hence it can be analytically continued to $\xi ^*\to-\xi^* $,
giving

\be
  \int_{\,\cxs^n}d\mu (\zeta )\,e^{\pi (\zeta ^*\xi -\xi ^*\zeta) }
=e^{-\pi \xi ^*A\xi }.
\eno21)
 \ee
If $\zeta =\alpha +i\beta $ and $\xi =u+iv$ with $\alpha ,\beta ,u,
v\in\rl^n$, then

\be
 \int_{\rl^{2n}} d\mu (\alpha ,\beta )\,e^{2\pi i(\alpha v-\beta u)}
=e^{-\pi (uAu+vAv)}\equiv S(v,-u),
\eno22)
 \ee
showing that $S(v,-u)$ is  the Fourier transform of $d\mu (\alpha
,\beta )$.

If $A$ is merely positive--semidefinite, i.e., if it is singular, then
$d\mu $ still exists but is concentrated on the {\sl range\/}  of $A$, and $A
^{-1} $ makes sense as a map from this range to the orthogonal complement
of the kernel of $A$.
\skp

 Eq. (19) is a finite--dimensional version of eq. (16) (with $g=h$).  In going to
the infinite--dimensional case, two separate complications arise.    
 First of all, recall that in finite dimesions, the Fourier transform takes
funcions on $\rl^n$ to functions on the dual space, $(\rl^n)^*$ (section 1.1). 
One usually identifies these two spaces by choosing an inner product on
$\rl^n$,  e.g., the Euclidean inner product.  In the infinite--dimensional
case, it is tempting to extend Bochner's theorem by letting $\rl^n$ go to a
Hilbert space and looking for a measure on this space.  However, Segal
[1956a, 1958] has shown that the ensuing $d\mu $ cannot be a Borel
measure, since it is only finitely additive.  To obtain a Borel measure, the
domain of integration must be expanded to a space of distributions, its dual
then being a space of test functions.  Minlos' theorem (Gel'fand and Vilenkin
[1964]; Glimm and Jaffe [1981]) states that
 if a functional $S[g]$ defined  on the Schwartz space of test functions ${\cal
S}(\rl^s)$ satisfies  appropriate conditions (positive--definiteness,
normalization and continuity), a Borel probability measure $d\mu $ exists on
the dual space of tempered distributions ${\cal S}'(\rl^s)$ such that eq. (17) is
satisfied for all $g\in {\cal S}(\rl^s)$, the integration being over  ${\cal
S}'(\rl^s)$.  This resolves the problem of infinite dimansionality.  In our case,
however, the situation is further complicated by the fact that the space of
$f$'s over which we wish to integrate, even if it is enlarged, consists not of
free functions but of solutions of the Klein--Gordon equation.  This difficulty
can be overcome by first applying Minlos' theorem in the momentum space
representation, where 
$S[\hat g]\equiv\exp\left( -\|\hat g\|^2_{L^2(d\tilde p)} \right)$
satisfies the necessary conditions for $\hat g\in{\cal S}(\rl^s)$. This gives a
probability measure $d\tilde\mu(\fhat)$ on ${\cal S}'(\rl^s)$.  We then
define the  spaces

\ba{
 \kco\a\equiv \left\{g\z=\inop e^{-izp}\,\hat g({\bf p})\ \,|\,\hat g\in {\cal
S}\,\right\}\cr
 \kcp\a\equiv \left\{f\z=\inop e^{-izp}\,\fhat({\bf p})\ \,|\,\fhat\in {\cal
S}'\,\right\}.
 \cr}
 \eno23)
 \ea
These may be regarded as mutually dual, under the sesquilinear pairing 

\be
 \l\,f, g\,\r\equiv \l\,\fhat, \hat g\,\r\equiv \inop \,\overline{\fhat (p)}\,
\hat g(p),\quad \fhat\in {\cal S}',\, \hat g\in{\cal S}.
\eno24)
 \ee
Together with $\kc$, they form a ``triplet''

\be
 \kco\subset\kc\subset\kcp.
\eno25)
 \ee
We now  use the map $\fhat\mapsto f$ to transfer the measure from ${\cal
S}'$ to $\kcp$, obtaining a probability measure $d\mu $ on $\kcp$.  This
results, finally, in the resolution of unity 

\be
 \intfp \,|\,E^f\,\r\l\,E^f\,|\,=I,
\eno 26)
 \ee
where the integral converges, as usual, in the weak operator topology.
$d\mu $  is Gaussian in the sense that its restrictions to finite--dimensional
cylinder sets in $\kcp$ are all Gaussian measures.  It is, therefore, an
infinite--dimensional version of the Gaussian measure on $\cx^s$ which
gave the resolution of unity for the canonical coherent states in section 1.2. 

\skp 

One might well ask what is the point of insisting that the integration take
place over $\kcp$ rather than ${\cal S}'(\rl^s)$, the momentum space
representation.  One reason is esthetic:  The vectors $E^f$ combine the {\sl
finite--dimensional\/}  (particle) coherent--state representation  with the
{\sl infinite--dimensional\/}  (field) coherent--state representation. 
Another reason is that whereas the ``sample points'' $\fhat$ in 
${\cal S}'(\rl^s)$ are merely distributions, the elements $f$ in $\kcp$ are
holomorphic functions, since the decaying exponential $e^{-yp}$ dominates
the singular behavior of $\fhat$, just as  it did when 
$\fhat\in L^2(d\tilde p)$.  Note, however, that non--normalizable field
coherent states $E^f$ now enter the resolution of unity.  In fact, the Hilbert
space $\kc$ has measure zero with respect to $d\mu $, since 
$L^2(d\tilde p)$ has measure zero with respect to $d\tilde \mu $.  This is
remedied by the fact that only vectors $h, g$ in the test function space ${\cal
K}_0$ are now allowed in eq. (16).
\skp

With the resolution of unity provided by the field coherent states, the
inner product in $\hc$ can be represented by the {\sl functional integral\/} 

\ba{
 \langle\, \Phi \,|\,\Psi \,\rangle_\hc\a=\intfp \langle\, \Phi \,|\,E^f\,\rangle
\langle\, E^f\,|\,\Psi \,\rangle\cr
\a\equiv \intfp\,\overline{ \Phi [f]}\,\Psi [f].
 \cr}
 \eno27)
 \ea
\skp  
The above construction was for a neutral scalar field.  If $\phi $ were
{\sl charged,\/}  its coherent states would take the form

\be
 E^{f,\gbar}=e^{\phi ^*(f)+\phi (\gbar)}\,\Psi _0,
\eno28)
 \ee
where $\phi ^*(f)$ is as before and $\phi (\gbar)$ is an antiparticle creation
operator,

\be
 \phi (\gbar)\equiv \insm\phi (z)\,\gbar(z),
\eno29)
 \ee
which commutes with $\phi ^*(f)$. $\gbar(z)$ is a negative--energy solution
of the Klein--Gordon equation, holomorphic in $\tm$, or, equivalently, 
$\gbar(z)=\overline{ g(\zbar)}$ with $g\in\kcp$.  Thus we write
$\gbar\in\overline{\kcp }$.   The  resolution of unity for charged fields is
therefore

\be
\int_{\kcp\times\overline{\kcp} }d\mu (f,\gbar)
\,|\,E^{f, \gbar}\,\rangle\langle\, E^{f, \gbar}\,|\,=I_{\cal H},
\eno30)
 \ee
where $d\mu (f,\gbar)=d\mu (f)\,d\mu (\gbar)$ is the tensor product of
two Gaussian measures  defined as above.
 \skp
Finally, it is reasonable to ask whether coherent states
exist for an {\sl interpolating\/}  field, by analogy with the interpolating
particle coherent states studied in sections 5.3 and 5.6.  A {\sl necessary\/} 
condition would seem to be that the first half of the canonical commutation
relations  still be valid, i.e.,

\be
 [\phi (z), \phi (z')]=0
\eno31)
 \ee
for $z, z'\in\tp$, since one would like to find simultaneous eigenvectors of
$\phi (z)$ for all $z\in\tp$.  Recall that the extended {\sl free\/}  neutral
scalar field had the form

\be
 \phi (z)=\inop\left[\theta^{-izp}\,a(p) +\theta^{izp}\,a^*(p) \right]
\eno32)
 \ee
for arbitrary $z\in\cx^{s+1}$, and the commutation relation given by eq. (31)
was due to the polarization of the positive and negative frequencies into
$\tp$ and $\tm$, respectively.   If interactions are introduced, the positive--
and negative--frequency components get inextricably mixed together,
hence it is highly unlikely that the above commutation relation
survives.  However, a {\sl charged\/}  free field has the form

\be
 \phi (z)=\inop\left[\theta^{-izp}\,a(p) +\theta^{izp}\,b^*(p) \right],
\eno33)
 \ee
where $b^*$ commutes with $a$, hence

\be
 [\phi (z), \phi (z')]=0\qquad \forall z,z'\in\cx^{s+1} .
\eno34)
 \ee
I believe that {\sl this\/}  relation does have a chance of holding for
interpolating charged scalar fields.  It would be a consequence, for example,
of the physical requirement that the Lie algebra generated by  the
field has no operators which remove (or add) a double charge $2\ve$. This
commutation relation is the {\sl weaker half\/}  of the free--field canonical
commutation relations, the stronger half (which we do {\sl not\/}  assume)
being that $[\phi (z), \phi (z')^*]$ is a ``c--number,'' i.e. a multiple of the
identity.    If $[\phi (z), \phi (z')]=0$ for all $z$ and $z'$, then it makes
sense to look for common eigenvectors of all the $\phi (z)$'s, 
which would be coherent states of the  interpolating field.

\secskp

 \noindent {\bf Notes }
\def\rightheadline{\tenrm\hfil {\sl   Notes}\hfil\folio}  
\skp

\noindent Most of the results in sections 5.2--5.5 were announced in Kaiser
[1987b] and have been published in Kaiser [1987a].  An earlier attempt to
describe quantized fields in complex spacetime was made in Kaiser [1980b]
but was found  to be unsatisfactory.   The Analytic--Signal transform is
further studied in Kaiser [1990c]. 

\skp

Segal [1963b]  proposed a formulation of quantum field theory  in terms of
the symplectic geometry of the phase space of classical fields. 
 This phase space  corresponds, roughly,  to the space ${\cal K}_0'$ defined in
section 5.7.  An attempt to study quantized fields as (operator--valued)
functions on the Poincar\'e group--manifold has been made by Lur\c cat
[1964];  see also Hai [1969].  As mentioned in section 4.2, this manifold may
be regarded as an extended phase space which includes spin degrees of
freedom in addition to position-- and velocity coordinates.  In the context of
{\sl classical\/}  field theory, this point of view has been generalized to
curved spacetime by replacing the Poincar\'e group--manifold with the
orthogonal frame bundle over a Lorentzian spacetime (Toller [1978]).
 These efforts have not, however,
utilized holomorphy.  It may be interesting to expand the point of view
advocated here to a complex manifold containing the Poincar\'e
group--manifold  in order to account naturally for spin. This might result in
a ``total'' coherent--state representation  where the classical phase space 
coordinates range over $\tb$ and the spin phase space  coordinates range
over the Riemann sphere, as in section 3.5. 

\skp

 I owe special thanks to R.~F.~ Streater for  many important
comments and corrections in this chapter. 

\VE

\def\leftheadline{\tenrm\folio\hfil {\sl 6. Further Developments}\hfil} 
\def\rightheadline{\hfill\folio}

\headline={\ifodd\pageno\rightheadline\else\leftheadline\fi}
\def\be{$$}\def\ee{$$}\def\ba{$$\eqalign}\def\a{&}  
\def\eno{\eqno(}\def\ea{$$}

\centerline{\bf Chapter 6}\skp
\centerline{\bf FURTHER DEVELOPMENTS}
\vskip 3 cm
\noindent {\bf  6.1. Holomorphic Gauge theory}
\skp

\noindent In this section we give a  brief, not--very--technical but
hopefully intuitive, discussion of gauge theory and indicate how it may be
modified in order to make sense in complex spacetime.  This represents
work still in progress, and our account is accordingly incomplete. One
obstacle is the absence, so far, of a satisfactory Lagrangian formulation. 
This is due in part to the fact that fields in complex spacetime are
constrained since they can be derived from local fields in $\st$ (chapter
5).    Our treatment is as elementary as possible, with a geometrical
emphasis.  We   ignore global questions and work within a single chart.
 For more details, see Kaiser [1980a, 1981].
 \skp

Gauge theory is a natural, geometric way of introducing interactions.  It
applies some of the ideas of General Relativity to  quantum
mechanics and arrives at a class of theories which are generalizations of
classical electrodynamics, the latter being the simplest case.  The power of
Einstein's theory of gravitation owes much to  the fact that it actually
assumes {\sl less,\/}  initially,  than its predecessor,  Newton's theory.   By
dropping the assumption that spacetime is flat, we lose the ability to
transport tangent vectors from one point in spacetime to another, as is done
when differentiating a vector field or even finding the acceleration of a
particle moving in spacetime.  To regain it, we need a {\sl connection.\/} 
We need to know how a vector transforms in going from any given point
$x_0$ to a neighboring point along a curve $x\t$.  The tangent vector to the
curve at  the point  $x_0$ is 

\be
 X=\dot x^\mu \partial_\mu .
\eno1)
 \ee
(The partial--derivative operators $\partial_\mu =\partial/\partial x^\mu $
form a basis for the tangent space at $x_0$;  see Abraham and Marsden
[1978].)   An infinitesimal transport must have a linear effect, thus a
vector $Y$ at $x_0$ should change by 

\def\rightheadline{\tenrm\hfil {\sl 
 6.1. Holomorphic Gauge theory}\hfil\folio}

\be
 \delta Y=\Gamma (X) Y\,\delta t,
\eno2)
 \ee
where $\Gamma(X)$ is a linear transformation on the tangent space at
$x_0$. Furthermore, $\Gamma(X)$ must be linear in $X$ since the latter
is an infinitesimal (i.e., linearized) description of the curve at $x_0$.  If
$Y=Y^\nu \partial_\nu $, this gives

\be
\delta Y=\dot x^\mu Y^\nu \Gamma(\partial_\mu )\partial_\nu \,\delta t.
\eno3)
 \ee
Since $\Gamma (\partial_\mu )$ is a  linear transformation, we have

\be
 \delta Y=\dot x^\mu Y^\nu \Gamma^\kappa _{\mu \nu}\,\partial_\kappa 
\,\delta t,
\eno4)
 \ee
where $\Gamma^\kappa _{\mu \nu}$ are a set of (locally defined) funcions
on spacetime, known as the {\sl connection coefficients.\/}   This gives  the 
rate of change of $Y$  due to transport along $X$.  

Now suppose we are given a {\sl metric\/}  $g$ on spacetime.  (In Relativity,
the ``metric'' is indefinite, i.e. Lorentzian rather than Riemannian;  with this
understood, we continue to call it a metric.)    If two vectors are 
transported along a curve, then their inner product must not change  since
it is a scalar.   This gives a relation betweeen $\Gamma$ and $g$ which
determines the symmetric part (with respect to exchange of $X$ and $Y$)
of $\Gamma$.  The antisymmetric part is the {\sl torsion,\/}  which in the
standard theory is assumed to vanish.  Hence the metric uniquely
determines a torsionless connection known as the {\sl Riemannian
connection.\/}

  General Relativity (see Misner, Thorne and Wheeler [1970]) relates the 
Riemannian connection to  Newton's gravitational potential, thus giving
gravity  a geometric interpretation. This is reasonable, since the connection
determines the  acceleration of a particle moving freely (i.e., along a
geodesic) in spacetime, which is related to gravity.   By assuming less to
begin with, one discovers what must  necessarily  be added to even {\sl
compute\/}  the acceleration, and this additional structure turns out to
include  gravity!   In   Newton's theory, gravity must be added in an 
ad--hoc fashion.  The two theories coincide in the non--relativistic limit. 

 \skp

Now consider the wave function of a (scalar) quantum particle, for example 
a solution of the free Klein--Gordon equation, in {\sl real\/}  spacetime
$\st$.  Suppose   we drop the usually implicit assumption that $f $ can be
differentiated by simply taking the difference
between  its values at neighboring points.  Instead of regarding $f\x$ as a
complex {\sl number,\/}  we now regard it  as a one--dimensional
complex {\sl  vector\/}  attached to $x$, analogous to the tangent vectors
in Relativity.   This means assuming less structure to begin with, since $f\x$
now belongs to $\cx$ as a {\sl vector space\/}  rather than as an {\sl
algebra.\/}  The complex plane attached to $x$ will be denoted by $\cx_x$
and is called the {\sl fiber\/}  at $x$, and the set of all fibers is called a {\sl
complex line bundle\/}  over $\st$ (Wells [1980]).  To differentiate $f$, we
must know how it is affected by transport. The situation is similar to the
one above, only now $\Gamma(X)$ must be a $1\times 1$ complex matrix,
i.e. a complex number.  An infinitesimal transport gives

\be
 \delta f=\Gamma (X)f\,\delta t=\dot x^\mu \Gamma_\mu f\,\delta t,
\eno5)
 \ee
where $\Gamma_\mu \x$ is  a complex--valued function. The total
rate of change  along $X$ is therefore

\ba{
 \dot x^\mu \partial_\mu f+\dot x^\mu \Gamma_\mu f\equiv D_Xf,
 \cr}
 \eno6)
 \ea
where $D_X$ is called the {\sl
covariant derivative along $X$.\/}   Equivalently, the
differential change is

\be
 Df=df+\Gamma f,\quad \ \hbox{where}\  \quad \Gamma\equiv 
\Gamma_\mu\,dx^\mu .
\eno7)
 \ee
The 1--form $\Gamma$ is called the {\sl connection form.\/} $Df$ is the
sum of a ``horizontal'' part $df$ (which measures change due to the
dependence of $f$ on $x$) and a ``vertical'' part (which measures change
due to transport).  A {\sl gauge transformation\/}  is represented locally by
a multiplication by a variable phase factor, i.e. 

\be
 f\x\mt e^{i\chi  \x}\,f\x.
\eno8)
 \ee
This is a linear map on each fiber $\cx_x$, which corresponds in Relativity
to the linear map on tangent spaces induced by a coordinate
transformation. In fact, since there is  no longer any natural way to identify
distinct fibers, a gauge transformation {\sl is\/}  a coordinate
transformation of sorts.
 We therefore require that $Df$ be invariant under gauge
transformations, which implies that $\Gamma$ transforms as

\be
 \Gamma\mt \Gamma-id\chi  .
\eno9)
 \ee
Suppose now that we try to complete the analogy with Relativity by
deriving $\Gamma$ from a Hermitian metric  on the fibers
such that the scalar product of two vectors remains invariant under
transport.  If we assume the metric to be positive--definite, it must have
the form

\be
 (f\x, g\x)=\bar f\x\,h\x\, g\x,
\eno10)
 \ee
where $h\x$ is a positive function.  It will suffice to consider the inner
product of $f\x$ with itself, i.e. the quantity

\be
 \rho \x\equiv (f\x, \,f\x)=\overline{ f\x}\,h\x \,f\x.
\eno11)
 \ee
We  require that $\rho $ be invariant under transport.  This means that $(f,
f)$ changes only due to its dependence  on $x$, i.e.

\be
 (Df,\, f)+(f, \,Df)=d(f, f).
\eno12)
 \ee
It follows that 

\ba{
\overline{ \Gamma}\,h+h\Gamma=dh,
 \cr}
 \eno13)
 \ea
which constrains the real part of $\Gamma$ but leaves the imaginary part
arbitrary.  Writing $\Gamma= R+iA$, where $R$ and $A$ are real 1--forms, 
we have

\be
 2R=d \log h.
\eno14)
 \ee
The real part $R$ of the connection can be transformed away by defining 
$\ftil=h^{1/2}f$ and  $\tilde h\equiv 1$, which gives $\tilde \rho =\rho $
and

\be
 Df=h^{-1/2}\,(d\ftil+iA\ftil)\equiv h^{-1/2}\,(d+\tilde \Gamma)\ftil.
\eno15)
 \ee
Since $\rho =\overline{\ftil }\,\ftil$, we may as well assume from the
outset that $\rho =|f|^2$ and $\Gamma=iA$ is purely imaginary. 
\skp
\noindent {\sl Note:\/}  The mapping $f\mt\ftil$  is {\sl not\/}  a gauge
transformation in the usual sense; in the standard gauge theory the metric
is assumed to be constant ($h\x\equiv 1$), hence only phase translations
are allowed. This corresponds to having already transformed away $R$.   It
turns out that in phase space, it will be natural to admit non--constant
metrics.

 \skp

To make the Klein--Gordon equation invariant under gauge
transformations,  we now replace $\partial _\mu $ by
$D_\mu=\partial_\mu +iA_\mu $.  The result is

\be
(  D^\mu D_\mu+m^2) f=( \partial^\mu +iA^\mu )
( \partial_\mu +iA_\mu )f+m^2f=0.
\eno16)
 \ee
This equation was known (even before gauge theory) to be a relativistically
covariant description of a Klein--Gordon particle in the presence of the
  electromagnetic field determined by the {\sl vector potential\/} 
$A_\mu \x$.  Hence the connection, which describes a geometric property of
the  the complex line bundle, acquires a  physical significance with respect
to electrodynamics, just as did the connection $\Gamma$ with respect to
gravitation.   When $f$ is differentiated in the usual way,  it is
unconsciously assumed that the connection vanishes.  Coupling to an
electromagnetic field then has to be put in ``by hand,''  through the
substitution $\,\partial_\mu \to \partial_\mu +iA_\mu $, which is known as
the  {\sl minimal coupling\/} prescription.  Gauge theory gives this ad--hoc
prescription a geometric interpretation.  But note that in this case, the fiber 
metric $h\x$ did {\sl not\/}  determine the connection.  This is due to the
complex structure:  $iAf$ cancels in the inner product  because it is
imaginary.   The electromagnetic field generated by the potential $A$ is
given by the 2--form

\be
 F=dA,
\eno17)
 \ee
which in fact measures the non--triviality of the connection form $A$:  if
$F=0$, then $A$ is closed and therefore (locally) exact, i.e. it is due purely to
a choice of gauge.  This is analogous in Relativity to choosing an
accelerating coordinate system, which gives the illusion of gravity.

\skp
 Note the complementary nature of the two theories:  in Relativity, the skew
part of the connection, which is the torsion, is assumed to vanish.  In gauge
theory, the inner product becomes Hermitian, and the symmetric and
antisymmetric parts correspond to its real and imaginary parts,
respectively.  It is the {\sl real\/}  part of the connection which is assumed
to vanish in gauge theory.  Were $\Gamma$ required to be real, it could in
fact be transformed away as above, giving rise to no gauge field.
That is, only the  {\sl  trivial\/}   part of the connection  is determined by
the metric.  The non--trivial (imaginary) part is arbitrary. 
 We will see that when  the theory is extended to complex
spacetime, the metric {\sl does\/}  determine a non--trivial connection.

 \skp
Gauge theory usually begins with a Lagrangian invariant under phase
translations $f\mt e^{i\phi }f$, which form the group $U(1)$.  The equations
satisfied by $f$ and $A_\mu$ are derived using variational principles (see
Bleeker [1981]).  There is  a natural generalization where $f\x$ is an
$n$--dimensional complex vector.  In that case,
the group of phase tanslations is replaced by a non--abelian group $G$,
usually a subgroup of  $U(n)$.  $G$ is called the {\sl gauge group,\/} 
and the correponding gauge theory is said to be non--abelian or of the
Yang--Mills type.  Such theories have in recent years been applied with
great success to   the two remaining known interactions (aside from gravity
and electromagnetism), namely the {\sl weak\/}  and the {\sl strong\/}
forces, which involve nuclear matter (Appelquist et al. [1987]). 

\skp
Let us now see how non--abelian gauge theory may be extended to
{\sl complex\/}  spacetime. (This will include the abelian case of
electrodynamics when $n=1$.)   Consider a  field $f$ on  complex  spacetime,
say on the double tube $\tb$,  whose values are $n$--dimensional complex
vectors.  The set of all possible values at $z\in\tb$ is a complex vector space
$F_z\approx \cx^n$ called the {\sl fiber\/}  at $z$.    The collection of all
fibers is called a {\sl vector  bundle.\/}  We assume that this bundle is
holomorphic (Wells [1980]), so that holomorphic sections $z\mt f\z\in F_z$,
represented locally by holomorphic vector--valued functions, make sense. 
Upon transport along a curve $z\t$ having the complex tangent vector $Z$,
$f$ changes by

\be
 \delta f=\theta (Z)f\,\delta t,
\eno18)
 \ee
where  $\theta(Z)$ is a linear map on each fiber.
The total differential change   is 

\be
 Df=\left(  d+\theta\right)f.
\eno19)
 \ee 
If $z=x-iy$, then

\be
 d=\partial+\bar\partial,
\eno20)
 \ee
where 

\ba{
 \partial\a=dz^\mu \,\partial_\mu 
=\h\,dz^\mu\left({ \partial\over \partial x^\mu }+
i { \partial\over \partial y^\mu }\right) \cr
\bar\partial\a=d\zbar^\mu \,\bar\partial_\mu 
=\h\,d\zbar^\mu \left({ \partial\over \partial x^\mu }-
i { \partial\over \partial y^\mu }\right) .
 \cr}
 \eno21)
 \ea
  Since a general tangent vector has the form

\be
 Z=Z_\mu \,\partial_\mu +Z_\mb\,\bar\partial_\mu ,
\eno22)
 \ee
we have

\be
 \theta (Z)=\theta _\mu \,dz^\mu +\theta _\mb\,d\zbar^\mu ,
\eno23)
 \ee
where $\theta _\mu =\theta (\partial_\mu )$ and $\theta _\mb=\theta
(\bar\partial_\mu )$.

Let us now try again to derive the connection  from a fiber metric, as we
have failed to in the case of real spacetime.  A positive--definite metric on
the fibers $F_z$ must have the form

\be
 (f\z, \,g\z)= f\z^*\, h\z \, g\z
\eno24)
 \ee
where $h\z$ is a positive--definite matrix.  Again, it will suffice to consider
the (squared) fiber norm

\be
 \rho \z=(f\z, \, f\z).
\eno25)
 \ee
We define a {\sl holomorphic  gauge transformation\/} to be a map of the
form
\ba{
 f\z\a\mt f'\z=\chi  \z^{-1} \, f\z,\cr
h \z\a\mt h'=\chi  \z^*  \,h\z \,\chi  \z,
 \cr}
 \eno26)
 \ea
where $\chi  \z$ is an invertible $n\times n$ matrix--valued
holomorphic  function.  Clearly $\rho $ is invariant under holomorphic  gauge
transformations.  The  corresponding  gauge group acting on a single fiber
$F_z$   is  the general linear group $G=GL(n, \cx)$, which includes the usual
gauge group $U(n)$.  However, analyticity correlates the values of $\chi \z$
at different fibers.  Invariance of the inner product  under transport gives

\be
 (Df, f)+(f, Df)=d(f,f),
\eno27)
 \ee
from which we obtain the matrix equation

\be
 \theta ^*h+h\theta =dh=\db h+\d h.
\eno28)
 \ee
As in the case of real spacetime, this only determines the Hermitian part
(relative to the metric) of $\theta $.  But if we make the {\sl Ansatz\/} 

\be
 h\theta =\d h,
\eno29)
 \ee 
then the resulting connection $\theta =h ^{-1} \d h$ satisfies the above
constraint.

We now show that in general, the connection $\theta $ is non--trivial, i.e.
cannot be transformed away by a holomorphic gauge transformation. Under
such a  transformation, $\theta $ becomes

\ba{
\theta '\a=(\chi ^*h\chi )^{-1} \d (\chi ^*h\chi) \cr
\a=\chi ^{-1} h ^{-1} \d h \chi +\chi ^{-1} \d \chi \cr
\a=\chi ^{-1} \theta \chi +\chi ^{-1} \d \chi,
 \cr}
 \eno30)
 \ea
since $\d \chi ^*=0$ by analyticity.  It follows that 

\be
 D'f'\equiv (d+\theta ')f'=\chi ^{-1} D(\chi f'),
\eno31)
 \ee
hence
\be
( D')^2f'=\chi ^{-1} D^2(\chi f').
\eno32)
 \ee
If $\theta $ were trivial, then for some gauge we would have $\theta '=0$,
hence  $(D')^2=d^2=0$, so $D^2=0$.  But

\ba{
D^2f\a=
(d+\theta )(d+\theta)\,f=d(\theta \,f)+\theta\w df+\theta\w\theta f\cr
\a=(d\theta )\,f+\theta \w \theta \,f\equiv \Theta \,f,
 \cr}
 \eno33)
 \ea
where the 2--form

\be
 \Theta =d\theta +\theta \w \theta =\bar\partial \theta +\partial \theta +
\theta \w \theta
\eno34)
 \ee
is the {\sl curvature form\/} of the connection $\theta $, analogous to the
Riemann curvature tensor in Relativity.   Using the matrix equation 

\be
 dh ^{-1} =-h^{-1} dh\cdot h ^{-1}
\eno35)
 \ee
  and $\d^2=\db^2=\d\db+\db\d=0$, we find

\ba{
\partial \theta +\theta \w\theta =\left( -h ^{-1} \partial h\cdot h ^{-1} 
\right)\w \partial h+\left(h ^{-1} \partial h  \right)\w\left(  
h ^{-1} \partial h  \right)=0.
 \cr}
 \eno36)
 \ea
This is an {\sl integrability condition\/}  for $\theta $,
being a consequence of the fact that $\theta $ can be derived from $h$.  One
could say that $h$ is a ``potential'' for $\theta $.  Therefore the quadratic
term cancels in eq. (34) and the curvature form reduces to 

\be
 \Theta =\bar\partial \theta .
\eno37)
 \ee
Hence if $h$ is such that  $\db \left(h ^{-1}  \d h  \right) \ne 0$, then
the connection is non--trivial.  The form $\Theta $ is the complex spacetime
version of a {\sl Yang--Mills field,\/}  and $\theta $ corresponds to the
Yang--Mills potential.

 For $n=1$, $\theta $ and $\Theta $ are the complex
spacetime versions of the electromagnetic potential and the electromagnetic
field, respectively.  Since $h\z$ is a positive function, it may be written as

\be
 h\z=e^{-\phi \z}
\eno38)
 \ee
where $\phi \z$ is real.  Then

\be
 \theta =-\d\phi ,\qquad \Theta =-\db\d\theta .
\eno39)
 \ee
To relate $\theta $ and $\Theta $ to the electromagnetic potential $A$ and
the electromagnetic field $F$, one performs a non--holomorphic gauge
transformation similar to that in eq.(15):  let

\be
 \tilde f\z=e^{-\phi \z /2}f\z,\qquad \tilde h\z\equiv 1.
\eno40)
 \ee
Then the transformed potential becomes purely imaginary,

\be
 \tilde\theta =-{i \over 2}{\partial \phi  \over \partial y^\mu }dx^\mu ,
\eno41)
 \ee
giving

\be
 A_\mu \z=-\h{\partial \phi  \over \partial y^\mu }
\eno42)
 \ee
for the complex spacetime version of the electromagnetic potential.  Note
that although $ A_\mu \z$ is a pure gradient in the $y$--direction, the
corresponding electromagnetic field is {\sl not\/} trivial, since

\ba{
 F_{\mu \nu }\z\a\equiv  {\partial A_\mu  \over \partial x^\nu }-
 {\partial A_\nu   \over \partial x^\mu  }  \cr
\a=\h\left[ {\partial ^2\phi  \over \partial x^\mu \partial y^\nu } -
{\partial ^2\phi  \over \partial x^\nu  \partial y^\mu  } \right]
 \cr}
 \eno43)
 \ea 
need not vanish, in general.
\skp
Incidentally, there is an intriguing similarity between the inner product
using the fiber metric,

\be
 \l f\,|\,g\r=\int_\sigma d\mu \z\, \overline{ f\z}\,e^{-\phi \z}\,g\z,
\eno44)
 \ee
and that in Onofri's holomorphic coherent states representation  (sec. 3.4, eq.
(62)).  Note that our $\Theta $ coincides with
Onofri's symplectic form $-\omega $.  This possible connection
remains to be explored.
\skp

\noindent {\sl Remarks.\/} 
\item{1.  }  The   relation between the Yang--mills potential $A$ and the
Yang--Mills field $F$  in {\sl real\/} spacetime  is $F=dA+A\w A$, which is
{\sl quadratic\/}  in the non--abelian case ($n>1$), since the wedge product
$A\w A$ involves matrix multiplication.  In our case,  however, the
connection satisfies the  integrability condition  given by eq. (36), hence
the quadratic term cancels and the relation becomes {\sl linear,\/} just as it
is normally in the abelian case.  The non--holomorphic gauge transformation
$f\mt \tilde f$ in eq. (40) can be generalized to the  non--abelian case as
follows:  Since $h\z$ is a postive matrix, it can be written as
$h\z=k\z^*k\z$, where $k\z$ is an $n\times n$ matrix.  ($k\z$ need not be
Hermitian;  the holomorphic case $\db k=0$ corresponds to a ``pure gauge''
field, i.e. $\Theta =0$.)  Setting $\tilde f\z=k\z f\z$ and $\tilde h\z\equiv 1$
brings us to the {\sl unitary gauge,\/} where the new gauge transformations
are given by unitary matices $\tilde f\z\mt U\z \tilde f\z$.  This amounts to
a {\sl reduction\/}  of the gauge group from $GL(n, \cx)$ to $U(n)$.  In the
unitary gauge, the relation between the connection and the curvature
becomes non--linear, as it is in the usual Yang--Mills theory.  See Kaiser
[1981] for details.
\skp

 \item{2. } $\Theta $ has a symmetric real
part and an antisymmetric imaginary part.  The antisymmetric part  
corresponds to the usual Yang--Mills field,  whereas  the symmetric part
does not seem to have an obvious counterpart in real spacetime.  

\skp

\item{3. }  In complex differential geometry, $\theta =h ^{-1} \d h$ is
known as the {\sl canonical connection of type\/}  (1, 0) determined by
$h$. (Wells [1980]). The  functions $f$ are assumed to be (local
representations of) holomorphic sections of the vector  bundle.  We do not
make this assumption, since it appears that analyticity may be lost in the
presence of interactions. However,  it is possible that $f\z$ is holomorphic,
and it is the non--holomorphic gauge transformation $f\mt \tilde f$ which
spoils the analyticity.  That is, the non--analytic part of the theory may be
all contained in the fiber metric $h$.

\secskp

\noindent {\bf  6.2  Windowed X--Ray Transforms: Wavelets Revisited}
\def\rightheadline{\tenrm\hfil {\sl 
6.2  Windowed X--Ray Transforms: Wavelets Revisited}\hfil\folio}  

\skp

\noindent In this final section we generalize the idea behind the
Analytic--Signal transform (section 5.2) and arrive at an $n$--dimensional
version of the Wavelet transform (chapters 1 and 2).  In a certain sense,
relativistic wave functions and fields in complex spacetime  may be
regarded as {\sl generalized wavelet transforms\/} of their counterparts in
real spacetime.  This is related to the fact, mentioned earlier, that 
relativistic windows shrink in the direction of motion,  due to the
Lorentz contraction associated with the hyperbolic geometry of
spacetime.  This contraction is like the compression associated with
wavelets.  

Other generalizations of the wavelet transform to more than one dimension
 have been studied (see Malat [1987]), but they are usually
obtained from the one--dimensional one by taking tensor products, hence
are not natural with respect to symmetries of $\rl^n$ (such as rotations or
Lorentz transformations), and consequently do not lends themselves  to
analysis by group--theoretic methods.  The generalization proposed here,
which we call a {\sl windowed X--Ray transform,\/}  does not assume any
prefered directions, hence it respects all symmetries of $\rl^n$.  For
example, the transforms of functions over real spacetime $\st$ will transform
naturally under the Poincar\'e group. If these functions form a
representation  space for ${\cal P}_0$ (whether irreducible, as in the case of
a free particle, or reducible, as in the case of systems of interacting fields), 
then so do their transforms. \skp

Let us start directly with the windowed X--Ray transform (Kaiser [1990b]). 
Fix a window
 function  $h{:}\,\rl\to\cx$, which will play a role similar to a ``basic
wavelet.''    For a given (sufficiently well--behaved) function $f$ on $\rl^n$,
define $f_h$ on $\rl^n\times\rl^n$ by

\be
 f_h(x,y)=\inr  dt\,\overline{h(t) }\,f(x+ty).
\eno1)
 \ee
Note that for $n=1$ and $y\ne 0$, 

\ba{
f_h(x,y)\a=|y|^{-1} \int dt'\,\overline{ h\left({ t'-x\over y}  \right)}\,f(t')\cr
\a=|y|^{-1/2}\,(Wf)(x,y),
 \cr}
 \eno2)
 \ea
where $Wf$ is the usual Wavelet transform based on the affine group
$A=\rl\sdp\rl^*$, defined in section 1.6.  On the other hand, for arbitrary
$n$ but

\be
 h(t)={ 1\over 2\pi (1-it)},
\eno3)
 \ee
$f_h$ coincides with the Analytic--Signal transform defined in section 5.2. 
The name ``windowed  X--Ray transform'' derives from the fact that for
the choice $h\equiv 1$, $f_h(x, y)$ is simply the integral of $f$ along the
line $x\t=x+ty$, and if $|y|=1$ this is known as the {\sl X--Ray transform\/} 
(Helgason [1984]), due to its applications in tomography (Herman [1979]).

Returning to $n$ dimensions and an arbitrary window  function $h\t$, we
wish to know, first of all, whether and how $f$ can be reconstructed from
$f_h$.  Note that the transform is trivial for $y=0$, since 

\be
 f_h(x,0)=f(x)\int dt \,\overline{h\t },
\eno4)
 \ee
so we rule out $y=0$ and let $y$ range over $\rl^n_*\equiv \rl^n\backslash
\{0\}$.    Note also that $f_h$ has  the following {\sl dilation property\/} for
$a\ne 0$:

\be
 f_h(x, ay )=\int dt\,|a |^{-1} \overline{ h(t/a)}\,f(x+ty)
= f_{h_a }(x,y) ,
\eno5)
 \ee
where $h_a (t)\equiv |a|^{-1} \,h(t/a)$.

 To reconstruct $f$, begin by formally substituting the Fourier representation 
of $f$ into $f_h$:

\ba{
f_h(x,y)\a=\int dt\,\int d^np\,e^{-2\pi ip(x+ty)}\,
\overline{ h(t)}\,\fhat(p)\cr
\a=\int d^np\,e^{-2\pi ipx}\,\overline{ \hat h(py)}\,\fhat(p)\cr 
\a\equiv \langle\, \hat h_{x,y} \,|\, \fhat\,\rangle_{L^2}=
\langle\,  h_{x,y} \,|\, f\,\rangle_{L^2},
 \cr}
 \eno6)
 \ea 
where $\hat h_{x,y}$ is defined by

\be
 \hat h_{x,y}(p)=e^{2\pi i px}\,\hat h(py), 
\eno7)
 \ee
so that 

\be
 h_{x,y}(x')=\int d^np\,e^{-2\pi ip(x'-x)}\,\hat h(py).
\eno8)
 \ee
 Note that $\hat h_{x,y}$, and hence also $h_{x,y}$, is not
 square--integrable for $n>1$, since its modulus is constant along
directions orthogonal to $y$.\footnote*{So far, we have  {\sl not\/}  assumed
any metric structure on $\st$.  Recall (sec. 1.1) that the natural domain of
$\fhat(p)$ is the {\sl dual\/} space $(\rl^n)^*$.  ``Orthogonal''
here means that $p(y)=0$ as a linear functional.  This will be important 
when  $\rl^n$ is spacetime  with its Minkowskian metric.}  This means that
eq. (6) will not make sense for arbitrary $f\in L^2(\rl^n)$.  We therefore
assume, initially, that $f$ is a test function, say it belongs to the Schwartz
space ${\cal S}(\rl^n)$.  Then  eq. (6) makes sense with $\langle\,   \,|\, 
\,\rangle$ as the (sesquilinear) pairing between distributions and test
functions.  If $h$ is also sufficiently well--behaved, then we can substitute 

\be
 \hat h(py)=\int d\nu \,\hat h(\nu )\,\delta (py-\nu )
\eno9)
 \ee
and change the order of integration in eq. (8),  obtaining

\ba{
 h_{x,y}(x')=\int d\nu \,\hat h(\nu)\int d^np\,e^{-2\pi
ip(x'-x)}\,\delta (py-\nu).
 \cr}
 \eno10)
 \ea

To reconstruct $f$, we look for a  resolution of unity in terms of the
vectors $h_{x,y}$ (chapter 1).   That is,  we need a measure $d\mu (x,y)$ on
$\rl^n\times \rl^n_*$ such that 

\be
 \int d\mu (x,y)\,|f_h(x,y)|^2=\|f\|^2_{L^2}.
\eno11)
 \ee
For then the map $T{:}\,f\mapsto f_h$ is an isometry onto its range in
$L^2(d\mu )$, and polarization gives

\be
\langle\,g\,|\,f\,\rangle =\langle\, Tg\,|\,Tf\,\rangle=
 \langle\, g\,|\, T^*Tf\,\rangle,
\eno12)
 \ee
showing that $f=T^*(Tf)$  in $L^2(\rl^n)$, which is the desired
reconstruction formula.  There are various ways to obtain a 
resolution of unity, since $f$ is actually overdetermined by $f_h$, i.e. giving
the values of $f_h$ on all of $\rl^n\times \rl^n_*$ amounts to
``oversampling,''  so $f_h$ will have to satisfy a consistency
condition.  We have seen several examples of this in the study  of the
windowed Fourier transform (section 1.5) and the   one--dimensional wavelet
transform (section 1.6 and chapter 2), where {\sl discrete\/}  subframes
were obtained starting with a continuous resolution of unity.  However, for
$n>1$, there are other options than discrete subframes, as we will see.
 In the spirit of the one--dimensional wavelet transform, our first resolution
of unity will involve an integration over  all of $\rl^n\times \rl^n_*$.  Note
that

\be
 f_h(x,y)=\left( \overline{ \hat h(py)} \,\fhat\right)\check{\ }(x),
\eno13)
 \ee
so Plancherel's theorem gives

\be
 \int d^nx\,|f_h(x,y)|^2=\int d^np\,|\hat h(py)|^2\,|\fhat(p)|^2.
\eno14)
 \ee
We therefore need a measure $d\rho (y)$ on $\rl^n_*$ such that 

\be
 H(p)\equiv \int d\rho (y)\,|\hat h(py)|^2\equiv 1 
\quad \hbox{for almost all }p. 
\eno15)
 \ee
The solution is simple:  Every $p\ne 0$ can be transformed to $q\equiv (1,
0,  \ldots  0)$ by a  {\sl dilation and rotation\/} of $\rl^n$.  That is, the orbit
of $q$ (in Fourier space) under dilations and rotations is all of  $\rl^n_*$. 
Thus we choose $d\rho $ to be invariant under rotations and dilations, which
gives

\be
 d\rho (y)=N|y|^{-n} d^ny,
\eno16)
 \ee
where $N$ is a normalization constant and $|y|$ is the Euclidean norm of
$y$.  Then for $p\ne 0$,

\ba{
H(p)\a=H(q)=N\int |y|^{-n} d^ny\,|\hat h(y_1)|^2\cr
\a=N\int dy_1\,|\hat h(y_1)|^2\,
\int { dy_2 \cdots dy_n\over (y_1^2+ \cdots+ y_n^2)^{n/2}}.
 \cr}
 \eno17)
 \ea
But a straightforward computation gives 

\be
 \int { dy_2  \cdots dy_n\over (y_1^2+ \cdots +y_n^2)^{n/2} }=
{ \pi ^{n/2}\over |y_1|\,\Gamma(n/2)}.
\eno18)
 \ee
This shows that  the measure $d\mu (x,y)\equiv d^nx \,d\rho (y)$ gives a
resolution of unity if and only if  

\be
 c_h\equiv \int { d\xi \over |\xi |}\,|\hat h (\xi )|^2<\infty,
\eno19)
 \ee
which is precisely the {\sl admissibility condition \/} for the usual
(one--dimensional) Wavelet transform (section 1.6).  If $h$ is admissible, the
normalization constant is given by

\be
N={ \Gamma(n/2)\over \pi  ^{n/2} \,c_h}
\eno20)
 \ee
and the reconstruction formula is then

\be
 f(x')=(T^*Tf)(x')=N\int { d^nx\,d^ny\over |y|^n}\,
\overline{h_{x,y}(x')}\,f_h(x,y). 
\eno21)
 \ee
The {\sl sense\/}  in which this formula holds depends, of course on the
behavior of $f$.  The class of possible $f$'s, in turn, depends on the choice
of $h$.  A rigorous analysis of these questions is not easy, and will not be
attempted here.  Note that in spite of the factor $|y|^n$ in the denominator,
there is no problem at $y=0$ since 

\be
 f_h(x,0)=\overline{ \hat h(0)}\,f(x)=0
\eno22)
 \ee
by the admissibility condition.  The behavior of $f_h$ for small $y$ can be
analized using the dilation property, since eq. (5) implies that for $\lambda
>0$,

\ba{
 f_h(x, y/\lambda )=\int dt\,\lambda \overline{ h(\lambda t)}\,f(x+ty).
 \cr}
 \eno23)
 \ea
Thus if  $h\t$ decays rapidly, say if 

\be
 \lambda h(\lambda t)\to 0\quad \ \hbox{as}\  \quad \lambda \to\infty,
\eno24)
 \ee
then we expect $f_h(x, y/\lambda )\to 0$ as $\lambda \to \infty$.

\skp
Since eq. (11) holds for admissible $h$, we can now  allow
$f\in L^2(\rl^n)$.  We would like to characterize the {\sl range\/} $\Re_T$ 
of the map $T{:}\ f\mt f_h$ from $L^2(\rl^n)$ to $L^2(d\mu )$.
The relation 

\be
 f_h(x, y)=\l h_{x, y}\,|\,f\r_{L^2}
\eno25)
 \ee
shows that $ h_{x, y}$ acts like an {\sl evaluation map\/} taking $f_h\in
L^2(d\mu )$ to its ``value'' at $(x, y)$.    These linear maps on $\Re_T$
are, however, not bounded if $n>1$, since then $h_{x, y}$ is not
square--integrable.  (In general, the ``value'' of $f_h$ at a point may be
undefined.)  Hence $\Re_T$ is {\sl not\/}  a reproducing--kernel Hilbert
space (chapter 1).  But in any case, the distributional kernel

\ba{
 K(x, y;\, x', y')\a\equiv \l h_{x, y}\,|\,h_{x', y'}\r\cr
\a=\int d^np\,e^{-2\pi i p(x-x')}\,\overline{ \hat h(py)}\hat h(py')
 \cr}
 \eno26)
 \ea
represents the orthogonal projection from $L^2(d\mu )$ onto
$\Re_T$.  Thus a given function in $L^2(d\mu )$ belongs to
$\Re_T$ if and only if it satisfies the {\sl consistency condition\/} 

\be
 g(x, y)=\int d\mu (x', y')\,K(x, y;\, x', y')\,g(x', y'),
\eno27)
 \ee
where the integral is the symbolic representation  of the action of $K$ as a
distribution.  

\sp

\noindent {\sl Remarks.\/} 
\skp

\item{1.} For $n=1$, the reconstruction formula is identical with the one for
the continuous one--dimensional wavelet transform\/ $Wf$, since by 
Eq. (2),
\be
 \int { dx\,dy\over |y|}\,|f_h(x, y)|^2=\int { dx\, dy\over y^2}\,
\left| (Wf)(x, y) \right|^2.
\eno28)
 \ee

\sp

\noindent
\item{2.}   In deriving the resolution of unity and the related
reconstruction formula, we have tacitly identified $\rl^n$ as a Euclidean
space, i.e. we have equipped it with the Euclidean metric and identified the
pairing $px$ in the Fourier transform as the inner product.  The exact place
where this assumption entered was in using the rotation group plus dilations
to obtain $\rl^n_*$ from the single vector $q$, since rotations presume a
metric. 

\vskip .5cm

\noindent
Having established $f_h$ as a generalization of the one--dimensional wa\-ve\-let 
transform, let us now investigate it in its own right.  First, note that
for $n=1$ there were only two simple types of candidates for generalized
frames of wavelets:  (a) all continuous translations and dilations of the basic
wavelet, or (b) a discrete subset thereof.  For $n>1$, any choice of a discrete
subset  of vectors $h_{x, y}$ spoils the  invariance under continuous
symmetries such as rotations, and it is therefore not obvious how to use
the above group--theoretic method to find discrete subframes.  In fact, the
discrete subsets $\{(a^m, na^mb)\}$ which gave frames of wavelets in
section 1.6 and chapter 2 do {\sl not\/}  form subgroups of the affine group.
One of the advantages of using tensor products of one--dimensional
wavelets is that they do generate discrete frames for $n>1$, though
sacrificing symmetry.   However,  other options exist for choosing 
generalized (continuous) subframes when $n>1$, and one may adapt one's
choice to the problem at hand.   Such choices fall  between the two extremes
of using {\sl all\/}  the vectors $h_{x, y}$ and merely summing over a
discrete subset, as seen in the examples below.

\skp

\centerline {\sl 1.  The X--Ray Transform} 
\skp

\noindent The usual X--Ray transform is obtained by choosing $h\t\equiv
1$, which is not admissible in the above sense;  hence the above ``wavelet''
reconstruction fails.  The reason is easy to see:  Note that now $f_h$ has the
following symmetries:

\ba{
f_h(x, ay)\a=|a| ^{-1} f_h(x, y)\quad \forall \, a\in\rl^*\cr
f_h(x+s\a y, y)=f_h(x,y)\quad \forall s\in\rl  .
 \cr}
 \eno29)
 \ea
Together, these equations state that $f_h$ depends only on the 
line of integration and not on the way it is parametrized.
The first equation shows that integration over all $y\ne 0$ is unnecessary
as well as undesirable, and it suffices to integrate over the unit sphere 
$|y|=1$.  The second equation shows that for a given $y$, it is (again)
unnecessary and undesirable to integrate over all $x$, and  it suffices to
integrate over the hyperplane orthogonal to $y$.  The set of all such $(x,y)$
does, in fact, correspond to the set of all lines in $\rl^n$, and the
corresponding set of $h_{x,y}$'s forms a continuous frame which gives the
usual reconstruction formula for the X--Ray transform (Helgason [1984]). 
The moral of the story is that sometimes, inadmissibility in the ``wavelet''
sense carries a message:  Reduce the size of the frame.
\skp

\centerline {\sl 2. The Radon Transform \/} 
\skp
\noindent Next, choose $\nu \in\rl$ and 

\be
 h_\nu \t=e^{-2\pi i\nu t}.
\eno30)
 \ee
Like the previous function, this one is inadmissible, hence the\break
``wavelet'' reconstruction fails.  Again, this can be corrected by
understanding the reason for inadmissibility.   Eq. (6) now gives

\be
 f_{h_\nu }(x, y)=\int d^np\, e^{-2\pi ipx}\,\delta (py-\nu )\,\fhat(p).
\eno31)
 \ee
For any $a\ne 0$, we have

\be
 f_{h_\nu }(x, ay)=|a|^{-1} f_{h_\omega }(x, y),
\eno32)
 \ee
where $\omega =\nu /a$. Hence it suffices to restrict the $y$--integration
to the unit sphere, provided we also integrate over $\nu \in\rl$.  Also, 
for any $\tau \in\rl$,

\ba{
f_{h_\nu }(x+\tau y, y)\a=\int d^np\, e^{-2\pi ipx}\,e^{-2\pi i\tau py}\,
\delta (py-\nu )\,\fhat(p)\cr
\a=e^{-2\pi i\tau \nu }\,f_{h_\nu }(x, y).
 \cr}
 \eno33)
 \ea
Fixing $x=0$,  the function 

\be
 (R\fhat)(y, \nu )=f_{h_\nu }(0, y)
\eno34)
 \ee
is called the {\sl Radon transform \/}  of $\fhat$ (Helgason [1984]).  It may 
be regarded as being defined on the set of all hyperplanes in 
the Fourier space $(\rl^n)^*$, and $\fhat$ can be reconstructed by integrating
over the set of these hyperplanes.

\skp

\centerline{ \sl  3.  The Fourier--Laplace Transform}
\skp
\noindent Now consider

\be
 h(t)={ 1\over 2\pi i(t-i)}
\eno35)
 \ee
which gives rise to the Analytic--Signal transform.  (We have adopted  a
slightly different sign convention than is sec. 5.2.  Also, note that we have
re--inserted a factor of $2\pi $ in the exponent in the Fourier transform,
which simplifies the notation.)
 Then $\hat h$ is the exponential step function (sec. 5.2)

\be
 \hat h(\xi )=\theta (\xi )\,e^{-2\pi \xi }=\theta ^{-2\pi \xi },
\eno36)
 \ee
and  eq. (6) reads

\ba{
f_h(x,y)\a=\int d^np\,\theta (py)\, e^{-2\pi ip(x-iy)  }\,\fhat (p)\cr
\a=\int_{M_y}d^np\,e^{2\pi ip(x-iy)  }\,\fhat (p),
 \cr}
 \eno37)
 \ea
where $M_y$ is the half--space $\{p\,|\,py>0\}$.  This is the 
{\sl Fourier--Laplace transform\/}  of $\fhat$ in $M_y$.  For $n=1$ and
$y>0$, it reduces to the usual Fourier--Laplace transform.

This $h$, too, is not admissible.  $f(x)$ can be recovered simply by letting
$y\to 0$,  and $f_h(x,y)$ may be regarded as a {\sl regularization\/}  of
$f(x)$.  If the support of $\fhat$ is contained in some closed convex cone
$\Gamma^*\subset (\rl^n)^*$, then $f_h(x, y)\equiv f(x-iy)$ is  holomorphic  in
the tube ${\cal T}_\Gamma$ 
over the cone $\Gamma$ {\sl dual\/}  to $\Gamma^*$, i.e.

\ba{
\Gamma\a=\{y\in\rl^n\,|\,py>0\quad \forall\, p\in \Gamma^*\}\cr
{\cal T}_\Gamma\a=\{x-iy\in\cx^n\,|\,y\in \Gamma\}.
 \cr}
 \eno38)
 \ea
(Note that no metric has been assumed.)
In that case, $f\x$ is a {\sl boundary value\/}  of $f(x-iy)$.
This forms the background for the theory of Hardy spaces (Stein and Weiss
[1971]). We have encountered a similar situation when $\rl^n$ was
spacetime ($n=s+1$), $\Gamma^*=\overline{V}_+$, and $f\x$ was a
positive--energy solution  of the Klein--Gordon equation;  
then $\Gamma=V'$ and  ${\cal T}_\Gamma=\tp$.
But in that
case, $f\x$ was not in $L^2(\rl^{s+1})$ due to the conservation of probability. 
There it was unnecessary and undesirable to integrate $|f\z|^2$ over all of
$\tp$ since it was determined by its values on any phase space $\sigma_+
\subset \tp$, and reconstruction
 was then achieved by integrating over   $\sigma_+$  (chapter 4). 

 \skp

As seen from these examples,  the  windowed X--Ray transform has
the  remarkable feature of being related to most of the ``classical''  integral
transforms:  The X--Ray,  Radon  and  Fourier--Laplace transforms. Since the
Analytic--Signal transform is a close relative of the  multivariate 
Hilbert transform  $H_y$ (sec. 5.2), we may also add $H_y$ to this
collection. 

\VE

\def\rightheadline{\tenrm\hfil {\sl  References}\hfil\folio}  
\def\leftheadline{\tenrm\folio\hfil \sl References\hfil} 

\parindent=40pt

\skp
\centerline {\bf REFERENCES } 
\skp\skp

\noindent Abraham, R. and Marsden, J. E.
\item{[1978]} {\sl Foundations of Mechanics,\/} Benjamin/Cummings,
Reading, MA.

\noindent Abramowitz, M. and Stegun, L. A., eds.
\item{[1964]} {\sl Handbook of Mathematical Functions,\/} National Bureau
of Standards, Applied Math. Series {\bf  55}.

\noindent Ali, T. and Antoine, J.--P.
\item{[1989]} Ann. Inst. H. Poincar\'e {\bf  51}, 23.

\noindent Applequist, T. Chodos, A. and Freund, P. G. O., eds.
\item{[1987]} {\sl Modern Kaluza--Klein Theories,\/} Addison--Wesley,
Reading, MA.

\noindent Aronszajn, N.
\item{[1950]} Trans. Amer. Math. Soc. {\bf  68}, 337.

\noindent Aslaksen, E. W. and Klauder, J. R.
\item{[1968]} J. Math. Phys. {\bf  9}, 206.

\item{[1969]} J. Math. Phys. {\bf  10}, 2267.

\noindent Bargmann, V.
\item{[1954]} Ann. Math. {\bf  59}, 1.

\item{[1961]} Commun. Pure Appl. Math. {\bf 14}, 187.

\noindent Bargmann, V., Butera, P., Girardello, L. and Klauder, J. R.
\item{[1971]} Reps. Math. Phys. {\bf  2}, 221.

\noindent Barut, A.O. and Malin, S.
\item{[1968]} Revs. Mod. Phys. {\bf  40}, 632.

\noindent Battle, G.
\item{[1987]} Commun. Math. Phys. {\bf  110}, 601.

\noindent Berezin, F. A.
\item{[1966]} {\sl The Method of Second Quantization,\/} Academic Press,
New York.

\noindent Bergman, S. 
\item{[1950]} {\sl The Kernel Function and Conformal Mapping,\/}  Amer.
\hfil\break Math. Soc.,  Providence.

\noindent Bialinicki--Birula, I. and Mycielski, J.
\item{[1975]} Commun. Math. Phys. {\bf  44}, 129.

\noindent Bleeker, D.
\item{[1981]} {\sl Gauge Theory and Variational Principles,\/}
Addison-- Wesley, Reading, MA.

\noindent Bohr, N. and Rosenfeld, L.
\item{[1950]} Phys. Rev. {\bf  78}, 794.

\noindent Bott, R.
\item{[1957]} Ann. Math. {\bf  66}, 203.

\noindent Carey, A. L.
\item{[1976]} Bull. Austr. Math. Soc. {\bf  15}, 1.

\noindent Daubechies, I.
\item{[1988a]}  ``The wavelet transform, time--frequency localization
and signal analysis,'' to be published in IEEE Trans. Inf. Thy.

 \item{[1988b]}  Commun.  Pure  Appl.  Math. XLI, 909.

\noindent De Bi\`evre, S.
\item{[1989]} J. Math. Phys. {\bf  30}, 1401.

\noindent Dell'Antonio, G. F.
\item{[1961]} J. Math. Phys. {\bf  2}, 759.

\noindent Dirac, P. A. M.
\item{[1943]} {\sl  Quantum Electrodynamics,\/} Commun. of the
Dublin\hfil\break  Inst. for Advanced Studies series A, \#1.

\item{[1946]} {\sl Developments in Quantum Electrodynamics,\/} Commun.
of the Dublin Inst. for Advanced Studies series A, \#3.

\noindent Dufflo, M. and Moore, C. C.
\item{[1976]} J. Funct. Anal. {\bf  21}, 209.

\noindent Dyson, F. J.
\item{[1956]} Phys. Rev. {\bf  102}, 1217.

\noindent Fock, V. A.
\item{[1928]} Z. Phys. {\bf 49}, 339.

\noindent Gel'fand, I. M., Graev, M. I. and Vilenkin, N. Ya.
\item{[1966]} Generalized Functions, vol 5, Academic Press, New York.

\noindent Gel'fand, I. M. and Vilenkin, N. Ya.
\item{[1964]} Generalized Functions, vol 4, Academic Press, New York.

\noindent Gerlach, B.,  Gomes, D. and Petzold, J.
\item{[1967]} Z. Physik {\bf  202}, 401.

\noindent Glauber, R. J.
\item{[1963a]} Phys. Rev. {\bf  130}, 2529.

\item{[1963b]} Phys. Rev. {\bf  131}, 2766.

\noindent Glimm, J. and Jaffe, A.
\item{[1981]} {\sl Quantum Physics,\/} Springer, New York.

\noindent Green, M. W., Schwarz, J. H. and Witten, E.
\item{[1987]} {\sl Superstring Theory,\/} vols. I, II, Cambridge Univ. Press,
Cambridge.

\noindent Greenberg, O. W.
\item{[1962]} J. Math. Phys. {\bf  3}, 859.

\noindent Guillemin, V. and Sternberg, S.
\item{[1984]} {\sl Symplectic Techniques in Physics,\/}  Cambridge
Univ.   \hfil\break Press, Cambridge.

\noindent Hai, N. X.
\item{[1969]} Commun. Math. Phys. {\bf  12}, 331.

\noindent Halmos, P.
\item{[1967]} {\sl A Hilbert Space Problem Book,\/} Van Nostrand, London.

\noindent Hegerfeldt, G. C.
\item{[1985]} Phys. Rev. Lett. {\bf  54}, 2395.

\noindent Helgason, S.
\item{[1962]} {\sl Differential Geometry and Symmetric Spaces,\/} \hfill\break
Academic Press, NY.

\item{[1978]} {\sl Differential Geometry, Lie Groups, and Symmetric
Spa\-ces,\/} Academic Press, New York.

\item{[1984]} {\sl Groups and Geometric Analysis,\/}
 Academic Press, New York.

\noindent Henley, E. M. and Thirring, W.
\item{[1962]} {\sl Elementary Quantum Field Theory,\/}  \hfill\break
McGraw--Hill, New York.

\noindent Herman, G. T., ed.
\item{[1979]} {\sl Image Reconstruction from Projections:  Implementation \hfill\break
and Applications,\/} Topics in Applied Physics {\bf  32}, Springer
New York.

\noindent Hermann, R.
\item{[1966]} {\sl Lie Groups for Physicists,\/} Benjamin, New York.

\noindent Hille, E.
\item{[1972]}  Rocky Mountain J. Math. {\bf  2}, 319.

\noindent In\"on\"u, E. and Wigner, E. P.
\item{[1953]} Proc. Nat. Acad. Sci., USA, {\bf  39}, 510.

\noindent Itzykson, C. and Zuber, J.--B.
\item{[1980]} {\sl Quantum Field Theory,\/} McGraw--Hill, New York.

\noindent Jost, R.
\item{[1965]} {\sl The General Theory of Quantized Fields,\/} Amer. Math.
Soc., Providence.

\noindent Kaiser, G.
\item{[1974]}  ``Coherent States and Asymptotic Representations,''  in
Proc. of Third Internat. Colloq. on Group--Theor. Meth. in Physics, Marseille.

\item{[1977a]} `` Relativistic Coherent--State Representations,'' in
\hfil\break
 {\sl Group--Theoretical Methods in Physics,\/} R. Sharp and B. Coleman, eds.,
Academic Press, New York.

\item{[1977b]} J. Math. Phys. {\bf  18}, 952.

\item{[1977c]} {\sl Phase--Space Approach to Relativistic Quantum
Mechanics,\/} Ph. D. thesis, Mathematics Dept, Univ.  of Toronto.

\item{[1977d]} ``Phase--Space Approach to Relativistic Quantum\hfil\break
Mechanics,'' in proc. of Eighth Internat. Conf. on General Relativity,
Waterloo, Ont.

\item{[1978a]} J. Math. Phys. {\bf  19}, 502.

\item{[1978b]} ``Hardy spaces associated with the Poincar\'e group,''  in {\sl
Lie Theories and Their Applications, \/} W. Rossmann, ed., \hfill\break
Queens Univ. Press, Ont.

\item{[1978c]} ``Local Fourier Analysis and Synthesis,'' Univ. of Lowell
preprint (unpublished).

\item{[1979]} Lett. Math. Phys. {\bf  3}, 61.

\item{[1980a]} ``Holomorphic gauge theory,'' in {\sl Geometric Methods
in\hfil\break  Mathematical Physics,\/}  G. Kaiser and J. E. Marsden, eds., 
Lecture Notes in Math. {\bf  775},  Springer, Heidelberg.

\item{[1980b]} ``Relativistic quantum theory in complex spacetime,'' in {\sl
Differential Geometrical Methods in Mathematical  Phys\-ics,\/} \hfill\break
P.L.  Garcia, A. $\!$Perez--Rendon and J.--M. $\!$Souriau, eds., \hfill\break
Lecture Notes in Math. {\bf  836},  Springer, Berlin.

\item{[1981]} J. Math. Phys. {\bf  22}, 705.

\item{[1984a]} ``A sampling theorem for signals in the joint
time-- frequency domain,''  Univ. of Lowell preprint (unpublished).

\item{[1984b]} ``On the relation between the formalisms of 
Prugove${\rm\check c}$ki and Kaiser,'' Univ. of Lowell preprint,  
presented as a poster at the VII-th International Congress of Mathematical
Physics, in {\sl Mathematical  Physics VII,\/}  W. E. Brittin, K. E.
Gustafson and W. Wyss, eds., North--Holland, Amsterdam

\item{[1987a]} Ann. Phys. {\bf  173}, 338.

\item{[1987b]} ``A phase--space formalism for quantized fields,'' in
\hfil\break {\sl Group--Theoretical Methods in Physics,\/} R. Gilmore, ed., \hfill\break
World Scientific, Singapore.

\item{[1990a]} ``Algebraic Theory of Wavelets. I:  Operational Calculus and
Complex Structure,'' Technical Repors Series    \#17,  Univ. of Lowell.

\item{[1990b]} ``Generalized Wavelet Transforms. I: The Windowed  X--Ray
Transform,''  Technical Repors Series    \#18,  Univ. of Lowell.

\item{[1990c]} ``Generalized Wavelet Transforms. II:  The Multivariate \hfill\break
Analytic--Signal Transform,''  Technical Repors Series 
  \#19,  Univ. of Lowell.

\noindent Kirillov, A. A.
\item{[1976]} {\sl Elements of the Theory of Representatins,\/} 
Springer--\hfil \break  Verlag, Heidelberg.

\noindent Klauder, J. R. 
\item{[1960]} Ann. Phys. {\bf  11}, 123.

\item{[1963a]} J Math. Phys. {\bf  4}, 1055.

\item{[1963b]} J Math. Phys. {\bf  4}, 1058.

\noindent Klauder, J. R. and Skagerstam, B.--S., eds.
\item{[1985]} {\sl Coherent States,\/}  World Scientific, Singapore.

\noindent Klauder, J. R. and Sudarshan, E. C. G.
\item{[1968]} {\sl Fundamentals of Quantum Optics,\/} Benjamin, New York.

\noindent Kobayashi, S. and Nomizu, K.
\item{[1963]} {\sl Foundations of Differential Geometry,\/}  vol. I,
Interscience, New York.

\item{[1969]} {\sl Foundations of Differential Geometry,\/}  vol. II,
\hfil\break  Interscience, New York.

\noindent Kostant, B.
\item{[1970]} ``Quantization and Unitary Representatins,'' in {\sl Lectures in
Modern Analysis and Applications,\/} R.~M.~Dudley et al., eds. Lecture
Notes in Math. {\bf  170}, Springer, Heidelberg.

\noindent Lemari\'e, P. G. and Meyer, Y.
\item{[1986]} Rev. Math. Iberoamericana {\bf  2}, 1.

\noindent Lur\c cat,  F.
\item{[1964]} Physics {\bf  1}, 95.

\noindent Mackey, G.
\item{[1955]} {\sl The Theory of Group--Representations,\/} Lecture Notes,
Univ. of Chicago.

\item{[1968]} {\sl Induced Representations of Groups and Quantum
Mechanics,\/} Benjamin, New York.

\noindent Mallat, S.
\item{[1987]} ``Multiresolution approximation and wavelets,'' \hfill\break
preprint, GRASP Lab., Dept. of Computer and Information Science, Univ. of
Pennsylvania, to be published in Trans. Amer. Math. Soc.

\noindent Marsden, J. E.
\item{[1981]} {\sl Lectures on Geometric Methods in Mathematical
Physics,\/}  Notes from a Conference held at  Univ. of Lowell,
CBMS--NSF Regional Conference Series in Applied Mathematics vol. {\bf 37},
SIAM, Philadelphia, PA.

\noindent Meschkowski, H.
\item{[1962]} {\sl Hilbertsche R\"aume mit Kernfunktion,\/}
Springer, Heidelberg.

\noindent Messiah, A.
\item{[1963]} {\sl Quantum Mechanics,\/} vol. II, North--Holland,
Amsterdam.

\noindent Meyer, Y.
\item{[1985]} ``Principe d'incertitude, bases hilbertiennes et alg\`ebres \hfill\break
d'op\'erateurs,''   S\'eminaire Bourbaki {\bf  662}.

\item{[1986]} ``Ondelettes et functions splines,'' S\'eminaire EDP, \hfill\break
Ecole Polytechnique, Paris, France (December issue).

\noindent Misner, C., Thorne, K. S. and Wheeler, J. A.
\item{[1970]} {\sl Gravitation,\/} W.~H.~Freeman, San Francisco.

\VE  

\noindent Nelson, E.
\item{[1959]} Ann. Math. {\bf  70}, 572.

\item{[1973a]} J. Funct. Anal. {\bf  12}, 97.

\item{[1973b]} J. Funct. Anal. {\bf  12}, 211.

\noindent Neumann, J. von
\item{[1931]} Math. Ann. {\bf  104}, 570.

\noindent Newton, T. D. and Wigner, E.
\item{[1949]} Rev. Mod. Phys. {\bf  21}, 400.

\noindent Onofri, E.
\item{[1975]} J. Math. Phys. {\bf  16}, 1087.

\noindent Papoulis, A. 
\item{[1962]} {\sl The Fourier Integral and its Applications,\/} 
McGraw--Hill.

\noindent Perelomov, A. M.
\item{[1972]} Commun. Math. Phys. {\bf  26}, 222.

\item{[1986]} {\sl Generalized Coherent States and Their Applications,\/} 
\hfill\break Springer, Heidelberg.

\noindent Prugove$\check {\rm c}$ki, E. 
\item{[1976]} J. Phys. {\bf  A9}, 1851.

\item{[1978]} J. Math. Phys. {\bf  19}, 2260.

\item{[1984]} {\sl Stochastic Quantum Mechanics and Quantum
Spacetime,\/} D.~Reidel, Dordrecht.

\noindent Robinson, D. W.
\item{[1962]} Helv. Physica Acta {\bf  35}, 403.

\noindent Roederer, J. G.
\item{[1975]} {\sl Introduction to the Physics and Psychophysics of
Music,\/} Springer, New York.

\noindent Schr\"odinger, E.
\item{[1926]} Naturwissenschaften {\bf  14}, 664.

\noindent Segal, I. E.
\item{[1956a]} Trans. Amer. Math. Soc. {\bf  81}, 106.

\item{[1956b]} Ann. Math. {\bf  63}, 160.

\item{[1958]}  Trans. Amer. Math. Soc. {\bf  88}, 12.

\item{[1963a]} Illinois J. Math. {\bf  6}, 500.

\item{[1963b]} {\sl Mathematical Problems of Relativistic Physics,\/} Amer.
\hfil \break Math. Soc., Providence.

\item{[1965]} Bull. Amer. Math. Soc. {\bf  71},  419.

\noindent Simms, D. J. and Woodhouse, N. M. J.
\item{[1976]}  {\sl Lectures on Geometric Quantization,\/} Lecture Notes in
Phys\-ics {\bf  53}, Sprin\-ger Verlag, Heidelberg. 

\VE

\noindent \'Sniatycki, J.
\item{[1980]} {\sl Geometric Quantization and Quantum mechanics,\/}  
\hfil\break  Springer, Heidelberg.

\noindent Souriau, J.--M.
\item{[1970]} {\sl Structure des Syst\`emes Dynamiques,\/} Dunod, Paris.

\noindent Stein, E.
\item{[1970]} {\sl Singular Integrals and Differentiability Properties
of\hfil\break  Functions,\/} Princeton Univ. Press, Princeton.

\noindent Strang, G.
\item{[1989]}  SIAM Rev. {\bf 31}, 614.

\noindent Streater, R. F.
\item{[1967]} Commun. Math. Phys. {\bf  4}, 217.

\noindent Streater, R. F. and Wightman, A. S.
\item{[1964]} {\sl PCT, Spin \& Statistics, and All That,\/} Benjamin, New York.

\noindent Thirring, W. 
\item{[1980]} {\sl Quantum Mechanics of Large Systems,\/}
Springer, New York.

\noindent Toller, M.
\item{[1978]} Nuovo Cim. {\bf  44B}, 67.

\noindent Unterberger, A.
\item{[1988]} Commun. in Partial Diff. Eqs. {\bf  13}, 847.

\noindent Varadarajan, V. S.
\item{[1968]} {\sl Geometry of Quantum Theory,\/} vol. I, D. Van Nostrand,
Princeton, NJ.

\item{[1970]} {\sl Geometry of Quantum Theory,\/} vol. II, D. Van Nostrand,
Princeton, NJ.

\item{[1974]} {\sl Lie Groups, Lie Algebras, and their Representations,\/}
\hfil\break Prentice--Hall, Englewood Cliffs, NJ.

\noindent Warner, F.
\item{[1971]} {\sl Foundations of Differentiable Manifolds and Lie
Groups,\/} \hfill\break
Scott, Foresman and Co., London.

\noindent Wells, R. O.
\item{[1980]} {\sl Differential Analysis on Complex Manifolds,\/}
Springer, New York.

\noindent Weyl, H.
\item{[1919]} Ann. d. Physik {\bf  59}, 101.

\item{[1931]} {\sl The Theory of Groups and Quantum Mechanics,\/} 
translated by H. P. Robertson,   Dover, New York 

\noindent Wightman, A. S.
\item{[1962]} Rev. Mod. Phys. {\bf  34}, 845.

\noindent Yosida, G.
\item{[1971]} {\sl Functional Analysis,\/} Springer, Heidelberg.

\VE

\noindent Young, R. M.
\item{[1980]} {\sl An Introduction to Non--Harmonic Fourier Series,\/} 
Academic Press, New York.

\noindent Zakai, M.
\item{[1960]} Information and Control {\bf  3}, 101.

\bye